\documentstyle[12pt,latexsym,amssymb,graphicx,axodraw,epsfig,epsf,rotating,color,graphics,cite]{article}

\newcommand{\mste}{m_{\tilde{t}_1}}
\newcommand{\mstz}{m_{\tilde{t}_2}}

\newcommand{\msusy}{M_{\rm SUSY}}
\newcommand{\msq}{m_{\tilde{q}}}

\newcommand{\msbar}{$\overline{\rm{MS}}$}
\newcommand{\oaas}{{\cal O}(\alpha\alpha_s)}
\newcommand{\ogmzmts}{{\cal O}(G_F^2\mt^6)}
\newcommand{\cp}{{\cal CP}}
\newcommand{\twol}{two-loop}
\newcommand{\onel}{one-loop}
\newcommand{\fh}{{\tt FeynHiggs}}
\newcommand{\mh}{m_h}

\newcommand{\mt}{m_{t}}
\newcommand{\mgl}{m_{\tilde{g}}}
\newcommand{\Stop}{\tilde{t}}
\newcommand{\tst}{\theta_{\tilde{t}}}

\newcommand{ \bea }{\begin{eqnarray}}
\newcommand{ \eea }{\end{eqnarray}}

\newcommand{ \NPB    }[3]{Nucl. Phys.      {\bf B#1}, #3 (#2)}
\newcommand{ \PLB    }[3]{Phys. Lett.      {\bf B#1}, #3 (#2)}
\newcommand{ \PLBold }[3]{Phys. Lett.      {\bf #1B}, #3 (#2)}
\newcommand{ \PRD    }[3]{Phys. Rev.       {\bf D#1}, #3 (#2)}
\newcommand{ \PRL    }[3]{Phys. Rev. Lett. {\bf #1}, #3 (#2)}
\newcommand{ \ZPC    }[3]{Z. Phys.         {\bf C#1}, #3 (#2)}
\newcommand{ \EPJC   }[3]{Eur. Phys. J.    {\bf C#1}, #3 (#2)}

\newcommand{ \centeron }[2]{{\setbox0=\hbox{#1}\setbox1=\hbox{#2}\ifdim
                             \wd1>\wd0\kern.5\wd1\kern-.5\wd0\fi \copy0
                             \kern-.5\wd0\kern-.5\wd1\copy1\ifdim\wd0>\wd1
                             \kern.5\wd0\kern-.5\wd1\fi}}
\newcommand{ \ltap }{\;\centeron{\raise.35ex\hbox{$<$}}
                     {\lower.65ex\hbox{$\sim$}}\;}
\newcommand{ \gtap }{\;\centeron{\raise.35ex\hbox{$>$}}
                     {\lower.65ex\hbox{$\sim$}}\;}

\newcommand{ \slashchar }[1]{\setbox0=\hbox{$#1$}   
   \dimen0=\wd0                                     
   \setbox1=\hbox{/} \dimen1=\wd1                   
   \ifdim\dimen0>\dimen1                            
      \rlap{\hbox to \dimen0{\hfil/\hfil}}          
      #1                                            
   \else                                            
      \rlap{\hbox to \dimen1{\hfil$#1$\hfil}}       
      /                                             
   \fi}                                             %

\newcommand{ \NI         }{ \tilde{N}_1 }
\def\G{ \tilde{G} }
\newcommand{ \lR         }{ { \tilde \ell}_R }
\newcommand{ \tauu       }{ { \tilde \tau}_1 }
\newcommand{\sss}{\scriptscriptstyle}

\newcommand{\Z}{Z^{\sss 0}}  
\newcommand{\epem}{e^{\sss +}e^{\sss -}}   

\newcommand{\stopu}{\tilde{t}_1}
\newcommand{\stopd}{\tilde{t}_2}

\RequirePackage{epsfig}
\RequirePackage{wrapfig}

\catcode`@=11
\def\citer{\@ifnextchar
[{\@tempswatrue\@citexr}{\@tempswafalse\@citexr[]}}

\def\@citexr[#1]#2{\if@filesw\immediate\write\@auxout{\string\citation{#2}}\fi
  \def\@citea{}\@cite{\@for\@citeb:=#2\do
    {\@citea\def\@citea{--\penalty\@m}\@ifundefined
       {b@\@citeb}{{\bf ?}\@warning
       {Citation `\@citeb' on page \thepage \space undefined}}%
\hbox{\csname b@\@citeb\endcsname}}}{#1}}
\catcode`@=12

\newcommand{\ie}{{\it i.e.\/}\ }

\newcommand{\gsim}{\raisebox{-0.13cm}{~\shortstack{$>$ \\[-0.07cm] $\sim$}}~}
\newcommand{\rpv}{$\mathrm{\not\!R_p}$}
\def\fb{{\rm fb}}

\newcommand{\st}{\tilde{t}}

\newcommand{\sq}{\tilde{q}}
\newcommand{\sqb}{\bar{\tilde{q}}}
\newcommand{\gl}{\tilde{g}}
\newcommand{\gau}{\tilde{\chi}}
\newcommand{\MS}{\mbox{$\overline{\rm MS}$}}
\newcommand{\MSSM}{\mbox{$\MSSM}$}

\newcommand{\TeV}{\mbox{Te$\!$V}}
\newcommand{\GeV}{\mbox{Ge$\!$V}}
\newcommand{\tb}{\mbox{$\tan\beta$}}
\newcommand{\tgb}{\mbox{$\tan\beta$}}
\newcommand{\nn}{\noindent}

\newcommand{\zp}[3]{{Z.\ Phys.} {\bf #1} (19#2) #3}
\newcommand{\np}[3]{{Nucl.\ Phys.} {\bf #1} (19#2)~#3}
\newcommand{\pl}[3]{{Phys.\ Lett.} {\bf #1} (19#2) #3}
\newcommand{\pr}[3]{{Phys.\ Rev.} {\bf #1} (19#2) #3}
\newcommand{\prl}[3]{{Phys.\ Rev. Lett.} {\bf #1} (19#2) #3}

\newcommand{\ra}{\rightarrow}

\newcommand{\beq}{\begin{eqnarray}}
\newcommand{\eeq}{\end{eqnarray}}
\newcommand{\bq}{\begin{equation}}
\newcommand{\eq}{\end{equation}}
\newcommand{\be}{\begin{equation}}
\newcommand{\ee}{\end{equation}}

\begin{document}
\renewcommand{\thefootnote}{\fnsymbol{footnote}}

\def\dofig#1#2{\epsfxsize=#1\centerline{\epsfbox{#2}}}
\def\dofigs#1#2#3{\centerline{\epsfxsize=#1\epsfbox{#2}%
   \hfil\epsfxsize=#1\epsfbox{#3}}}

\begin{center}

\vspace*{1.2cm}

{\Large\sc \bf THE SUSY WORKING GROUP: }

\vspace*{0.4cm}

{\Large\sc \bf Summary Report} 

\vspace*{1.cm}

Conveners: \\[0.2cm] 
{\sc S.~Abdullin$^1$, M.~Drees$^{2,3}$, H.--U.~Martyn$^4$, G.~Polesello$^5$ }

\vspace*{0.6cm}

Subgroup Coordinators: \\[0.2cm]
{\sc S.~Ambrosanio$^6$, H.~Dreiner$^7$,
R.M.~Godbole$^8$, J.~Wells$^{6,9}$ }

\vspace*{0.6cm}

Contributing members: \\[0.2cm]
{\sc 
P.~Chiappetta$^{10}$,
D.~Choudhury$^{11}$, 
A.K.~Datta$^{11}$, 
A.~Deandrea$^6$, 
O.J.P.~\'Eboli$^2$,
N.~Ghodbane$^{12}$,
S.~Heinemeyer$^{13}$,
V.~Ilyin$^{14}$, 
T.~Kon$^{15}$,
S.~Kraml$^{16}$, 
Y.~Kurihara$^{17}$,
M.~Kuroda$^{18}$,
L.~Megner$^{19}$,
B.~Mele$^{20}$,
G.~Moreau$^{21}$,
B.~Mukhopadyaya$^{11}$, 
E.~Nagy$^{22}$,
S.~Negroni$^{22}$,
K.~Odagiri$^7$,
F.E.~Paige$^{23}$,
E.~Perez$^{24}$,
S.~Petrarca$^{20}$,
P.~Richardson$^{25}$,
A.~Rimoldi$^{26}$,
S.~Roy$^{11}$,
M.H.~Seymour$^7$,
M.~Spira$^{27}$,
J.M.~Virey$^{10,28}$,
F.~Vissani$^{13}$,
G.~Weiglein$^6$

}

\vspace*{1.cm}

{\small

$^1$ ITEP, Moscow, Russia.\\
$^2$ IFT, Univ. Estadual Paulista, Rua Pamplona 145, S\~ao Paulo,
Brazil. \\
$^3$  Physik Department, TU M\"unchen, James Franck Str., D--85748 Garching, 
Germany. \\
$^4$ I. Physikalisches Institut, RWTH Aachen, Germany. \\
$^5$ INFN, Sezione di Pavia, Via Bassi 6, Pavia, Italy. \\
$^6$ CERN Theory Group, CH-1211 Gen\`eve 23, Switzerland.\\
$^7$  Rutherford Appleton Laboratory, Chilton, Didcot, Oxon OX11 OQX, U.K. \\
$^8$ Centre for Theoretical Studies, Indian Institute of Science,
Bangalore 560 012, India. \\
$^9$ Physics Department, University of California, Davis, CA 95616. \\
$^{10}$  Centre de Physique Th\'eorique, UPR7061, CNRS-Luminy,
F-13288 Marseille, France. \\
$^{11}$ Mehta Research Institute, Chhatnag Road, Jhusi, Allahabad 211 
019, India. \\
$^{12}$ Inst. de Physique Nucl\'eaire de Lyon,
43 Bd du 11 novembre 1918, 69622 Villeurbanne, France. \\
$^{13}$ DESY--Theorie, Notkestrasse 85, D--22603 Hamburg, Germany. \\
$^{14}$ INP MSU, Moscow, Russia. \\
$^{15}$ Faculty of Engineering, Seikei University, Tokyo, Japan. \\
$^{16}$ Inst. f. Hochenergiephysik d. \"Oster. Akademie
d. Wissenschaften, A-1050 Wien, Austria. \\
$^{17}$ KEK, Tsukuba, Japan. \\
$^{18}$ Institute of Physics, Meiji-Gakuin University, Tokyo, Japan. \\
$^{19}$ Physics Dept. Frescati, KTH, S-104 05 Stockholm, Sweden. \\
$^{20}$ INFN - Sezione di Roma I, and Dip. di Fisica, Univ. ``La
Sapienza'', Rome, Italy. \\
$^{21}$ Service de Physique Th\'eorique, 
CEA-Saclay, F91191, Gif-sur-Yvette, Cedex France.\\
$^{22}$ Centre de Physique des Particules, Universit\'e de la 
M\'editerran\'ee, F-13288 Marseille, France.\\ 
$^{23}$ Physics Department, Brookhaven National Laboratory, 
Upton, NY 11973. \\
$^{24}$ Service de Physique des Particules, DAPNIA,
CEA-Saclay, F91191, Gif-sur-Yvette, France. \\
$^{25}$ Physics Dept., Theoretical Physics, Univ. of Oxford, UK. \\
$^{26}$ CERN, EP--Division, CH-1211 Gen\`eve, Switzerland. \\
$^{27}$ II.\ Inst. f\"ur Theoretische Physik, Univ. Hamburg,
D--22761 Hamburg, Germany. \\
$^{28}$ Universit\'e de Provence, Marseille, France.
}

\vspace*{1cm}

{\it Report of the SUSY working group for the Workshop \\[0.1cm]
``Physics at TeV Colliders", Les Houches, France 8--18 June 1999.}

\end{center} 

\newpage


\vspace*{1cm} 
\begin{center}
{\bf \large CONTENTS} 
\end{center} 

\vspace*{0.8cm} 

\nn {\bf 0. SYNOPSIS}  \\ \vspace*{3mm}

\nn {\bf 1. General SUSY} \\[0.2cm]

\oddsidemargin 5mm

1a) {\it SUSY particle production at hadron colliders} \\[0.2cm]
\hspace*{12mm}
M. Spira \\

1b) {\it Comparison of exact matrix element calculations with
ISAJET and PYTHIA in case \\ \hspace*{12mm} of degenerate spectrum in the MSSM}
\\[0.2cm] \hspace*{12mm}
S. Abdullin, V. Ilyin and T. Kon \\

1c) {\it The finite width effect on neutralino production}
\\[0.2cm]  \hspace*{12mm}
T. Kon, Y. Kurihara, M. Kuroda and K. Odagiri \\

1d) {\it Width effects in slepton production: $e^+e^- \rightarrow
\tilde{\mu}_R^+ \tilde{\mu}_R^-$ } \\[0.2cm]  \hspace*{12mm}
H.-U. Martyn. \\

1e) {\it Radiative effects on squark pair production at $e^+e^-$
colliders} \\[0.2cm] \hspace*{12mm}
M. Drees, O.J.P. \'Eboli, R.M. Godbole and S. Kraml. \\

1f) {\it Spin correlations and phases for SUSY particle searches
at $e^+e^-$ colliders} \\[0.2cm] \hspace*{12mm}
N. Ghodbane. \\

\vspace*{3mm}

\nn {\bf 2. $R-$parity violation} \\[0.2cm]

2a) {\it The three lepton signature from resonant sneutrino
production at the LHC} \\[0.2cm]
\hspace*{12mm}
G. Moreau, E. Perez and G. Polesello. \\

2b) {\it Resonant slepton production at the LHC}
\\[0.2cm] \hspace*{12mm}
H. Dreiner, P. Richardson and M.H. Seymour. \\

2c) {\it $\tilde{\chi}_1^0$ reconstruction in mSUGRA models with
$R-$parity violation} \\[0.2cm] \hspace*{12mm}
L. Megner and G. Polesello. \\

2d) {\it Supersymmetry with $R-$parity violation at the LHC:
discovery potential from single \\ \hspace*{12mm} top production}
\\[0.2cm]  \hspace*{12mm}
F. Chiappetta, A. Deandrea, E. Nagy, S. Negroni, G. Polesello and
J.M. Virey. \\

2e) {\it Probing $R \hspace{-3mm} /$ couplings through indirect
effects on Drell--Yan production at the LHC}
\\[0.2cm]  \hspace*{12mm}
D. Choudhury and R.M. Godbole. \\

2f) {\it Neutrino masses, $R-$parity violating supersymmetry and
collider signals} \\[0.2cm] \hspace*{12mm}
A.K. Datta, B. Mukhopadhyaya and S. Roy. \\

\vspace*{3mm}

\nn {\bf 3. Gauge--mediated supersymmetry breaking}  \\[0.2cm]

3a) {\it Introduction to GMSB phenomenology at TeV colliders},
\\[0.2cm] \hspace*{12mm} S. Ambrosanio. \\

3b) {\it The light Higgs spectrum in GMSB models} \\[0.2cm]
\hspace*{12mm} S. Ambrosanio, S. Heinemeyer and G. Weiglein. \\

3c) {\it Measuring the SUSY breaking scale at the LHC in the slepton
NLSP scenario of \\ \hspace*{12mm} GMSB models} \\[0.2cm] \hspace*{12mm}
S. Ambrosanio, B. Mele, S. Petrarca, G. Polesello and A. Rimoldi. \\

\vspace*{3mm}

\nn {\bf 4. Anomaly mediated SUSY breaking at the LHC} \\[0.2cm] \hspace*{5mm}
F.E. Paige and J.D. Wells. \\

\vspace*{3mm}

\newpage


\begin{center}
{\large\sc {\bf SYNOPSIS}} 
\end{center}

\vspace*{0.1cm}

Supersymmetry remains the best motivated extension of the Standard
Model, since it stabilizes the gauge hierarchy against quadratically
divergent radiative corrections. However, this argument by itself does
not yet tell us very much about the way supersymmetry is realized in
Nature. In particular, it tells us nothing about the mechanism that
breaks supersymmetry, except that sparticles masses should not greatly
exceed 1 TeV (with the possible exception of first and second
generation sfermions; this possibility will not be pursued here). One
can be completely ``agnostic'' and simply parameterize SUSY breaking
by soft terms. This approach has largely been taken by the first and
second subgroup of this working group. Alternatively, one can explore
the predictions of specific models, as has been done by subgroups three
(on gauge mediated SUSY breaking) and four (on anomaly mediated SUSY
breaking).

An additional complication arises because supersymmetric models allow
new, gauge--invariant interactions that have no equivalent in the
standard model; this is true even if one sticks to the minimal
particle content. These interactions break $R-$parity. Some
implications of this have been investigated in subgroup two. 

Some aspects of SUSY phenomenology have already reached quite a high
level of maturity. Radiative corrections have been computed for total
cross sections and partial widths; their impact on kinematic
distributions is now being explored. Finite width effects formally
also belong to higher orders in perturbation theory. Finally, in some
cases it is important to keep spin correlations through the entire
sparticle production and decay process. Some investigations along
these lines are described in contributions from subgroup one; they are
(possibly with minor modifications) relevant for all SUSY models.

Another area of current interest is the determination of soft breaking
parameters, in particular of sparticle masses. Here a realistic
assessment of the capabilities of various experiments requires
detailed Monte Carlo simulations. Several studies of this kind are
described in contributions from subgroups one, two and four.

\smallskip
\nn Manuel Drees (for the Working Group). \bigskip

\noindent {\bf Acknowledgements}: \smallskip
\noindent We thank the organizers of this workshop for giving us the
opportunity to attend a true renaissance meeting, stimulating for body
and mind. We are particularly indebted to Jean-Philippe Guillet et
al., and the Lapp computer engineers, for providing state--of--the--art
computing devices. We also wish to express our gratitude to the staff
of the Les Houches center, whose efficient and friendly service
greatly contributed to the pleasant memories we took away from the
workshop.

\newpage


\setcounter{footnote}{0}

\begin{center}
{\large \bf SUSY particle production at hadron colliders}

\vspace*{3mm}

{\sc M. SPIRA\footnote{Heisenberg Fellow}}

\end{center}

\vspace*{3mm}

\begin{abstract}
The determination of the full SUSY QCD corrections to the
production of squarks, gluinos and gauginos at hadron colliders is reviewed.
The NLO corrections stabilize the theoretical predictions of
the various production cross sections significantly and lead to sizeable
enhancements of the most relevant cross sections. We discuss the
phenomenological consequences of the results on present and future
experimental analyses.
\end{abstract}

\section{Introduction}
The search for Higgs bosons and
supersymmetric particles is among the most important endeavors of present and
future high energy physics.  The novel colored particles, squarks and gluinos,
and the weakly interacting gauginos can be searched for at the upgraded
Tevatron, a $p\bar p$ collider with a c.m.\ energy of 2 TeV, and the
LHC, a $pp$ collider with a c.m.\ energy of 14 TeV.  Until now the search at
the Tevatron has set the most stringent bounds on the colored SUSY particle
masses.  At the 95\% CL, gluinos have to be heavier than about 180 GeV, while
squarks with masses below about 180 GeV have been excluded for gluino masses
below $\sim 300$ GeV \cite{bounds}.  Stops, the scalar superpartners of the
top quark, have been excluded in a significant part of the MSSM parameter
space with mass less than about 80 GeV by the LEP and Tevatron experiments
\cite{bounds}.  Finally charginos with masses below about 90 GeV have been
excluded by the LEP experiments, while the present search at the Tevatron is
sensitive to chargino masses of about 60--80 GeV with a strong dependence on
the specific model \cite{bounds}. Due to the negative search at LEP2
the lightest neutralino $\tilde
\chi_1^0$ has to be heavier than about 30 GeV in the context of SUGRA
models \cite{bounds}. In the $R$-parity-conserving MSSM,
supersymmetric particles can only be produced in pairs.  All supersymmetric
particles will decay to the lightest supersymmetric particle (LSP), which is
most likely to be a neutralino, stable thanks to conserved $R$-parity.  Thus
the final signatures for the production of supersymmetric particles will
mainly be jets, charged leptons and missing transverse energy, which is
carried away by neutrinos and the invisible neutral LSP.

In Section 2 we shall summarize the
details of the calculation of the NLO QCD corrections, as described in
Refs.\ \citer{sqgl_sp,gaunlo} for
the case of $\sq \sqb$ production.  The evaluation of the full SUSY QCD
corrections splits into two pieces, the virtual corrections, generated by
virtual particle exchanges, and the real corrections, which originate from
gluon radiation and the corresponding crossed processes with three-particle
final states.

In Section 3 we shall consider the production of squarks and gluinos except
stops \cite{sqgl_sp}.  We assume the light-flavored squarks to be mass
degenerate, which is a
reasonable approximation for all squark flavors except stops, while the light
quarks ($u,d,s,c,b$) will be treated as massless particles.
The production of stop pairs requires the inclusion of mass splitting and
mixing effects \cite{stops} and
will be investigated in Section 4. In Section 5 we will summarize the
results for the production of charginos and neutralinos at
NLO \cite{gaunlo}. The calculation of the LO cross sections has been
performed a long time
ago \cite{lo}.  Since in most of the cases the [unphysical] scale
dependence of the LO quantities
amounts up to about 50\%, the determination of the NLO corrections is necessary
in order to gain a reliable theoretical prediction, which can be used in
present and future experimental analyses.

\section{SUSY QCD corrections}

\subsection{Virtual corrections}
\begin{figure}[hbt]
\begin{center}
\begin{picture}(120,80)(20,10)

\Gluon(0,20)(50,20){-3}{5}
\Gluon(0,80)(50,80){3}{5}
\Gluon(50,50)(75,80){-3}{4}
\DashLine(100,20)(50,20){5}
\DashLine(50,80)(100,80){5}
\DashLine(50,20)(50,80){5}
\put(-15,78){$g$}
\put(-15,18){$g$}
\put(75,48){$g$}
\put(105,18){$\sqb$}
\put(105,78){$\sq$}

\end{picture}
\begin{picture}(120,80)(-40,10)

\Gluon(0,20)(50,20){-3}{5}
\Gluon(0,80)(50,80){-3}{5}
\Gluon(75,20)(75,80){-3}{5}
\Line(75,80)(75,20)
\ArrowLine(75,20)(50,20)
\ArrowLine(50,20)(50,80)
\ArrowLine(50,80)(75,80)
\DashLine(100,20)(75,20){5}
\DashLine(75,80)(100,80){5}
\put(-15,78){$g$}
\put(-15,18){$g$}
\put(40,48){$q$}
\put(85,48){$\gl$}
\put(105,18){$\sqb$}
\put(105,78){$\sq$}

\end{picture}  \\
\caption[]{\label{fg:virt} \it Typical diagrams of the virtual corrections.}
\end{center}
\end{figure}
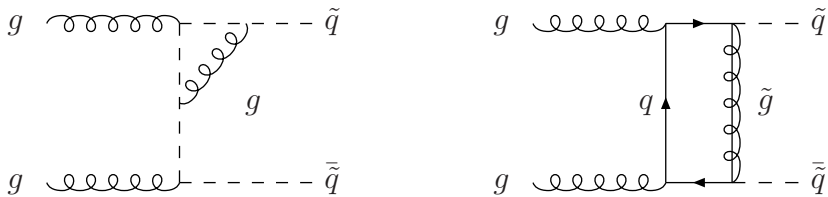
The one-loop virtual corrections are built up by gluon, gluino, quark and
squark exchange
contributions [see Fig.~\ref{fg:virt}]. They have to be contracted with the
LO matrix elements. The calculation of the one-loop contributions has been
performed in dimensional regularization, leading to the extraction of
ultraviolet, infrared and collinear singularities as poles in
$\epsilon = (4-n)/2$. For the chiral $\gamma_5$ coupling we have used the naive
scheme, which is well justified in the present analysis at the one-loop
level\footnote{We have explicitly checked that the results obtained
with a consistent $\gamma_5$ scheme are identical to the one with the naive
scheme.}.
We have explicitly checked that after summing all virtual corrections no
quadratic divergences are left over, in accordance with the general property
of supersymmetric theories. The renormalization has been performed by
identifying the squark and gluino masses with their pole masses, and defining
the strong
coupling in the $\overline{\rm MS}$ scheme including five light flavors in the
corresponding $\beta$ function. The massive particles, i.e.\ squarks, gluinos
and top quarks, have been decoupled by subtracting their contribution at
vanishing momentum transfer \cite{decouple}. In dimensional regularization,
there is a mismatch between the gluonic degrees of freedom [d.o.f. = $n-2$] and
those of the gluino [d.o.f. = $2$], so that SUSY is explicitly broken. In
order to restore SUSY in the physical observables in the massless limit, an
additional finite counter-term is required for the renormalization of the novel
$\sq \gl \bar q$ vertex \cite{count}.

\subsection{Real corrections}
\begin{figure}[hbt]
\begin{center}
\setlength{\unitlength}{1pt}
\begin{picture}(120,80)(0,10)

\Gluon(0,20)(50,50){-3}{5}
\Gluon(0,80)(50,50){3}{5}
\Gluon(25,65)(75,95){3}{5}
\DashLine(50,50)(100,80){5}
\DashLine(100,20)(50,50){5}
\put(-15,78){$g$}
\put(-15,18){$g$}
\put(80,93){$g$}
\put(105,18){$\sqb$}
\put(105,78){$\sq$}

\end{picture}
\begin{picture}(170,80)(-30,10)

\Gluon(0,20)(50,50){-3}{5}
\ArrowLine(0,80)(50,50)
\ArrowLine(50,50)(100,50)
\DashLine(100,50)(150,80){5}
\Gluon(100,50)(125,35){-3}{3}
\Line(100,50)(125,35)
\ArrowLine(125,35)(150,10)
\DashLine(125,35)(150,35){5}
\put(-15,78){$q$}
\put(-15,18){$g$}
\put(102,30){$\gl^*$}
\put(155,78){$\sq$}
\put(155,33){$\sqb$}
\put(155,8){$q$}

\end{picture}  \\
\setlength{\unitlength}{1pt}
\caption[]{\label{fg:real} \it Typical diagrams of the real corrections.}
\end{center}
\end{figure}
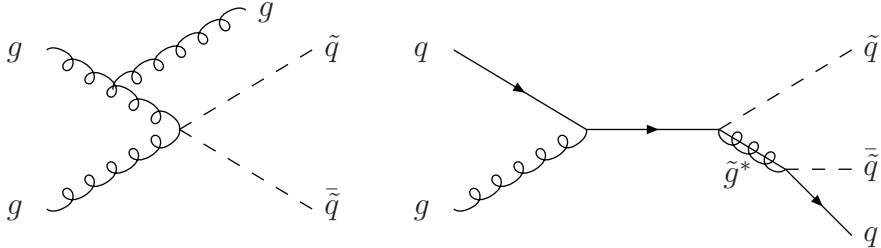
The real corrections originate from the radiation of a gluon in all possible
ways and from the crossed processes by interchanging the gluon of the final
state against a light (anti)quark in the initial state.  The phase-space
integration
of the final-state particles has been performed in $n=4-2\epsilon$ dimensions,
leading to the extraction of infrared and collinear singularities as poles in
$\epsilon$.  After evaluating all angular integrals and adding the virtual and
real corrections, the infrared singularities cancel.  The left-over collinear
singularities are universal and are absorbed in the renormalization of the
parton densities at NLO.  We defined the parton densities in the conventional
$\overline{\rm MS}$ scheme including five light flavors, i.e.\ the squark,
gluino and top quark contributions are not included in the mass factorization.
Finally we end up with an ultraviolet, infrared and collinear finite partonic
cross section.

However, there is an additional class of physical singularities, which have to
be regularized. In the second diagram of Fig.~\ref{fg:real} an
intermediate $\sq \gl^*$ state is produced, before the [off-shell] gluino splits
into a $q\sqb$ pair. If the gluino mass is larger than the common squark mass,
and the partonic c.m.\ energy is larger than the sum of the squark and gluino
masses, the intermediate gluino can be produced on its mass-shell. Thus the
real corrections to $\sq \sqb$ production contain a contribution of $\sq \gl$
production. The residue of this part corresponds to $\sq \gl$ production with
the subsequent gluino decay $\gl \to \sqb q$, which is already contained
in the LO cross section
of $\sq \gl$ pair production, including all final-state cascade decays.
This term has to be subtracted in order to derive a well-defined production
cross section. Analogous subtractions emerge in all reactions: if the gluino
mass is larger than the squark mass, the contributions from $\gl \to \sq \bar
q, \sqb q$ have to be subtracted, and in the reverse case the contributions of
squark decays into gluinos have to subtracted.

\section{Production of Squarks and Gluinos}
\begin{figure}[hbt]
\vspace*{-1.0cm}

\hspace*{2.0cm}
\epsfxsize=12cm \epsfbox{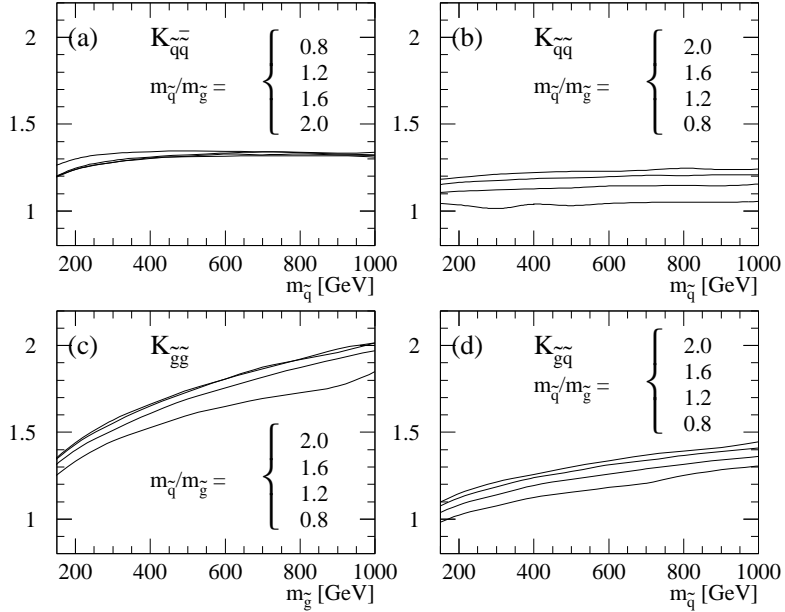}
\vspace*{-0.5cm}

\caption[]{\label{fg:kfac} \it K factors of the different squark and gluino
production cross sections at the LHC [$\sqrt{s} = 14$ TeV].
Parton densities: GRV94 with $Q=m$. Top mass: $m_t=175$ GeV.}
\end{figure}

\begin{figure}[hbt]
\vspace*{-0.5cm}

\hspace*{2.0cm}
\epsfxsize=10cm \epsfbox{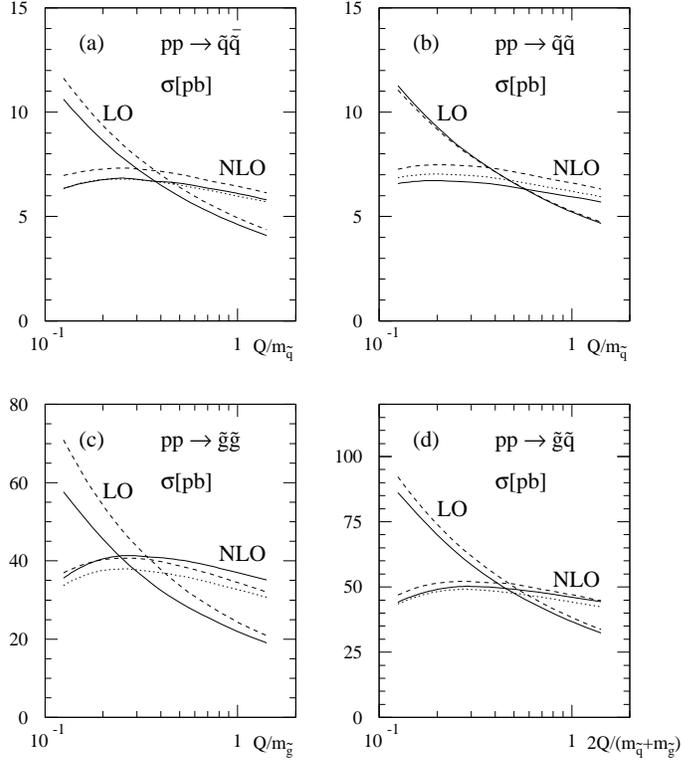}
\vspace*{-0.5cm}

\caption[]{\label{fg:scale} \it Scale dependence of the
total squark and gluino production cross sections at the LHC in LO
and NLO. Parton densities: GRV94 (solid), CTEQ3 (dashed) and MRS(A')
(dotted); mass parameters: $m_{\sq}=600$ GeV, $m_{\gl}=500$ GeV and $m_t=175$
GeV.}
\end{figure}

\begin{figure}[hbt]
\vspace*{-2.5cm}

\hspace*{-0.0cm}
\begin{turn}{-90}%
\epsfxsize=10cm \epsfbox{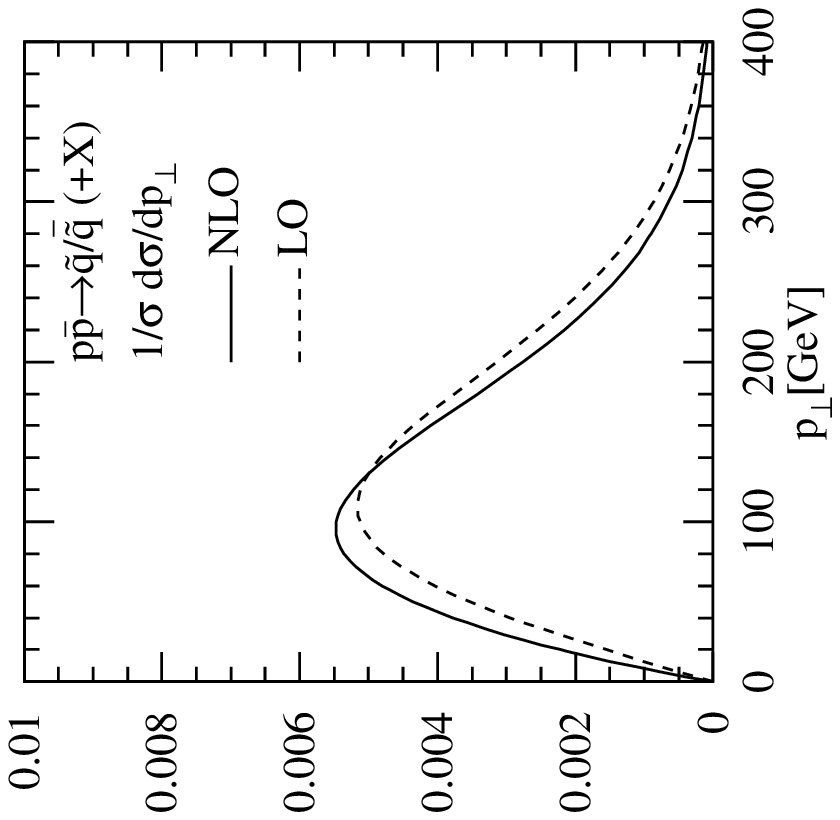}
\end{turn}

\vspace*{-10.05cm}

\hspace*{6.0cm}
\begin{turn}{-90}%
\epsfxsize=10cm \epsfbox{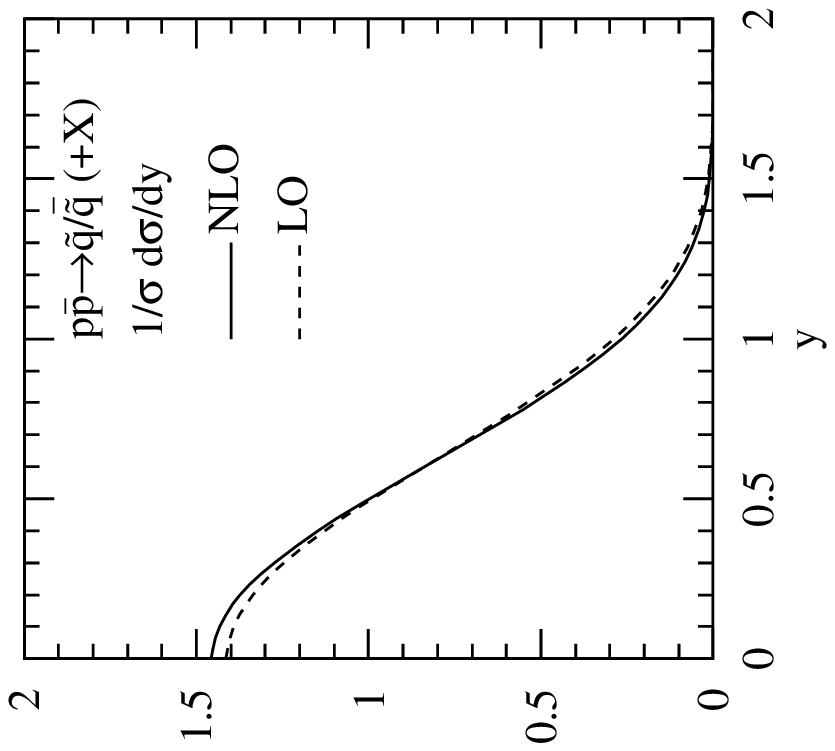}
\end{turn}
\vspace*{-2.8cm}

\caption[]{\label{fg:pty} \it Normalized transverse-momentum and rapidity
distributions of $p\bar p\to \sq \sqb + X$ at the upgraded Tevatron
[$\sqrt{s}=2$ TeV] in LO (dotted)
and NLO (solid). Parton densities: CTEQ4L (LO) and
CTEQ4M (NLO) with $Q=m$; mass parameters: $m_{\sq}=250$ GeV, $m_{\gl}=300$ GeV
and $m_t=175$ GeV.}
\end{figure}

Squarks and gluinos can be produced via $pp, p\bar p \to \sq \sqb, \sq \sq,
\sq \gl, \gl \gl$ at hadron colliders.
The hadronic squark and gluino production cross sections can be obtained from
the partonic ones by convolution with the corresponding parton densities.  We
have performed the numerical analysis for the upgraded Tevatron and the
LHC. For the
natural renormalization/factorization scale choice $Q=m$, where $m$ denotes
the average mass of the final-state SUSY particles, the SUSY QCD corrections
are large and positive, increasing the total cross sections by 10--90\%
\cite{sqgl_sp}. This is shown in Fig.~\ref{fg:kfac}, where the K factors,
defined as the ratios of the NLO and LO cross sections, are presented as a
function of the corresponding SUSY particle mass for the LHC.
We have investigated the residual scale dependence in LO and NLO, which is
presented in Fig.~\ref{fg:scale}.  The inclusion of the NLO corrections
reduces the LO scale dependence by a factor 3--4 and reaches a typical level
of $\sim 15\%$, which serves as an estimate of the remaining theoretical
uncertainty.  Moreover, the dependence on different sets of parton
densities is rather weak and leads to an additional uncertainty of $\sim 15\%$.
In order to quantify the effect of the NLO corrections on the
search for squarks and gluinos at hadron colliders, we have extracted the SUSY
particle masses corresponding to several fixed values of the production cross
sections.  These masses are increased by 10--30 GeV at the Tevatron and
10--50 GeV at the LHC, thus enhancing the present and future bounds on the
squark and gluino masses significantly.
Finally we have evaluated the QCD-corrected transverse-momentum and rapidity
distributions for all different processes.  As can be inferred from
Fig.~\ref{fg:pty}, the modification of the normalized distributions in NLO
compared to LO is less than about 15\% for the transverse-momentum
distributions and much less for the rapidity distributions.  Thus it is a
sufficient approximation to rescale the LO distributions uniformly by the K
factors of the total cross sections.

\section{Stop Pair Production}
At LO only pairs of $\st_1$ or pairs of $\st_2$ can be produced at hadron
colliders.  Mixed $\st_1 \st_2$ pair production is only possible at NLO and
beyond.  However, we have estimated that mixed stop pair production is
completely suppressed by several orders of magnitude and can thus safely be
neglected \cite{stops}. The evaluation of the QCD corrections proceeds along
the same lines as in the case of squarks and gluinos. The strong coupling and
the parton densities have been defined in the \MS~scheme with 5 light flavors
contributing to their scale dependences, while the stop masses are
renormalized on-shell.
\begin{figure}[hbt]
\vspace*{-2.5cm}

\hspace*{-0.0cm}
\begin{turn}{-90}%
\epsfxsize=10cm \epsfbox{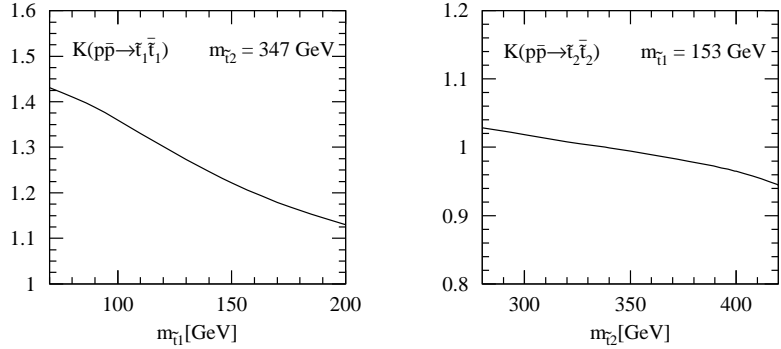}
\end{turn}
\vspace*{-2.5cm}

\caption[]{\label{fg:kst} \it K factor of the light stop
production cross sections at the upgraded Tevatron [$\sqrt{s}=2$ TeV].
Parton densities: CTEQ4L (LO)
and CTEQ4M (NLO) with $Q=m_{\st_1}$. Top mass: $m_t=175$ GeV.}
\end{figure}
The QCD corrections increase the cross sections by up to
about 40\% [see Fig.~\ref{fg:kst}] and thus lead to an increase of
the extracted stop masses from
the measurement of the total cross section.
\begin{figure}[hbt]
\vspace*{0.5cm}

\hspace*{3.5cm}
\epsfxsize=8cm \epsfbox{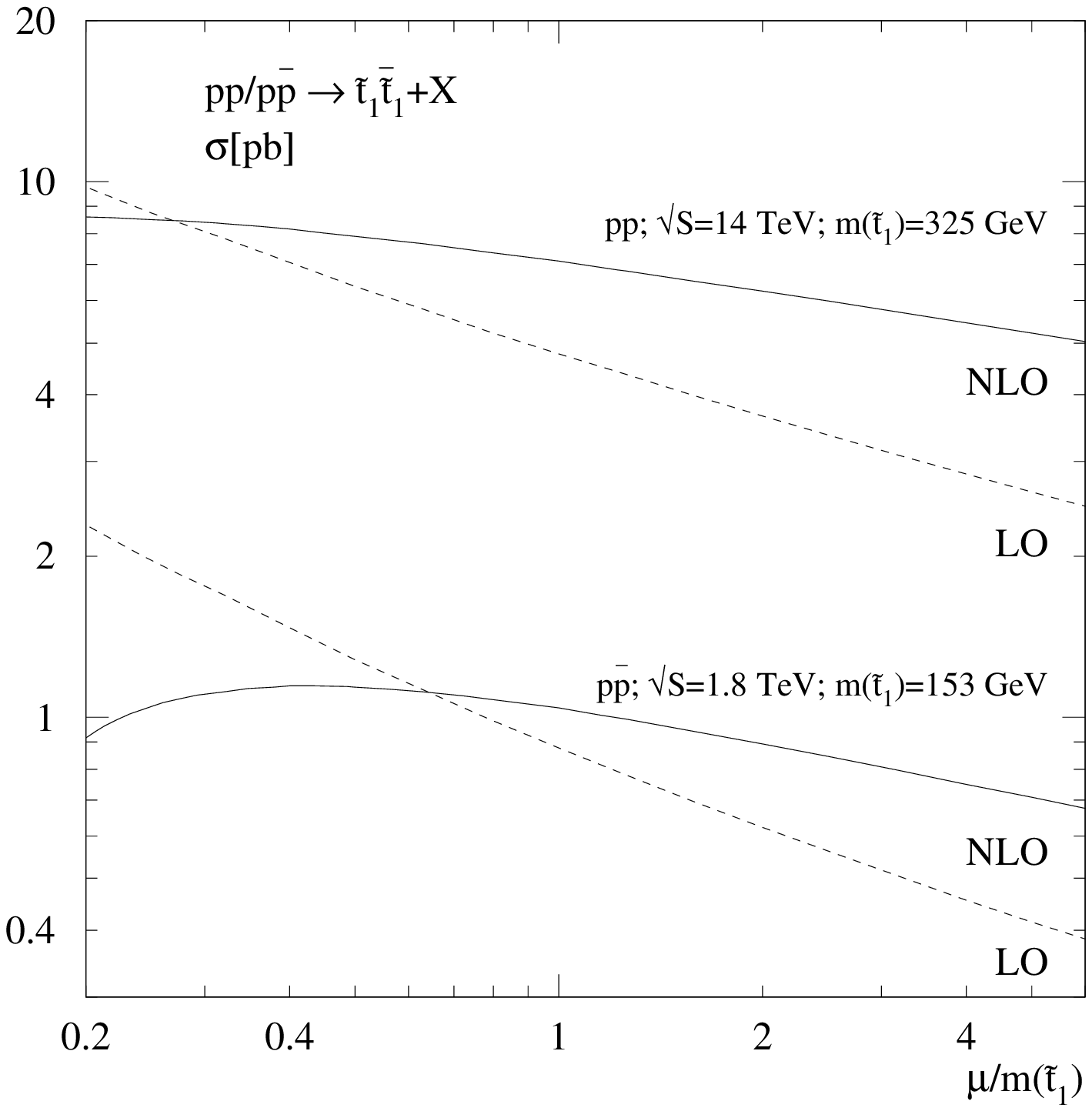}
\vspace*{-0.0cm}

\caption[]{\label{fg:scst} \it Scale dependence of the
total light stop production cross sections at the Tevatron and LHC in LO
and NLO. Parton densities: CTEQ4L (LO) and CTEQ4M (NLO);
Top mass: $m_t=175$ GeV.}
\end{figure}
Moreover, as in the squark/gluino
case the scale dependence is strongly reduced [see Fig.~\ref{fg:scst}] and
yields an estimate of about
15\% of the remaining theoretical uncertainty at NLO. At NLO the virtual
corrections depend on the stop mixing angle, the squark, gluino and second stop
masses. However, it turns out that these dependences are
very weak and can safely be neglected as can be inferred from
Fig.~\ref{fg:stopprod}.
\begin{figure}[hbt]
\vspace*{-0.0cm}

\hspace*{3.5cm}
\epsfxsize=8cm \epsfbox{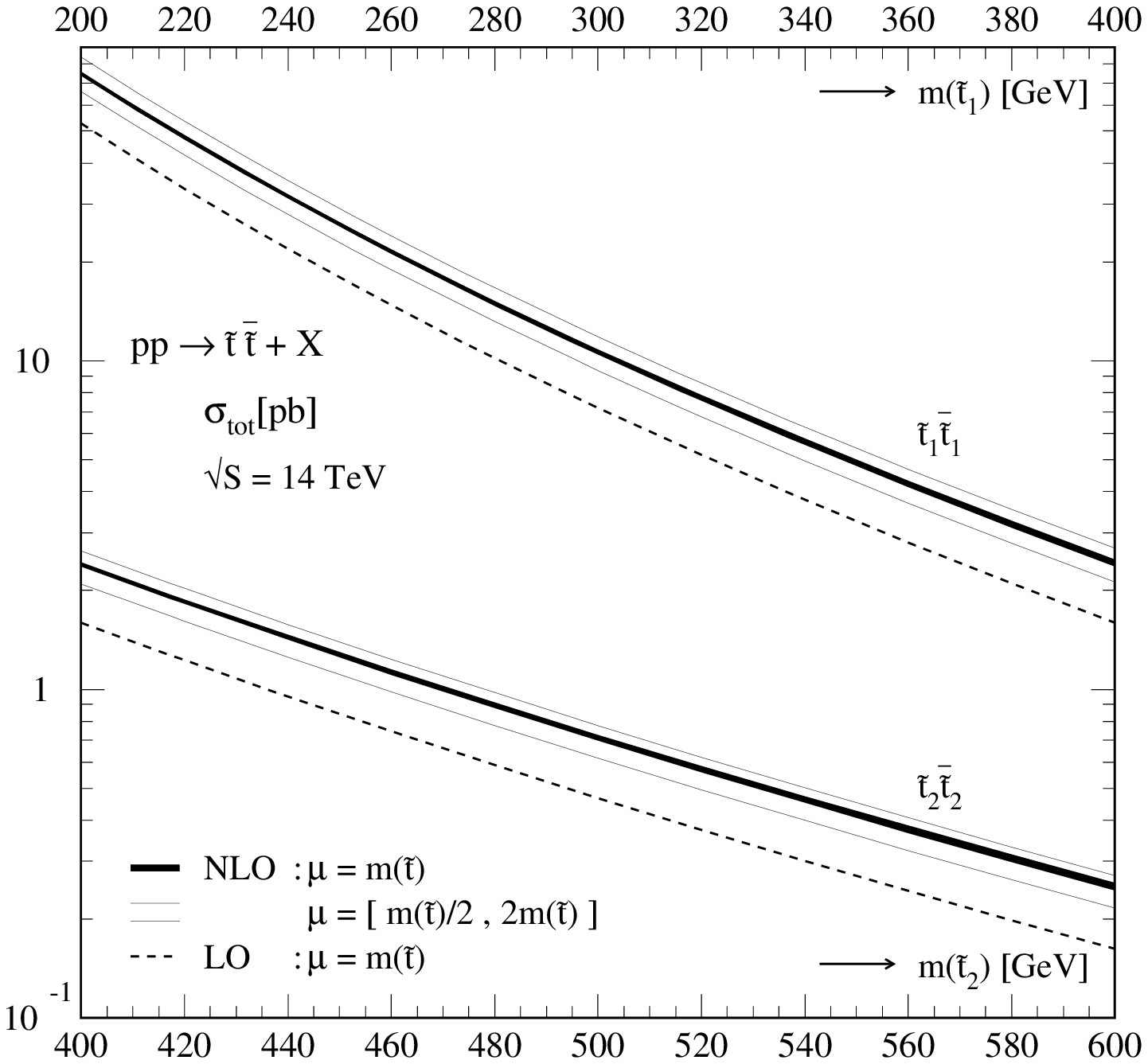}
\vspace*{-0.0cm}

\caption[]{\label{fg:stopprod} \it Production cross sections of the
light and heavy stop states at the LHC at LO [dashed] and NLO
[solid]. The thickness of the NLO curves represents the dependence of
the cross sections on the stop mixing angle and the squark and gluino
masses. The shaded NLO bands indicate the theoretical uncertainties due
to the scale dependence within $m_{\st}/2 < \mu < 2m_{\st}$.
Parton densities: CTEQ4L (LO) and CTEQ4M (NLO); Top mass: $m_t=175$ GeV.}
\end{figure}

\section{Chargino and Neutralino Production}
The production cross sections of charginos and neutralinos depend on several
MSSM parameters, i.e.\ $M_1, M_2, \mu$ and $\tgb$ at LO \cite{lo}. The cross
sections are sizeable for chargino/neu\-tra\-lino masses below about 100 GeV at
the upgraded Tevatron and less than about 200 GeV at the LHC. Due to the
strong dependence on the MSSM parameters the
extracted bounds on the chargino and neutralino masses depend on the
specific region in the MSSM parameter space \cite{bounds}. The outline of
the determination
of the QCD corrections is analogous to the previous cases of
squarks, gluinos and stops. The resonance contributions due to $gq \to \gau_i
\sq$ with $\sq \to q \gau_j$ have to be subtracted in order to avoid double
counting with the associated production of gauginos and strongly interacting
squarks and gluinos. The parton densities have been defined with 5 light
flavors contributing to their scale evolution in the \MS~scheme, while the
$t$-channel squark masses have been renormalized on-shell.
\begin{figure}[hbt]
\vspace*{0.5cm}

\hspace*{0.5cm}
\epsfxsize=6cm \epsfbox{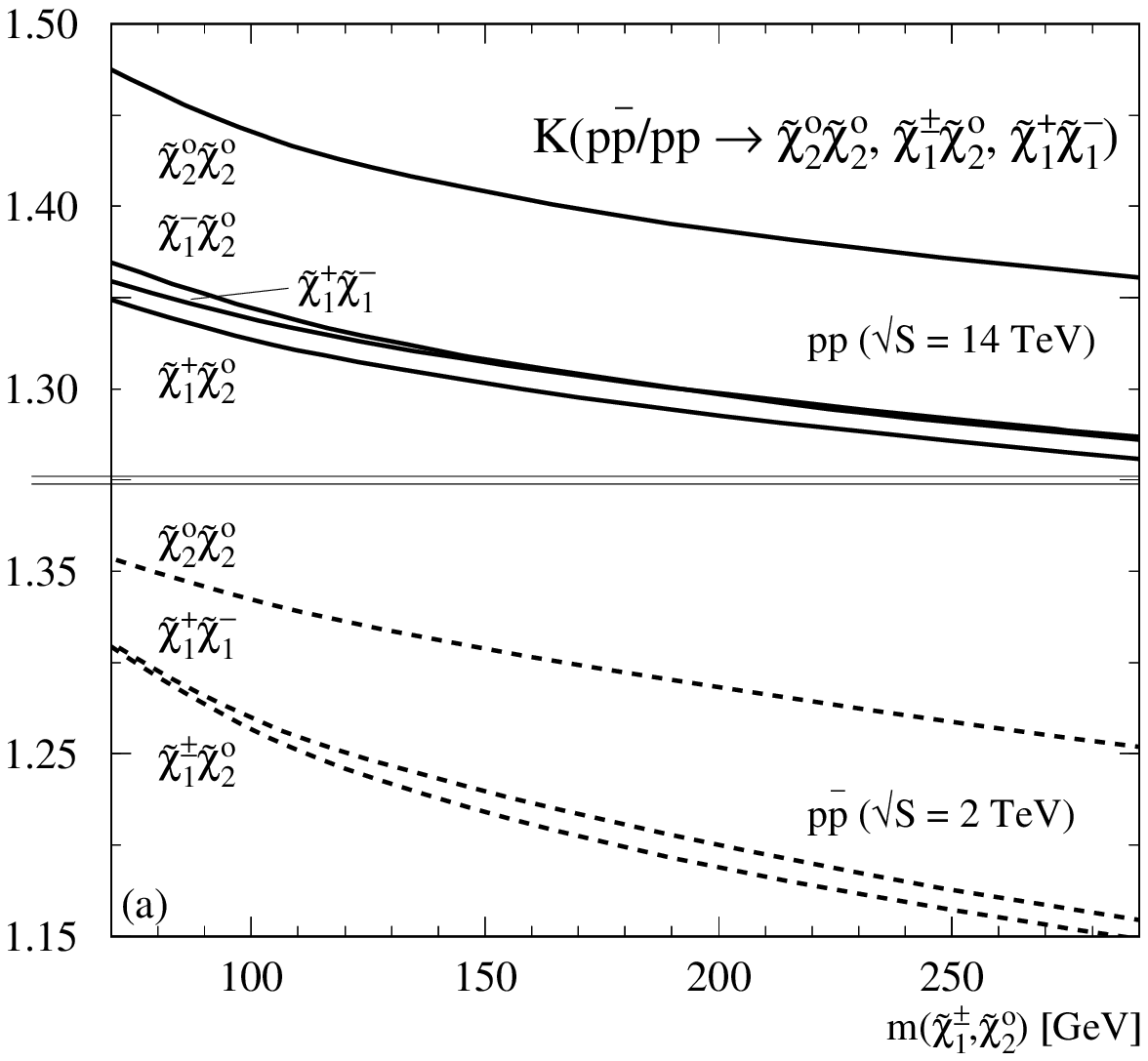}
\vspace*{-5.6cm}

\hspace*{7.2cm}
\epsfxsize=7.65cm \epsfbox{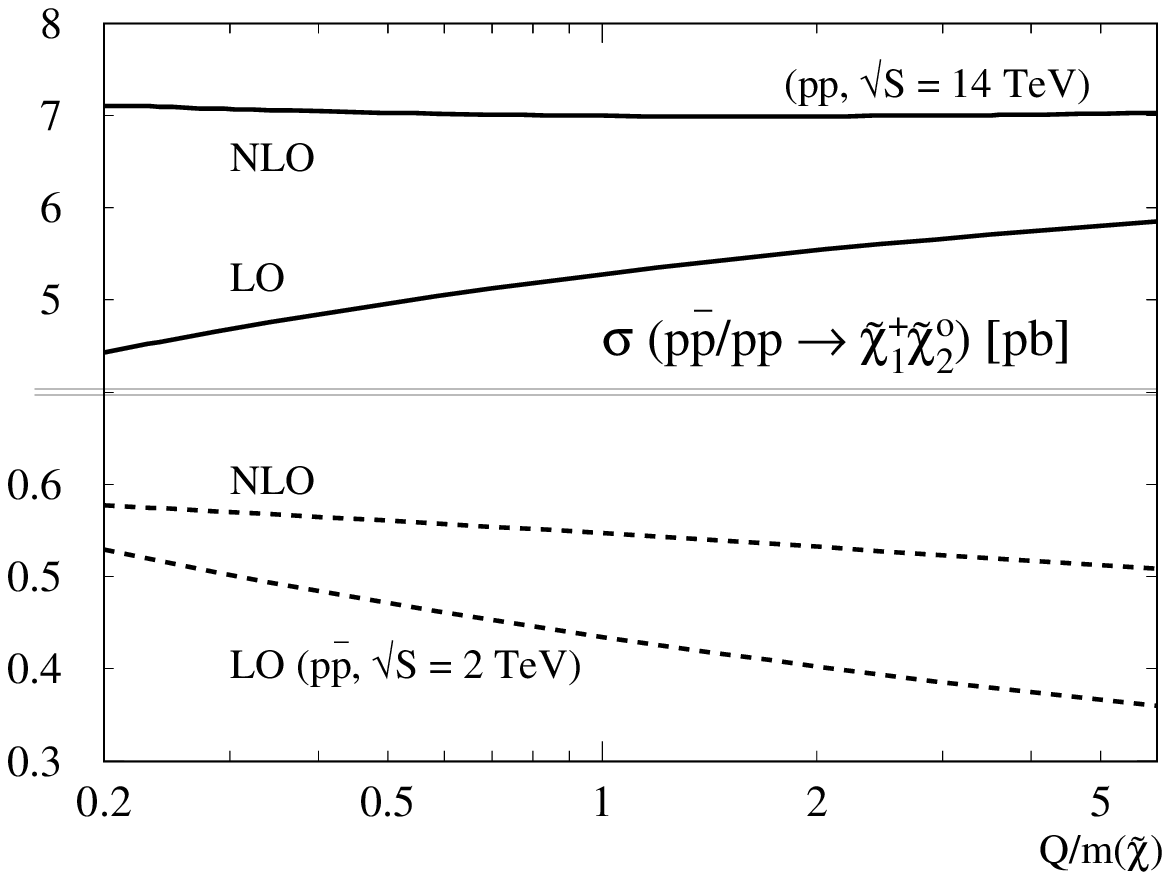}
\vspace*{-0.2cm}
 
\caption[]{\label{fg:kgausc} \it K factor and scale dependence of the $\gau^+_1
\gau^0_2$
production cross section at the Tevatron and LHC. Parton densities: CTEQ4L (LO)
and CTEQ4M (NLO) with $Q=m_{\st_1}$. Top mass: $m_t=175$ GeV;
SUGRA parameters: $M_0 = 100$ GeV, $A_0 = 300$ GeV, $\tgb = 4$, $\mu > 0$.}
\end{figure}
The QCD
corrections enhance the production cross sections of charginos and
neutralinos by about 10--40\% [see Fig.~\ref{fg:kgausc}]. The LO
scale dependence is reduced to about
10\% at NLO [see Fig.~\ref{fg:kgausc}], which signalizes a significant
stabilization of the thoretical prediction for the production cross sections
\cite{gaunlo}. The dependence of the
chargino/neutralino production cross sections on the specific set of parton
densities ranges at about 15\%.

\section{Conclusions}
In this work we have reviewed the status of SUSY particle production at
hadron colliders at NLO.  Most QCD corrections to the production processes
are known, thus yielding a nearly complete theoretical status.  There are
especially large QCD corrections to the production of gluinos, which
significantly increase the extracted bounds on the gluino mass from the
negative search for these particles at the Tevatron.  In all processes, which
are known at NLO, the theoretical uncertainties are reduced to about 15\%,
which should be sufficient for the upgraded Tevatron and the LHC\footnote{The
computer code PROSPINO \cite{prospino}
for the production of squarks,
gluinos and stops at hadron colliders is available at
{\tt http://www.desy.de/$\sim$spira}. The NLO production of gauginos and
sleptons will be included soon.}. \\

\noindent {\bf Acknowledgements} \\
I would like to thank W.\ Beenakker, R.\ H\"opker, M.\ Kr\"amer, M.\ Klasen,
T.\ Plehn and P.M.\ Zerwas for their collaboration and the organizers of
the Les Houches workshop for the invitation, the pleasant atmosphere and
financial support.

\setcounter{figure}{0}
\setcounter{table}{0}
\setcounter{section}{0}
\setcounter{equation}{0}
\setcounter{footnote}{0}
\clearpage

\begin{center}
{\large \bf
Comparison of exact matrix element calculations with ISAJET and PYTHIA
in case of degenerate spectrum in the MSSM }\\ 

\vspace*{3mm}
{\sc S.~ABDULLIN, V.~ILYIN and T.~KON }\\

\end{center}

\vspace*{3mm}

\begin{abstract}
We calculate the transverse momentum distribution of additional partons from
$pp\rightarrow \tilde{g}\tilde{g} + X$  with $m_{\tilde{g}}=1.2$ TeV at
$\sqrt{s}$=14 TeV. The calculations are carried out with exact $2\rightarrow
3$  matrix element generators CompHEP and GRACE, and widely used physics
generators ISAJET and PYTHIA. In the latter cases only $2\rightarrow 2$ 
subprocesses are included and additional partons originate mainly from initial
and final states radiation. The results obtained by CompHEP and GRACE are in
perfect agreement in the whole region of $p_T$ (10 -- 2000 GeV) of interest.
The comparison with $p_T$ distributions produced by ISAJET and PYTHIA does not
show significant difference. ISAJET tends to overshoot for $p_T$ values in the
region of 200-900 GeV, while PYTHIA's distribution has quite a good agreement
with exact ME calculations.

\end{abstract}


One of the main purposes of the LHC collider is to search for physics
beyond the Standard Model (SM), in particular to look for
superpartners of ordinary particles expected in SUperSYmmetric
extensions of the SM (SUSY). SUSY, if it exists, is expected to
reveal itself at the LHC firstly via an excess of $(multilepton +)$
$multijet$ + $E_{T}^{miss}$ final states compared to Standard Model
(SM) expectations \cite{reach_papers,sqgl_ab}. The bulk of these studies
was carried out in the framework of the Minimal SuperSymmetric Model
(MSSM), or the Supergravity-inspired subset of MSSM, minimal SUGRA
(mSUGRA) with universal gaugino mass and corresponding mass
hierarchy.

Unfortunately even in the framework of the MSSM one can expect
scenarios with non-universal gaugino mass leading to some
``pathological'' SUSY cases which are difficult to observe due to
a ``degenerate'' mass spectrum \cite{nkrasnik}.  For instance, it is
quite obvious that there are no visible objects in the SUSY event if
$m_{\tilde{\chi}_{1}^{0}} \approx m_{\tilde{g}} \approx m_{\tilde{q}}
= M_{SUSY}$. In this case one can rely mainly on observation of jet
activity which is additional to sparticle pair production. The idea is
that additional high $p_T$ recoil partons can provide a clear
$E_T^{miss} + jet(s)$ signature to observe the excess of SUSY events
over SM expectations.

Our preliminary estimates of the MSSM signal generated with ISAJET
\cite{ISAJET_ab} show that the limit of visibility (at the 5 $\sigma$
level) in the MSSM is about 1.2 -- 1.3 TeV with integrated luminosity
of 100 fb$^{-1}$ for $m_{\tilde{\chi}_{1}^{0}} \approx m_{\tilde{g}}
\ll m_{\tilde{q}}$, thus much smaller than the squark-gluino mass
reach ($\geq$ 2.5 TeV) in mSUGRA with the same integrated luminosity
\cite{sqgl_ab}.

It is well known that data bases in ISAJET and PYTHIA \cite{PYTHIA_ab}
contain matrix elements only for $2\rightarrow 2$ subprocesses. So,
additional partons can be produced only {\em via} initial and final
state radiation (ISR/FSR parton showering). If one discusses
production of hard partons, this mechanism is an approximation with
{\em a priori} unknown precision. From a general point of view this
approximation should underestimate the result obtained within the
perturbative QCD theory through evaluation of the complete set of
relevant Feynman diagrams. Thus, one might assume that there will be a
substantial number of events with high $p_T$ of the heavy SUSY system
recoiling against observable jet(s), larger than predicted by the
parton showering mechanism in ISAJET and PYTHIA. We use the packages
CompHEP \cite{CompHEP} and GRACE \cite{GRACE} to evaluate
$2\rightarrow 3$ matrix elements in leading order and calculate the
distributions. Finally we compare these results with ones obtained
with ISAJET and PYTHIA.

\section*{Calculational details}

To make a coherent calculation we choose a set of parameters as follows:
\begin{itemize}
 \item gluino mass $m_{\tilde{g}}$ = 1.2 TeV;
 \item common squark mass $m_{\tilde{q}}$ = 100 TeV (10 TeV for PYTHIA) to
neglect squark contributions;  
 \item CTEQ4M set of structure functions;
 \item $Q^2$ = $4m_{\tilde{g}}^2$ ( in PYTHIA
                   $Q^2=(m^2_{\bot1}+m^2_{\bot2})/2$); 
 \item NLO running strong coupling with normalization $\alpha_{S}(M_Z) = 0.118$,
       so $\alpha_{S}(2400 \ {\rm GeV}) = 0.0817$.
\end{itemize}
We use ISAJET 7.44 and PYTHIA 6.125 with multiple interactions
switched off. We treat the jet in ISAJET/PYTHIA not as the single
hardest final state parton, but use the final state parton list
(excluding the two gluinos) as input to an iterative cone jet finder
algorithm ($R_{cone}$ = 0.5) to find a group of partons forming
parton-level jets.

In CompHEP and GRACE we calculate exact tree level matrix elements for the
subprocesses $:$
$$ gg\to \tilde{g}\tilde{g}+g; \ \ 
gq(Q)\to \tilde{g}\tilde{g}+q(Q); \ \
qQ\to \tilde{g}\tilde{g}+g;$$

\noindent
which were convoluted with the CTEQ4m parton distributions to get LHC cross
sections and distributions.
The  $q$ ($Q$) terms stand for quarks (antiquarks) of the first five flavours:
$u$, $d$, $c$, $s$ and $b$.

We ask the final (additional) parton, gluon or quark, to radiate within the
typical ATLAS and CMS rapidity, $|\eta_j|<4.5$, and to be hard
enough, $E^j_T>10$ GeV. No cuts are applied on the gluino
pair.

\section*{Results and Conclusions}

The cross section of the process $pp\to \tilde{g}\tilde{g}+jet$ within cuts on
the jet turns out to be rather high,
about 120 fb, with the following contributions from different channels :
$$ \sigma_{gg} = 78.9 ~\mbox{fb}; \ \
\sigma_{gq(Q)} = 21.9 ~\mbox{fb}; \ \
\sigma_{qQ} = 20.0 ~\mbox{fb}.$$

\begin{figure}[hbtp]
\begin{center}
\centerline{\epsfig{figure=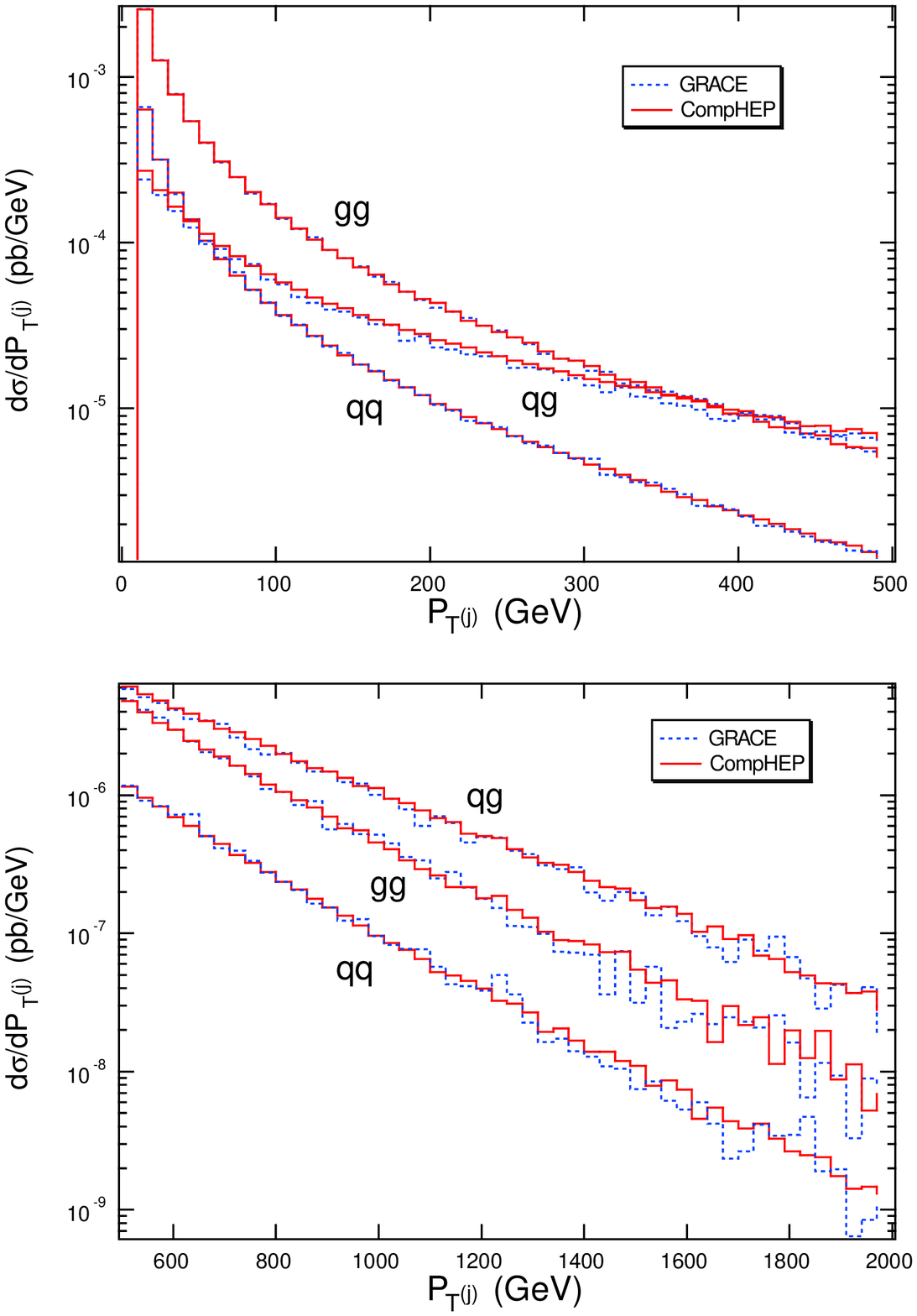,height=14cm,angle=0}}
\caption{Contribution of different channels to the $p_T$ distribution of
the additional jet in $pp\rightarrow\tilde{g}\tilde{g} + X$,
calculated with CompHEP and GRACE.
\label{fig:channels}}
\end{center}
\end{figure}

We found that the shapes of the $p_T$ distributions are quite
different in these three cases, see Fig.\ref{fig:channels}. Indeed,
one can expect that the $qQ$ channel produces less events than the
$gg$ contribution, due to the dominance of the gluon component in the
proton for small $x$. However, one can see that the $gq(Q)$ curve is
higher than the gluon-gluon one for $p_T>400$ GeV, being much smaller
even than $qQ$ at the soft end. Note that the set of topologies of
Feynman diagrams is the same in all three channels. So, the observed
difference connects surely with different interplay of contributions
of vector and spinor virtual particles.

\begin{figure}[hbtp]
\begin{center}
\centerline{\epsfig{figure=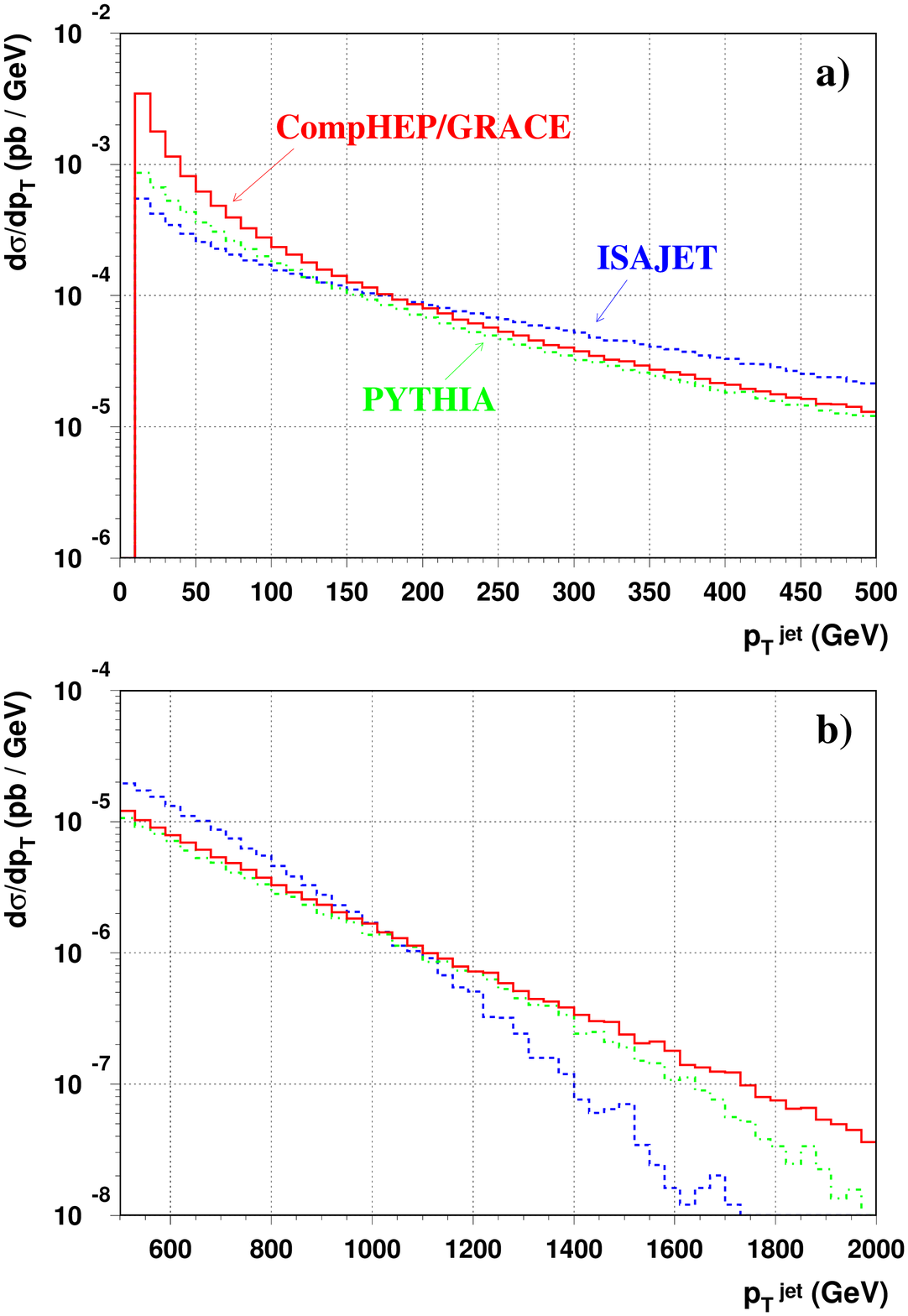,height=15cm,angle=0}}
\caption{$P_T$ distribution of the additional jet in 
$pp \rightarrow\tilde{g}\tilde{g} + X$, calculated with different
packages.
 \label{fig:compare}}
\end{center}
\end{figure}

In Fig.\ref{fig:compare} the $p_T$ distribution for $pp\to
\tilde{g}\tilde{g}+jet$ is given as a sum of the three channels
discussed above. The CompHEP/GRACE result is represented by solid
histograms, ISAJET by dashed histograms and PYTHIA by dot-dashed
ones. One can see that the PYTHIA distribution underestimates the {\em
exact ME} result smoothly. Of course this (almost constant) difference
can be explained (corrected) by a different setting of some QCD
switches. At the same time one can see a more pronounced deviation
from the {\em exact ME} in the shape of the ISAJET curve. In particular at
moderate $p_T$ ($\sim 200-900$ GeV) ISAJET overestimates the {\em
exact ME} result. One can also see that for very high transverse
momenta ($\geq$ 1200 GeV for ISAJET and 1400 GeV for PYTHIA) the {\em
parton showering} approximation is too crude.

Coming back to the initial idea to look at additional jet activity
with high $p_T$ recoiling against the heavy SUSY system one can
conclude that parton shower approximation works well enough up to very
high values of transverse momenta.  Therefore, unfortunately one can
not find room for sizable improvements of {\em parton shower}
predictions by using exact matrix elements in the process discussed.

Some interesting discrepancy is observed in the region of relatively small
transverse momenta, $p_T<150$ GeV, where ISAJET and PYTHIA produce much weaker
(by a factor of 2-3) jet activity than CompHEP/GRACE. However, this region is
sensitive to higher order corrections (in particular due to resummation of
large logarithms of $p_T/m_{\tilde{g}}$), thus a more careful analysis is
necessary to get reliable conclusions.


\setcounter{figure}{0}
\setcounter{table}{0}
\setcounter{section}{0}
\setcounter{equation}{0}
\setcounter{footnote}{0}
\clearpage

\def\bq{\begin{equation}}
\def\eq{\end{equation}}
\def\bqa{\begin{eqnarray}}
\def\eqa{\end{eqnarray}}
\def\bqb{\begin{eqnarray*}}
\def\eqb{\end{eqnarray*}}

\begin{center}

{\large \bf
     The finite width effect on neutralino production } \\

\vspace*{3mm}

{\sc T. KON, Y. KURIHARA, M. KURODA and K. ODAGIRI}

\end{center}

\vspace*{3mm}

\newdimen\nude\newbox\chek
\def\slash#1{\setbox\chek=\hbox{$#1$}\nude=\wd\chek#1{\kern-\nude/}}
\def\up#1{$^{\left.#1\right)}$}
\def\to{\rightarrow}
\def\dag{\dagger}
\def\llgm{\left\lgroup\matrix}
\def\rrgm{\right\rgroup}
\def\vectrl #1{\buildrel\leftrightarrow \over #1}
\def\partrl{\vectrl{\partial}}

It has been proposed in \cite{atlas} that neutralino mass differences 
can be determined fairly accurately through the $\mu$-pair invariant
mass distribution $d\sigma/d M_{\mu\mu}$ of the process
\bq
    pp \to \tilde\chi_1^+\tilde\chi_2^0X\to \tilde\chi_1^+ (\mu\tilde\mu)X
     \to \tilde\chi_1^+\mu (\tilde\chi_1^0 \mu)X,
\eq
as the distribution exhibits a sharp decrease at its
kinematic endpoint. When, however, the width of
the neutralino is taken into account, the smearing of the cross
section might make this observation obsolete and less appealing
for the determination of the neutralino mass.  Therefore, it is
important to know to what extent the width of neutralino makes
the mass determination less accurate.\par
We have investigated this problem by comparing the muon-pair
invariant mass distribution in the following three cases.\par
\noindent{(case~1)} Zero width approximation for $\tilde\chi_2^0$.
Only those diagrams which proceed via the chain decay (1) are taken into
account.\par
\noindent{(case~2)} Same as (1) but finite width of $\tilde\chi_2^0$
is used. Only those diagrams that proceed via resonances as shown in (1)
are
considered.\par
\noindent{(case~3)} Finite width of $\tilde\chi_2^0$  as well as the entire
set of diagrams (165 diagrams in unitary gauge) that creates
the final state  $\tilde\chi_1^+\tilde\chi_1^0\mu\mu$ are taken into
account.
The diagrams which are considered in case 3 but not in case 2 constitute
the background to this process.\par

\medskip
In the numerical evaluation of the muon pair distribution, we have used the
following set of SUSY parameters,
\bq
    \tan\beta =12,~~~~~M_2 = 200 GeV, ~~~~\mu = -500 GeV,
\eq
which results in the following masses for charginos and neutralinos,
\bqa
    m_{\tilde\chi_i^0} &=& ( 97, 197, 507, 511),\\
    m_{\tilde\chi_i^+} &=& ( 197, 514).
\eqa
In particular, in the present set of parameters we are considering,
$\tilde\chi_1^0$ is almost purely bino (the SUSY partner of the $U(1)$
gauge boson, 
$B_\mu$), while $\tilde\chi_2^0$ is almost purely neutral wino (the SUSY
partner of the
neutral $SU(2)_L$ gauge-boson, $W_3^\mu$).
As for the slepton masses, we have used the common mass
$m_{\tilde\ell}$ = 150 GeV.\par
Assuming that other SUSY particles are heavy so that $\tilde\chi_2^0$ decays
dominantly into $\ell \tilde\ell$,
we have evaluated the width of $\tilde\chi_2^0$ as
\bq
     \Gamma_{\tilde\chi_2^0} = 1.25GeV.
\eq
The process is described by the subprocess
\bq
    q \bar q' \to \tilde\chi_1^+\tilde\chi_2^0\to \tilde\chi_1^+
(\mu\tilde\mu)
     \to \tilde\chi_1^+\mu (\tilde\chi_1^0 \mu).
\eq
For the quark distribution functions, we have used
QTEC-3D \cite{pdflib}.\par

The numerical evaluation was performed by {\tt HERWIG} \cite{herwig_ko}
in case 1 and case 2,
while the case 3 was calculated by {\tt GRACE} \cite{grace}.\par
Results are shown in Fig.1(a) and (b),
where the distributions of the invariant mass
$M_{\mu\mu}$ in the second neutralino decay are shown.
In Fig.1(a),  the dashed line
represents the zero width approximation (case 1), the dotted line
represents the distribution for the finite $\tilde\chi_2^0$ width
but with only the resonance diagrams (case 2),
while the solid line
represents the distribution for the finite $\tilde\chi_2^0$ width and
full set of 
diagrams which give the final states
$\tilde\chi^+_1\tilde\chi_1^0\mu\mu$ (case 3).
We show a detailed comparison between case 2 and case 3 around the end
point region of $M_{\mu\mu}$ in Fig.1(b).
As one sees from the figures, the edge of the
distribution at the end point of the phase space remains, although it is
broadened by the finite width effect. As a result, the exact position of
the end point acquires an ambiguity of about 5 GeV, although the precision
of mass determination, which may additionally make use of
the shape of the distribution, can presumably be made much less
than this. A full analysis will need to take into account
the experimental resolutions and signal selection efficiency,
and this is beyond the scope of our study.
The total cross section of the signal process (1) is 0.015 pb.

\vspace*{3mm}
{\bf Acknowledgements}
One of the authors, K.O., would like to thank Drs. J. Kanzaki and
P. Richardson for their computing help.

\begin{figure}[hbtp]
\begin{center}
\centerline{\epsfig{figure=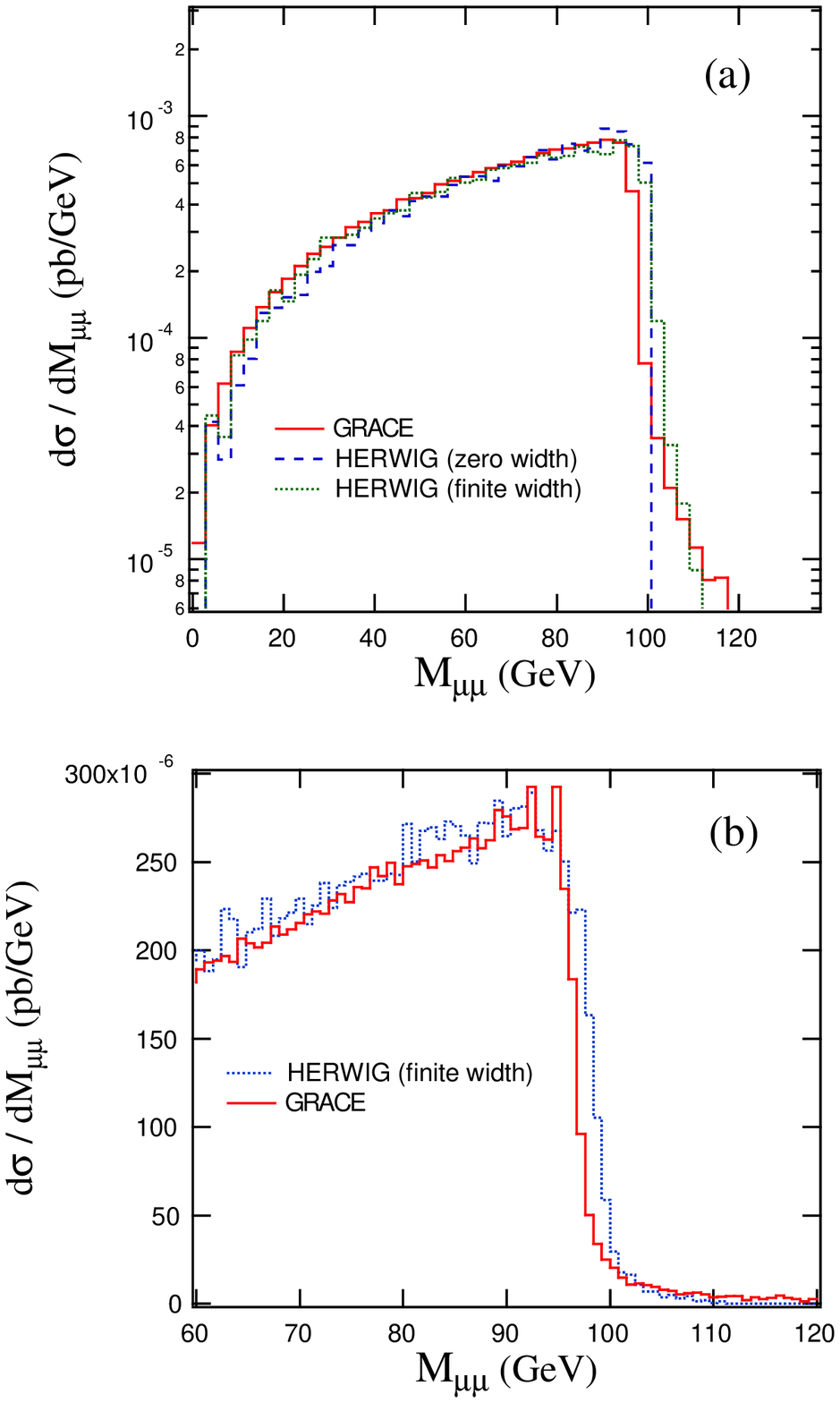,height=15cm,angle=0}}
 \caption{ Distribution of the invariant mass
$M_{\mu\mu}$ in the second neutralino decay from the
$\tilde\chi_1^+ \tilde\chi_2^0$
associated production. In (a) a comparison is made among the full
calculation
using {\tt GRACE} (continuous line, case 3),
the calculation with {\tt HERWIG} assuming a narrow
width and an isotropic decay (dashed line, case 1),
and another calculation using
{\tt HERWIG} incorporating a finite $\tilde\chi_2^0$
width (dotted line, case 2).
(b) corresponds to the detailed comparison between case 2 and case 3
around the end point region of $M_{\mu\mu}$.
 \label{fig:GH}}
\end{center}
\end{figure}
 
\setcounter{figure}{0}
\setcounter{table}{0}
\setcounter{section}{0}
\setcounter{equation}{0}
\setcounter{footnote}{0}
\clearpage


\def\MeV{{\rm MeV}}

\begin{center}
{\large \bf Width effects in slepton production
  $e^+e^- \to \tilde{\mu}^+_R \, \tilde{\mu}^-_R$} \\ 

\vspace*{3mm}
{\sc H.-U. MARTYN} \\
\vspace*{3mm}

\begin{abstract}
  A case study will be presented to investigate width effects
  in the precise determination of slepton masses at the 
  $e^+e^-$ {\sc Tesla} Linear Collider. 
\end{abstract}
\end{center}

\section{Introduction}

If supersymmetry will be discovered in nature a precise measurement of the 
particle spectrum will be very important in order to determine the underlying 
theory.
The potential of the proposed {\sc Tesla} Linear Collider~\cite{cdr} with its 
high luminosity and polarization of both $e^\pm$ beams will allow to obtain 
particle masses with an accuracy of $10^{-3}$ or better~\cite{lcphysics}.
At such a precision width effects of primary and secondary particles
may become non-negligible.

\begin{figure}[hbt]
\centering
\epsfig{file=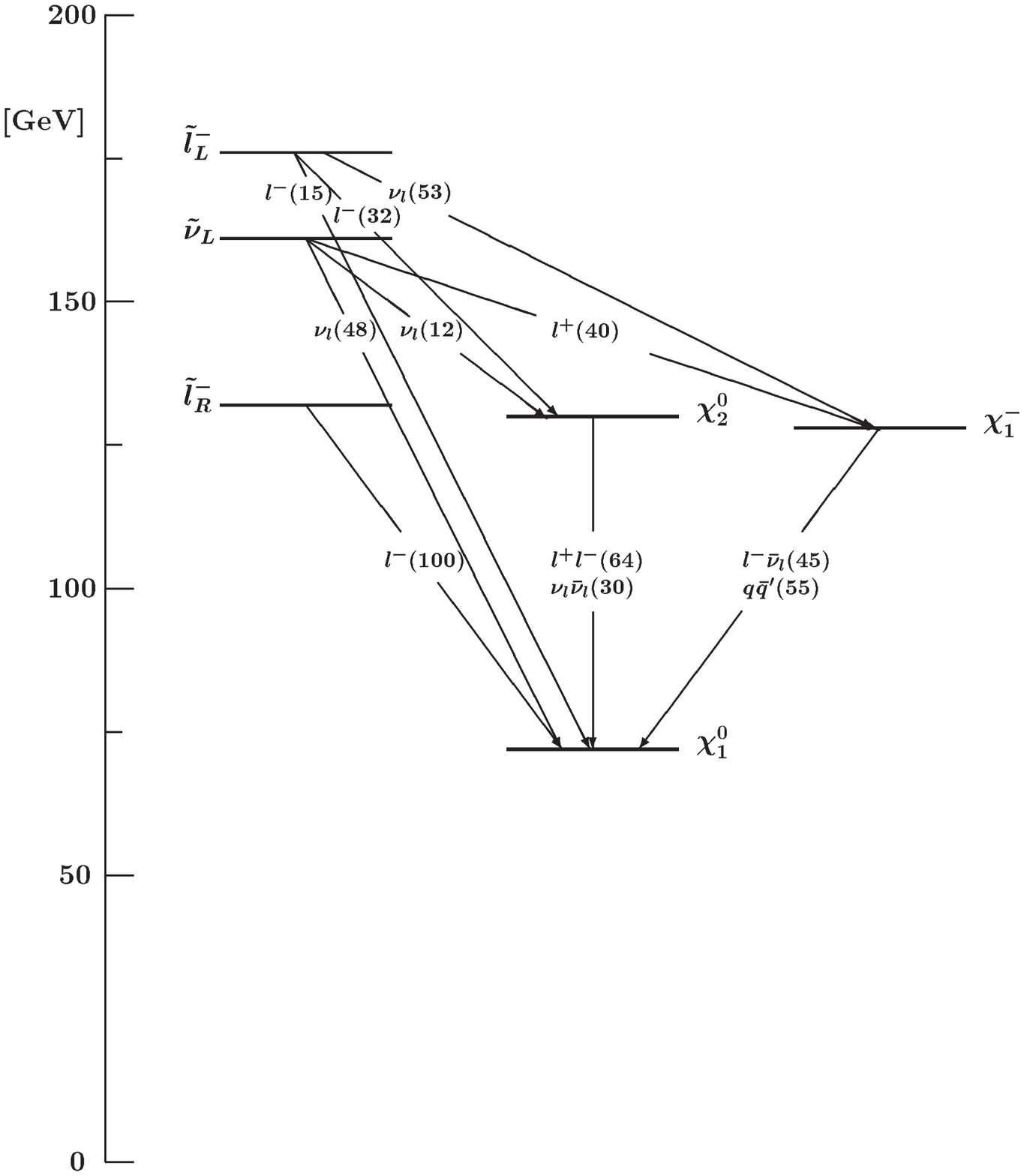,%
  bbllx=0pt,bblly=340pt,bburx=570pt,bbury=680pt,clip=,%
  height=9cm}
\caption{Mass spectrum and decay modes of sleptons and light gauginos
\label{fig:spectrum}}
\end{figure}

The present case study is based on a particular
$R$-parity conserving m{\sc SUGRA} scenario,
also investigated in the {\sc Ecfa/Desy} Study~\cite{ecfadesy},
with parameters
$  m_0 = 100~\GeV, \ m_{1/2} = 200~\GeV, \ A_0 = 0~\GeV, \
  \tan\beta = 3$ and ${\rm sgn}(\mu) > 0 $.
The particle spectrum is shown in fig.~\ref{fig:spectrum}.
Typical decay widths of the scalar leptons are expected to be 
$\Gamma \sim 0.3 - 0.5~\GeV$, 
while the widths of the light gauginos,
decaying into 3-body final states, are (experimentally) negligible.

This note presents, as an example, a simulation of right scalar muon
production 
\begin{eqnarray}
  e^-_R e^+_L & \to & \tilde{\mu}_R^- \, \tilde{\mu}_R ^+\ ,\\
              & \to & \mu^- \chi^0_1 \, \mu^+ \chi^0_1 \ . \nonumber
\end{eqnarray}
The analysis is based on the methods and techniques described in a
comprehensive study of the same {\sc Susy} spectrum~\cite{mb}.
The detector concept, acceptances and resolutions are taken from the
{\sc Tesla} Conceptual Design Report~\cite{cdr}.
Events are generated with the Monte Carlo program 
{\sc Pythia}~6.115~\cite{pythia_ma}, 
which includes the width of supersymmetric particles as well as
QED radiation and beamstrahlung~\cite{circe}.
It is assumed that both beams are polarized, right-handed electrons
to a degree of ${\cal P}_{e^-_R} = 0.80$ and left-handed positrons by
${\cal P}_{e^+_L} = 0.60$.
A proper choice of polarizations increases the cross section by a factor of 
$\sim 3$ and reduces the background substantially, e.g. by more than an 
order of magnitude for Standard Model processes.

\section{Mass determinations}

Scalar muons $\tilde{\mu}_R$ are produced in pairs via $s$ channel
$\gamma$ and $Z$ exchange and decay into an ordinary muon and a stable
neutralino $\chi_1^0$ (LSP), which escapes detection. The
experimental signatures are two acoplanar muons in the final state
with large missing energy and nothing else in the detector.  Simple
selection criteria~\cite{cdr} (essentially cuts on acollinearity angle
and missing energy) suppress background from $W^+W^-$ pairs and
cascade decays of higher mass {\sc Susy} particles and result in
detection efficiencies around $\sim 70\%$.

Two methods to determine the mass of $\tilde{\mu}_R$ will be discussed:
(i) a threshold scan of the pair production cross section, and
(ii) a measurement of the energy spectrum of the decay muons,
which simultaneously constrains the mass of the primary smuon
and the secondary neutralino.
The particle mass parameters given by the chosen {\sc Susy} model
are $m_{\tilde{\mu}_R} = 132.0~\GeV$,
$\Gamma_{\tilde{\mu}_R} = 0.310~\GeV$ and $m_{\chi^0_1} = 71.9~\GeV$.

\subsection{Threshold scan} 

\begin{figure}[htb]
  \centering \vspace{-.3cm}
  \epsfig{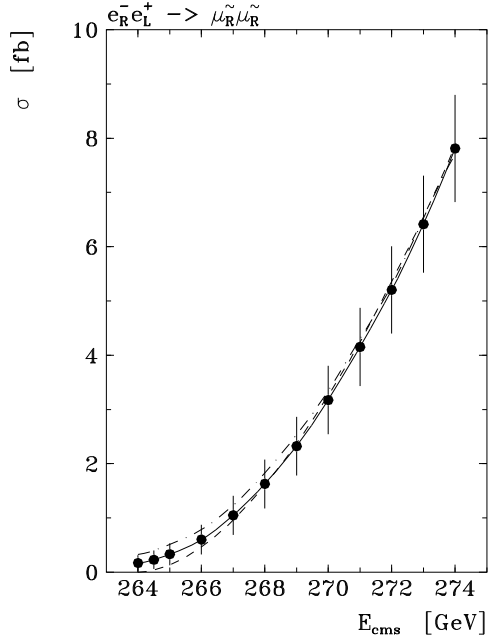}
  \vspace{-.2cm}
  \caption{ Observable cross section near threshold of the reaction
    $e^-_R e^+_L \rightarrow \tilde{\mu}_R \tilde{\mu}_R$
    including QED radiation and Beamstrahlung. 
    Curves assume a mass $m_{\tilde{\mu}_R} = 132.0~\GeV$ and width
    $\Gamma_{\tilde{\mu}_R}$ of $310~\MeV$ (full curve),
    $0~\MeV$ (dashed curve) and $620~\MeV$ (dashed-dotted curve).
    Measurements correspond to ${\cal L} = 10~\fb^{-1}$ per point.}
  \label{fig:scan}
\end{figure}

Cross section measurements close to production threshold are
relatively simple.  One essentially counts additional events with a
specific signature, here two oppositely charged, almost monoenergetic
muons, over a smooth background.  The cross section for slepton pair
production rises as $\sigma \propto \beta^3$, where $\beta = \sqrt{1 -
4\,m_{\tilde{\mu}_R}^2/s}$ is the velocity related to the
$\tilde{\mu}_R$ mass.  The excitation curve as a function of the cms
energy, including effects due to QED initial state radiation and
beamstrahlung, is shown in figure~\ref{fig:scan}.  The sensitivity to
the width $\Gamma_{\tilde{\mu}_R}$ is most pronounced close to the
kinematic production limit and diminishes with increasing energy.  A
larger width `softens' the rise of the cross section with energy.
Fits to various mass and/or width hypotheses are performed by
simulating measurements with a total integrated luminosity of
$100~\fb^{-1}$ distributed over 10 equidistant points around $\sqrt{s}
= 264 - 274~\GeV$.  The data may be collected within a few months of
{\sc Tesla} operation.

Taking the width from the model prediction $\Gamma_{\tilde{\mu}_R} = 310~\MeV$,
a fit to the threshold curve gives a statistical accuracy for the smuon mass
of $\delta m_{\tilde{\mu}_R} = 90~\MeV$. This error is considerably smaller
than the expected width.
A two-parameter fit yields
$m_{\tilde{\mu}_R} = 132.002~^{+0.170}_{-0.130}~\GeV$ and
$\Gamma_{\tilde{\mu}_R} = 311~^{+560}_{-225}~\MeV$. 
However, both parameters are highly correlated with a correlation coefficient 
of $0.95$. 
Finally, if one may fix the $\tilde{\mu}_R$ mass from another measurement, the
width can be determined to $\delta\Gamma_{\tilde{\mu}_R} = \pm 190~\MeV$.
It should be noted that the scan procedure and choice of energy points is by
no means optimized. 
Possibilities to reduce the correlations should be studied.

\subsection{Energy spectrum of $\mu^\pm$}

For energies far above threshold,  the kinematics of the decay chain 
of reaction (1) allows to identify and to reconstruct the
masses of the primary and secondary sparticles.
The isotropic decays of the scalar muons lead to a flat energy
spectrum of the observed final $\mu^\pm$ in the laboratory frame.
The endpoints of the energy distribution are related to the masses of the
$\tilde{\mu}_R$ and $\chi^0_1$ via
\begin{eqnarray}
  \frac{m_{\tilde{\mu}}^2-m_{\chi^0}^2}{2\,(E_{\tilde{\mu}}+p_{\tilde{\mu}})}
  \ \leq & E_\mu & \leq \
  \frac{m_{\tilde{\mu}}^2-m_{\chi^0}^2}{2\,(E_{\tilde{\mu}}-p_{\tilde{\mu}})} 
  \ .
\end{eqnarray}

\begin{figure}[htb]
  \begin{center}
  \hspace{-.5cm}
  \epsfig{file=martyn3a.eps,%
    bbllx=20pt,bblly=40pt,bburx=470pt,bbury=740pt,clip=,%
    angle=90,height=7cm}
  \epsfig{file=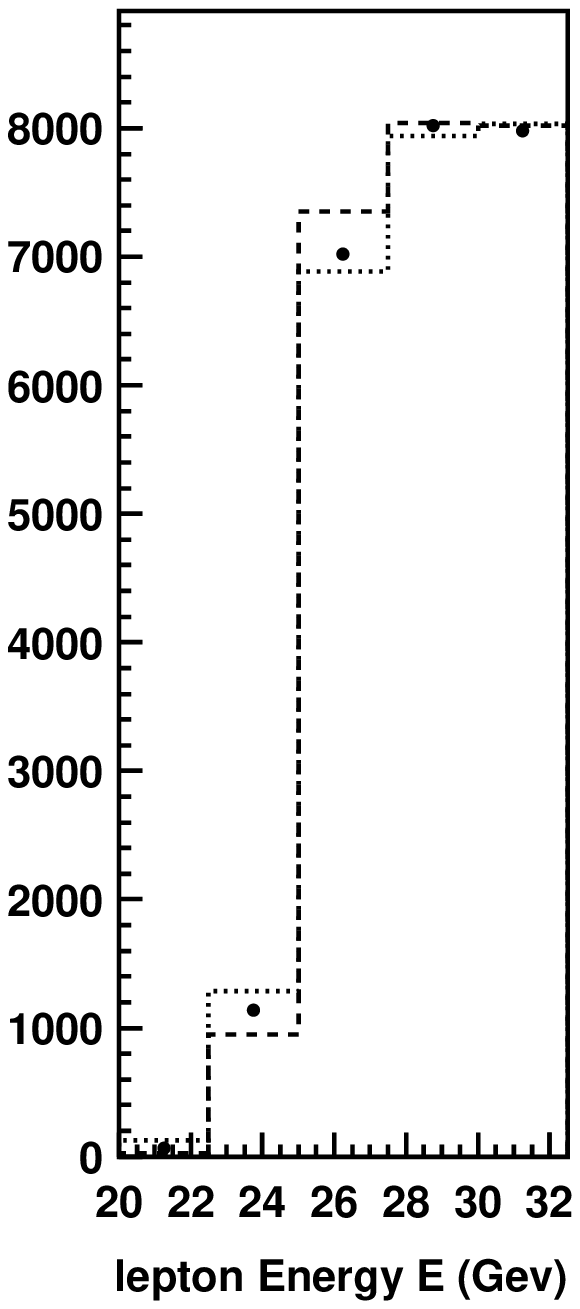,%
    bbllx=0pt,bblly=0pt,bburx=230pt,bbury=390pt,clip=,%
    angle=0,height=7cm}
  \end{center}
  \caption{{\bf Left:}
    Energy spectrum of di-muon events from the reaction
    $e^-_R e^+_L \to \tilde{\mu}_R \tilde{\mu}_R$ 
    and the background $e^-_R e^+_L \to \chi^0_1 \chi^0_2$
    at $\sqrt{s}=320~\GeV$ assuming ${\cal L} = 160~\fb^{-1}$.
    {\bf Right:} 
    Lower endpoint region of the $\mu$ energy spectrum illustrating 
    the effect of various widths $\Gamma_{\tilde{\mu}_R}$ of 
    $310~\MeV$ (dots), $0~\MeV$ (dashed) and $620~\MeV$ (dotted)
    using the tenfold luminosity.}
  \label{fig:sleptons}
\end{figure}

In practice the sharp edges of the energy spectrum will be smeared by effects
due to detector resolution, selection criteria and in particular
initial state radiation and beamstrahlung.
The results of a simulation at $\sqrt{s} = 320~\GeV$ assuming an integrated
luminosity of $160~\fb^{-1}$ are shown in figure~\ref{fig:sleptons}.
One observes a clear signal from $\tilde{\mu}_R$ pair production above a
small background of cascade decays $\chi^0_2 \to \mu^+ \mu^- \chi^0_1$
from the reaction $e^-_R e^+_L \to \chi^0_2 \chi^0_1$. 
Contamination from chargino or $W$ pair production is completely negligible.

A two-parameter fit to the $\mu$ energy spectrum yields masses of
$m_{\tilde{\mu}_R} = 132.0 \pm 0.3~\GeV$ and
$m_{\chi^0_1}      =  71.9 \pm 0.2~\GeV$.
The statistical accuracy is of the same size as the expected width of the
scalar muon.
Choosing a different width $\Gamma_{\tilde{\mu}_R}$ in the simulation modifies
essentially the $\mu$ energy spectrum at the low endpoint and has little impact
at higher energies. 
This is illustrated in figure~\ref{fig:sleptons}, right part, 
which compares the lower part of the spectrum with the predictions of width 
zero and twice the expected value.
The sharp rise is getting smeared out with increasing width.
With the anticipated luminosity of $160~\fb^{-1}$ it may be feasible to 
distinguish these cases.

\subsection{Production of other sparticles}

It should be noted that estimates on the sensitivity of width effects
in other slepton production channels can be obtained from the above results
by scaling the cross section and taking the branching ratios into final 
states into account. 
Thus one expects e.g. a gain by a factor of $\sim$2 for selectron
$\tilde{e}_R$ and sneutrino $\tilde{\nu}_e$ pair production.
For the higher mass chargino $\chi^\pm_2$ and neutralinos $\chi^0_{3,\,4}$
mass resolutions of $\sim 0.25 - 0.50~\GeV$ may be obtained from threshold
scans~\cite{mb}, where the cross sections rise as $\sigma \propto \beta$.
The corresponding widths are expected to be $\sim 2 - 5~\GeV$ 
(two-body decays in a gauge boson and gaugino) and have certainly to be
considered.

\section{Conclusions}

The high luminosity of {\sc Tesla} allows to study the production and
decays of the accessible {\sc Susy} particle spectrum.  Polarization
of both $e^-$ and $e^+$ beams is very important to optimize the signal
and suppress backgrounds.  A simulation of slepton production $e^+e^-
\to \tilde{\mu}_R^+ \tilde{\mu}_R^-$ shows that for precision mass
measurements with an accuracy of ${\cal O}(100~\MeV)$ the widths of
the primary particles have to be taken into account.  Finally, it is
worth noting that the anticipated mass resolutions from threshold
scans or lepton energy spectra can only be obtained if beamstrahlung
effects are well under control.

\setcounter{figure}{0}
\setcounter{table}{0}
\setcounter{section}{0}
\setcounter{equation}{0}
\setcounter{footnote}{0}
\clearpage

\def\be{\begin{equation}}
\def\ee{\end{equation}}
\def\epem{\mbox{$e^+ e^-$}}
\def\tchi{\mbox{$\tilde \chi_1^0$}}
\def\mchi{\mbox{$m_{\tilde \chi}$}}
\def\sq{\mbox{$\tilde{q}$}}
\def\qp{\mbox{$\tilde{q} \tilde{q}^*$}}
\def\mq{\mbox{$m_{\tilde{q}}$}}
\def\msqmin{\mbox{$m_{\tilde{q},{\rm min}}$}}
\def\rs{\mbox{$\sqrt{s}$}}


\begin{center}
{\large \bf Radiative Effects on Squark Pair Production at
$e^+ e^-$ Colliders} \\
\vspace*{3mm}
{\sc M. DREES, O.J.P. \'EBOLI, R.M. GODBOLE and S. KRAML}\\

\end{center}

\vspace*{.3cm}
\begin{abstract}

We study the impact of various radiative effects on the kinematic
reconstruction of squark pair events at future \epem\ colliders. We
focus on the simplest case where both squarks decay directly into a
stable neutralino, but include both photon radiation off the initial
state and gluon radiation during the production and/or decay of the
squarks. These effects change the shapes of the distributions used to
determine the squark mass; they can therefore also introduce additional
systematic uncertainties.

\end{abstract}

\noindent



Among the advantages of \epem\ colliders over hadron colliders are the
fairly well defined center--of--mass energy in hard (annihilation)
events, and the comparative cleanliness of the environment. Taken
together, these properties make the kinematic reconstruction of \epem\
events much easier than that of comparable events at hadron
colliders. In particular, experiments at \epem\ colliders should be
able to measure the masses of new particles (with unsuppressed
electroweak couplings) with an error of 1\% or less \cite{degk1},
either through threshold scans, or through fitting kinematic
distributions of events well above threshold. The latter method is
more versatile, since the same data set can well contain several kinds
of ``new physics'' events, which can be separated from each other (and
from backgrounds) using kinematical cuts. Moreover, kinematical
reconstruction allows to determine not only the masses of the new
particles produced in the primary interaction, but also those of their
decay products.

However, in order to correctly interpret the information contained in
various distributions one needs an accurate model of the final
state. In particular, if one wants to achieve errors of 1\% or less,
various radiative effects have to be taken into account. These are of
special importance if the new particles and/or their decay products
have strong interactions, since a significant fraction of all signal
events will then contain hard gluons. Here we study the impact of
these radiative effects on the measurement of squark masses at \epem\
colliders. Squarks are the currently most plausible new particles that
have both strong and electroweak interactions.

In the last few years considerable progress has been made \cite{degk2}
in the accurate calculation of total cross sections for squark pair
production and of squark branching ratios. In particular, one--loop
corrections to these quantities from both ordinary QCD and SUSY QCD,
as well as from Yukawa interactions, are now known. These have been
used \cite{degk4} to estimate the error with which the squark mass can
be extracted from a measurement of the total squark pair production
cross section times branching ratio into a given final state. However,
in such a ``dynamical'' determination of the squark mass one has to
assume values for all the other input parameters that affect the cross
section times branching ratio. These include the gluino mass and, for
third generation squarks, also the masses and mixing angles of
squarks, charginos, neutralinos, and Higgs bosons \cite{degk2}.

In contrast, relatively little attention has been paid to the
kinematical determination of squark masses at \epem\ colliders. The
pioneering work by Feng and Finnell \cite{degk5} investigates the
usefulness of various kinematical distributions, and concludes that
experiments at a 500 GeV collider should be able to determine the mass
of 200 GeV squarks with an error of $\leq 0.5\%$ using just 20
fb$^{-1}$ of data, if all squarks decay directly into a massless
quark and an invisible (stable or long--lived) neutralino
\tchi. However, their study did not include any radiative
effects. Here we update their analysis by including initial state
radiation of photons, as well as the emission of hard gluons during
the production ($\epem \rightarrow \qp g$) and/or decay ($\tilde{q}
\rightarrow q \tchi g$) of the squarks. We also take larger squark
masses $(\mq \sim 300$ GeV) to account for recent experimental bounds
from the Tevatron \cite{degk6}, and a correspondingly higher
center--of--mass energy of 800 GeV.

For reasons of space we only briefly summarize the ingredients of our
analysis; details will be given elsewhere. We treat the emission of
photons off the initial state in the structure function formalism
\cite{degk8}. The differential cross section is then given by
\be \label{e_degk1}
d \sigma = \int dx_1 dx_2 f_{e|e}(x_1,\sqrt{s}) f_{e|e}(x_2,\sqrt{s}) 
d \hat{\sigma}(\hat{s} = x_1 x_2 s),
\ee
where $\hat{\sigma}$ is the cross section in the absence of ISR, and
\be \label{e_degk2} 
f_{e|e}(x,\sqrt{s}) = \beta \left[ (1-x)^{\beta-1} \left(1+
\frac{3}{4} \beta \right) - \frac{\beta}{2} (1+x) \right],
\ee
with $\beta = \frac {\alpha_{\rm em}} {\pi} \left( \log \frac{s}
{m_e^2} - 1 \right)$, is the leading--log resummed effective $e^\pm$
distribution function. Note that we do not include beamstrahlung
(which is expected to further smear out the peak in the \epem\
luminosity at $\hat{s} = s$), since it depends on details of
accelerator design. Moreover, we conservatively assume that all ISR
photons escape detection, even though eq.(\ref{e_degk2}) is strictly
valid only if there are no experimental constraints on the phase space
of the emitted photons. However, we will see that the main effect of
ISR is an overall reduction of the cross section by $\sim 15\%$;
kinematical distributions are little affected even if all ISR photons
are invisible.

We treat the emission of gluons during \qp\ production as described in
\cite{degk9}. In particular, we introduce a minimal gluon energy
$E_{g,{\rm min}}$ to regularize IR divergences. The final results will
not depend on the value of this regulator after contribution from \qp\
and $\qp g$ events have been added, if $E_{g,{\rm min}}$ is
sufficiently small.\footnote{Of course, virtual QCD corrections to
\qp\ production have to be added for this cancellation to
work.} In the numerical results presented below we take $E_{g,{\rm
min}}=1$ GeV. For $\sqrt{s}=800$ GeV and $\mq=300$ GeV this implies that
only 18\% of all squark pairs are produced together with a ``hard''
gluon.

The squared matrix element for $\tilde{q} \rightarrow q \tchi g$ can
be found in \cite{degk3}. We again have to include virtual QCD
corrections, in this case to $\tilde{q} \rightarrow q \tchi$ decays,
to cancel IR singularities. In this case we regularize these
singularities by introducing a finite gluon mass $m_g$, which we also
set to 1 GeV in our numerical examples. For our choice $\mq=300$ GeV,
$\mchi=50$ GeV this means that nearly 90\% of all squark decays
produce a ``hard'' gluon. Altogether more than 95\% of all $\epem
\rightarrow \qp \rightarrow q \bar{q} \tchi \tchi X$ events therefore
contain at least one ``hard'' gluon, as defined through our two IR
regulators. Note that, unlike for $\epem \rightarrow \qp$, virtual QCD
corrections to $\tilde{q} \rightarrow q \tchi$ decays introduce UV
divergences. These cancel only after including full SUSY--QCD
corrections \cite{degk3}, which depend on the mass of the gluino. We
take $m_{\tilde g} = 450$ GeV; the shapes of the kinematical
distributions we study here are almost independent of this choice.

Due to the emission of hard gluons we have up to 5 visible partons in
the final state\footnote{We allow gluon emission during the production
and both decays simultaneously, i.e. we include events with 2 or 3
gluons. These contributions are formally of NNLO. However, since
production and decays are independent processes, up to terms ${\cal
O}(\Gamma_{\tilde q}/\sqrt{s})$, other (as yet unknown) NNLO
contributions cannot cancel these known contributions.}. In the
numerical results presented below we group these into exactly 2 jets
(except for events with $<2$ partons in the acceptance region defined
by $|\cos(\theta)| \leq 0.9$, which we discard), using the $k_T$
clustering (``Durham'') algorithm \cite{degk10}. We have checked that
one obtains very similar results if one merges partons with a fixed
$y_{\rm cut}$ parameter, rather than a fixed number of final state
jets, and only uses the two hardest jets for the kinematical
analysis. In order to simulate experimental resolutions, we smear the
energies (but not directions) of all partons before jet merging, with
a Gaussian error given by $\delta(E) = 0.3/\sqrt{E} \oplus 0.01$ ($E$
in GeV). Finally, we apply a set of cuts in order to suppress
backgrounds. We require that the energy of each jet exceeds 15 GeV,
that the missing $p_T$ exceeds 56 GeV (taken from \cite{degk5} after
scaling up from $\sqrt{s}=500$ GeV to $800$ GeV), and that the
acoplanarity angle between the two jets exceed $30^\circ$.

In Figs. 1a,b we show the resulting distribution in the jet energy
(1a) and in the variable \msqmin\ (1b) defined below. In the absence
of cuts, radiative effects and energy smearing, the jet energy
distribution should be constant (flat) between two kinematical
endpoints, and zero elsewhere. The cuts distort this simple shape,
producing a small peak near the lower edge of the spectrum; this is
shown by the solid curve in Fig.~1a. The histograms show
``experimental'' distributions, based on 3,500 events before cuts; for
a single $q=2/3$ $SU(2)-$singlet ($\tilde{u}_R$) squark, this would
require an integrated luminosity of about 205 fb$^{-1}$ (including ISR
and QCD corrections, and assuming all squarks decay directly into $q
\tchi$); the total cross section is 17.0 (11.7) fb before (after)
cuts.\footnote{The total cross section for $\tilde{u}_L$ pair
production for the same mass is 26.8 fb. However, in most models one
expects $SU(2)$ doublet squarks to predominantly decay into charginos
and heavier neutralinos, rather than directly into the LSP.} The
dotted histogram still does not include any radiative effects, but
includes a finite energy resolution as described above; clearly this
has little effect on the shape of the spectrum. As mentioned earlier,
including ISR (dashed histogram) reduces the cross section, but again
does not distort the shape of the spectrum very much; in particular,
there is only little ``leakage'' beyond the nominal endpoints. In
contrast, including gluon emission (solid histogram) does change the
shape of the distribution. In particular, there are now quite a few
entries below the lower nominal endpoint. In most of these events one
of the quarks falls out of the acceptance region, so that one of the
jets is entirely made up of gluons. QCD corrections also increase the
total cross section before (after) cuts by 24\% (42\%)\footnote{Note
that QCD corrections markedly increase the acceptance of the cuts. It
is therefore also important to include their effect on the {\em
kinematics} when trying to extract \mq\ from measurements of the \qp\
production cross section.}

\begin{figure}[h]
\vspace*{-1cm}
\hspace*{-14mm}
\dofigs{2.7in}{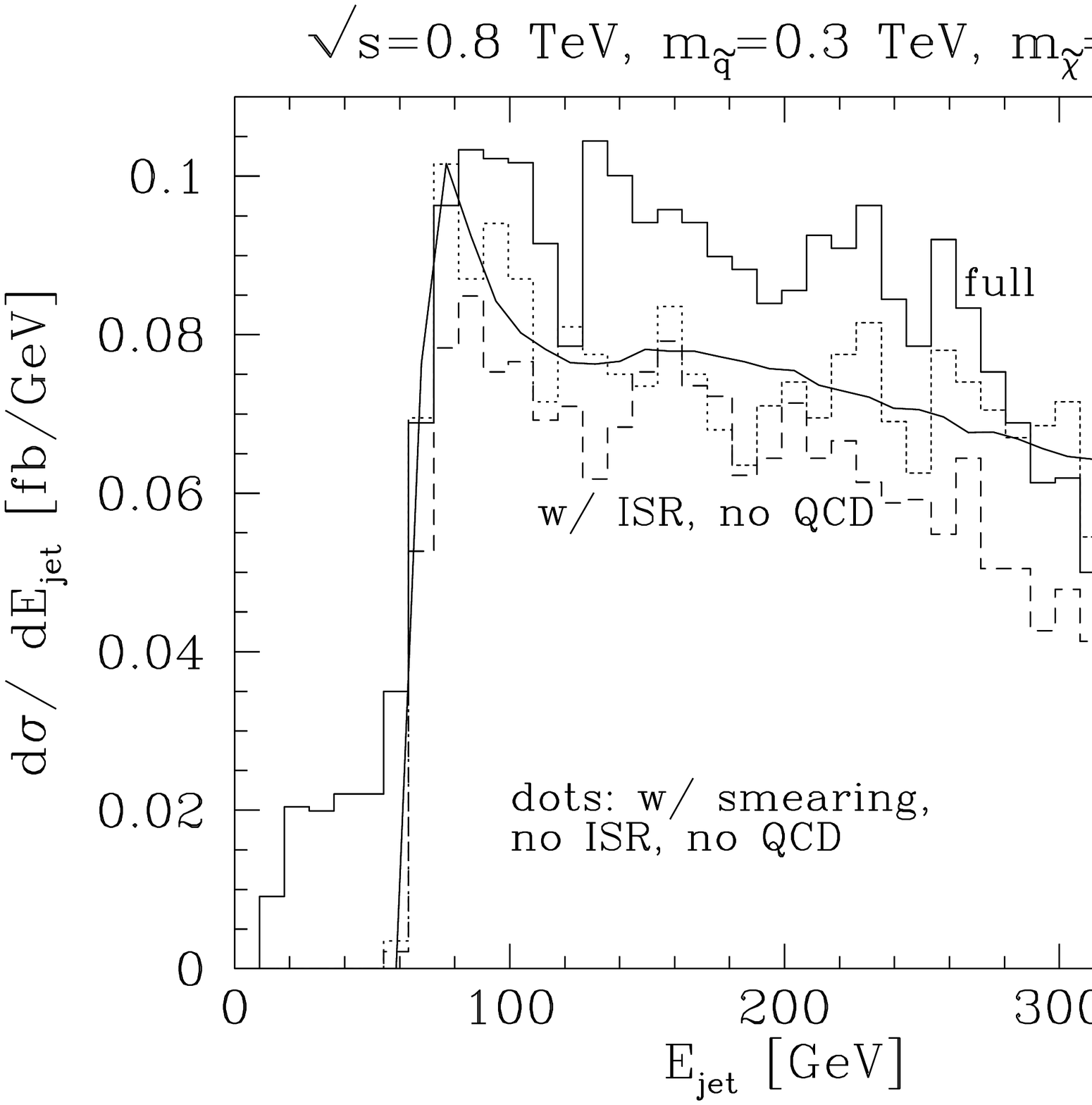}{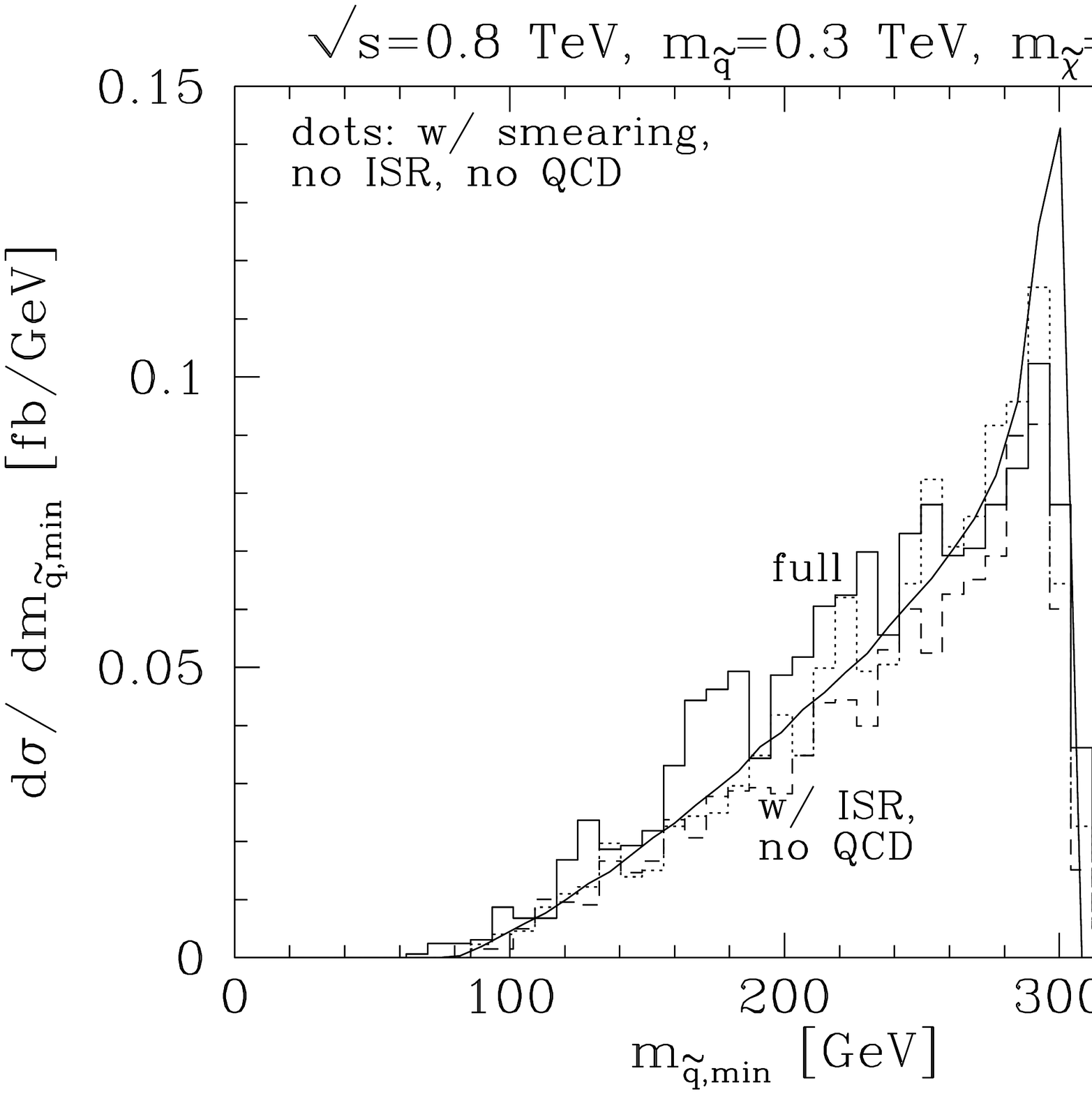}
\vspace*{-0.5cm}

\caption
{\small The jet energy (left) and \msqmin\ (right) distributions after
cuts. The solid curves show the ``ideal'' distributions in the absence
of energy smearing and radiative effects. In the dotted, dashed and
solid histograms, resolution smearing, ISR, and gluon radiation are
``switched on'' successively. See the text for further details.}
\end{figure}

The variable \msqmin\ plotted in Fig.~1b has been introduced in
\cite{degk5}. It is the minimal possible squark mass, {\em if} one
assumes a fixed value of $\sqrt{s}$ (no ISR), and {\em if} the final
state consists of exactly two massless quarks and two neutralinos with
equal (and known) mass.\footnote{The neutralino mass is expected to be
known from analyses of data at lower energy, e.g. from chargino pair
events.} None of these assumptions hold in our case. In fact, in some
cases the reconstruction described in \cite{degk5} is impossible,
i.e. some trigonometrical function acquires a value exceeding unity; we
discard these events. We nevertheless find that radiative effects
introduce only a modest amount of ``leakage'' beyond the nominal
endpoint of the distribution, which is at \mq. Note, however, that
these effects significantly broaden the distribution, i.e. many events
have migrated to lower values of \msqmin\ (compare the solid and dotted
histograms). Since events in the peak contribute most to the
determination of \mq\ \cite{degk5}, this broadening is expected to
increase the error on \mq.

Fig.~2 shows that a fit of the \msqmin\ distribution nevertheless
yields a smaller statistical error for \mq\ than a fit of the $E_{\rm
jet}$ distribution does. In this figure we compare a mock data set for
$\mq=300$ GeV, based on an integrated luminosity of just 50 fb$^{-1}$
(852 events before cuts), with ``template'' distributions, which have
been computed for 13 different values of \mq, leaving all other input
parameters unchanged, and applying the cuts described above. This
gives 13 different values of $\chi^2$; the open (filled) squares have
been computed from the $E_{\rm jet}$ (\msqmin) distributions. A
parabola is then fitted to $\chi^2(\mq)$. The minimum of this parabola
gives the ``measured'' value $m_{\tilde{q},0}$ of \mq, while the
($1\sigma$) error is computed from $\chi^2(m_{\tilde{q},0}\pm \delta
\mq) = \chi^2_{\rm min} + 1$. Note that the $\chi^2$ values are only
based on the {\em shapes} of the distributions, i.e. the ``data'' have
been normalized to the template before computing $\chi^2$ for a given
assumed value of \mq. Fig.~2 therefore shows the results of purely
kinematical determinations of \mq.

\begin{figure}[htb]
\vspace*{-.5cm}
\begin{center}
\mbox{
\epsfig{file=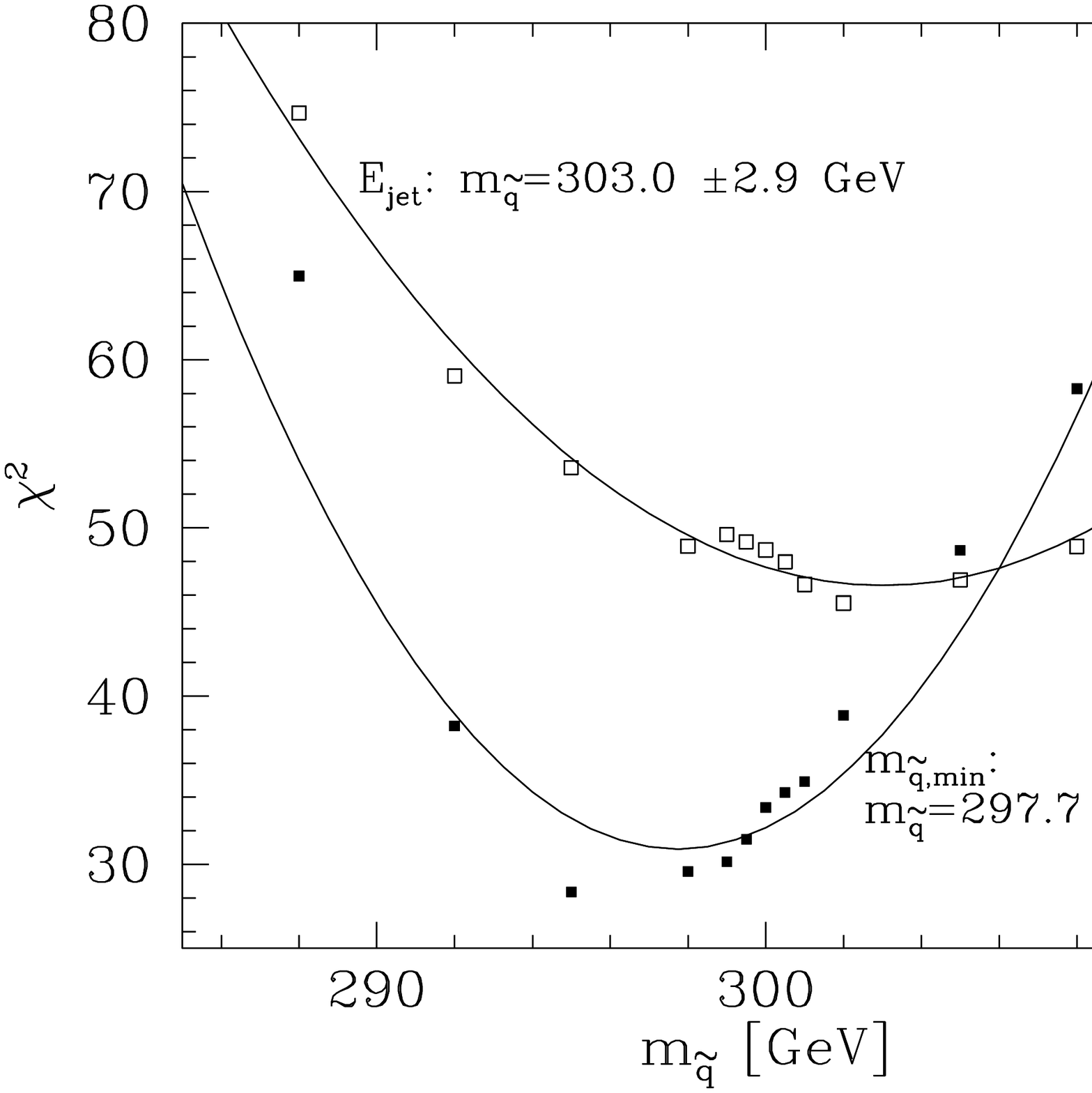,height=10cm} }
\end{center}
\vspace*{-0.5cm}

\caption
{\small
The $\chi^2$ values computed by comparing a mock data set with 13
different template distributions. The ``data'' are for $\mq=300$ GeV,
while the templates are for values of \mq\ between 288 and 312
GeV. The open (full) boxes are for the $E_{\rm jet}$ (\msqmin)
distributions, and the curves are parabolic fits to the corresponding
functions $\chi^2(\mq)$. These parabolas determine the ``measured''
\mq, including its error.}
\end{figure}

We see that even after radiative effects have been included, a fit of
the \msqmin\ distribution determines the squark mass with a
statistical error of well under 1\% already with 50 fb$^{-1}$ of data,
which corresponds to only about 1 month of running time for the
planned TESLA collider \cite{degk11}. However, radiative effects also
introduce possible new systematic errors, e.g. due to the choice of
scale in $\alpha_s$. (In our calculation we used $\sqrt{\hat{s}}$ for
the corrections to \qp\ production, and \mq\ in $\tilde{q}$ decays.)
Moreover, we haven't included any hadronization effects yet. Note that
the lighter scalar top eigenstate $\tilde{t}_1$, which is likely to be
the lightest, and hence most easily accessible squark, might well
hadronise before it decays \cite{degk9}. In addition, the massless
quarks and gluons in the final state will hadronise into jets with
finite masses. Finally, the error on the assumed LSP mass also
propagates into the error on \mq\ \cite{degk5}. We plan to investigate
these issues in a future publication.


\subsection*{Acknowledgements}
This work was supported in part by Conselho Nacional de
Desenvolvimento Cient\'{\i}fico e Tecnol\'ogico (CNPq), by
Funda\c{c}\~ao de Amparo \`a Pesquisa do Estado de S\~ao Paulo
(FAPESP), by Programa de Apoio a N\'ucleos de Excel\^encia
(PRONEX), by ``Fonds zur F\"orderung der wissenschaftlichen 
Forschung'' of Austria, project no. P13139--PHY, by the Department of 
Science and Technology (India), and by the US National Science Foundation 
under NSF grant INT-9602567.


\setcounter{figure}{0}
\setcounter{table}{0}
\setcounter{section}{0}
\setcounter{equation}{0}
\setcounter{footnote}{0}
\clearpage

\def\Journal#1#2#3#4{{#1} {\bf #2}, #3 (#4)}

\begin{center}

{\large \bf Spin correlations and phases for SUSY particle searches
 at $e^+e^-$ colliders } \\
\vspace*{3mm}

N. GHODBANE \\
\vspace*{3mm}
\end{center}

\section{Introduction}
Since one expects high luminosities for the next generation of linear
colliders (e.g. $\sim 500 fb^{-1}$ for the TESLA project), one can use
beam polarization to reduce the standard model backgrounds and use the
polarization dependence of the cross sections to study specific SUSY
parameters.  Moreover, as it has been stressed by several authors
\cite{gudi:spincorr}, spin correlations play a major role in the
kinematic distributions of final particles. For this reason, we
upgraded the \texttt{SUSYGEN} Monte Carlo generator\footnote{A
description of the Monte Carlo generator can be found in the Higgs
working group report.} \cite{nous:susygen3}.

Here, we firstly study the spin correlations effects in the gaugino
and the stau searches. Then, in a second part, we show how the newly
reconsidered CP violating phases arising from MSSM can affect the
chargino searches.

\section{Beam polarization and spin correlations}

As it has been stressed by several papers \cite{gudi:spincorr} the
study of the angular distribution of $e^\pm$ produced in
$e^+e^-\to\tilde{\chi}^0_2\tilde{\chi}^0_1\to \tilde{\chi}^0_1
\tilde{\chi}^0_1 e^+ e^- $ will give valuable information concerning
the neutralino nature and then enable MSSM parameter extraction. Here
we consider the two scenarios ({\bf{A}}: $M_2=78$ GeV$/c^2$,
$\mu=-250$ GeV$/c^2$, $\tan\beta=2$) and ({\bf{B}}: $M_2=210$
GeV$/c^2$, $\mu=-60$ GeV$/c^2$, $\tan\beta=2$). In model {\bf{A}},
$\tilde{\chi}^0_1$ is Bino like ($m_{\tilde{\chi}^0_1}=42.5 \mbox{
GeV}/c^2, \tilde{\chi}^0_1=+0.98\tilde{B} + 0.17\tilde{W}^3 -
0.09\tilde{H}^0_1 + 0.04\tilde{H}^0_2$) whereas $\tilde{\chi}^0_2$ is
Wino like ($m_{\tilde{\chi}^0_2}= 91.9 \mbox{ GeV }/c^2,
\tilde{\chi}^0_2 = +0.14\tilde{B} - 0.95\tilde{W}^3 -
0.28\tilde{H}^0_1 + 0.05\tilde{H}^0_2$). In model {\bf{B}}, the two
lightest neutralinos are Higgsino like $(m_{\tilde{\chi}^0_1} =
55.1\mbox{ GeV}/c^2, \tilde{\chi}^0_1 = +0.16\tilde{B} -
0.09\tilde{W}^3 - 0.78\tilde{H}^0_1 - 0.60\tilde{H}^0_2)$ and
$(m_{\tilde{\chi}^0_2}=88.9\mbox{ GeV }/c^2, \tilde{\chi}^0_2 =
+0.21\tilde{B} - 0.24\tilde{W}^3 + 0.62\tilde{H}^0_1 -
0.71\tilde{H}^0_2)$. For each one of these two models, we considered
the common sfermions mass at the GUT scale, $m_0$, being equal to 80
GeV$/c^2$ and 200 GeV/$c^2$.

The right side of figure \ref{fig:beampolcorr} illustrates the effect
of spin-correlations in angular distributions of $e^-$, decay product
of $\tilde{\chi}^0_2$. One can notice that the angular distribution of
the final leptons depends strongly on the sfermion mass parameter
$m_0$. For small $m_0$ of inclusion of spin correlations gives an
effect $\sim 20 \%$ in the angular distributions. For higher selectron
masses, the total cross section is smaller, but the spin correlation
effects appear to be more important $\sim 30\%$. If the two
neutralinos are Higgsino like, the effects are negligible ($\sim
0.7\%$) (see figure \ref{fig:beampolcorr2}). Moreover these angular
distributions depend strongly on the center of mass energy.
\begin{center}
\vskip -0.5cm
\begin{figure}[h!]
\includegraphics[width=7cm,height=7cm]{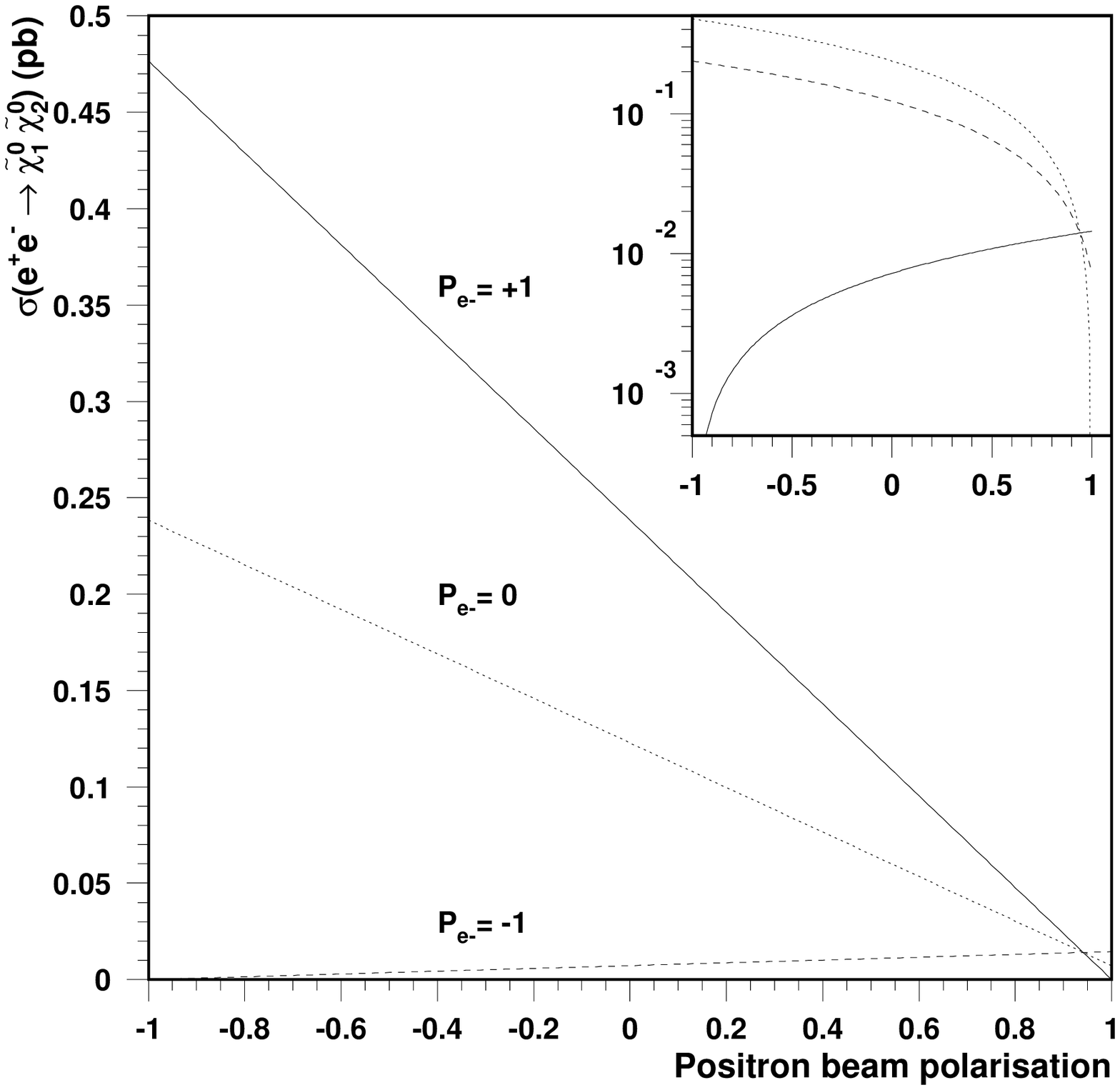}
\includegraphics[width=7cm,height=7cm]{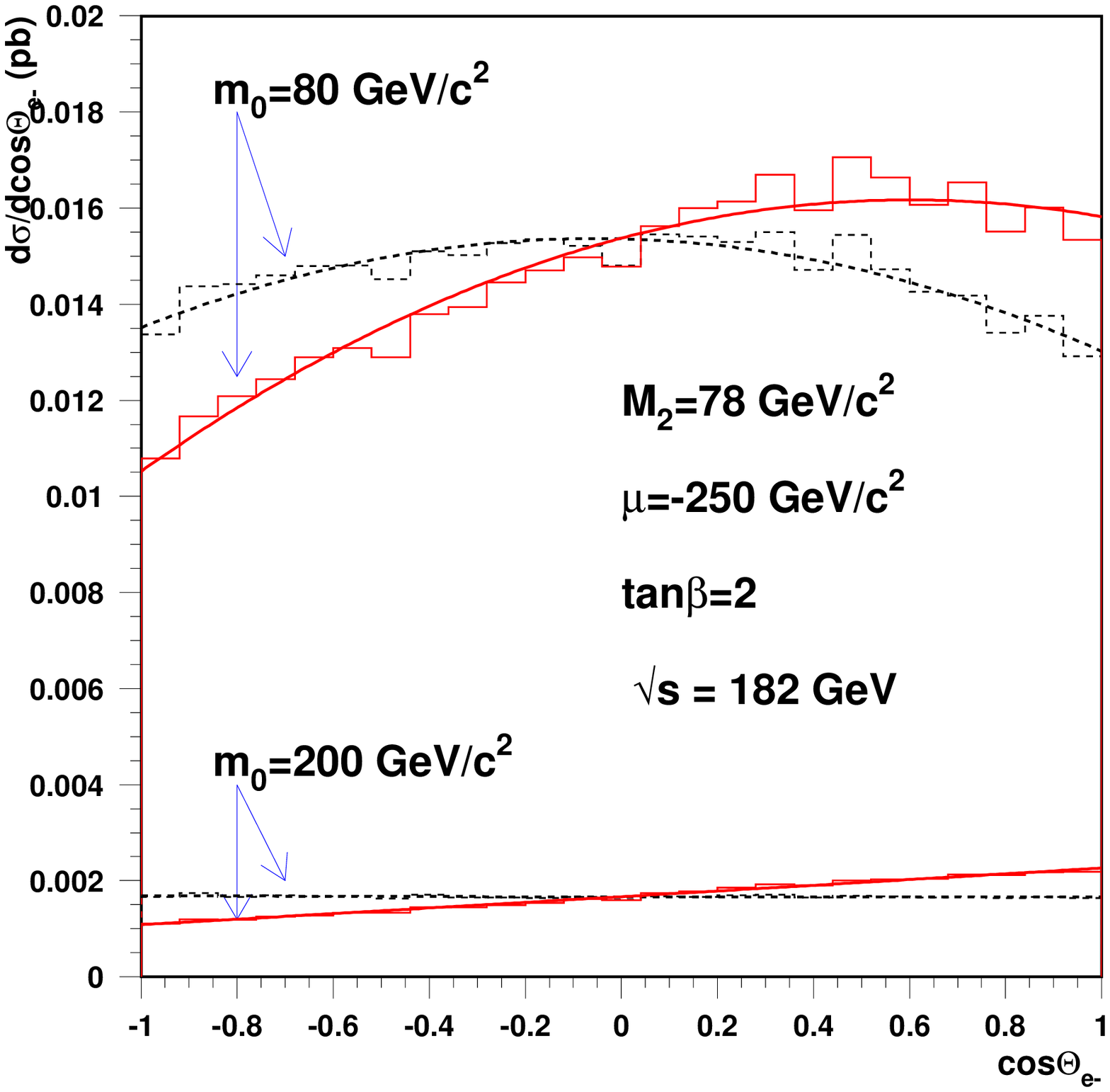}
\vskip -0.5cm
\caption{The figure on the left side shows the evolution of the cross
section associated to the production of $\tilde{\chi}^0_1
\tilde{\chi}^0_2$ for different beam polarizations. The figure on the
right side shows the $d\sigma / d\cos\theta$ distribution for the
final $e^-$ decay product of the $\tilde{\chi}^0_2$ with (solid line)
and without spin correlations (dashed line) for two assumptions on the
$m_0$ parameter (scenario {\bf{A}}).}
\label{fig:beampolcorr}
\end{figure}

\begin{figure}[h!]
\vskip -0.5cm
\includegraphics[width=7cm,height=7cm]{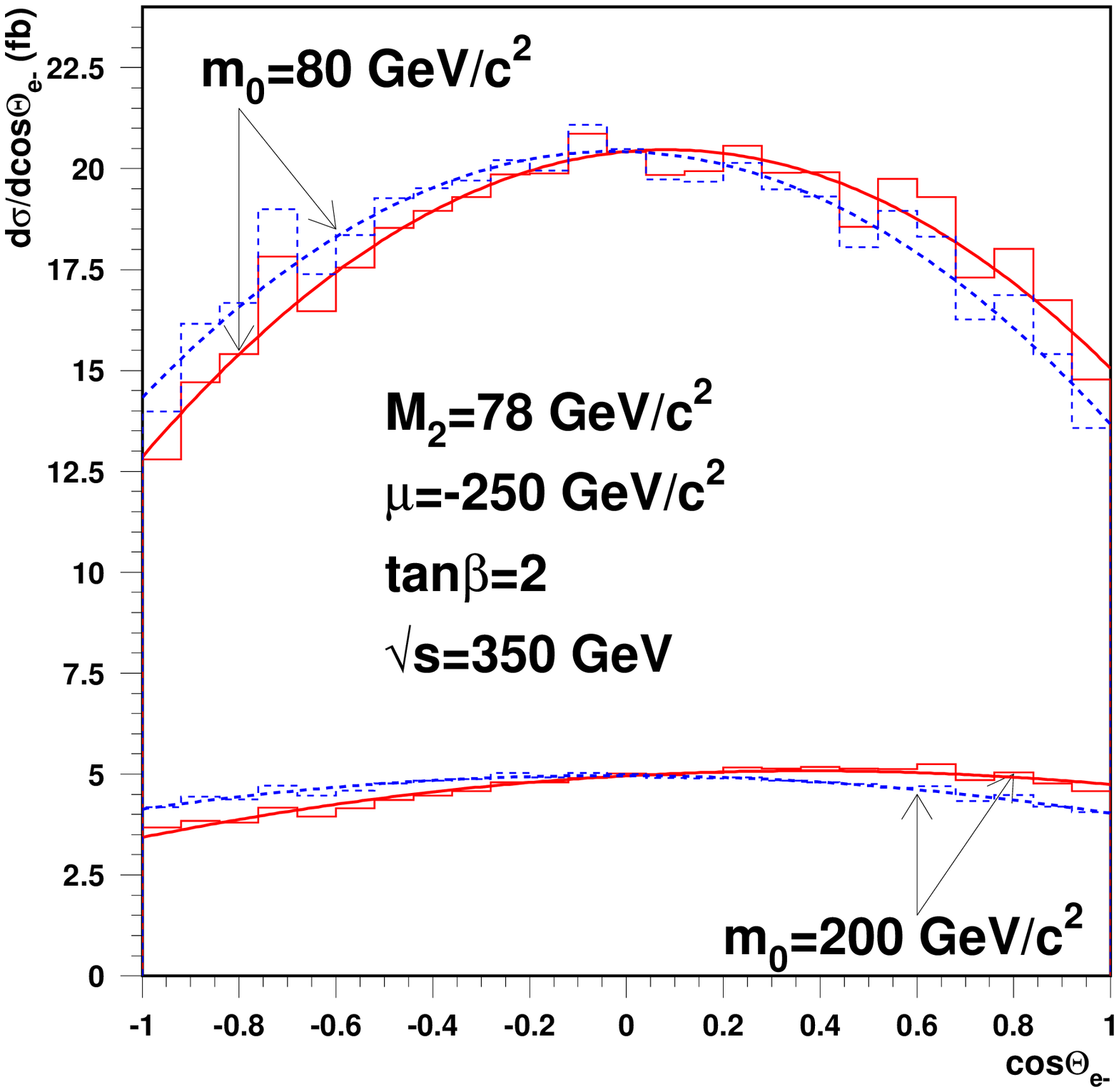}
\includegraphics[width=7cm,height=7cm]{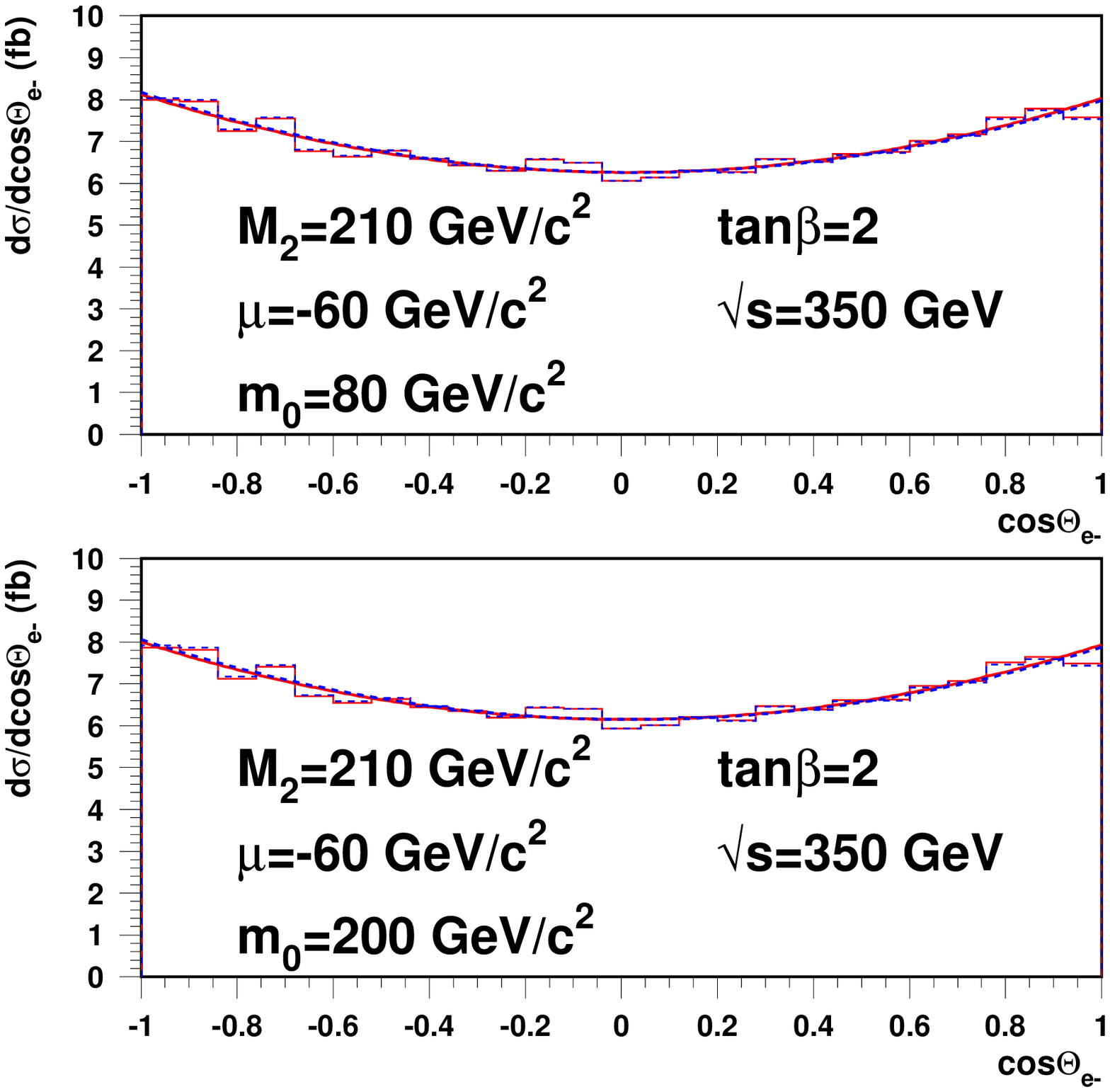}
\vskip -0.5cm
\caption{The figure on the left side shows the $d\sigma / d\cos\theta$
distribution for the final $e^-$ decay product of the
$\tilde{\chi}^0_2$ with (solid line) and without spin correlations
(dashed line) for two assumptions on the $m_0$ parameter assuming the
neutralino being bino like (scenario {\bf{A}}). The Higgsino like
case (scenario {\bf{B}} and figure on the right) shows no strong
dependence on the spin correlations.}
\label{fig:beampolcorr2}
\end{figure}
\end{center}

Recent studies \cite{Nojiri:stau} have shown that $\tau$ polarization
effects yield valuable information for the MSSM parameters e.g for
$\tan\beta$, the nature of the LSP and the mixing angle
$\theta_{\tilde{\tau}}$.

\begin{figure}[h!]
\begin{center}
\vskip -0.5cm
\includegraphics[width=7cm,height=7cm]{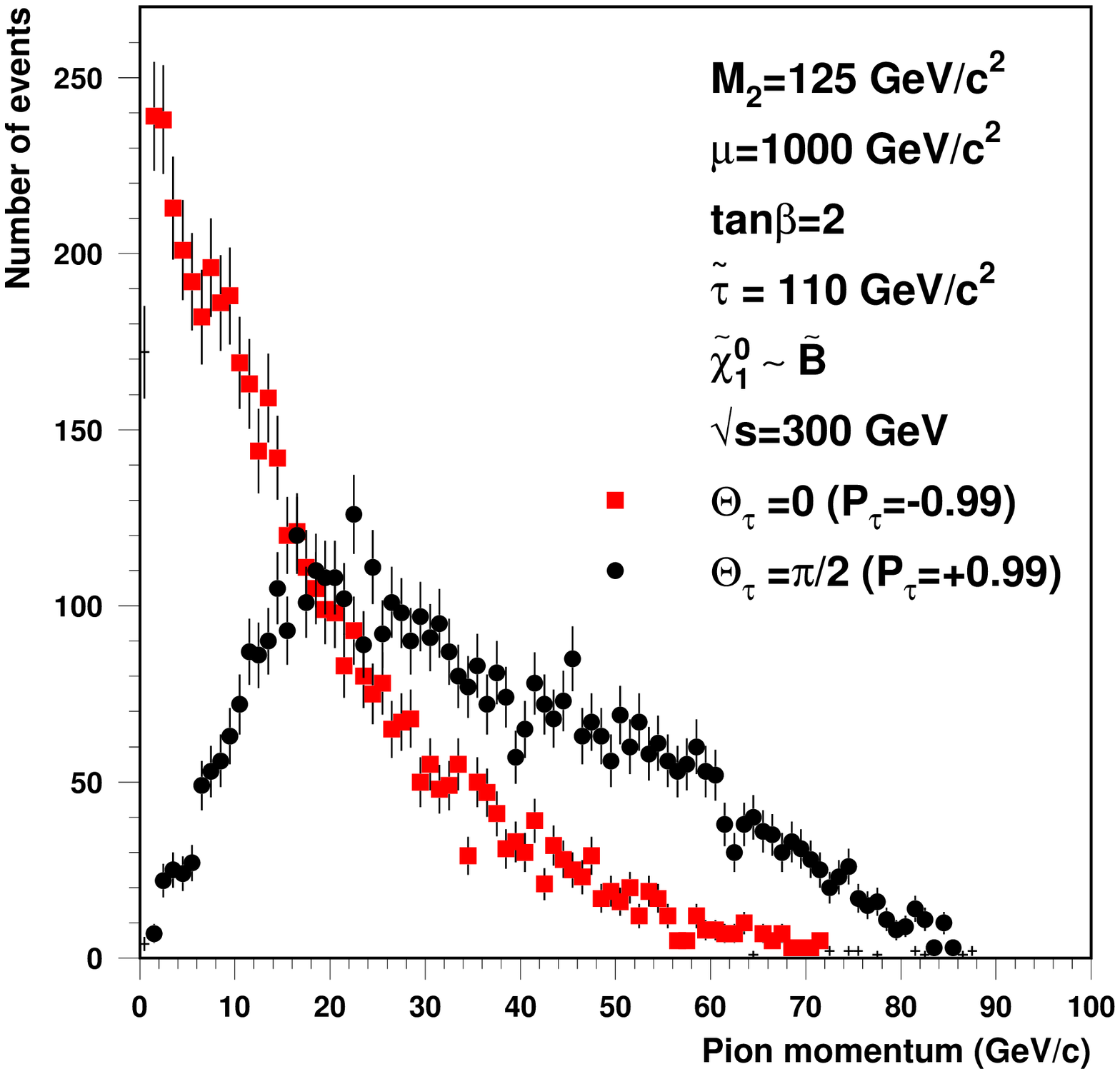}
\includegraphics[width=7cm,height=7cm]{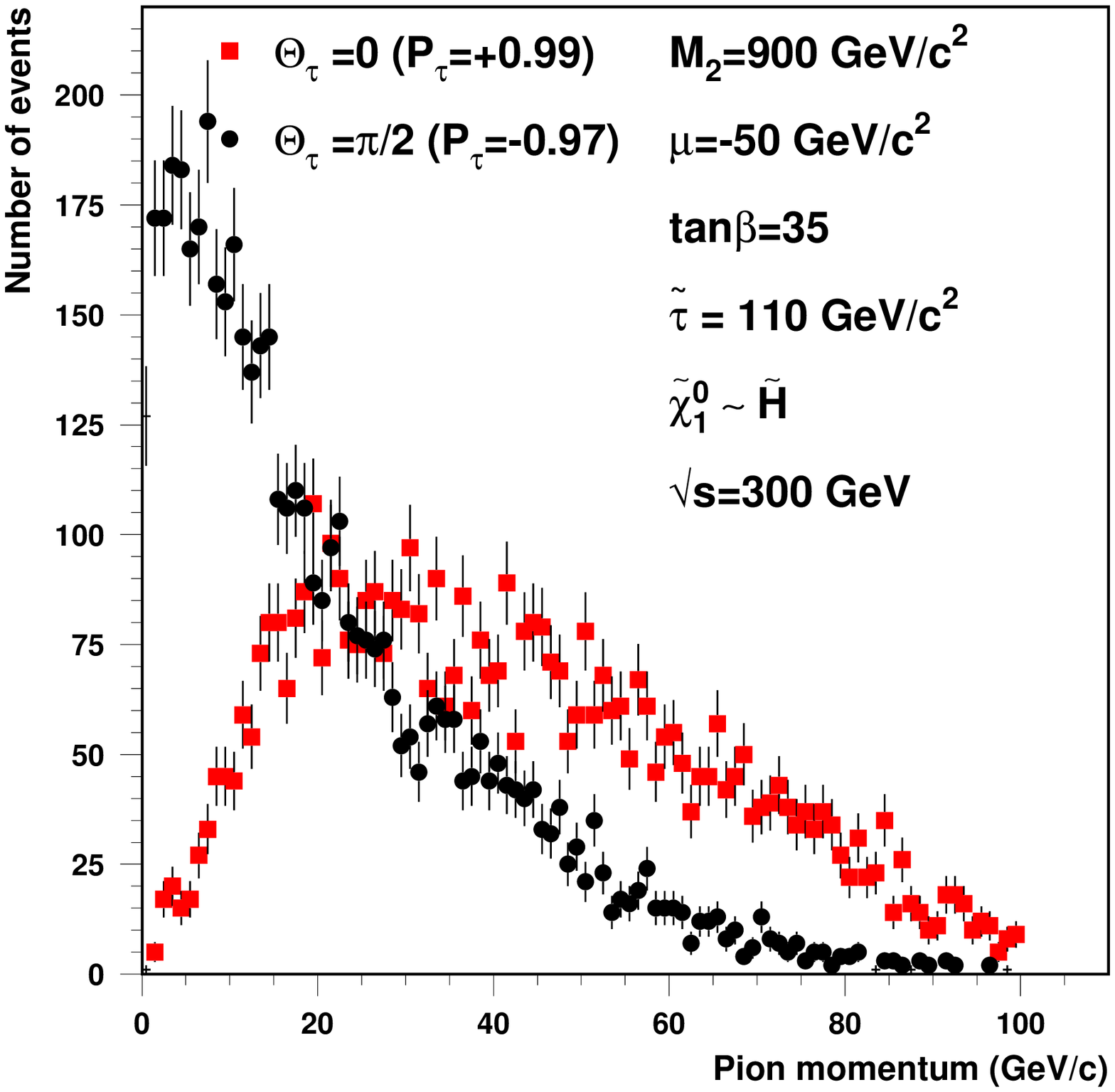}
\vskip -0.5cm
\caption{Momentum distributions of $\pi$, decay product of $\tau$
produced in the process $e^+e^- \to \tilde{\tau}_1^+ \tilde{\tau}_1^-
\to \tau^+ \tilde{\chi}^0_1 \tau^- \tilde{\chi}^0_1$. The
distributions have been plotted for two hypotheses concerning the stau
mixing angle $\theta_{\tilde{\tau}}$ ($\tilde{\tau}_L$ and
$\tilde{\tau}_R$) and for two hypotheses concerning the neutralino
$\tilde{\chi}^0_1$ nature.}
\label{fig:pionmomentum}
\end{center}
\end{figure}

Figure \ref{fig:pionmomentum} shows the momentum distribution of pions
produced by $e^+e^- \to \tilde{\tau}_1^+ \tilde{\tau}_1^- \to \tau^+
\tilde{\chi}^0_1 \tau^- \tilde{\chi}^0_1 \to \pi^+ \nu_\tau
\tilde{\chi}^0_1 \pi^- \bar{\nu}_\tau \tilde{\chi}^0_1$. The
distributions have been plotted assuming two scenarios for the stau
mixing angle $\Theta_{\tilde{\tau}}=0$ and $\pi$, and
$\tilde{\chi}^0_1$ nature (Bino, Higgsino). One can see that the final
particle distributions will give access to the tau polarization
${\mathcal{P}}_\tau$, to $\tan\beta$, and to $\tilde{\chi}^0_1$ nature
and through them, to the MSSM parameters.

\section{Phases in supersymmetry searches}

In the MSSM, there are new potential sources of CP non
conservation\cite{poko:phases}. Complex CP violating phases can arise
from several parameters present in the MSSM Lagrangian: the Higgs
mixing mass parameter $\mu$, the gauginos masses $M_i$, the trilinear
couplings $A_i$.  Experimental constraints on these CP violating
phases come from the electric dipole moment of the electron and the
neutron \cite{EDM_E_EXP}.

\begin{figure}[h!]
\vskip -0.5cm
\includegraphics[width=7cm,height=7cm]{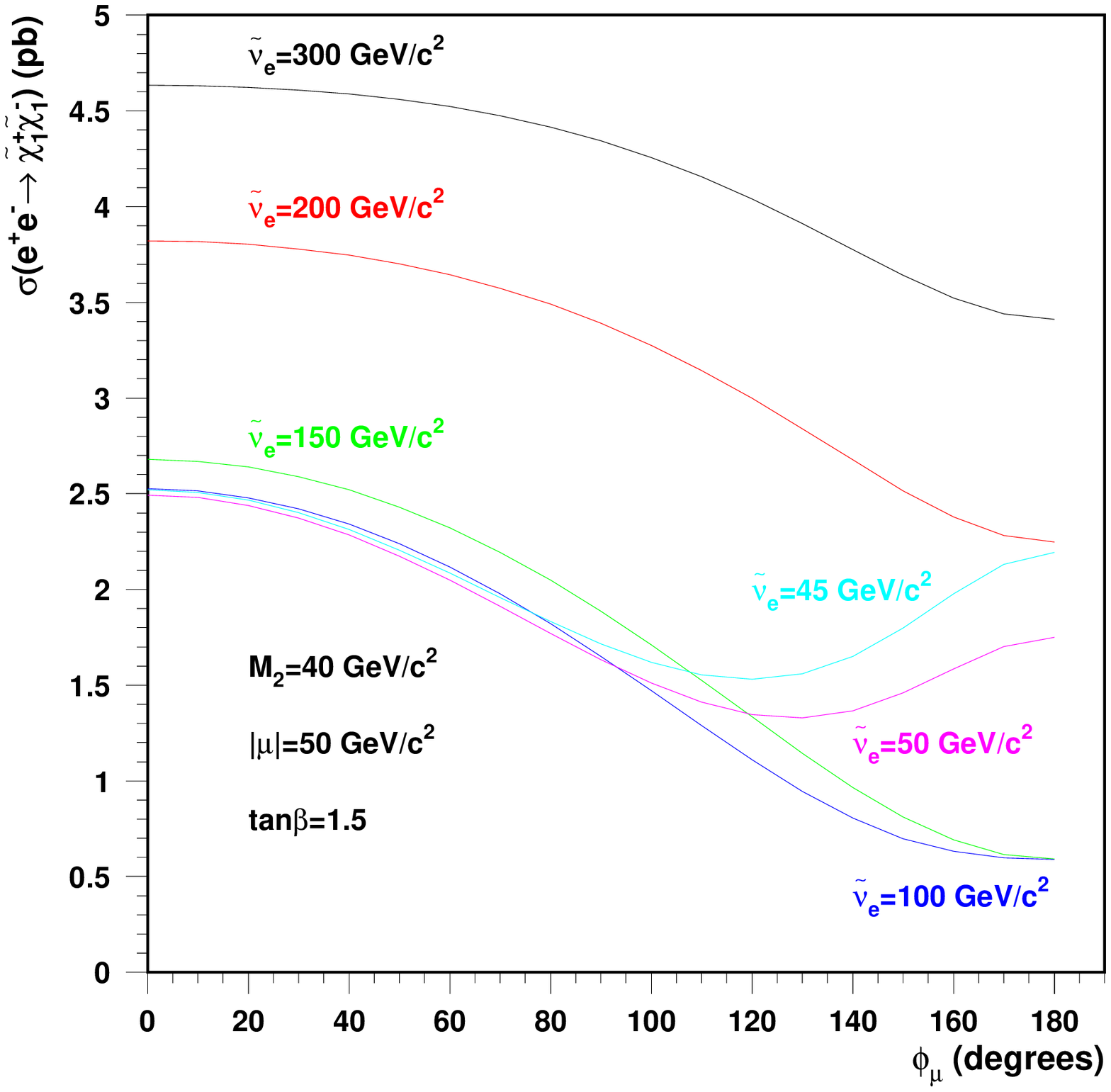}
\includegraphics[width=7cm,height=7cm]{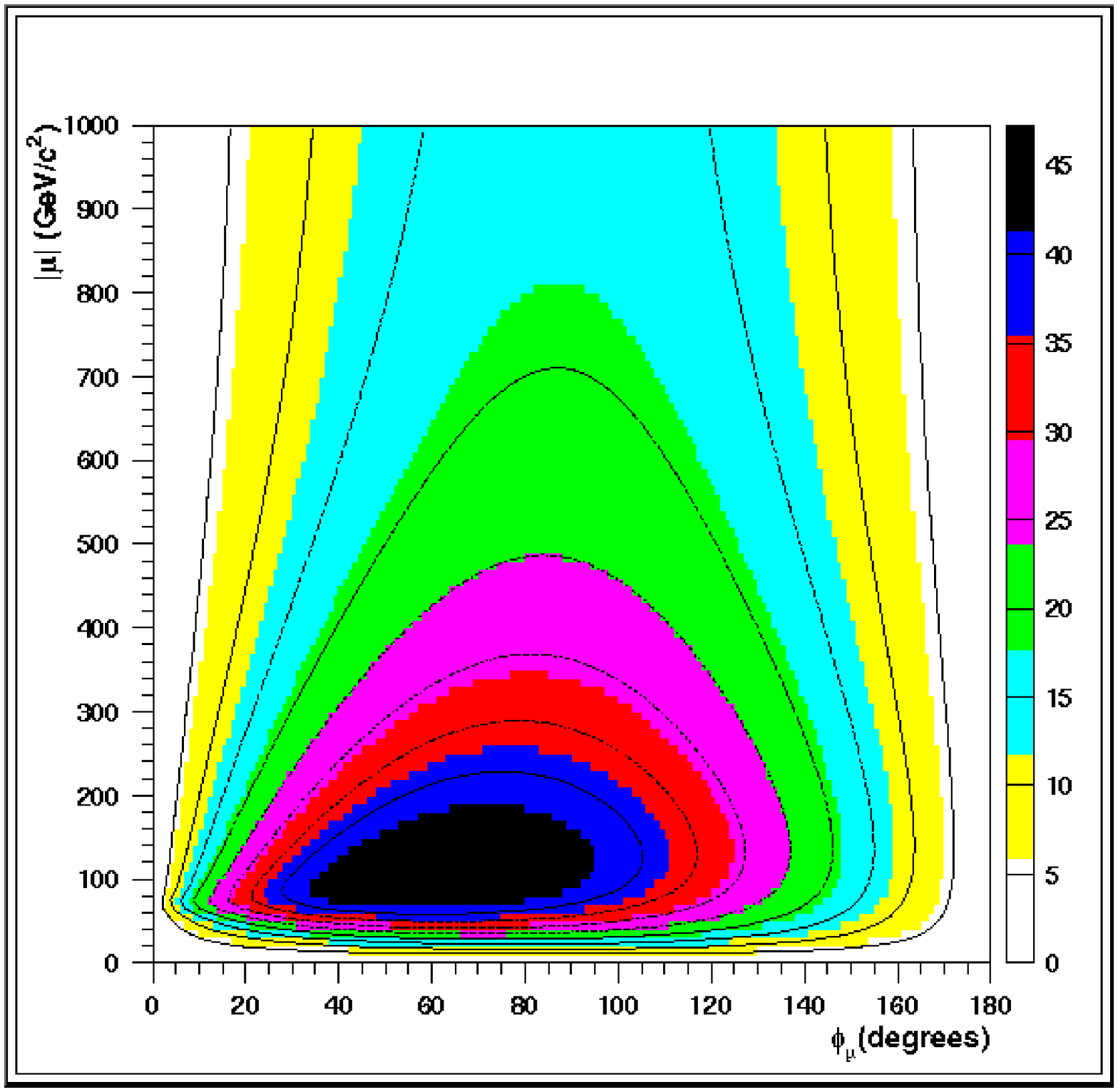}
\vskip -0.5cm
\caption{The figure on the left side shows the evolution of the cross section for $e^+e^-\to \tilde{\chi}^+_1 \tilde{\chi}^-_1$ in terms of the $\phi_\mu$ phase ($\mu=|\mu|e^{i\phi_\mu}$) for a center of mass energy of 189 GeV. The figure on the right side shows the evolution of the dipole electric moment divided by the experimental upper limit ($E_e^{exp}<4.3\cdot 10^{-27} e\cdot cm $) in terms of $\mu=|\mu|e^{i\phi_\mu}$ parameter.}
\label{fig:phases}
\end{figure}

Figure \ref{fig:phases} (left side) shows the chargino pair production
cross section variation in terms of $\phi_\mu$, the phase associated
to the $\mu$ parameter, for several values of the sneutrino mass
$m_{\tilde{\nu}_e}$ and for a value of $\tan\beta=1.5$. One sees that
there is a local minimum between the two extreme values
($\Phi_{\mu}=0,\pi$), which are tested at LEP searches ($\mu>0,
\mu<0$), but also that it is not so deep as to raise doubts on the
exhaustiveness of ``phaseless'' searches. Further the electric dipole
moment experimental upper limit ($E_e^{exp}<4.3\cdot 10^{-27} e\cdot
cm$) \cite{EDM_E_EXP} will constrain these phases (figure on the right
side) and rule out many scenarios for which the smallest cross section
for chargino pair production is obtained for a $\phi_\mu$ parameter
different from $0$ and $\pi$.

At the linear collider, one cannot neglect this strong dependence of
the cross section on phases.

\setcounter{figure}{0}
\setcounter{table}{0}
\setcounter{section}{0}
\setcounter{equation}{0}
\setcounter{footnote}{0}
\clearpage

\def\chione{\tilde \chi_1^0}
\def\chionepm{\tilde \chi_1^\pm}
\def\snm{{\tilde\nu_\mu}}
\def \susyq {supersymmetric }
\def\l {\lambda }

%

\begin{center}

{\large \bf The three-leptons signature from resonant sneutrino
  production at the LHC} \\
\vspace*{3mm}

{\sc G. MOREAU, E. PEREZ and G. POLESELLO} \\

\end{center}

\vspace*{3mm}
\begin{abstract}
The resonant production of sneutrinos at the LHC via the
R-parity violating couplings $\l ' _{ijk} L_i Q_j D^c_k$
is studied through its three-leptons signature.
A detailed particle level study of signal and background
is performed using a fast simulation of the ATLAS detector.
Through the full reconstruction of
the cascade decay, a model-independent and precise measurement
of the masses of the involved sparticles can be performed.
Besides, this signature can be detected in a large part of the SUSY
parameter space and for wide ranges of values of several $\l ' _{ijk}$
coupling constants.
\end{abstract}

\section{Introduction}

In extensions of the Minimal Supersymmetric Standard Model (MSSM)
where the so-called R-parity symmetry is violated, the superpotential
contains some additional trilinear couplings which offer the
opportunity to singly produce \susyq (SUSY) particles as resonances.
The analysis of resonant SUSY particle production allows an easier
determination of the these R-parity violating (\rpv) couplings than
the displaced vertex analysis for the Lightest Supersymmetric Particle
(LSP) decay, which is difficult experimentally especially at hadronic
colliders.

In this paper, we study the sensitivity provided by the ATLAS detector
at the LHC on singly produced charginos via the $\l'_{211}$ coupling,
the main contribution coming from the resonant process $ p p \to
\tilde \nu_{\mu} \to \tilde \chi^{\pm}_1 \mu^{\mp}$.  At hadron
colliders, due to the continuous energy distribution of the colliding
partons, the resonance can be probed over a wide mass range.  We have
chosen to concentrate on $\l ' _{ijk} L_i Q_j D^c_k$ interactions
since $\l '' _{ijk} U_i^cD_j^cD_k^c $ couplings lead to multijet final
states with large QCD background.  Besides, we focus on $\l'_{211}$
since it corresponds to first generation quarks for the colliding
partons and it is not severely constrained by low energy experiments:
$\lambda_{211}^{\prime} < 0.09$ (for $\tilde m= 100$~GeV)
\cite{Drein}.  We consider the cascade decay leading to the
three-leptons signature, namely $\tilde \chi^{\pm}_1 \to \tilde
\chi^0_1 l^{\pm}_p \nu_p$ (with $l_p=e,\mu$), $\tilde \chi^0_1 \to \mu
u \bar d, \ \bar \mu \bar u d$.  The main motivation lies in the low
Standard Model background for this three-leptons final state.  The
considered branching ratios are typically of order $B(\tilde
\chi^{\pm}_1 \to \tilde \chi^0_1 l^{\pm}_p \nu_p) \approx 22\%$ (for
$m_{\tilde l},m_{\tilde q},m_{\tilde \chi^0_2}>m_{\tilde
\chi^{\pm}_1}$) and $B(\tilde \chi^0_1 \to \mu u d) \sim 40\% - \sim
70\%$.

\section{Mass reconstruction}
\label{secana}

The clean final state, with only two hadronic jets, three leptons and
a neutrino allows the reconstruction of the $\tilde\nu$ decay chain
and the measurement of the $\tilde \chi^0_1$, $\tilde \chi^{\pm}_1$
and $\tilde \nu_{\mu}$ masses. We perform full analysis for the
following point of the MSSM: $M_1=75$~GeV, $M_2=150$~GeV,
$\mu=-200$~GeV, $\tan \beta=1.5$, $A_t=A_b=A_{\tau}=0$, $m_{\tilde
f}=300$~GeV and for $\lambda_{211}^{\prime}$=0.09.  For this set of
MSSM parameters, the mass spectrum is: $m_{\chione}=79.9$~GeV,
$m_{\tilde \chi_1^{\pm}}=162.3$~GeV and the total cross-section for
the three-leptons production is 3.1~pb, corresponding to $\sim 100000$
events for the standard integrated luminosity of 30~fb$^{-1}$ expected
within the first three years of LHC data taking.

The single chargino production has been calculated analytically and
implemented in a version of the SUSYGEN MonteCarlo \cite{susygen}
modified to include the generation of $pp$ processes.  The generated
signal events were processed through the program ATLFAST
\cite{ATLFAST}, a parameterized simulation of the ATLAS detector
response.

First, we impose the following loose selection cuts in order to select
the considered final state and to reduce the Standard Model (SM)
background (see Section \ref{secback}): (a) Exactly three isolated
leptons with $p_T^1>20$~GeV, $p_T^{2,3}>10$~GeV and $|\eta|<2.5$, (b)
At least two of the three leptons must be muons, (c) Exactly two jets
with $p_T>15$~GeV, (d) The invariant mass of any $\mu^+\mu^-$ pair
must lie outside $\pm6.5$~GeV of the $Z$ mass.

\begin{figure}[htb]
\begin{center}
\vspace*{-8mm}
\dofigs{0.5\textwidth}{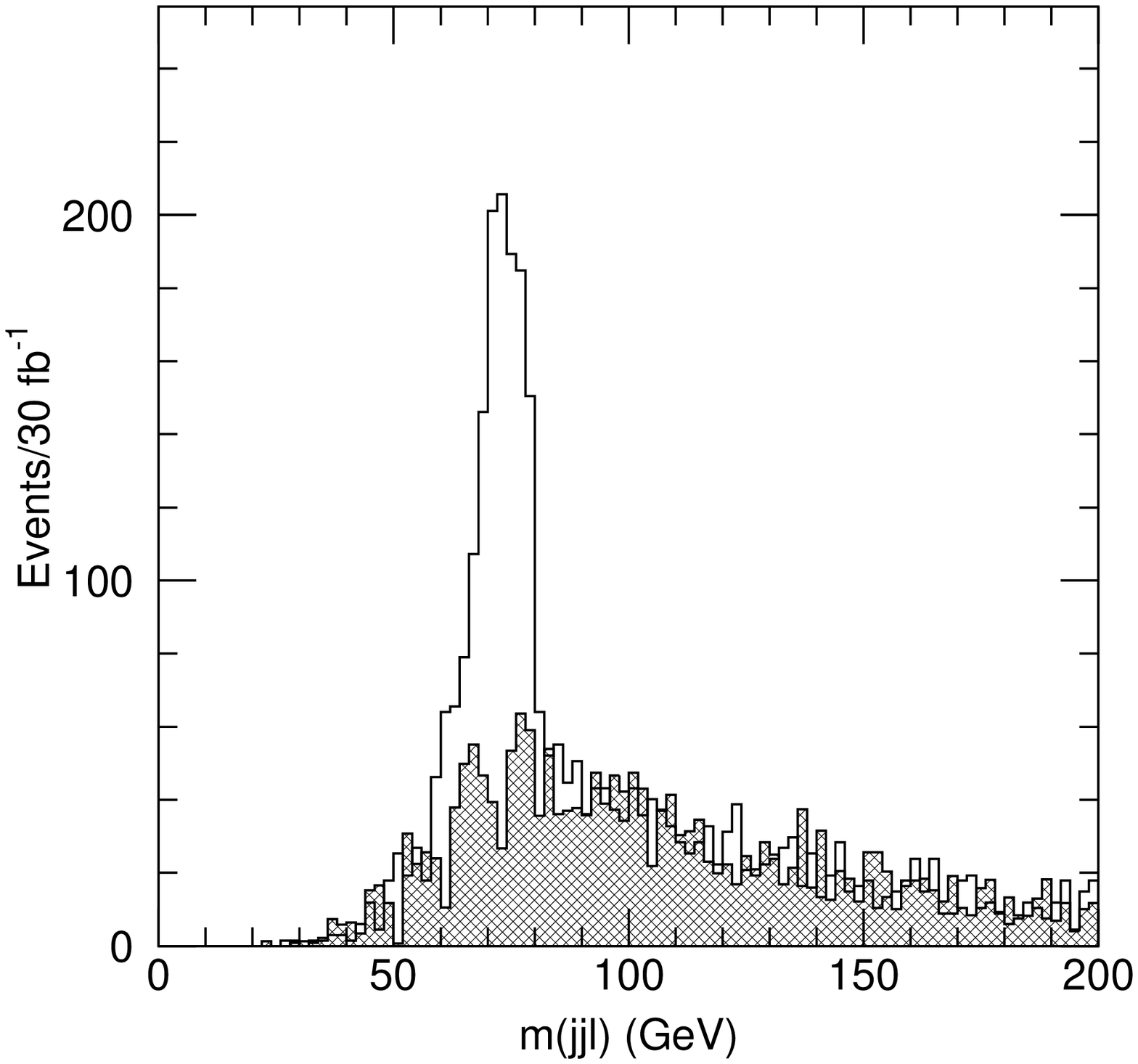}{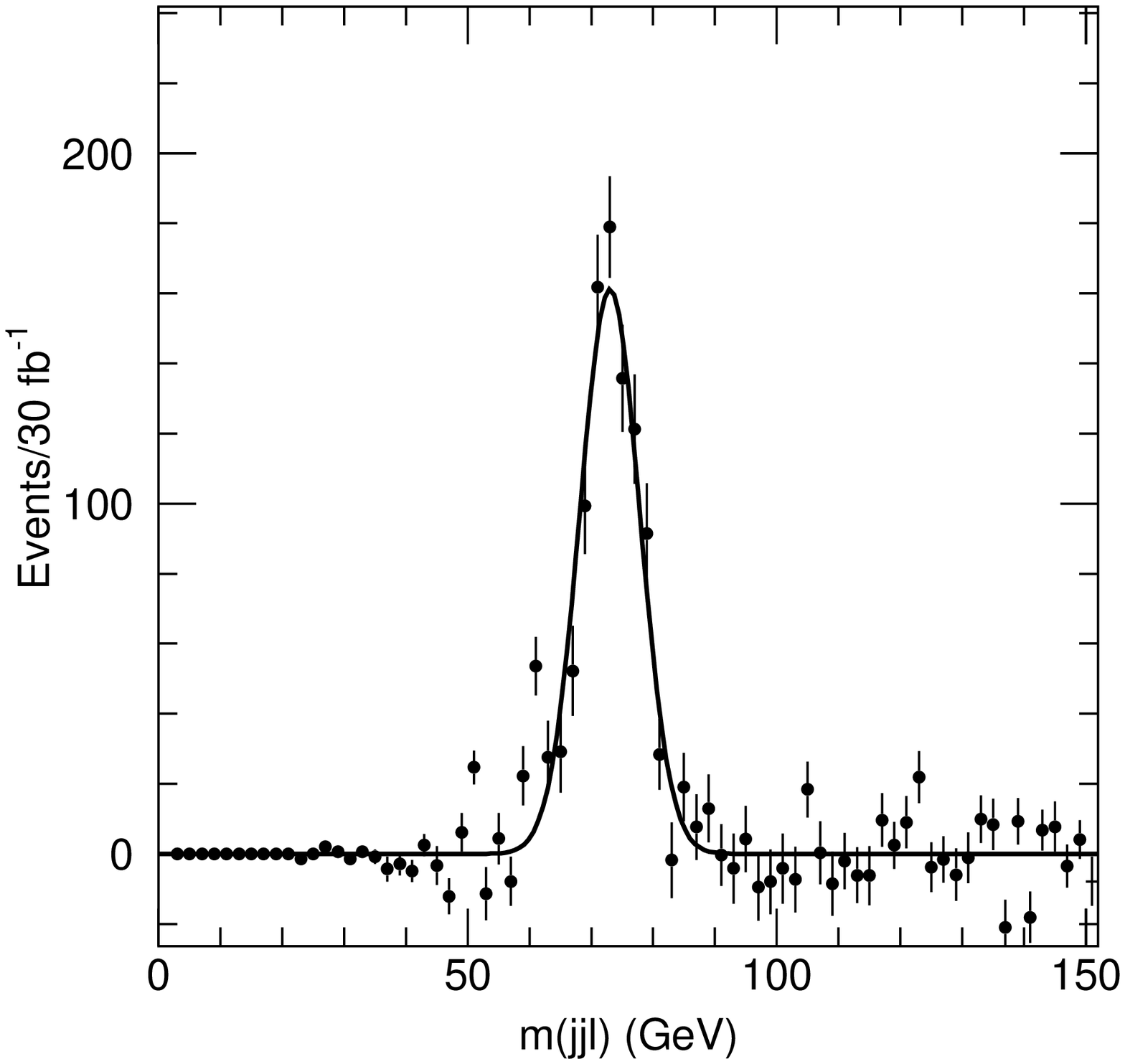}
\caption{\em $\mu$-jet-jet invariant mass for events in configuration (1)
(see text) before (left) and after (right) background subtraction.}
\label{figchi01}
\end{center}
\end{figure}

The three leptons come in the following flavor-sign configurations (+
charge conjugates): (1) $\mu^- e^+\mu^+$ (2) $\mu^- e^+\mu^-$ (3)
$\mu^-\mu^+\mu^+$ (4) $\mu^-\mu^+\mu^-$, where the first lepton comes
from the $\snm$, the second one from the $W$ and the third one from
the $\chione$ decay.  As a starting point for the analysis, we focus
on configuration (1) where the muon produced in the $\chione$ decay is
unambiguously identified as the one with the same sign as the
electron.  The distribution of the $\mu$-jet-jet invariant mass
exhibits a clear peak over a combinatorial background, shown on the
left side of Figure~\ref{figchi01}.  After combinatorial background
subtraction (right of Figure~\ref{figchi01}) an approximately Gaussian
peak is left, from which the $\chione$ mass can be measured with a
statistical error of $\sim 100$~MeV.  The combinatorial background is
due to events where one jet from $\chione$ decay is lost and a jet
from initial state radiation is used in the combination, and its
importance is reduced for heavier sneutrinos or neutralinos.  Once the
position of the $\chione$ mass peak is known, the reconstructed
$\chione$ statistics can be increased by also considering signatures
(2), (3) and (4), and by choosing as the $\chione$ candidate the
muon-jet-jet combination which gives invariant mass nearest to the
peak measured previously using events sample (1).  For further
reconstruction, we define as $\chione$ candidates the $\mu$-jet-jet
combinations with an invariant mass within 12~GeV of the measured
$\chione$ peak, yielding a total statistics of 6750 events for
signatures (1) to (4) for an integrated luminosity of 30~fb$^{-1}$ .

\begin{figure}[h!]
\begin{center}
\vspace*{-8mm}
\dofigs{0.5\textwidth}{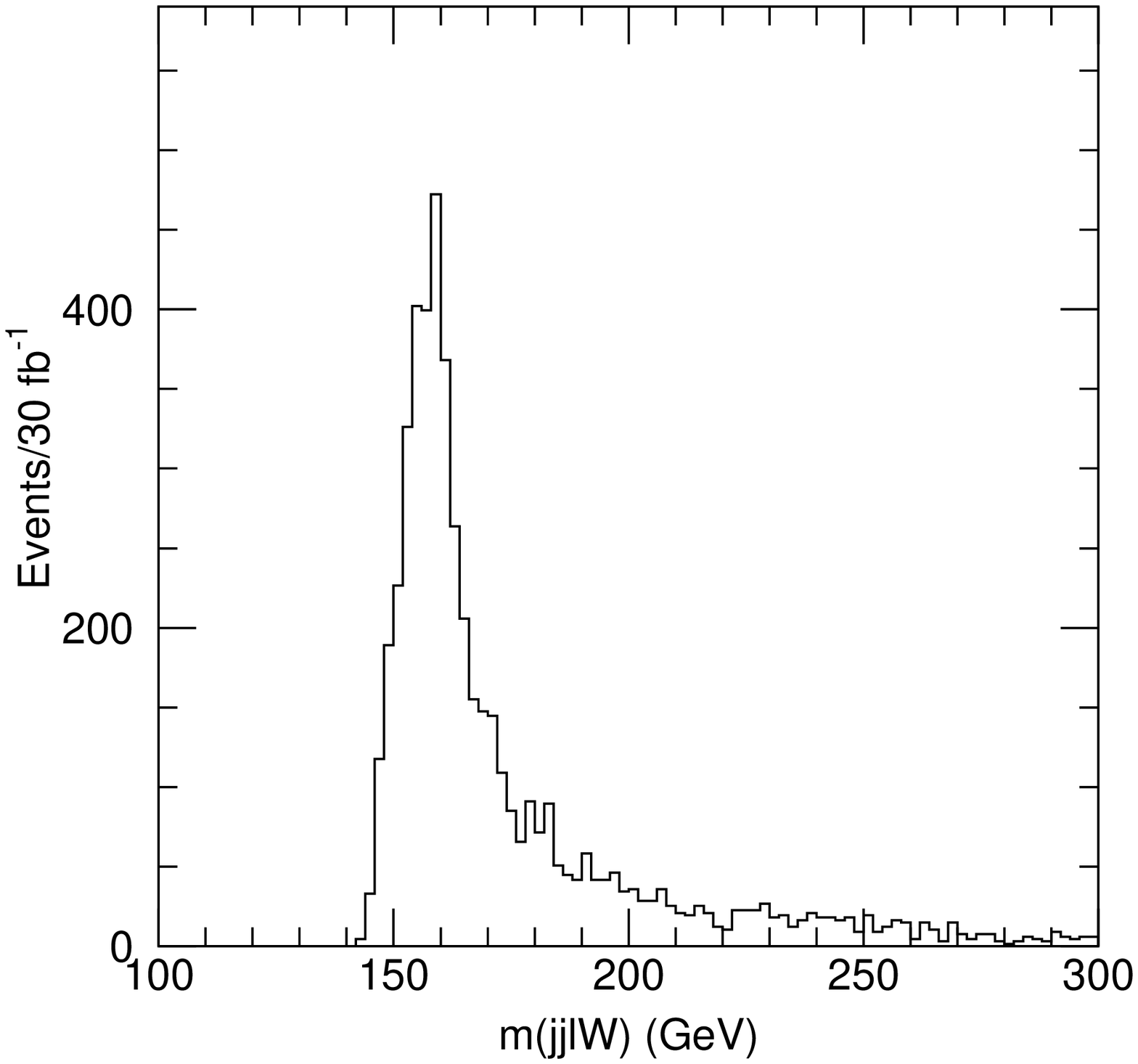}{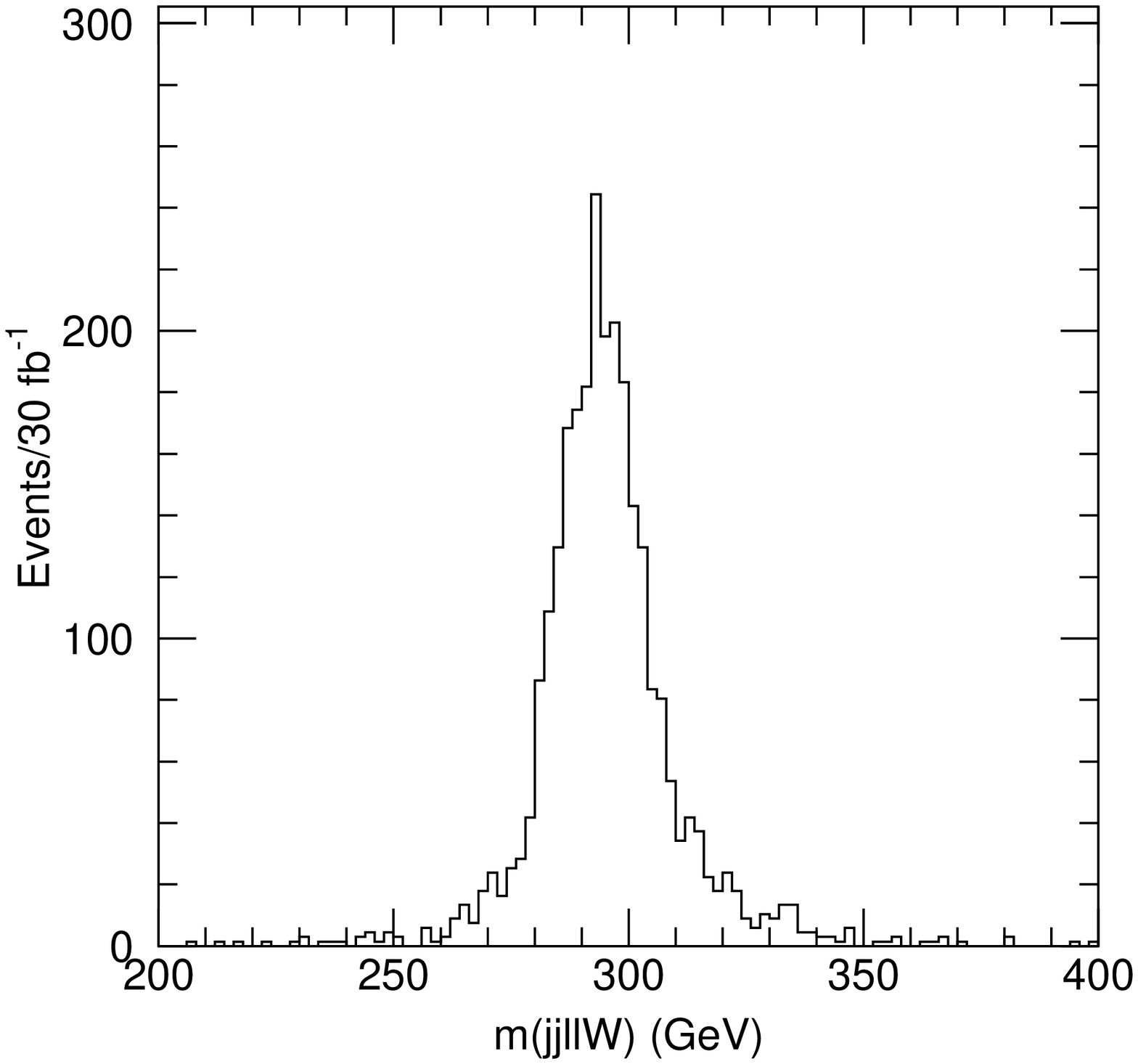}
\caption{\em Left: Invariant mass of the $W$ with the 
$\chione$ candidate. Right: Invariant mass of the third lepton in the event with the
$\chionepm$ candidate.}
\label{figsl}
\end{center}
\end{figure}

For $\chionepm$ reconstruction we consider only configurations (1) and
(2), for which the charged lepton from $W$ decay is unambiguously
identified as the electron. The longitudinal momentum of the neutrino
from the $W$ decay is calculated from the missing transverse momentum
of the event ($p_T^{\nu}$) and by constraining the electron-neutrino
invariant mass to the $W$ mass.  The resulting neutrino longitudinal
momentum has a twofold ambiguity.  We therefore build the invariant
$W-\chione$ mass candidate using both solutions for the $W$ boson
momentum.  The observed peak, represented on the left side of
Figure~\ref{figsl}, can be fitted with a Gaussian shape with a width
of $\sim 6$~GeV.  Only the solution yielding the $\chionepm$ mass
nearer to the measured mass peak is retained, and the $\chionepm$
candidates are defined as the combinations with an invariant mass
within 15~GeV of the peak, corresponding to a statistics of 2700
events.

Finally, the sneutrino mass is reconstructed by taking the invariant
mass of the $\chionepm$ candidate and the leftover muon
(Figure~\ref{figsl}, right).  The $\tilde\nu$ mass peak has a width of
$\sim10$~GeV and 2550 events are counted within 25~GeV of the measured
peak.

\vspace*{-4mm}

\vspace*{-4mm}

\section{Analysis reach}

\subsection{Standard Model background}
\label{secback}

We consider the following SM processes for the evaluation of the background 
to the three-leptons signature:
(1) $\bar tt$ production, followed by $t\rightarrow Wb$, where the two $W$
and one
of the $b$ quarks decay leptonically,
(2) $WZ$ production, where both bosons decay leptonically,
(3) $Wt$ production,
(4) $Wbb$ production,
(5) $Zb$ production.
These backgrounds were generated with the PYTHIA Monte Carlo \cite{PYTHIA},
and the ONETOP parton level generator \cite{ONETOP}, and passed through the
ATLFAST package \cite{ATLFAST}.

We apply to the background events the loose selection cuts described
in Section \ref{secana}, and in addition we reject the three same-sign
muons configurations which are never generated by our signal.  The
background to the sneutrino decay signal is calculated by considering
the events with a $\mu$-jet-jet invariant mass in an interval of $\pm
15$~GeV around the $\chione$ peak measured for the signal.  In order
to optimize the signal to background ratio only events containing
three muons (configurations (3) and (4)), which are less likely in the
Standard Model, are considered.  In each event two combinations,
corresponding to the two same-sign muons, can be used for the
$\chione$ reconstruction.  Both configurations are used when counting
the number of events in the peak. In most cases, however, the
difference in mass between the two combinations is such that they do
not appear in the same peak region. \par

\subsection{Supersymmetric background}

The pair production of SUSY particles through standard
$R_p$-conserving processes represents another source of background.  A
study based on the HERWIG 6.0 MonteCarlo \cite{herwig} has shown that
all the SUSY events surviving the cuts described in Section
\ref{secback} are mainly from $ pp \to \tilde \chi + X$ reactions
($\tilde \chi$ being either a chargino or a neutralino and $X$ any
other SUSY particle), and that the SUSY background decreases as the
$\tilde \chi^{\pm}$ and $\tilde \chi^0$ masses increase.  This
behavior is due to the combination of two effects: the $\tilde \chi +
X$ production cross-section decreases with increasing $\tilde \chi$
mass, and the probability of losing two of the four jets from the
decays of the two $\chione$ in the event becomes smaller as the
$\tilde \chi^{\pm}$ and $\tilde \chi^0_1$ masses increase.  The SUSY
background is only significant for $\chionepm$ masses lower than
$200$~GeV.

Besides, it can be assumed that the $\chione$ mass will be derived
from inclusive $\chione$ reconstruction in SUSY pair production as
shown in \cite{TDR} and \cite{lmgp}.  Hence, even in the cases where a
significant $\chione$ peak can not be observed above the SUSY
background, we can proceed to the further steps in the kinematic
reconstruction. The strong kinematic constraint obtained by requiring
both the correct $\chione$ mass and a peak structure in the
$\chione-W$ invariant mass will then allow to separate the single
sneutrino production from other SUSY processes.

Therefore, only the Standard Model background is considered in the
evaluation of the analysis reach presented below.

\subsection{Reach in the mSUGRA parameter space}

\begin{figure}[h!]
\begin{center}
\vspace*{-8mm}
\dofigs{0.5\textwidth}{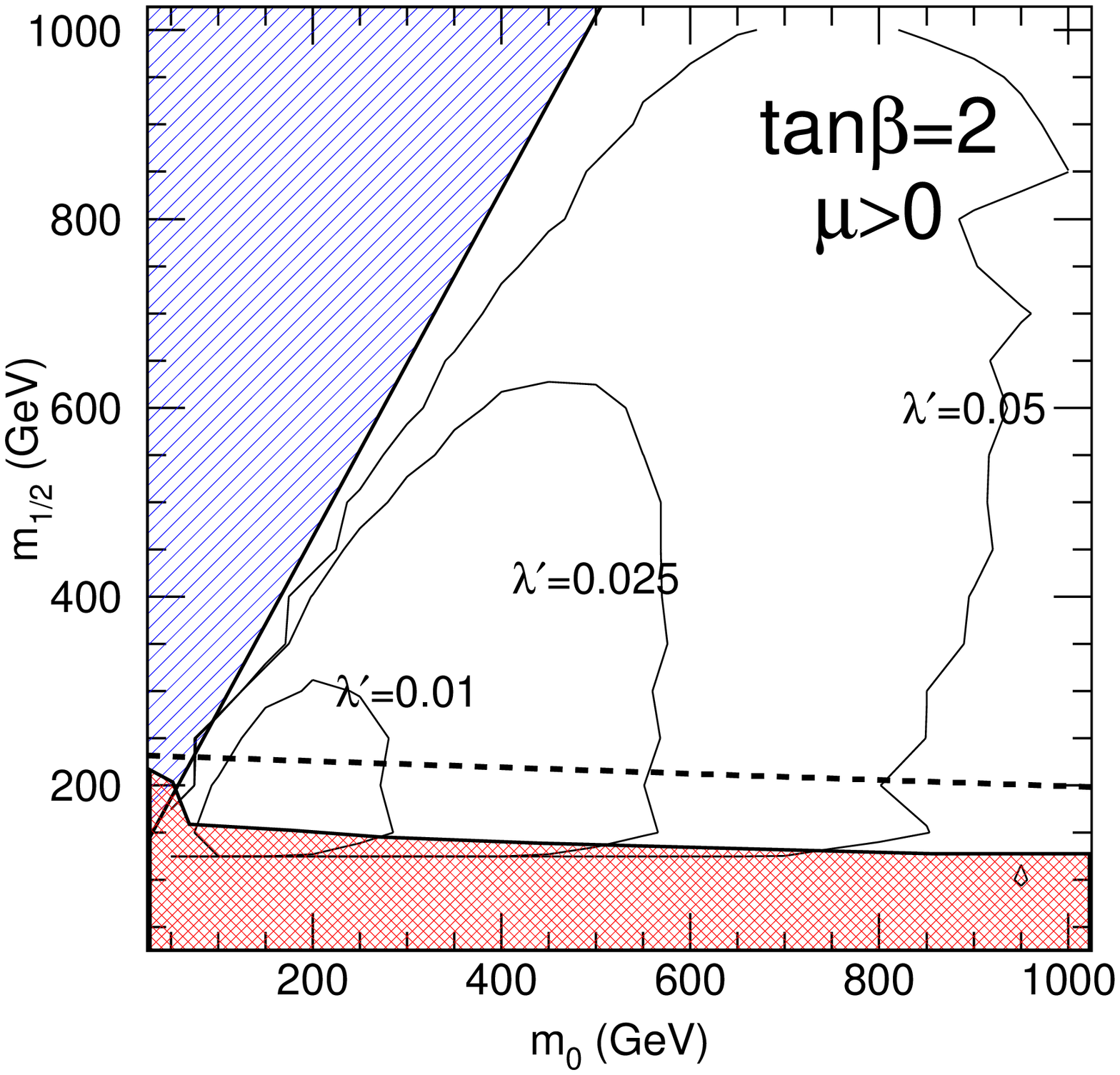}{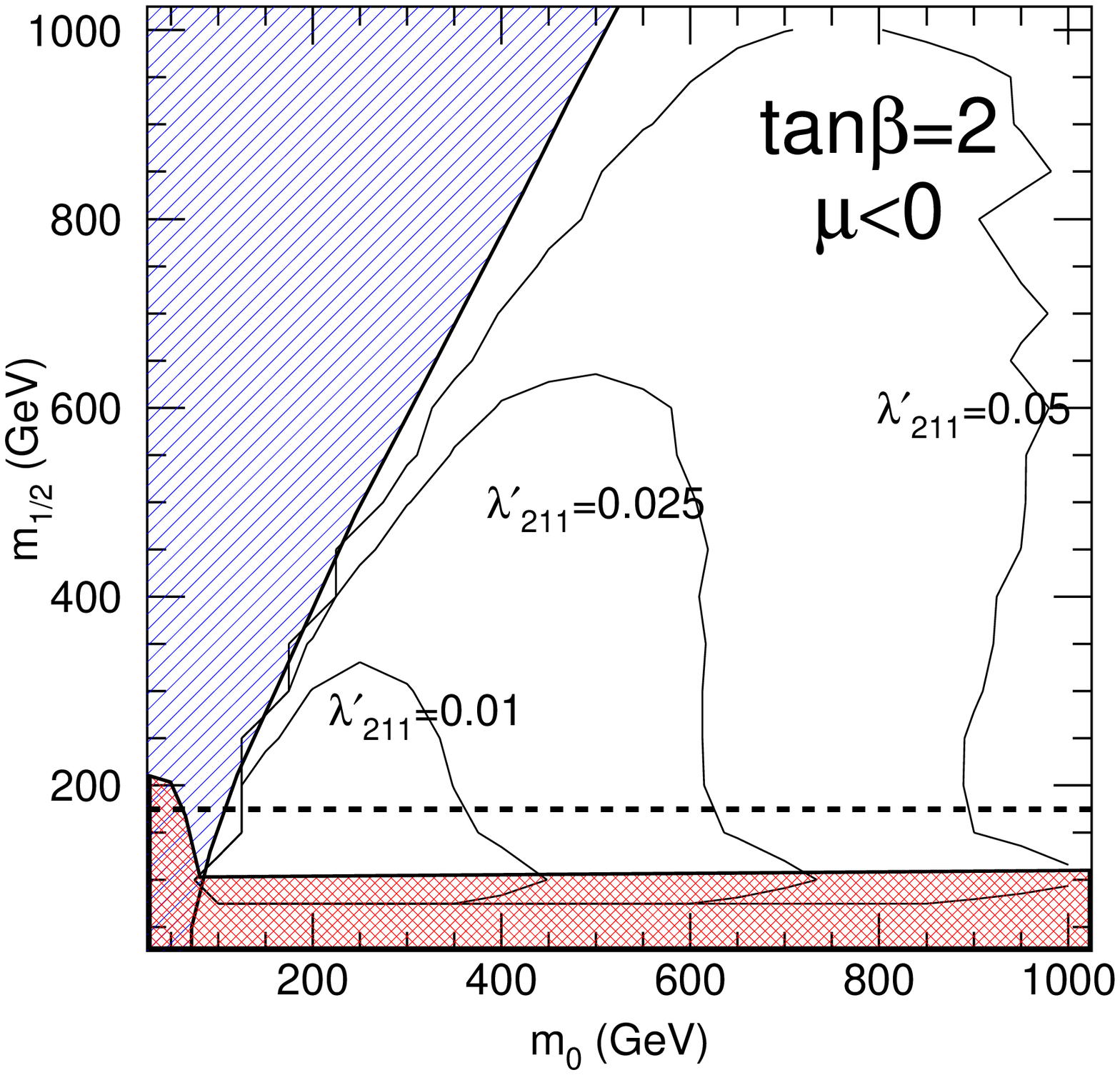}
\caption{\em $5\sigma$ reach in the $m_0-m_{1/2}$ plane
for $A_0=0$, $\tan\beta=2$, $\mu<0$ (left) and $\mu>0$ 
(right) and three
different choices of the $\lambda^\prime_{211}$ coupling, with an integrated
luminosity of 30~fb$^{-1}$ at the LHC. A signal
of at least ten events is required.
The hatched region at the upper
left corresponds to $m_{\tilde\nu}<m_{\tilde\chi^{\pm}_1}$.
The cross-hatched region for low $m_{1/2}$ gives the kinematical
limit for the discovery of $\chionepm$ or $\tilde l$ by LEP
running at $\sqrt{s}=196$~GeV \cite{aleph2}.
The dotted line shows the
region below which the $\chionepm$ decays to a virtual W.}
\label{plreach}
\end{center}
\end{figure}

In Figure~\ref{plreach}, we show the regions of the $m_0-m_{1/2}$
plane where the signal significance exceeds 5~$\sigma$ (${S \over
\sqrt B}>5$ with $S=Signal$ and $B= SM \ Background$) after the set of
cuts described in Section \ref{secback} has been applied, within the
mSUGRA model.  The full mass reconstruction analysis of Section
\ref{secana} is possible only above the dashed line parallel to the
$m_0$ axis. Below this line the decay $\chionepm\to\chione W^{\pm}$ is
kinematically closed, and the $W$ mass constraint can not be applied
to reconstruct the neutrino longitudinal momentum.

\vspace*{-4mm}

The basic feature in Figure~\ref{plreach} is a decrease of the
sensitivity on $\lambda^\prime_{211}$ as $m_0$ increases.  This is due
to a decrease of the partonic luminosity as $m_{\tilde\nu}$ increases.
The sensitivity on $\lambda^\prime_{211}$ is also observed to decrease
as $m_{\chionepm}$ approaches $m_{\tilde\nu}$.  There are two
reasons. First, in this region the phase space factor of the decay
$\tilde\nu\to\chionepm\mu^{\mp}$ following the resonant sneutrino
production is suppressed, thus reducing the branching fraction.
Secondly, as the $\tilde \nu_{\mu}$ and the $\chionepm$ become nearly
degenerate the muon from the decay becomes on average softer, and its
$p_T$ can fall below the analysis requirements.  In the region
$m_{\chionepm}>m_{\tilde\nu}$, shown as a hatched region in the upper
left of the plots, the resonant sneutrino production contribution
vanishes and there is essentially no sensitivity to
$\lambda^\prime_{211}$.  Finally, the the sensitivity vanishes for low
values of $m_{1/2}$.  This region, below the LEP 200 kinematic limit
for $\chionepm$ detection, corresponds to low values of the $\chione$
mass.  In this situation the two jets from the $\chione$ decay are
soft, and one of them is often below the transverse momentum
requirement, or they are reconstructed as a single jet.

For high $\tan\beta$, the three-lepton signature is still present, but
it may be produced through the decay chain
$\chionepm\to\tilde\tau_1\nu_{\tau}$, followed by
$\tilde\tau_1\to\tau\chione$. The full kinematic reconstruction
becomes very difficult, but the signal efficiency is essentially
unaffected, as long as the mass difference between the lightest
$\tilde\tau$ and the $\chione$ is larger than $\sim 50$~GeV.  For a
smaller mass difference the charged lepton coming from the $\tau$
decay is often rejected by the analysis cuts.

\section{Conclusion}

In conclusion we have shown that if minimal supersymmetry with
R-parity violation is realized in Nature, the three-leptons signature
from resonant sneutrino production will be a privileged channel for
precisely measuring sparticle masses in a model-independent way as
well as for testing a broad region of the mSUGRA parameter space.

This signature can lead to a high sensitivity to the
$\lambda^{\prime}_{211}$ coupling and should also allow to probe an
unexplored range of values for many other \rpv\ couplings of the type
$\lambda^\prime_{1jk}$ and $\lambda^\prime_{2jk}$.

\setcounter{figure}{0}
\setcounter{table}{0}
\setcounter{section}{0}
\setcounter{equation}{0}
\setcounter{footnote}{0}
\clearpage

\newcommand{\newc}{\newcommand}
\newc{\lam}{\lambda}
\newc{\cht}{\tilde{\chi}}
\newc{\dnt}{\tilde{d}}
\newc{\mr}{\mathrm}
\newc{\eg}{{\it e.g.\,}}
\newc{\barr}{\begin{eqnarray}}
\newc{\earr}{\end{eqnarray}}

\begin{center}
{\large \bf Resonant slepton production at the LHC }\\
\vspace*{3mm}
{\sc H. DREINER\footnote{E-mail address: dreiner@v2.rl.ac.uk},
P. RICHARDSON\footnote{E-mail address: p.richardson1@physics.ox.ac.uk},
and M. H. SEYMOUR\footnote{E-mail address: M.Seymour@rl.ac.uk} } \\

\end{center}
\vspace*{3mm}

\begin{abstract}
We consider the resonant production of charged sleptons at the LHC via
R-parity violation (\rpv\ ) followed by gauge decays to a charged
lepton and a neutralino which then decays via \rpv. This gives a
signature of two like-sign charged leptons. In the simulation we
include the full hadronization via Monte Carlo programs. We find a
background, after cuts, of $5.1\pm2.5$ events for an integrated
luminosity of $10fb^{-1}$.  A preliminary study of the signal suggests
that couplings of $2\times10^{-3}$ for a smuon mass of $223$\GeV\  and
smuon masses of up to  $540$\GeV\  for couplings of $10^{-2}$ can be
probed.

\end{abstract}

%
%
\section{Introduction}

In R-parity violating (\rpv) models the single resonant production of
charged sleptons in hadron-hadron collisions is possible. The most
promising channels for the discovery of these processes, at least with
small \rpv\   couplings, involve the gauge decays of these resonant
sleptons. In particular if we consider the production of a charged
slepton, this can then decay to give a neutralino and a charged
lepton, \ie the process
\be
  \mr{u} + \mr{\bar{d}} \longrightarrow \mr{\tilde{\ell}^+} 
	      		\longrightarrow \mr{\ell^+} + \mr{\cht^0}.
\ee
In addition to this $s$-channel process there are $t$-channel
processes involving squark exchange.  The neutralino decays via the
crossed process to give a charged lepton, which due to the Majorana
nature of the neutralino can have the same charge as the lepton from
the slepton decay. We therefore have a like-sign dilepton signature
which we expect to have a low Standard Model background.

\section{Backgrounds}
\label{sec:back}
  The dominant Standard Model backgrounds to this process come from
\begin{itemize}
\item 	Gauge boson pair production, \ie production of ZZ or WZ 
followed by leptonic decays of the gauge bosons with some of the
leptons not being detected.

\item 	$\mr{t\bar{t}}$ production. Either the t or $\mr{\bar{t}}$ 
        decays semi-leptonically, giving one lepton. The second top
        decays hadronically. A second lepton with the same charge can
        be produced in a semi-leptonic decay of the bottom hadron
        formed in the hadronic decay of the second top, \ie
\barr
 \mr{t}	&\ra&  \mr{W^+ b} \ra \mr{e^{+}\bar{\nu_{e}} b},\nonumber \\
 \mr{\bar{t}}	&\ra& \mr{W^{-}\bar{b}}	\ra \mr{q\bar{q}\bar{b}},\quad 
 \mr{\bar{b}} \ra \mr{e^{+}\bar{\nu_{e}}\bar{c}}.
\earr

\item 	$\mathrm{b\bar{b}}$ production. If either of these quarks 
        hadronizes to form a $\mr{B^0_{d,s}}$ meson this can mix to
        give a $\mr{\bar{B}^0_{d,s}}$.  This means that if both the
        bottom hadrons decay semi-leptonically the leptons will have
        the same charge as they are both coming from either b or
        $\mr{\bar{b}}$ decays.

\item 	Single top production. A single top quark can be produced together 
        with a $\mr{\bar{b}}$ quark by either an $s$-  or $t$-channel W
        exchange. This can then give one charged lepton from the top
        decay, and a second lepton with the same charge from the decay
        of the meson formed after the b quark hadronizes.

 \item Non-physics backgrounds. There are two major sources: (i) from
 misidentifying the charge of a lepton, \eg in Drell-Yan production, and 
 (ii) from incorrectly identifying an isolated hadron as a lepton. This
 means that there is a major source of background from W production
 with an additional jet faking a lepton.
\end{itemize}

Early studies of like-sign dileptons at the LHC
\cite{Dreiner:1994ba} only studied the backgrounds 
from heavy quark production. It was found that by imposing cuts on the
transverse momentum and isolation of the leptons the heavy quark backgrounds
could be significantly reduced. However more recent studies of the
like-sign dilepton production at the LHC \cite{Baer:1996va}
and the Tevatron
\cite{Matchev:1999nb,Baer:1999bq}
suggest that a major source of background to like-sign dilepton
production is from gauge boson pair production and
from fake leptons. Here we will consider the backgrounds from gauge boson
pair production as well as heavy quark production.
The study of the non-physics backgrounds (\eg 
fake leptons) requires a full simulation of the detector and it is
therefore beyond the scope of our study. In particular the background
from fake leptons cannot be reliably calculated from Monte Carlo
simulations and must be extracted from data
\cite{Matchev:1999nb}.
We can use the differences between the \rpv\  signature we are considering and
the MSSM signatures considered in \cite{Baer:1996va}
to reduced the background from gauge boson pair production.

\begin{figure}
\begin{center}
\vspace*{-3mm}
\includegraphics[angle=90,width=0.48\textwidth]{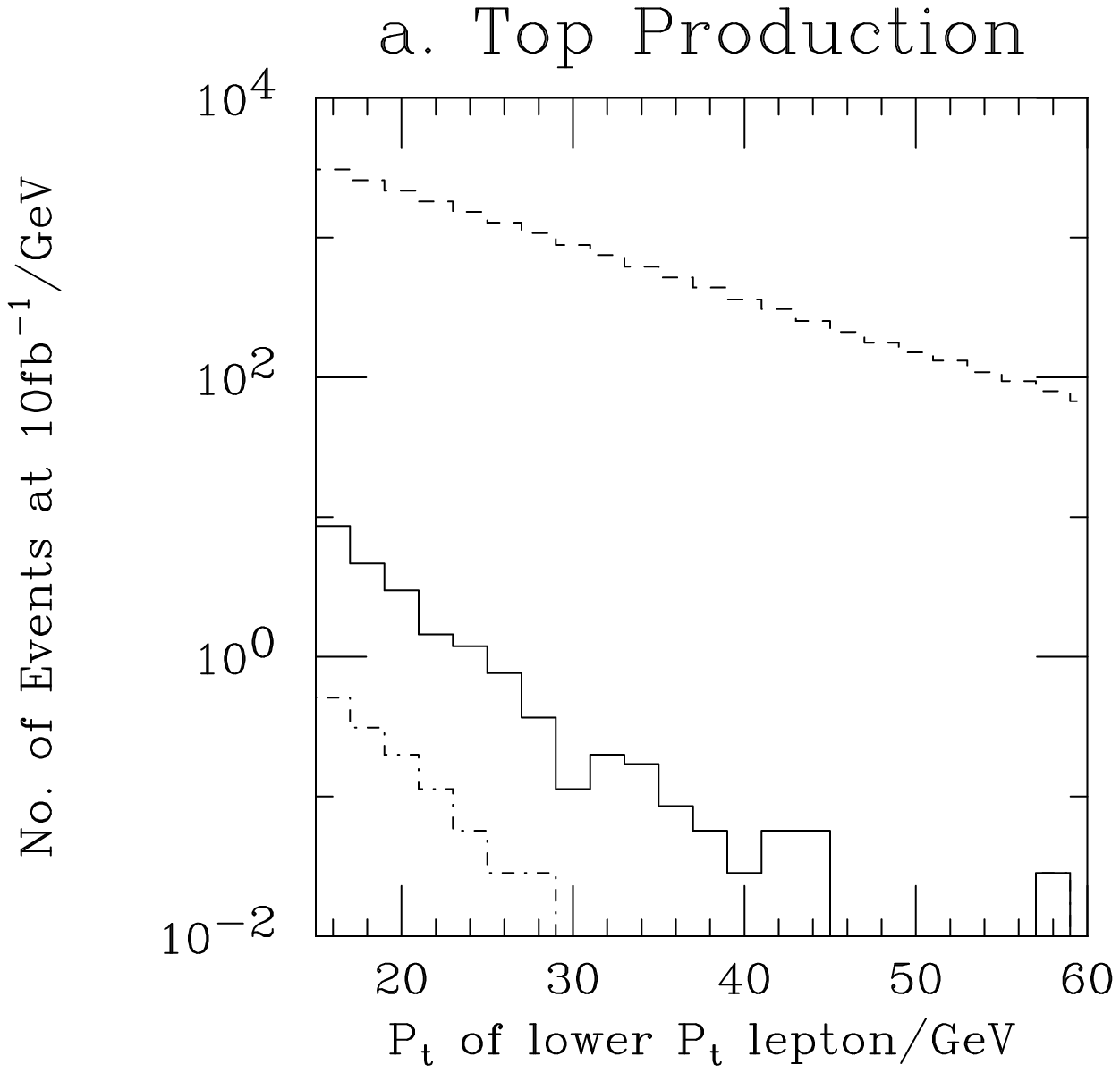}
\hfill
\includegraphics[angle=90,width=0.48\textwidth]{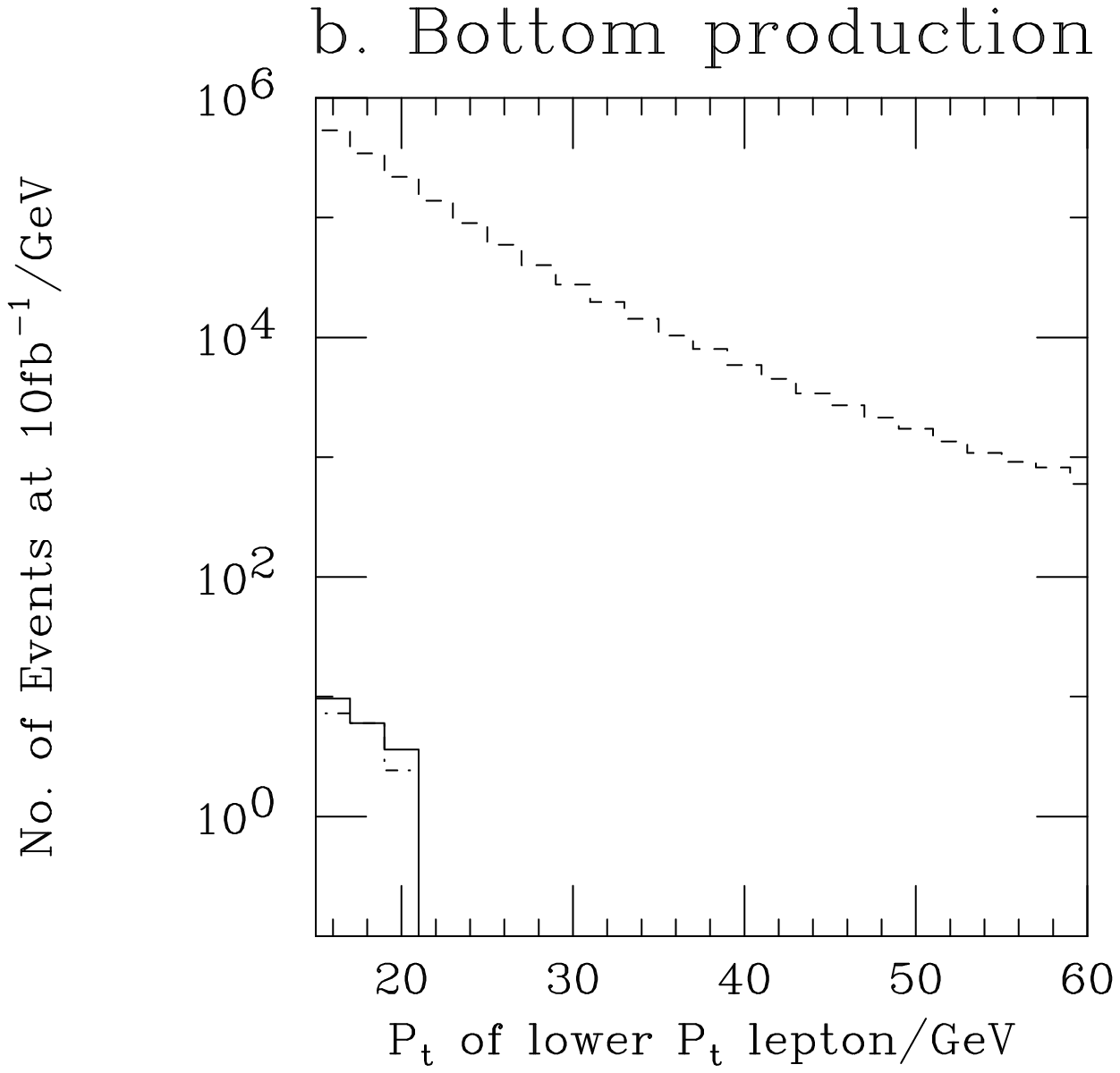}
\caption{Effect of the isolation cuts on the $\mr{t\bar{t}}$ and 
$\mr{b\bar{b}}$ 
backgrounds. The dashed line gives the background before any cuts, the solid
line shows the effect of the isolation cut described in the text.
The dot-dash line gives the effect of all the cuts.}
\label{fig:heavyiso}
\end{center}
\end{figure}

We impose the following cuts
\begin{itemize}

\item A cut on the transverse momentum of the like-sign 
leptons $p_T>40$~\GeV.

\item  An isolation cut on the like-sign leptons so that the transverse 
energy in a cone of radius $R = \sqrt{\Delta\phi^2+\Delta\eta^2} = 0.4$
about the direction of each lepton is less than $5$~\GeV.

\item  A cut on the transverse mass,
 	$M^2_T =
 	2|p_{T_{\ell}}||p_{T_\nu}|(1-\cos\Delta\phi_{\ell\nu})$, where
 	$p_{T_{\ell}}$ is the transverse momentum of the charged
 	lepton, $p_{T_\nu}$ is the transverse momentum of the
 	neutrino, assumed to be all the missing transverse momentum in
 	the event, and $\Delta\phi_{\ell\nu}$ is the azimuthal angle
 	between the lepton and the neutrino, \ie the missing momentum
 	in the event. We cut out the region where  $60\mr{\GeV} < M_T <
 	85\mr{\GeV} $.

\item   A veto on the presence of a lepton in the event with the same flavor
        but opposite charge (OSSF) as either of the leptons in the
        like-sign pair if the lepton has $p_T>10$~\GeV\  and which passes the
        same isolation cut as the like-sign leptons.

\item   A cut on the missing transverse energy, $E^T_{\mathit{miss}}
<20$~\GeV\ .
\end{itemize}

\begin{figure}
\begin{center}
\vspace*{-3mm}
\includegraphics[angle=90,width=0.4\textwidth]{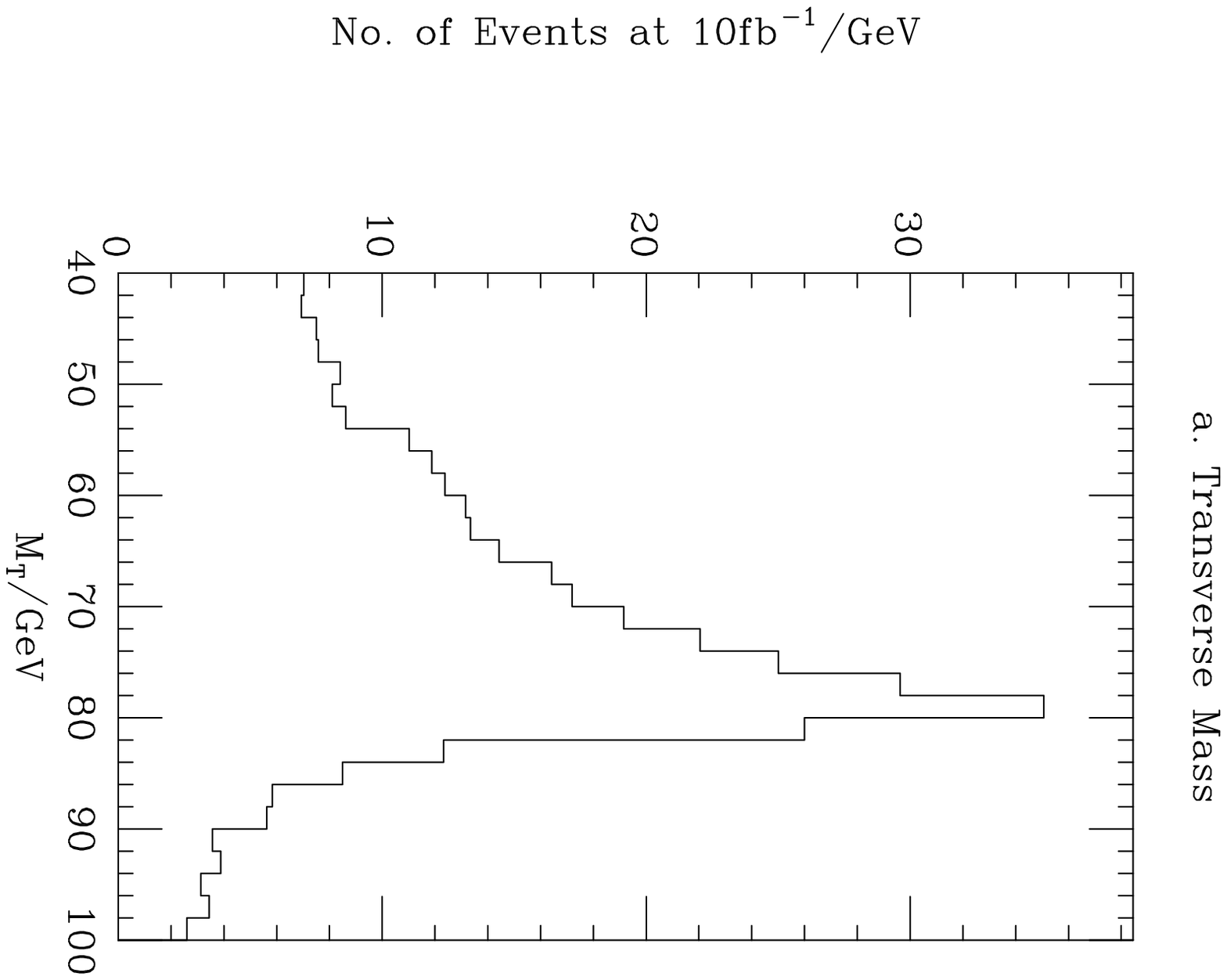}
\hfill
\includegraphics[angle=90,width=0.4\textwidth]{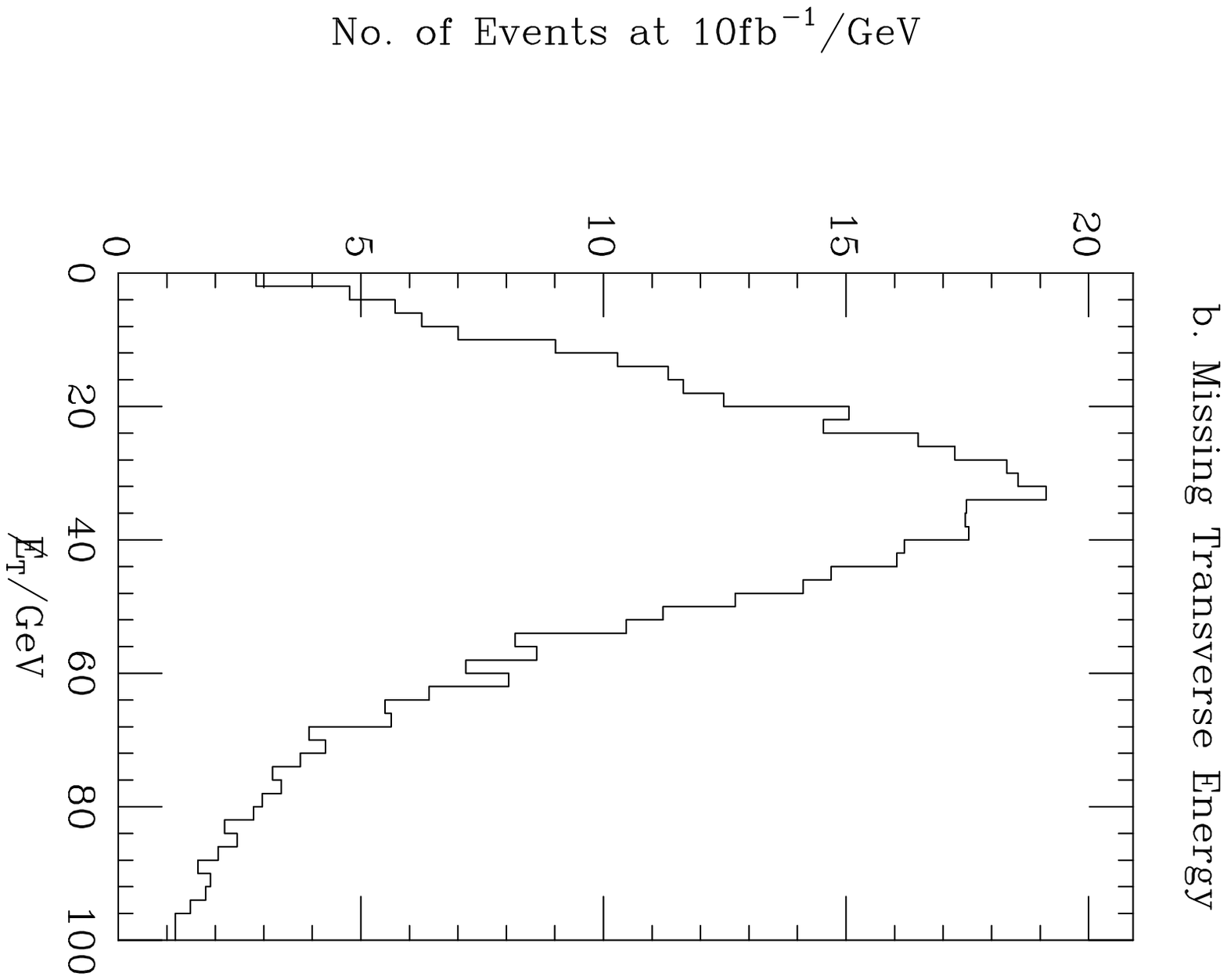}
\caption{Transverse mass and missing transverse energy in WZ events}
\label{fig:WZ}
\end{center}
\end{figure}
\vspace*{-3mm}

\begin{figure}
\begin{center}
\includegraphics[angle=90,width=0.48\textwidth]{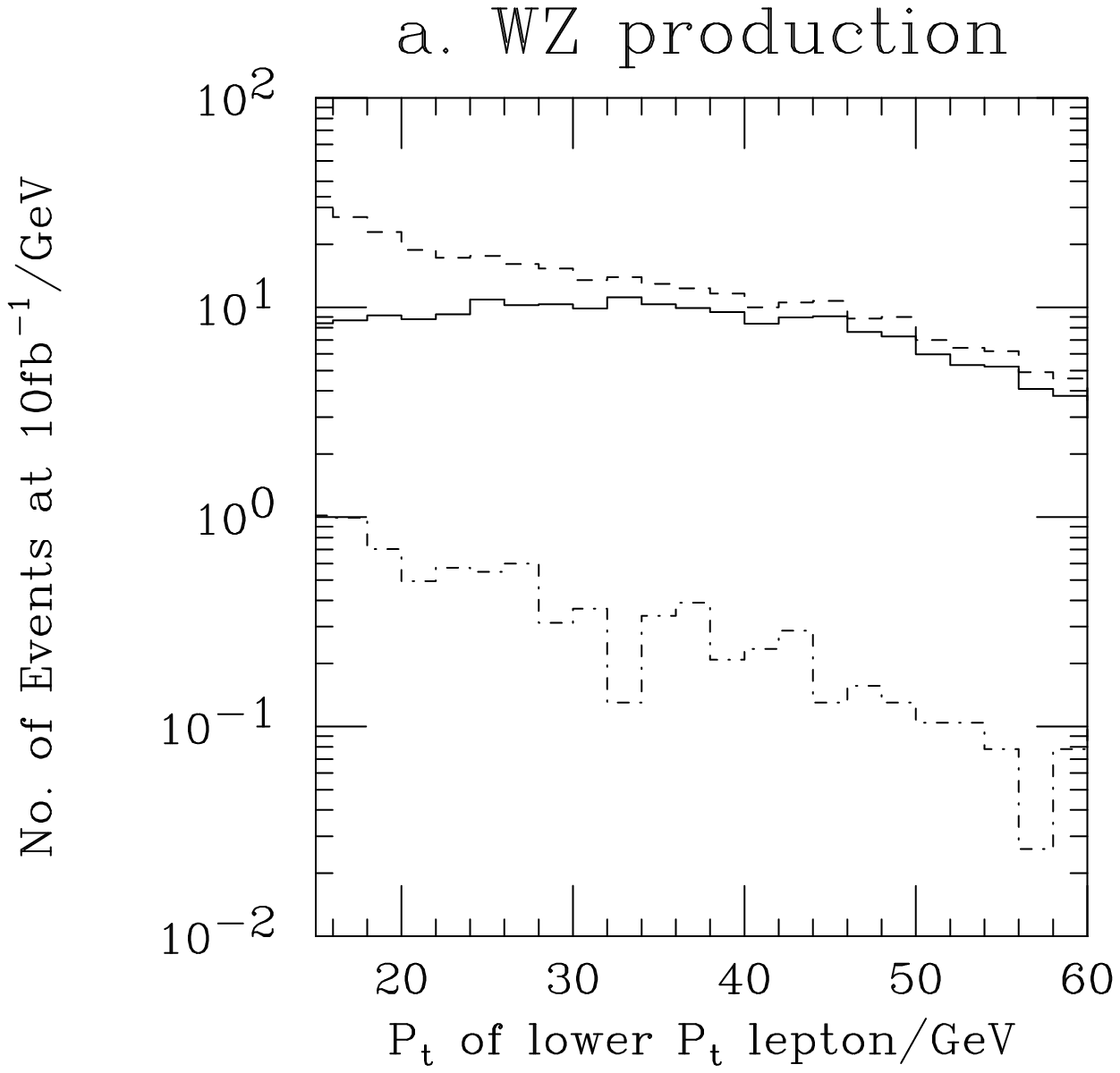}
\hfill
\includegraphics[angle=90,width=0.48\textwidth]{dreiner3a}
\caption{Effect of the isolation cuts on the WZ and ZZ 
backgrounds. The dashed line gives the background before any cuts, the solid
line shows the effect of the isolation cut described in the text.
The dot-dash line gives the effect of all the cuts.}
\label{fig:WZiso}
\end{center}
\end{figure}

While these cuts were chosen to reduce the background we have not attempted
to optimize them.
The first two cuts are designed to reduce the background from heavy
quark production.  As can be seen in Fig.\,\ref{fig:heavyiso}, these
cuts reduce this background by several orders of magnitude. The
remaining cuts are designed to reduce the background from gauge boson
pair, in particular WZ, production which is the major source of
background after the imposition of the isolation and $p_T$ cuts. The
transverse mass cut is designed to remove events with leptonic W
decays as can be seen in Fig.\,\ref{fig:WZ}a. The veto on the presence
of OSSF leptons is designed to remove events where one lepton from the
dilepton pair comes from the leptonic decay of a Z boson. The missing
transverse energy cut again removes events with leptonic W decays,
this is mainly to reduce the background from WZ production, as seen in
Fig.\,\ref{fig:WZ}b. The effect of these cuts on the heavy quark and
gauge boson pair backgrounds are shown in
Figs.\,\ref{fig:heavyiso}~and~\ref{fig:WZiso}, respectively.

The backgrounds from the various processes are summarized in
Table~\ref{tab:back}.  The simulations of the $\mr{b\bar{b}}$,
$\mr{t\bar{t}}$ and single top production were performed using
HERWIG6.1 \cite{Corcella:1999qn}. The simulations of
gauge boson pair production used PYTHIA6.1 \cite{Sjostrand:1994yb}. 
The major contribution to the background comes from WZ production; the
major contribution to the error comes from $\mr{b\bar{b}}$.  For the
$\mr{b\bar{b}}$ simulation we have required a parton-level cut of
$40$~\GeV\  on the transverse momentum of the bottom quarks. This
should not affect the results provided we impose a cut of at least
$40$~\GeV\  on the $p_T$ of the leptons. We also forced the B meson
produced to decay semi-leptonically. In events where there was one
$\mr{B^0_{d,s}}$ meson this meson was forced to mix, if there was more
than one $\mr{B^0_{d,s}}$ then one of the mesons was forced to mix and
the others forced to not mix. Even with these cuts it is impossible to
simulate the full luminosity with the resources available, due to the
large cross section for $b\bar{b}$ production. This gives the large
error on the estimate of this background.
  
\begin{table}
\begin{center}
\begin{tabular}{|c|c|c|c|}
\hline
Background Process 	& \multicolumn{3}{c|}{Number of Events} \\
\cline{2-4}
 	& After $p_T$ cut & After isolation and $p_T$ cuts & After all cuts \\
\hline
WW 		& $2.8\pm0.6$ 		& $0.0\pm0.1$ 	& $0.0\pm0.1$ 	\\
\hline
WZ 		& $226\pm3$		& $189\pm3$ 	& $4.1\pm0.5$ 	\\
\hline
ZZ 		& $50.4\pm0.9$		& $40.6\pm0.8$	& $0.9\pm0.1$ 	\\
\hline
$\mr{t\bar{t}}$ & $(4.8\pm0.3)\times10^3$ & $0.34\pm0.14$ & $0.06\pm0.06$ \\
\hline
$\mr{b\bar{b}}$ & $(5.69\pm0.8)\times10^4$  & $0.0\pm2.4$ & $0.0\pm2.4$   \\
\hline
Single Top 	& $11.5\pm0.3$ 		 &$0.0\pm0.008$	& $0.0\pm0.008$ \\
\hline
Total 	 	& $(6.2\pm0.8)\times10^4$ & $230\pm4$ 	& $5.1\pm2.5$ \\
\hline
\end{tabular}
\end{center}
\caption{Backgrounds to like-sign dilepton production at the LHC. The 
numbers of events are based on an integrated luminosity of $10\
\mr{fb}^{-1}$. We used the cross sections from the Monte Carlo
simulation for $\mr{b\bar{b}}$ and single top production, the
next-to-leading order cross section for gauge boson pair production
from \cite{Campbell:1999ah} and the next-to-leading order with
next-to-leading-log resummation cross section from
\cite{Bonciani:1998vc} for $\mr{t\bar{t}}$ production. We estimate an  error 
on the cross section from the effect of varying the scale between
half and twice the hard scale, apart from gauge boson pair production where
we do not have this information for the next-to-leading order cross section.
The error on the number of events is 
then the error in the cross section and the statistical error from the 
simulation added in quadrature.}
\label{tab:back}
\end{table}

\section{Signal}

We used HERWIG6.1 \cite{Corcella:1999qn} to simulate
the signal.  This version includes the resonant slepton production,
including the $t$-channel diagrams, and the R-parity violating decay
of the neutralino including a matrix element for the decay
\cite{Dreiner:1999qz}. We will only consider first generation quarks
as the cross sections for processes with higher generation quarks are
suppressed by the parton distributions. There are upper bounds
on the \rpv\  couplings from low energy experiments. The bound on
${\lam'}_{111}$ from neutrino-less double beta decay
\cite{Allanach:1999ic,Hirsch:1995zi} is very
strict so we consider muon production via the coupling
${\lam'}_{211}$, which has a much weaker bound,
\be
{\lam'}_{211} < 0.059 \times \left(\frac{M_{\dnt_R}}{100 \mr{GeV}}\right),
\ee
  from the ratio $R_\pi=\Gamma(\pi\ra e\nu)/\Gamma(\pi\ra \mu\nu)$
  \cite{Allanach:1999ic,Barger:1989rk}.
 
\begin{figure}
\begin{center}
\includegraphics[angle=90,width=0.48\textwidth]{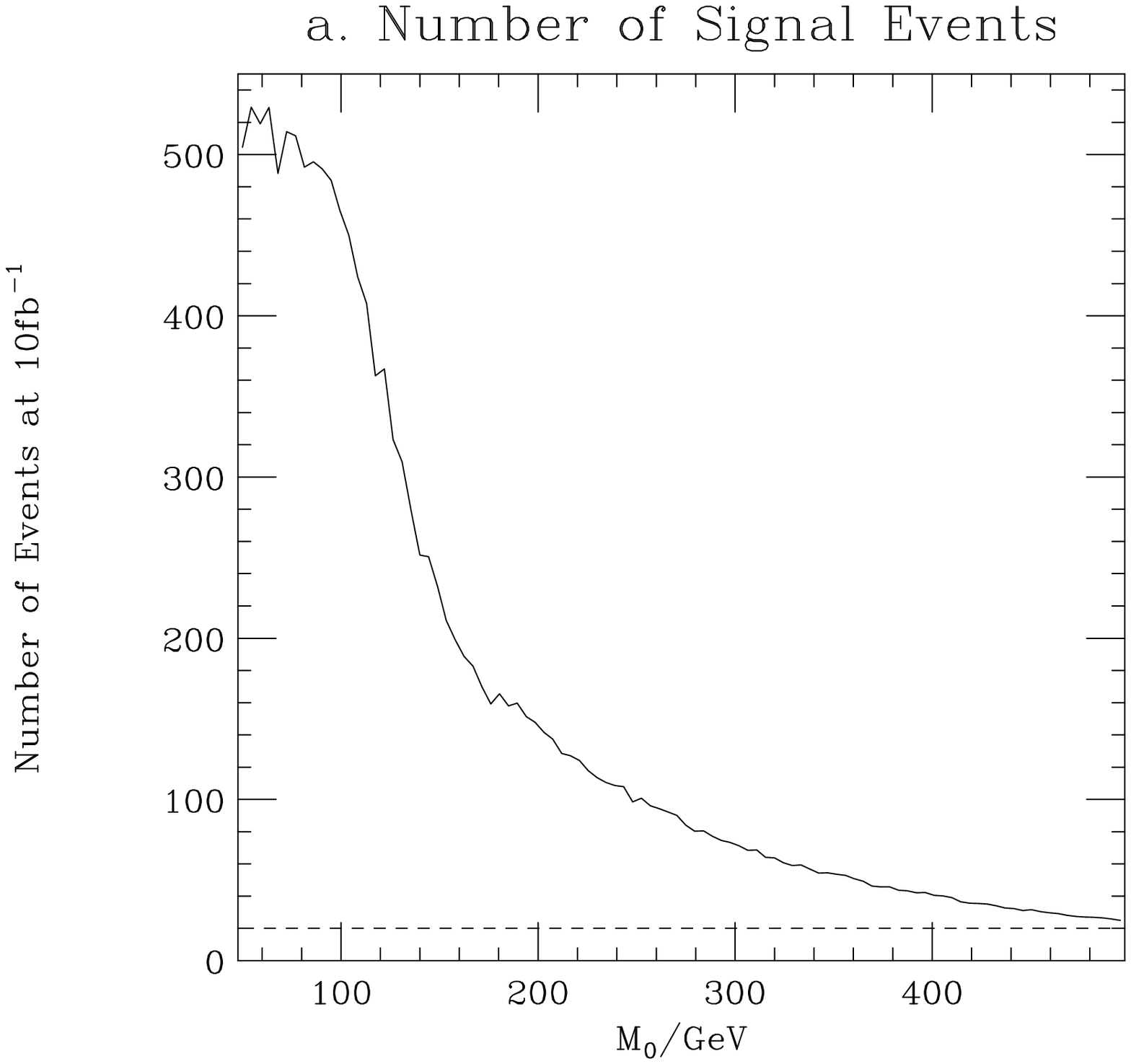}
\hfill
\includegraphics[angle=90,width=0.48\textwidth]{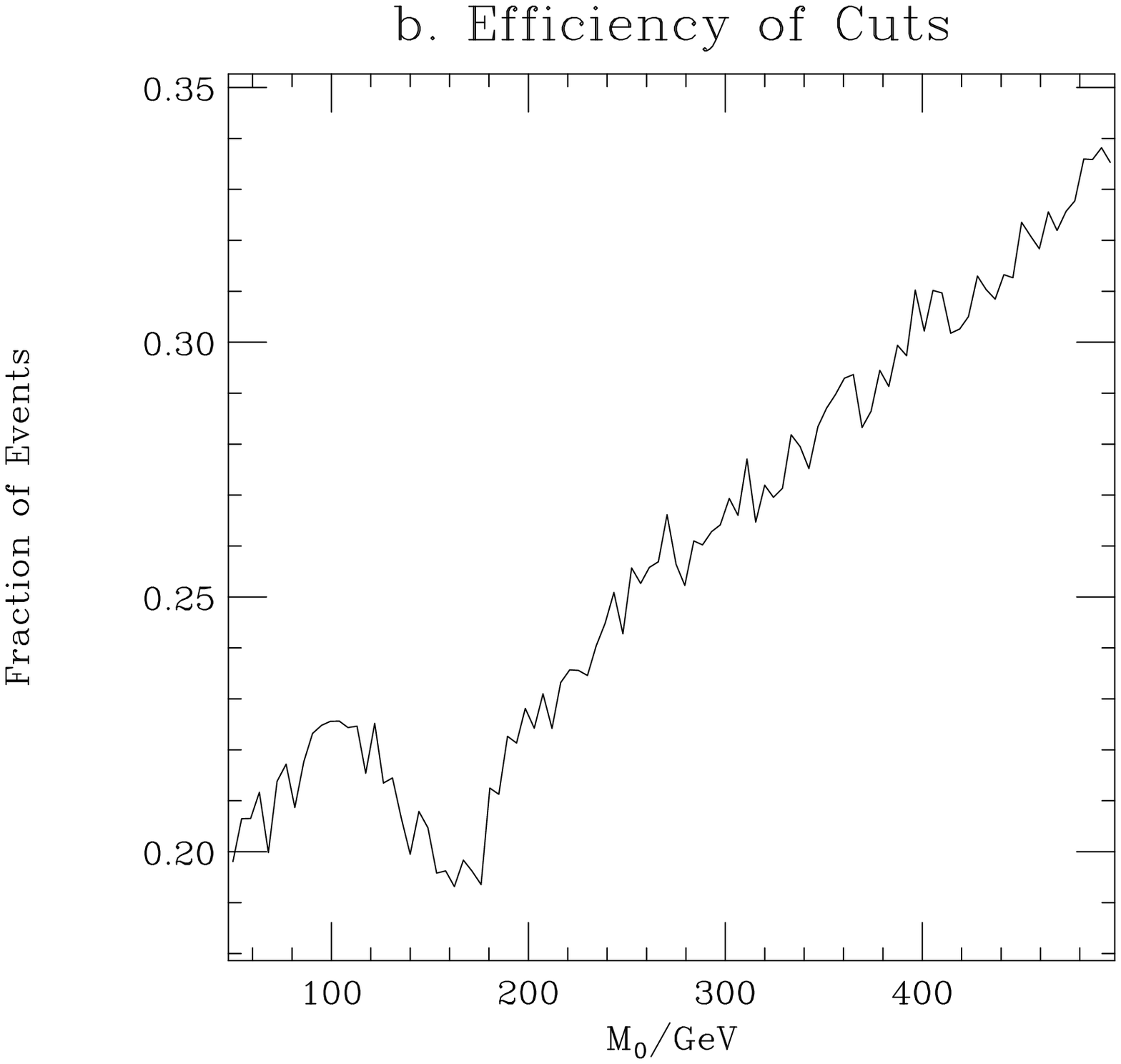}
\caption{Number of signal events passing the cuts and the efficiency for
$M_{1/2}=300$\GeV, $A_0=300$\GeV, $\tan\beta=2$, $\mr{sgn}\mu=+$, with the 
\rpv\  coupling ${\lam'}_{211}=0.01$. The dashed line gives the number of
events needed for a $5\sigma$ discovery.}
\label{fig:signal}
\end{center}
\end{figure}
 
We have performed a scan in $M_0$ using HERWIG with the following
SUGRA parameters, $M_{1/2}=300$\GeV, $A_0=300$\GeV, $\tan\beta=2$,
$\mr{sgn}\mu=+$, and with the \rpv\  coupling ${\lam'}_{211}=0.01$. The
number of events which pass the cuts given in Section~\ref{sec:back}
are shown in Fig.\,\ref{fig:signal}a, while the efficiency of the
cuts, \ie the fraction of the signal events which have a like-sign
dilepton pair passing the cuts, is shown in Fig.\,\ref{fig:signal}b.
The dip in the efficiency between $140\mr{\GeV}<M_0<180\mr{\GeV}$ is
due to the resonant production of the second lightest neutralino
becoming accessible. Just above threshold the efficiency for this
channel is low due to the low $p_T$ of the lepton produced in the
slepton decay.

If we conservatively take a background of $7.6$ events, \ie 1$\sigma$
above the central value of our calculation, a 5$\sigma$ fluctuation of
the background would correspond to 20 events, using Poisson statistics.
This is given as a dashed line in Fig.\,\ref{fig:signal}a. As can be
seen for a large range of values of $M_0$ resonant slepton production
can be discovered at the LHC, for $\lam'_{211}=0.01$. The production
cross section depends quadratically on the \rpv\  Yukawa coupling and
hence it should be possible to probe much smaller couplings for small
values of $M_0$.

\begin{figure}
\begin{center}
\includegraphics[angle=90,width=0.48\textwidth]{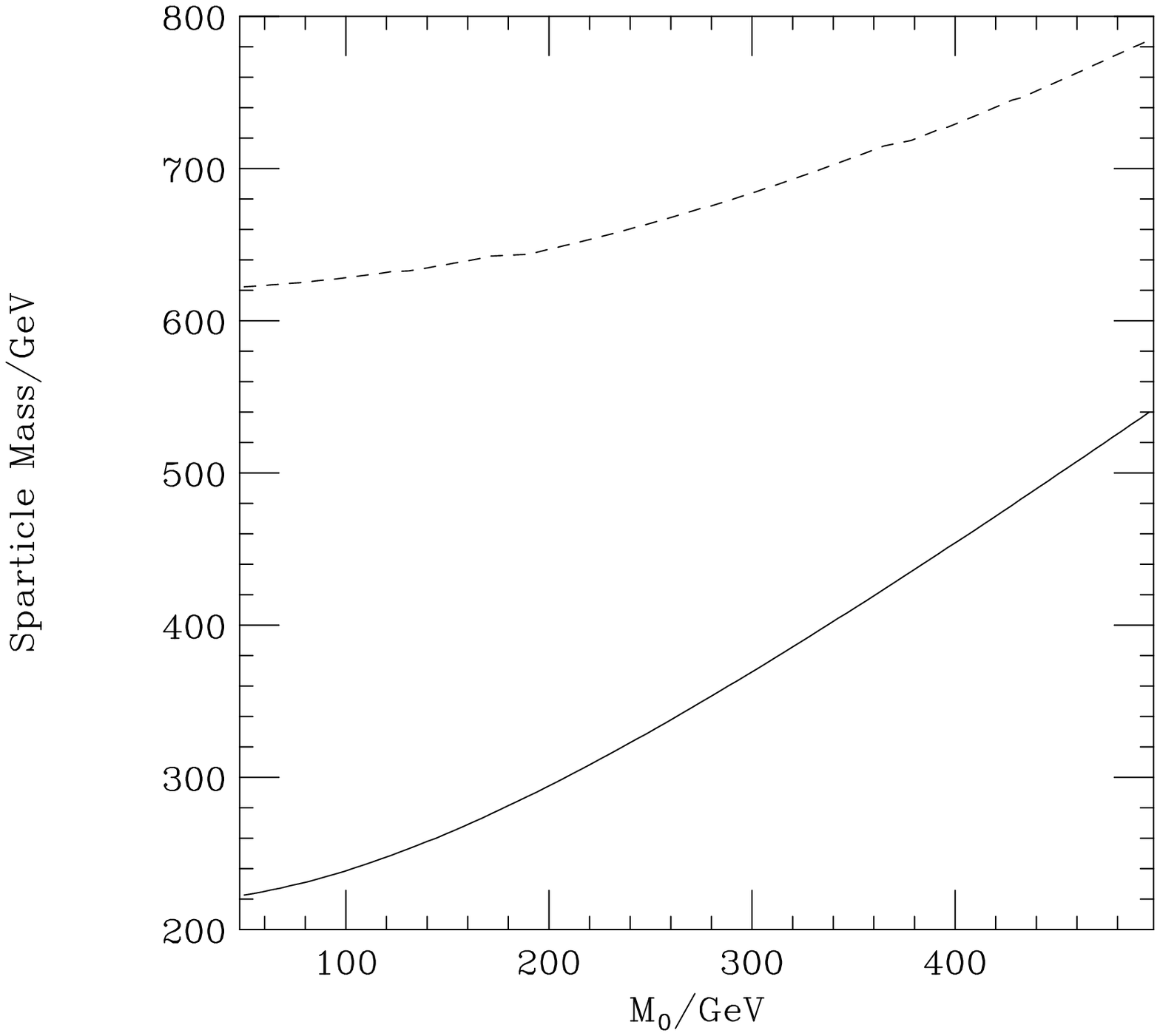}
\caption{Masses of the left smuon, solid line, and the right sdown,
dashed line, as a function of $M_0$ for 
$M_{1/2}=300$\GeV, $A_0=300$\GeV, $\tan\beta=2$, $\mr{sgn}\mu=+$.}
\label{fig:mass}
\end{center}
\end{figure}

As can be seen in Fig.\,\ref{fig:mass}, at this SUGRA point the sdown
mass varies between $622$\GeV\  at $M_0=50$\GeV\  and $784$\GeV\  at
$M_0=500$\GeV. The corresponding limit on the coupling ${\lam'}_{211}$
varies between 0.37 and 0.46.  We can probe couplings of ${\lam'}_{211}
=2\times10^{-3}$ for $M_0=50$\GeV\  which corresponds to a smuon mass
of $223$\GeV, and at couplings of ${\lam'}_{211}=10^{-2}$ we can probe
values of $M_0$ up to $500$\GeV, \ie a smuon mass of $540$\GeV. This
is more than an order of magnitude smaller than the current upper
bounds on the \rpv\  coupling given above for these values of $M_0$.
This is a greater range of couplings and
smuon masses than can be probed at the Tevatron 
\cite{Dreiner:1998gz}. The backgrounds are higher at 
the LHC but this is compensated by the higher energy and luminosity
leading to significantly more signal events.

\section{Conclusions}

  We have considered the backgrounds to like-sign dilepton production at
  the LHC and find a background after cuts of $5.1\pm2.5$ events for 
  an integrated luminosity of $10fb^{-1}$. This means,
  taking a conservative estimate of the background of 7.6 events, that
  20 events would correspond to a $5\sigma$ discovery. For a full
  analysis however, non-physics backgrounds must also be considered.

  A preliminary study of the signal suggests that an efficiency for detecting
  the signal in excess of 20\% can be achieved over a range of points in SUGRA
  parameter space. At the SUGRA point studied this means we can probe \rpv\  
  couplings of $2\times10^{-3}$ for a smuon mass of $223$\GeV\   and up to
  smuon masses of $540$\GeV\  for couplings of $10^{-2}$, and higher masses 
  for larger couplings.

  A more detailed scan of SUGRA parameter space for this signal remains to be
  performed.

\setcounter{figure}{0}
\setcounter{table}{0}
\setcounter{section}{0}
\setcounter{equation}{0}
\setcounter{footnote}{0}
\clearpage

\def\chione{\tilde \chi_1^0}
\def\l {\lambda }
\def \bt {\bar t }
\def \ud {{1 \over 2} }

\begin{center}

{\large \bf $\chione$ reconstruction in mSUGRA models with 
$R-$parity breaking $LQD$ term}\\
\vspace*{3mm}
{\sc L. MEGNER and G. POLESELLO} \\

\end{center}

\vspace*{3mm}
\begin{abstract}

The reconstruction of the $\chione$ LSP through the
decay to lepton-jet-jet is studied in the framework 
of the mSUGRA model. 
A detailed particle level analysis is performed 
on events generated with the HERWIG 6.0 Monte Carlo
and passed through the ATLAS fast simulation program ATLFAST.
The extraction of the $\chione$ mass peak and techniques 
for subtracting the combinatorial background are 
demonstrated on four example points in the mSUGRA parameter space.\\

\end{abstract}
\section[intro]{Introduction}
The $R-$parity violating (\rpv) extension of the MSSM contains the following  
additional terms in the superpotential,
which are trilinear in the quark and lepton superfields,  
\begin{eqnarray}
W_{R-odd}=\sum_{i,j,k} \bigg (\ud \l _{ijk} L_iL_j E^c_k+
\l ' _{ijk} L_i Q_j D^c_k+ \ud \l '' _{ijk} U_i^cD_j^cD_k^c   \bigg ), 
\label{super}
\end{eqnarray}
where $i,j,k$ are flavor indices.
The presence of R-violating terms will mainly manifest itself in two
ways: by production of single sparticles, or  by the fact
that the LSP is not stable and decays to standard model particles.
See \cite{Drein_me} for a recent review on the subject.

The reconstruction of the $\chione$ from its decay products in SUSY
particle pair production offers the possibility of detecting
$R-$parity violation for values of the $R$-violating couplings too
small to yield a significant cross-section for single sparticle
production.  The $\chione$ decay can be studied for all three coupling
types, $\l$, $\l '$ and $\l ''$ \cite{TDR_me} \cite{andy}.  The
hypothesis of a dominating $\lambda^{\prime}$ is the most favorable
case for the kinematic reconstruction of SUSY cascade decays, if the
decay of the $\chione$ into a charged lepton has a sizeable branching
fraction.  In this case, the charged lepton can be used as the
initiator of the mass reconstruction, yielding a reduced combinatorial
background with respect to three-jet case, and all three particles
from the $\chione$ decay are reconstructed in the detector, allowing
to fully reconstruct the $\chione$ mass peak.  In a pioneering work
\cite{connors} it was explicitly demonstrated that with appropriate
selection criteria on jets and leptons a clear $\chione$ peak could be
observed in the lepton-jet-jet invariant mass over an acceptable
combinatorial background. That work was based on $\chione$ decaying
with 100$\%$ branching fraction to $eqq'$, with the momenta of the
decay products distributed according to the phase space.  The present
study is based on a detailed implementation of the $R-$parity
violating decays in the HERWIG 6.0 Monte Carlo including matrix
elements for the $\chione$ decay \cite{herwig_me}.  We aim to to
verify if the $\chione$ peak can be observed for a broad range of SUSY
models, and to develop techniques for evaluating and subtracting the
combinatorial background, as a starting point for performing precision
measurements of the parameters of the underlying model.  \par

\section{The SUSY model}

We work in the framework of the minimal Supergravity inspired model
(mSUGRA) as implemented in the ISASUSY Monte Carlo \cite{isajet_me}.  The
SUSY events were generated with version 6.0 of the HERWIG Monte Carlo
\cite{herwig_me}, which implements all the $R-$parity conserving
two-to-two SUSY processes.  This version of HERWIG also includes the
simulation of all $R-$parity violating sparticle decays
\cite{drein_me}. The produced events were then passed through
ATLFAST, a fast simulation of the ATLAS detector \cite{ATLFAST_me}.

We assumed for this study a single dominant R-violating coupling
$\lambda^{\prime}_{111}$=0.01. The results obtained are independent of
the precise value of the $\lambda^{\prime}$ coupling, as long as it
induces a prompt $\chione$ decay.  The results are valid for all the
couplings of the type $\lambda^{\prime}_{ijk}$ with $i=1,2$. They
should be taken with some care if $j$ or $k$ is equal to three, since
in this case a $b$ quark is among the $\chione$ decay products.

In models with a single dominant $\lambda^{\prime}$ term $\chione$
has two decay modes: $\chione \to l^{\pm}qq^{\prime}$ and $\chione
\to\nu q\bar q$.  The relative branching ratio of the two modes is a
function of the parameters of the model.  The mode with a charged
lepton is the better mode for reconstructing the $\chione$ mass,
because there is no neutrino which escapes detection.  We have
verified \cite{gplm} that the charged lepton mode has a significant
branching fraction over all of the mSUGRA parameters space allowed by
present experimental searches. \par

\section{$\chione$ reconstruction in mSUGRA models}

\subsection{Introduction}

The possibility to extract a $\chione$ mass peak is determined by:
\begin{itemize} 
\item
the number of produced $\chione$ 
for which the decay $\chione\to l$-jet-jet with $l=e,\mu$  
can be completely reconstructed in the detector;
\item
the combinatorial background, both from SUSY events 
and from Standard Model processes.
\end{itemize}
These factors depend on the SUSY model considered, in particular the
$\chione$ mass, the squark and gluino masses, and the jet and lepton
multiplicity in SUSY events.  Our aim is to determine if the
reconstruction can be performed for the full parameter space. For this
reason we chose to perform a detailed study for a few representative
points in parameter space, trying to span a range as broad as possible
of the parameters listed above.  A convenient choice are the models
already studied in detail for R-conserving mSUGRA \cite{TDR_me}, which
present very varied phenomenologies, and also give the possibility to
cross-check the results of our analysis with previous detailed
studies. We have therefore selected points 1, 4 and 5, and we have
added a low mass point, which we call 7, in a region with low
branching fraction ($\sim 27\%$) for the decay $\chione \to
l^{\pm}qq^{\prime}$.  The mSUGRA parameters of the studied points are
given in Table~\ref{tabpoints}.

\begin{table}[htb]
\begin{center}
\begin{tabular}{|l|r|r|r|r|r|}
\hline
Point &  $m_0$  &    $m_{1/2}$  &    $A_0$   &  $\tan\beta$ &  sgn($\mu$) \\
\hline
    1   &   400   &     400 &   0 &  2.0 & + \\
    4 & 800 &  200 &  0 &  10.0 &  +  \\
    5 & 200 &  300 & 300 & 2.1 & +   \\
    7 & 200 &  200& 0 & 2.0 & +  \\
\hline
\end{tabular}
\caption{\em Parameters of the example mSUGRA points studied.}
\label{tabpoints}
\end{center}
\end{table}

\subsection{Selection criteria for jets}

In order to extract a significant $\chione$ peak the selection
criteria need to be optimized separately for each addressed model.  At
this level, the important parameter is not the purity of the obtained
$\chione$ peak, but its significance, and the possibility of precisely
measuring the peak position.

\begin{figure}[p]
\begin{center}
\dofig{\textwidth}{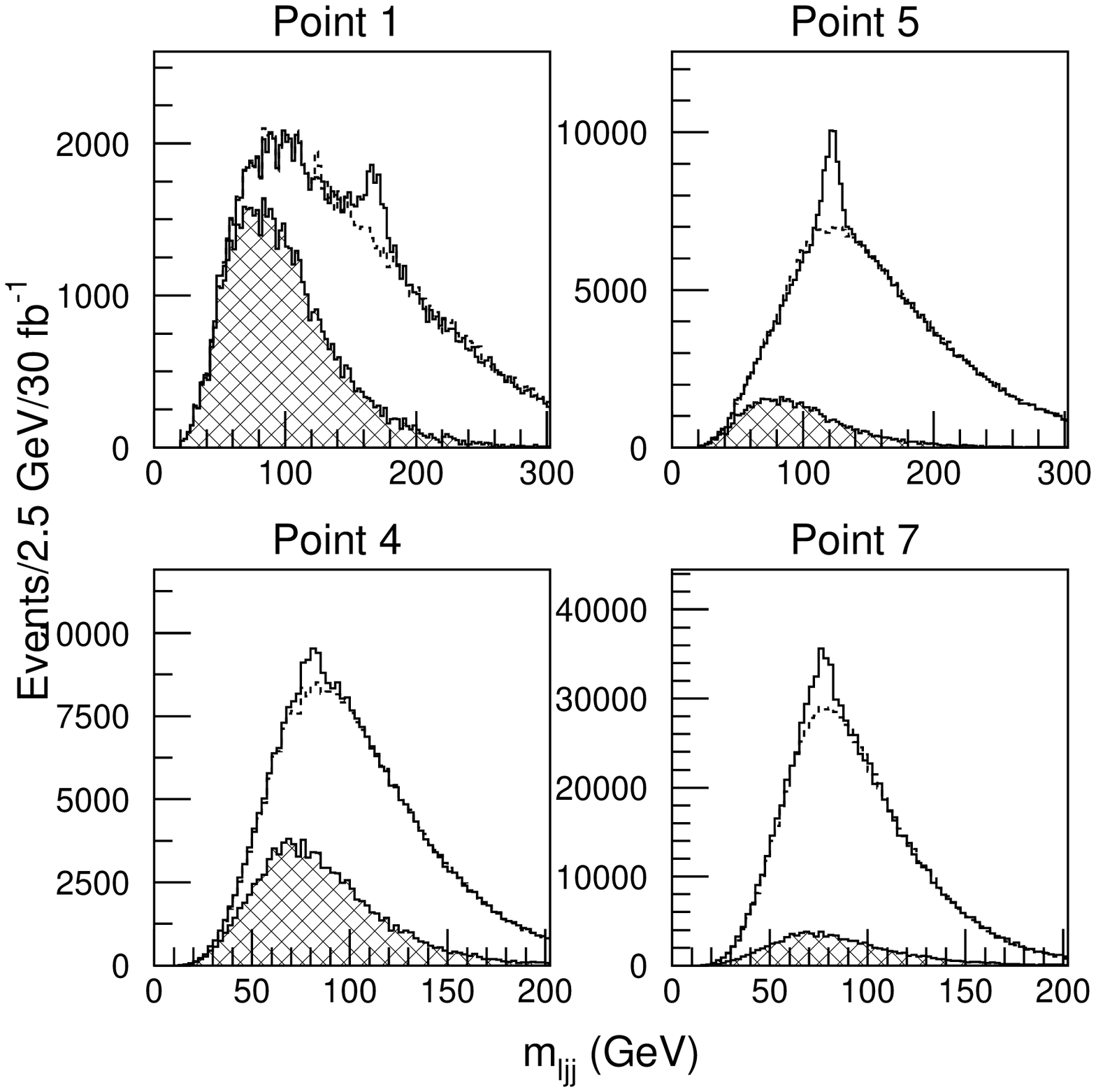}
\caption{\em Invariant mass of the accepted $m_{ljj}$ 
combinations for the sum of the SUSY signal and the SM $t\bt$ 
background for  the four mSUGRA points studied (solid line).
The background from $t\bt$ production is shown as a hatched
histogram. The dashed lines show the evaluation
of the combinatorial background performed with the `combined' 
method described in the text.
The statistics shown correspond to an integrated
luminosity of 30~fb$^{-1}$. }
\label{pointall}
\end{center}
\end{figure}

An initial event selection was  first made in order to reject the 
Standard model background. The events were required to have:
\begin{itemize}
\item at least 6 jets with p$_{\rm{t}} >$ 15 GeV
\item at least one lepton with p$_{\rm{t}} >$ 20 GeV in point 4 and
point 7, and at least two leptons with  p$_{\rm{t}} >$ 10 GeV in point
1 and point 5. 
\end{itemize}
The reason for requiring two leptons in point 1 and 5 is that a
stronger $t\bt$ background rejection is needed, since the SUSY
cross-section for these points is lower.  These are preliminary cuts
and will be further optimized in the final analysis \cite{gplm}.  For
the $\chione$ reconstruction the jets with the highest momentum were
not considered, since there are likely to come from the start of the
decay chain.  The 3rd to the 8th jet in decreasing momentum scale were
considered in point 1 and 5 and the 5th to the 8th jets in point 4 and
7. The $\chione$ was reconstructed starting from an identified
electron, and the invariant mass was calculated for the
electron-jet1-jet2 combinations such that $\Delta R$(electron-j1)$<2$,
$\Delta R$(electron-j2)$<2$, $\Delta R$(j1-j2)$<2$ with $\Delta
R=\sqrt{\Delta\eta^2+\Delta\phi^2}$, where $\eta$ is the
pseudorapidity, and $\phi$ the angle in the plane perpendicular to the
beam axis.  In figure~\ref{pointall} the invariant mass distribution
obtained in this way is shown in solid lines for the four points.  A
statistically significant peak, corresponding to the $\chione$ mass is
seen in all four plots.  The hatched histograms superimposed on the
plots are the $t\bt$ background.  This important background can easily
be evaluated and subtracted by performing the $\chione$ reconstruction
procedure using the wrong flavor of the lepton as described in the
`wrong flavor' technique below.  In the next section we concentrate
on the more difficult combinatorics from SUSY events.
 
\subsection{Subtraction of combinatorial background}
In order to estimate the shape of the combinatorial background 
two different techniques were combined.

In the `wrong flavor' technique, the combinatorial background is
evaluated using all of the lepton-jet-jet combinations satisfying the
cuts for the $\chione$ candidate, but using the leptons with the wrong
flavor, i.e. electrons if the $\chione$ decays into muons and vice
versa.

In the `mixed events' technique, the idea is to construct
lepton-jet-jet combinations and make sure that not all objects in such
a combination come from the same $\chione$, i.e. the combination does
not contribute to the signal. In order to make sure that this is the
case we use two different events (respectively 1 and 2).  The
directions ($\eta$,$\phi$) of the three objects (lepton, jet, jet) and
the momenta of two of the objects (lepton-jet or jet-jet) are taken
from one of the candidate $\chione$ combinations in event 1. The
momentum of the third object is taken from a combination in event 2.

These two approaches have complementary strengths and weaknesses.  The
`wrong flavor' approach correctly subtracts all of the combinatorial
background not involving one of the actual leptons from $\chione$
decay, whereas the `mixed events' background approximately accounts
for events where only two of the three particles come from a $\chione$
decay.\par

In order to combine the strengths of the two methods the background is
estimated through a linear combination of the backgrounds calculated
with the two previous methods. The relative weight of the two
components is a function both of the characteristics of the SUSY model
and of the applied cuts. In each case it is therefore necessary to
optimize the parameters of the linear combination.  A prescription
which was found to give reasonable results is:
\begin{itemize}
\item
Normalize the backgrounds estimated with the wrong lepton technique and with 
mixing events technique to the measured mass spectrum separately. This 
normalization is done in a region outside of the $\chione$ peak.
\item
Evaluate the relative weight of the two different background components
by performing a least-square fit to the observed $m_{ljj}$ mass distribution
outside of the peak. 
\end{itemize}

\begin{figure}[p]
\begin{center}
\dofig{\textwidth}{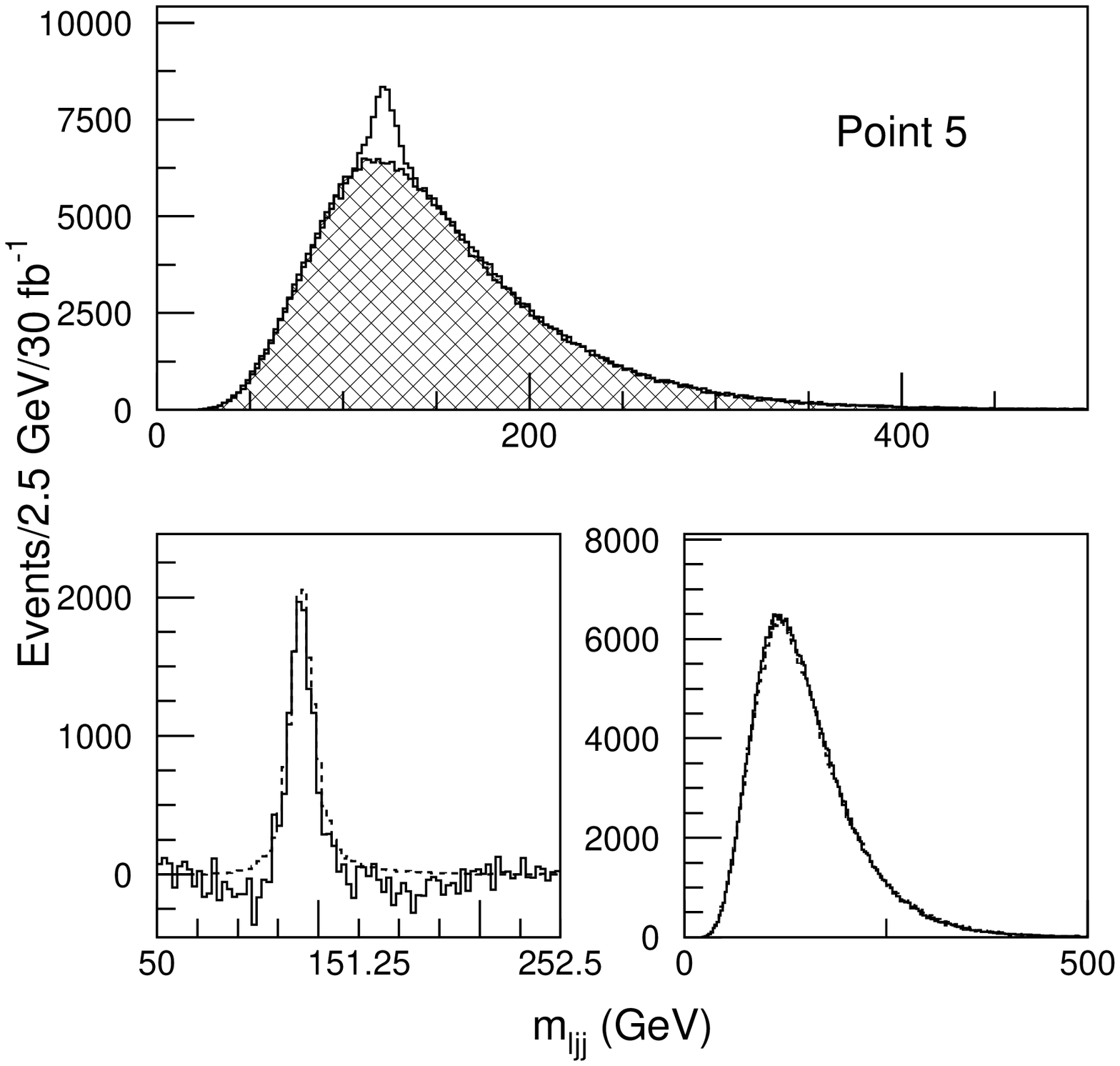}
\caption{\em Point 5: Invariant mass of the $m_{ljj}$ combinations 
for SUSY events. 
In the upper plot the mass distribution for the 
combinatorial background evaluated with the `combined' method (hatched)
is shown superimposed on the observed signal$+$background
distribution (solid line).  In the lower plot the mass distributions
for signal  (left) and background (right) are shown separately.
The actual signal/background distributions are shown as dashed lines. 
The solid lines are the corresponding distributions evaluated 
with the `combined' method. The statistics shown corresponds to an integrated 
luminosity of 30~fb$^{-1}$. } 
\label{p5all}
\end{center}
\end{figure}

The relative weight factors obtained exhibit some sensitivity to the
choice of the mass windows over which the least square fit is
performed.  The `real' background shape is best reproduced when the
lower limit of the mass window is as near as possible to the peak
position.  Thus the estimate of the peak purity is affected by a
systematic uncertainty, whereas the peak position has been checked to
be independent of the choice of the subtraction technique. The sum of
the background calculated with this combined method and the $t\bt$
background is shown for all the four points as dashed lines in
figure~\ref{pointall}. In all four cases the distribution for the
combinatorial background is nicely reproduced. \\ We show in
Figure~\ref{p5all} the results for Point 5 in order to illustrate the
power of this method. In the upper plot the invariant mass
distribution for all the accepted electron-jet-jet combinations in
SUSY events (solid line) is shown.  The combinatorial background
evaluated with the `combined' method is superimposed as a hatched
histogram.  The lower left plot shows the signal peak after background
subtraction (full line); the actual signal from $\chione$ decay is
superimposed (dashed line).  Likewise, the full line in the lower
right plot is the background distribution estimate obtained with the
`combined method', and the dashed line is the mass distribution for
actual background combinatorics. Distributions showing the same
excellent agreement are obtained for all four considered models.

\section{Conclusions}

We have performed a reconstruction of $\chione$ for the decay $\chione
\to l^{\pm}qq^{\prime}$ for different scenarios within the mSUGRA
model with an $R-$parity violating $\lambda^{\prime}$ term. We have
shown that it is possible to define simple cuts on the jet and lepton
multiplicity and topology which allow to observe a statistically
significant peak in the lepton-jet-jet invariant mass distribution, at
the position of the $\chione$ mass. We have discussed a method to
subtract the combinatorial background.  Using this subtraction the
$\chione$ mass and the number of reconstructed $\chione$s can be
precisely measured. This technique works effectively for all four
points considered in this analysis. The mass of the reconstructed
$\chione$ can be used as a starting point for the reconstruction of
particles further up in the decay chain.

\setcounter{figure}{0}
\setcounter{table}{0}
\setcounter{section}{0}
\setcounter{equation}{0}
\setcounter{footnote}{0}
\clearpage

\begin{center} 
{\large\bf Supersymmetry with $R$ parity violation at the LHC:
discovery potential from single top production} \\
\vspace*{3mm}
 
{\sc P. CHIAPPETTA, A. DEANDREA, E. NAGY, S. NEGRONI, G. POLESELLO and
J.M. VIREY}\\ 
 
\end{center}
\vspace*{0.3cm} 

\begin{abstract}
We study single top production through $R$-parity violating 
Yukawa type couplings at the LHC. 
We consider all $2\rightarrow 2$ partonic processes at tree-level, 
including interference terms. The calculated $2\rightarrow 2$ partonic 
cross sections are implemented in PYTHIA to generate complete 
particle final states. The generated events are processed through 
a fast particle level simulation of the ATLAS detector. We take into account 
all important SM backgrounds and study the signal-to-background 
ratio as a function of the initial partonic states, the exchanged 
sparticle mass and width, and of the value of the Yukawa couplings.
\end{abstract}
\vspace{0.4cm}
 
\noindent
The feasibility of single top quark production via squark and slepton 
exchanges  
to probe several combinations of $R$ parity violating couplings at
hadron colliders has been studied \cite{lavori}. 
According to those studies, the LHC is better 
at probing the $B$ violating couplings $\lambda^{\prime \prime}$
whereas the Tevatron and the LHC have a similar sensitivity to $\lambda'$
couplings. We perform a complete and detailed  
study including all signal channels using a Monte Carlo  
generator based on Pythia 6.1 \cite{sjostrand}, taking into account all the 
backgrounds and including the ATLAS detector  
response using ATLFAST 2.0 \cite{atlfast_de}.    

The $R$-parity violating parts of the Lagrangian that contribute to single  
top production are:  
\be 
L_{\not {\! R}} =  \lambda'_{ijk} {\tilde e}^i_L {\bar d}^k_R u^j_L -    
\lambda^{''}_{ijk} ({\tilde d}^k_R {\bar u}^i_L d^j_L + {\tilde d}^j_R  
({\bar d}^k_L)^c  u^i_L) + h.c.  
\label{eq2} 
\ee 
The superscript $c$ corresponds to charge conjugation. 
There are altogether 27 and 9 $\lambda'_{ijk}$ and $\lambda''_{ijk}$ Yukawa 
couplings, respectively. 
The most suppressed couplings are $ \lambda'_{111}$,  
$ \lambda'_{133}$, $\lambda^{''}_{112}$,   
$\lambda^{''}_{113}$  (see \cite{gdr} for detailed up to date reviews of the
existing bounds). In order to fix
the kinematical variables, the reaction we consider is 
\be 
u_i(p_1) + d_j (p_2) \rightarrow t(p_3) + b(p_4) \;\; , 
\label{eq3} 
\ee 
the $p_k$ being the 4-momenta of the particles and the indices $i$ and 
$j$ refer to the generations of the $u$ and $d$-type quarks. 
 
For valence-valence (VV) or sea-sea (SS) subprocesses, the scalar
slepton exchange in the $\hat u$-channel is taken into account but
appears to be suppressed within our assumptions about the $\lambda'$
couplings and sfermion masses. The down type squark exchange in the
$\hat s$-channel squared amplitude is dominant.

Let us now consider the subprocesses involving valence-sea (VS)
quarks.  Concerning $R$ parity violating terms, slepton exchange in
the $\hat s$-channel and down type squark exchange in the $\hat
u$-channel contribute.  The dominant terms are the squared amplitude
due to $\tilde e$ exchange, and for initial quarks of the same
generation ($i=j$), the interference between $W$ and $\tilde d$. The
result is sensitive to the interference term only if the product of
$\lambda''$ couplings is large (around $10^{-1}$).  For subprocesses
involving quarks of different generations in the initial state the
situation is more complex and all amplitudes have to be taken into
account.

We have carried out a feasibility study to detect  
single top production through $R$-parity violation at the LHC by 
measuring the $l\nu bb$ final state using the following procedure. 
 
First, we have implemented the partonic $2 \to 2$ 
cross sections  
in the PYTHIA event generator. Providing PYTHIA with 
the flavors and momenta of the initial partons 
using a given parton distribution function (p.d.f.)\footnote{ 
We have used the CTEQ3L p.d.f.},
complete final states including initial and final state 
radiations and hadronization are generated.  
 
The events were processed through ATLFAST to simulate the 
response of the ATLAS detector. In particular the energy 
of electrons, photons and muons  was smeared according to the 
resolution of the relevant detector element in the
pseudorapidity range $|\eta | < 2.5$. Finally, 
a simple fixed cone algorithm (of radius $R=0.4$) was used to 
reconstruct the parton jets. 
The minimum transverse energy of a jet was set at 15 GeV. 
According to the expected b-tagging performance of the ATLAS detector
\cite{TDR_de} for low luminosity at the LHC we have assumed a $60\%$ b-tag 
efficiency for a factor 100 of rejection against light jets.  
 
The same procedure was applied to the SM background with the 
exception that we used besides PYTHIA also the ONETOP~\cite{onetop}  
event generator. 
 
The integrated luminosity for one year at low luminosity at the LHC is
taken to be 10 fb$^{-1}$. In Table~\ref{tb:sq_xsec} we display the
total cross section values for different initial parton flavors in
the case of exchanged squarks of mass of 600 GeV and of $R$-parity
conserving width $\Gamma_{R}$ = 0.5 GeV. We took for all $\lambda '' =
10^{-1}$, which yields a natural width of the squark which is smaller
than the experimental resolution. Table~\ref{tb:sl_xsec} contains the
same information for slepton exchange ( $\lambda ' = 10^{-1}$, for a
slepton of mass of 250 GeV and a width of $\Gamma_{R}$ = 0.5 GeV).
Other processes are not quoted because the small value of the limits
of their couplings prevents their detection.

\begin{table}[htb] 
\begin{center}
\vspace*{-4mm}
\begin{tabular}{|c|c|c|c|c|c|}
\hline
Initial partons   & $cd$    &  $cs$    &  $ub$   &  \multicolumn{2}{ c|}{$cb$} \\ 
\hline 
Exchanged particle   
&$\tilde s$ & $\tilde d$ &$ \tilde s$ & $ \tilde d$ & $ \tilde s$ 
\\ \hline 
Couplings    
& $\lambda''_{212}\lambda''_{332}$   & $\lambda''_{212}\lambda''_{331}$   
& $\lambda''_{132}\lambda''_{332}$   & $\lambda''_{231}\lambda''_{331}$   
& $\lambda''_{232}\lambda''_{332}$   \\ \hline 
Cross section in pb    
& 3.98  & 1.45   & 5.01  &  \multicolumn{2}{ c|} {0.659}\\
\hline
\end{tabular} 
\caption{Total cross-section in pb for squark exchange in the $\hat s$-channel 
for a squark of mass of 600 GeV assuming $\Gamma_{R}$ = 0.5 GeV.} 
\label{tb:sq_xsec} 
\end{center}
\end{table} 
\vspace*{-2mm}

In order to study the dependence of the signal on the mass and the
width of the exchanged particle we have fixed the couplings to
$10^{-1}$ and have chosen three different masses for the exchanged
squarks: 300, 600 and 900 GeV, respectively. For each mass value we
have chosen two different $\Gamma_R$: 0.5 and 20 GeV, respectively.
For the first case $\Gamma_{\mathrm{\not\!R_p}}$ dominates, whereas in
the last one, when $\Gamma_{tot} \approx \Gamma_R$, the single
top-production cross section decreases by a factor $\sim 10$.  We have
considered here the $ub$ parton initial state, since this has the
highest cross section value.
 
In order to study the dependence on the parton initial state we have fixed 
the mass of the exchanged squark to 600 GeV and its width with 
$\Gamma_R$ = 0.5 GeV and varied the initial state according to the 
first line of Table~\ref{tb:sq_xsec}. 
 
Finally, for the exchanged sleptons we have studied only one case, 
namely the $u\bar d$ initial state with a mass and width of the exchanged 
slepton of 250 GeV and 0.5 GeV, respectively. 
In each case we have generated about $10^5$ signal events. 

The irreducible backgrounds are single top production through a  
virtual $W$ (noted $W^*$), or through $W$-gluon fusion.  
$W$-gluon fusion is the dominant process (for a detailed study 
see \cite{stelzer}). 
A $Wbb$ final state can be obtained either in direct production 
or through $Wt$ or $t \bar t$ production.   
Finally, the reducible background consists of $W+$n$j$ events where two of the 
jets are misidentified as $b$-jets. 
 
For the $tb$ final state first we 
reconstruct the top quark. 
The top quark can be reconstructed from the $W$ and from one of the  
$b$-quarks in the final state, requiring that their invariant mass satisfy  
$150 \leq M_{Wb} \leq 200$ GeV. 
The $W$ can be in turn reconstructed from either of the two decay channels: 
$W \rightarrow u \bar d$, $W \rightarrow l \nu$.  
Here we have considered only the latter case which gives a better  
signature due to the presence of a high $p_t$ lepton 
and missing energy. The former case suffers from multi-jets event  
backgrounds.  
As we have only one neutrino, its longitudinal momentum  can be reconstructed 
by using the $W$ and top mass constraints. 
The procedure used is the following : \\ 
- we select events with two b-jets of $p_t \ge 40$ GeV, with one lepton of 
$p_t \ge 25$ GeV, with $E_t^{miss} \ge  35$ GeV  and with a jet multiplicity 
$\le 3$,\\ 
- we reconstruct the longitudinal component ($p_z$) of the neutrino by  
requiring $M_{l\nu}$ = $M_W$. 
This leads to an equation with twofold ambiguity on $p_z$.  \\ 
- More than 80$\%$ of the events have at 
least one solution for $p_z$. In case of two solutions, we calculate $M_{l\nu  
b}$ for each of the two b-jets and we keep the $p_z$ that minimizes  
$|M_{top}-M_{l\nu b}|$.\\ 
- we keep only events where $150  \le   M_{l\nu b} \le 200$ GeV.\\ 
Next, the reconstructed top quark is combined with the 
$b$ quark not taking part in the top reconstruction.
In order to reduce the $t\bar t$ background to a manageable level, 
we need to apply a strong jet veto on the third jet by requiring that its 
$p_t$ should be $\le 20$ GeV. 

The invariant mass distribution of the $tb$ final state after the cuts
described above is shown in Fig.~1.  Once an indication for a signal
is found, we count the number of signal ($N_s$) and background ($N_b$)
events in an interval corresponding to 2 standard deviations around
the signal peak for an integrated luminosity of 30 fb$^{-1}$. Then we
rescale the signal peak by a factor $\alpha$ such that $N_s/\sqrt{N_b}
= 5$.  By definition the scale-factor $\alpha$ determines the limit of
sensitivity for the lowest value of the $\lambda''$ ($\lambda'$)
coupling we can test with the LHC:
$\lambda''_{ijk}\cdot\lambda''_{lmn} \le 0.01\cdot \sqrt\alpha$.
 
\begin{figure}
\epsfxsize=7cm
\vspace*{-15mm}
\centerline{\epsffile{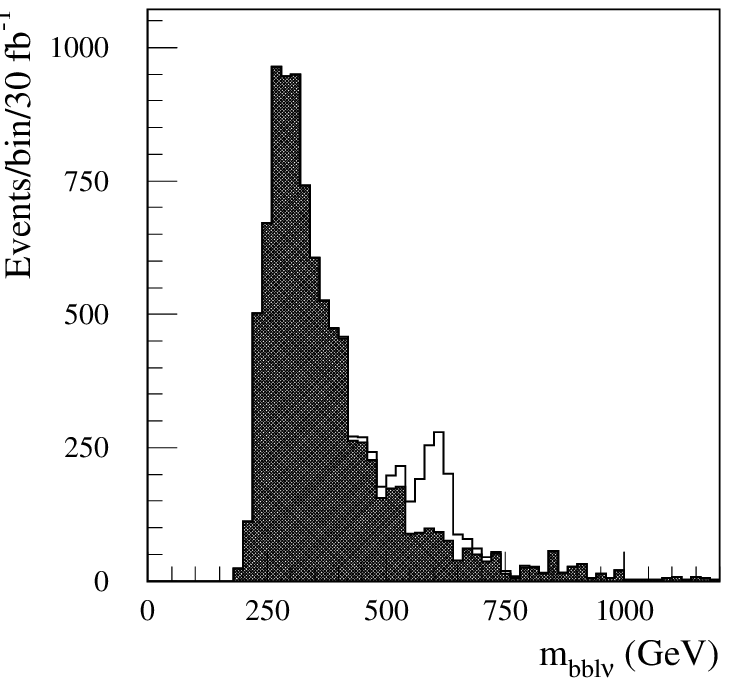}}
\noindent
{\bf Fig. 1} - {Invariant mass distribution of $l\nu bb$ for the signal and 
backgrounds (dashed histogram) after three years at LHC at low luminosity 
after having applied the cuts.}
\end{figure}

In Table~\ref{tb:lambda_ub} we show the limits obtained for the 
combinations of $\lambda''_{132}\lambda''_{332}$ for different masses 
and widths of the exchanged $\tilde s$-quark. Also shown are the current 
limits assuming a mass for $\tilde m_f$ = 100 GeV, 
the number of signal and background 
events, as well as the experimentally observable 
widths of the peak ($\Gamma_{exp}$). 

\begin{table} 
\begin{center} 
\vspace*{-10mm}
\begin{tabular}{|c|c|c|c|c|c|c|}
\hline
{$m_{\tilde s}$ (GeV) } & \multicolumn{2} {c|} {300}&\multicolumn{2} {c|} 
{600}&\multicolumn{2} {c|} {900} \\ \hline 
$\Gamma_R$ (GeV)  & 0.5 & 20 & 0.5 & 20 & 0.5 & 20 \\ \hline 
$N_s$            & 6300 & 250 & 703 & 69 & 161 & 22 \\ \hline 
$N_b$            & 4920 & 5640& 558 & 1056& 222& 215 \\ \hline 
$\Gamma_{exp}$ (GeV)    & 24.3 & 30.5 & 37.5 & 55.6 & 55.4 & 62.1 \\ \hline 
$\lambda''\times\lambda''$ & 2.36$\times10^{-3}$&1.21$\times10^{-2}$ 
& 4.10$\times10^{-3}$& 1.51 
$\times10^{-2}$& 6.09$\times10^{-3}$& 2.09$\times10^{-2}$\\
\hline
\end{tabular} 
\caption{Limits for the values of the $\lambda''_{132}\lambda''_{332}$ 
Yukawa couplings for an integrated luminosity of 30 fb$^{-1}$. For the 
other quantities see the text. The current limit is 6.25$\times10^{-1}$.}  
\label{tb:lambda_ub} 
\end{center}
\end{table}  

In Table~\ref{tb:lambda_600} we compile the sensitivity limit of the 
bilinear combination of the different Yukawa couplings one can 
obtain after 3 years of LHC run with low luminosity, if the exchanged 
squark has a mass of 600 GeV. 
 
\begin{table}[h!]
\begin{center}
\vspace*{-0mm}
\begin{tabular}{|c|c|c|c|c|c|}
\hline
Initial partons   & $cd$    &  $cs$    &  $ub$   
&  \multicolumn{2}{ c|}{$cb$} \\ 
\hline 
Exchanged particle   &$\tilde s$ & $\tilde d$ &$ \tilde s$ & $ \tilde d$ & $ \tilde s$ 
\\ \hline 
Couplings
& $\lambda''_{212}\lambda''_{332}$   & $\lambda''_{212}\lambda''_{331}$   
& $\lambda''_{132}\lambda''_{332}$   & $\lambda''_{231}\lambda''_{331}$   
& $\lambda''_{232}\lambda''_{332}$   \\ \hline 
$N_s$   & 660  & 236 & 703 & \multicolumn{2}{ c|} {96}  \\ \hline 
$N_b$   & \multicolumn{5}{ c|} {558}  \\ \hline 
$\Gamma_{exp}$ (GeV) & 38.5  & 31.3 & 37.5 &   \multicolumn{2}{ c|} {40.1} \\
\hline 
Limits on $\lambda''\times\lambda''$   & 4.26$\times10^{-3}$ 
& 7.08$\times10^{-3}$  
& 4.1$\times10^{-3}$ &  
\multicolumn{2}{ c|} {1.11$\times10^{-2}$}\\
\hline
\end{tabular} 
\caption{Limits on the Yukawa couplings for an exchanged squark of mass 
600 GeV assuming $\Gamma_R$ = 0.5 GeV, for an integrated luminosity 
of 30 fb$^{-1}$. Current limit is 6.25$\times10^{-1}$.} 
\label{tb:lambda_600} 
\end{center}
\end{table}

For the exchanged sleptons (cf Table~\ref{tb:sl_xsec}) 
we have calculated the sensitivity limit of the 
bilinear combination of the different Yukawa couplings 
only for the most favorable case, i.e. for the $u\bar d$
partonic initial state. 
We obtain 4.63$\times10^{-3}$ for the 
limits on $\lambda'_{11k}\lambda'_{k33}$ (in comparison with the limit of 
2.8$\times10^{-3}$ obtained by Oakes {\it et al.}). 

\begin{table}[h!] 
\begin{center} 
\vspace*{-0mm}
\begin{tabular}{|c|c|c|c|c|c|c|}
\hline
Initial partons & 
$u\bar d$ & $u\bar s$ & $c\bar d$ & $c\bar s$ & $u\bar b$ & $c\bar b$ \\
\hline
Couplings
& $\lambda'_{11k}\lambda'_{k33}$   & $\lambda'_{12k}\lambda'_{k33}$   
& $\lambda'_{21k}\lambda'_{k33}$   & $\lambda'_{22k}\lambda'_{k33}$   
& $\lambda'_{13k}\lambda'_{k33}$   & $\lambda'_{23k}\lambda'_{k33}$   
\\ \hline 
Cross section in pb    
& 7.05  & 4.45   & 2.31  &  1.07  & 2.64  & 0.525\\
\hline
\end{tabular} 
\caption{Total cross-section in pb for slepton exchange in the $\hat 
s$-channel for a slepton of mass of 250 GeV assuming $\Gamma_{R}$ = 0.5 GeV.} 
\label{tb:sl_xsec} 
\end{center}
\end{table} 

For those cases where the exchanged squark (slepton) can be 
discovered at the LHC we have made an estimate on the precision 
with which one can determine 
its mass. For this purpose, we have subtracted the background under the mass 
peak and fitted a Gaussian curve on the remaining signal. 
For the assumed value of the coupling constant, the error on 
the mass determination 
is dominated by the $1\%$ systematic uncertainty on the jet energy scale 
in ATLAS\cite{TDR_de}.

\setcounter{figure}{0}
\setcounter{table}{0}
\setcounter{section}{0}
\setcounter{equation}{0}
\setcounter{footnote}{0}
\clearpage

\def\ie{ {\it i.e.} }
\def\eg{ {\it e.g.} }
\def\etal{ {\it et al.} }
\def\dis{\displaystyle}
\def\mand{\qquad {\rm and} \qquad}
\newcommand{\eqn}[1]{(\ref{#1})}
\newcommand{\rpar}{$R$-parity}

\begin{center}
{\large \bf Probing \rpv\ couplings through indirect effects 
on Drell-Yan production at the LHC}
\\
\vspace*{3mm}
{\sc D. CHOUDHURY and  R.M. GODBOLE }\\

\end{center}
\vspace*{3mm}

\begin{abstract}
In this note we analyze the sensitivity to \rpv\ couplings at the LHC
through the measurement of dilepton pairs, taking $\lambda'_{211}$ as
example. We show that by exploiting the differences in the kinematic
distributions of the \rpv\ contributions from that of the SM case, we
can do significantly better than the bounds obtained from low energy
experiments. Further the possible LHC bound on the couplings does not
scale as the mass of the squark. Similar analysis can also be done to
assess the reach of the LHC to effects due to
leptoquarks/compositeness etc.
\end{abstract}

Within the Standard Model, electroweak gauge invariance ensures both
lepton number and baryon number conservation (at least in the
perturbative context). However, this is not so within the Minimal
Supersymmetric Standard Model (MSSM). In most standard treatments,
such interactions are avoided by the imposition of a discrete symmetry
called \rpar, which implies a conserved multiplicative quantum number,
$R\equiv (-1)^{3B+L+S}$, where $B$ is baryon number, $L$ is lepton
number, and $S$ is spin \cite{ff}.  All ordinary particles are \rpar\
even, while all superpartners are \rpar\ odd.  If \rpar\ is conserved,
superpartners must be produced in pairs and the lightest superpartner,
or LSP, is absolutely stable. Unfortunately, \rpar\ invariance of the
SUSY Lagrangian is an \textit{ad hoc} assumption and not derived from
any known fundamental principle.  Hence, it is of interest to consider
\rpar--violating extensions of the MSSM, especially since the
experimental signatures of low energy supersymmetry would then be
radically different. Curiously, \rpv\ interactions can improve the
agreement between theory and precision electroweak measurements, and
also offer explanations \cite{HERA_expl} for experimental anomalies
such as the high-$Q^{2}$ excess reported at HERA \cite{HERA}.
  
The most general \rpv\ terms in the superpotential consistent with 
Lorentz invariance, gauge symmetry, and supersymmetry 
are\footnote{We do not consider  here  the bilinear terms that mix lepton and 
Higgs superfields \protect\cite{hallsuz}.} 
%
\be
W_{\not{R}}  =  \lambda_{ijk} L_i L_j E^c_k +
                        \lambda'_{ijk} L_i Q_j D^c_k +
                        \lambda''_{ijk} U^c_i D^c_j D^c_k ,
\label{eq:superpot}
\ee
where $L_i$ and $Q_i $ are the $SU(2)$ doublet lepton and quark fields
and $E^c_i, U^c_i, D^c_i$ are the singlet  superfields.
The $UDD$ couplings violate baryon number while each 
of the other two sets violate lepton number. 
As our aim is to explore dilepton final states
in $pp$ collisions,  we shall explicitly forbid the $UDD$ 
interactions~\cite{leptopar} as an economical way to 
avoid unacceptably rapid proton decay.  

The remarkable agreement between low-energy experimental data and SM
expectations imply quite severe bounds on the strength of many \rpv\
operators \cite{BGH}.  Thus, within the context of colliders it is
natural to consider the production of supersymmetric particles to be
governed mainly by gauge interactions with \rpv\ being important only
in the subsequent decay\cite{collider_decays}. However, such analyses
are neither very sensitive to the actual size of the \rpv\ coupling
nor are they particularly useful for the case of heavy superpartners
(when the pair-production cross sections are smaller).  A
complementary approach\cite{b_c_s,Joanne_rizzo} is afforded by
processes like Drell-Yan production where the exchange of a heavy
squark is governed by the relevant \rpv\ coupling. In this note, we
refine the analysis of Ref.\cite{b_c_s} in the context of the LHC,
thereby extending its reach.

Expanding the superfield components in 
\eqn{eq:superpot}, we obtain the interaction Lagrangian that 
connects quarks to leptons:
\be
{\mathcal{L}}_{LQD} =
	\lambda^{\prime}_{ijk} 
      \left\{
	  \tilde{\nu}_{i\mathrm{L}} \bar{d}_{k\mathrm{R}} d_{j\mathrm{L}} 
	- \tilde{e}_{i\mathrm{L}}  \bar{d}_{k\mathrm{R}}  u_{j\mathrm{L}}
        + \tilde{d}_{j\mathrm{L}} \bar{d}_{k\mathrm{R}} \nu_{i\mathrm{L}}
	-  \tilde{u}_{j\mathrm{L}} \bar{d}_{k\mathrm{R}} e_{i\mathrm{L}}
        + \tilde{d}_{k\mathrm{R}}^{c} \nu_{i\mathrm{L}} d_{j\mathrm{L}} 
        - \tilde{d}_{k\mathrm{R}}^{c} e_{i\mathrm{L}} u_{j\mathrm{L}}  
     \right\} + 
  \mathrm{h.c.}
\label{eq:LQD}
\ee
With only one $\lambda^{\prime}_{ijk}$  being nonzero, the 
simplest processes (with an observable final state) 
to which ${\mathcal{L}}_{LQD}$ would contribute are
\begin{itemize}
\item $u_j \bar u_j \ra e_i^- e_i^+$ ($\tilde{d}_{k\mathrm{R}}$)
\item $d_k \bar d_k \ra e_i^- e_i^+$ ($\tilde{u}_{j\mathrm{L}}$)
\item $u_j \bar d_j \ra e_i^+ \nu_i$ ($\tilde{d}_{k\mathrm{R}}$)
\item $q_j \bar q'_j \ra q_j \bar q'_j$ ($\tilde{\nu}_{i\mathrm{L}},
					e_{i\mathrm{L}} $)
\end{itemize}	
with the particle in parentheses being exchanged in the $t$-channel. 
If we allow more than one \rpv\ coupling to be non-zero, more exotic 
final states would be possible. 
However, as simultaneous existence of more than one such coupling 
is disfavored from the data on flavor-changing neutral current 
processes\cite{products}, we shall not consider this possibility.  
Furthermore, the limits on $LQD$ couplings of muons with the 
first-generation quarks 
found in nucleon targets are weaker than those for electrons. 
Hence, in this note, we shall restrict ourselves to perhaps the 
most optimistic case, namely the study of a $\mu^+ \mu^-$ 
final state in the presence of a non-zero $\lambda'_{211}$. 
The corresponding details for the other cases will be presented 
elsewhere.

The signature we focus on  is a
$\mu^+ \mu^-$ {\em without} any missing transverse momentum. To ensure 
detectability, we demand that 
\be
	p_T(\mu^\pm) \geq 50 \GeV \mand \left| \eta(\mu^\pm) \right| < 3 
    \ .
   \label{eq:cuts}
\ee
The lowest order SM process leading to such a final state is the
Drell--Yan mechanism \ie $q \bar{q} \rightarrow (\gamma^\ast,
Z^\ast) \rightarrow \mu^+ \mu^-$. A non-zero $\lambda'_{211}$ induces
an additional $t$--channel diagram and the expressions for the total
cross sections can be found in Ref.\cite{b_c_s}. 
QCD corrections to the Drell--Yan process
have been calculated \cite{QCD} to the next-to-leading order and are
a function of the c.m. energy ($\sqrt{s}$) of the collider, the
structure functions used and the subprocess scale $M$.
The dependence on $M$ is marginal though and 
one may  approximate it by a scale--independent 
$K$-factor $1.12 $ (for the CTEQ4M\cite{CTEQ4} structure 
functions that we use). The same factor approximates the K-factor for
the dimuon mass distributions also for the range of dimuon masses
and rapidities that we use. 
In the absence of a calculation of the higher-order corrections 
to the \rpv\ contribution, we assume that $K$-factor for the 
full theory is the same as that within the SM.

To maximize the sensitivity, we need to consider the 
differential distributions 
especially since the event topology of the \rpv\ ``signal'' is quite different 
from that of the SM ``background''.  A convenient set of independent kinematical
variables is given by the dimuon invariant mass  $M$, the rapidity of the
$\mu^+ \mu^-$ pair $\eta_{\rm pair}$ and the difference of the 
individual rapidities $\Delta \eta$. 


\begin{figure}[htb]
\begin{center}
\epsfxsize=5.6cm
\epsffile{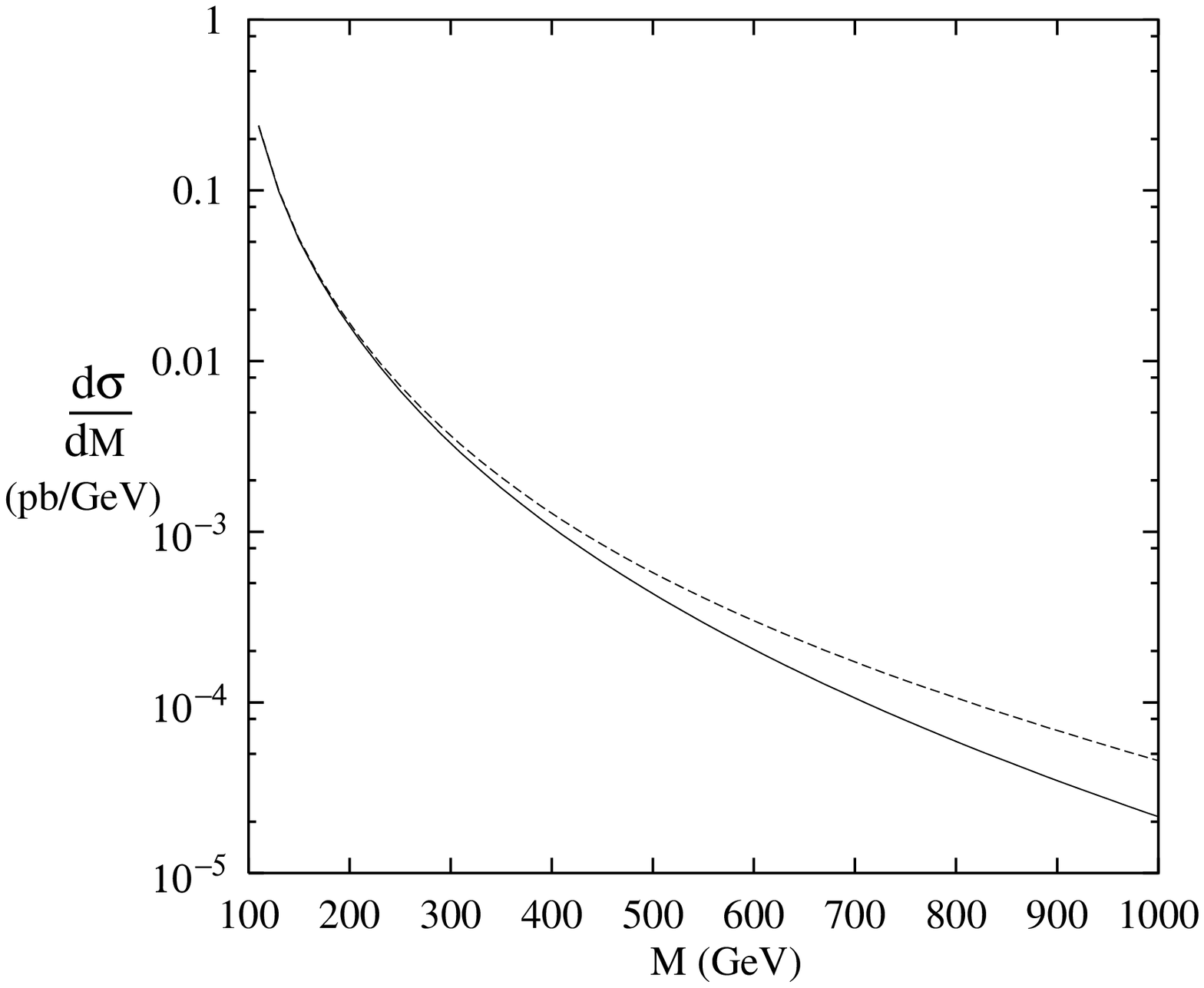}
\hspace*{-0.8cm}
\epsfxsize=5.6cm
\epsffile{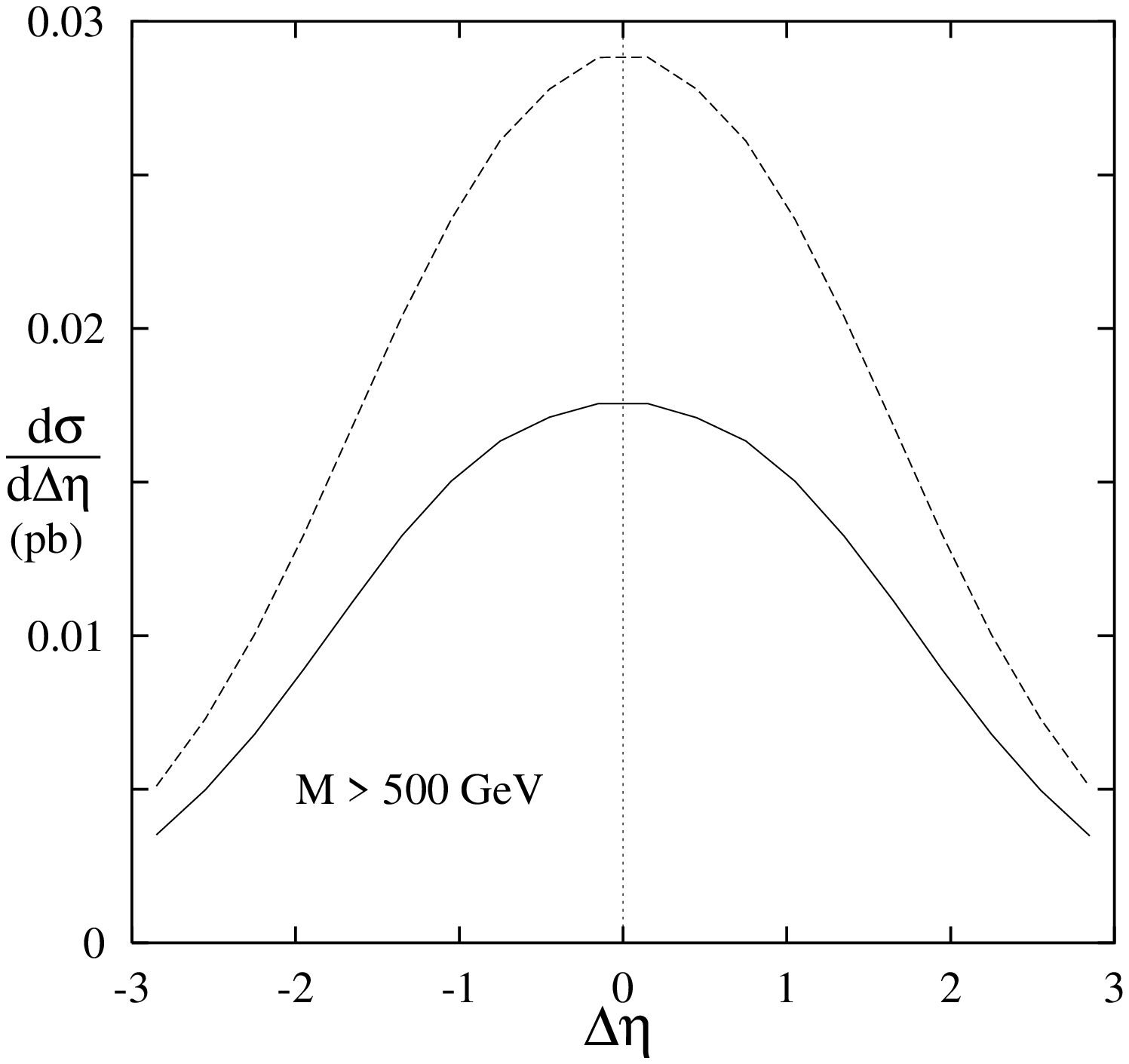}
\hspace*{-0.8cm}
\epsfxsize=5.6cm
\epsffile{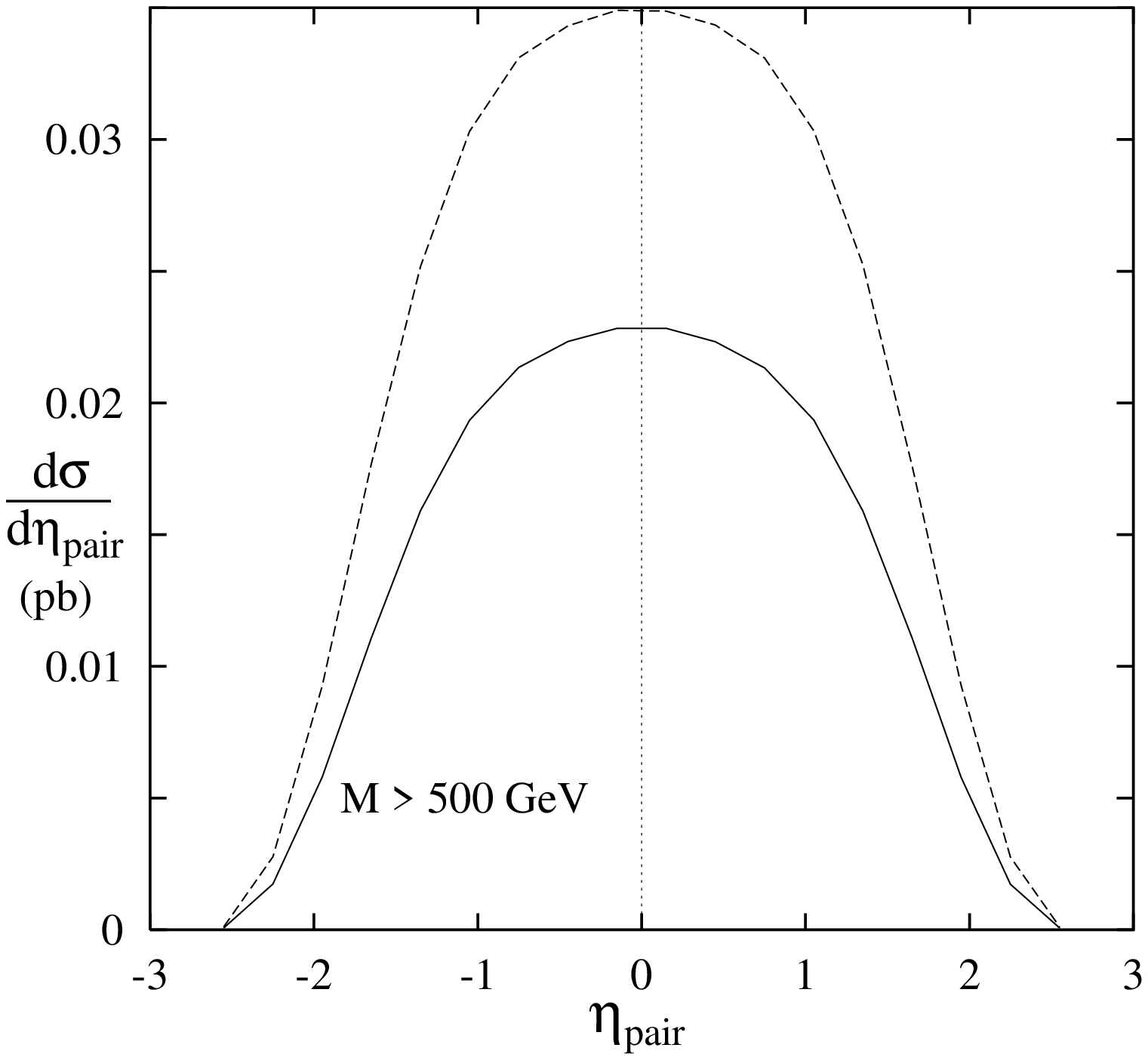}
\end{center}
\caption{\em The phase space distributions for the Drell-Yan process 
	at the LHC. The kinematical cuts are as in \protect\ref{eq:cuts}.
	For the rapidity distributions, an additional cut ($M > 500 \GeV$) 
	has been imposed. In each case, the lower curve corresponds to the 
	SM and the upper curve to 
	$\lambda'_{211} = 0.5$, $m_{\rm sq} = 800 \GeV$.
	}
    \label{fig:distribs}
\end{figure}


It is obvious that the contribution due to a heavy squark exchange 
would be non-negligible only for $M \gsim m_{\rm sq}$. Thus the 
relative deviations are pronounced only for large values of $M$ 
(see Fig.\ref{fig:distribs}). It has been demonstrated~\cite{dittmar}  
that the  parton luminosities  can be measured with high accuracy 
at the LHC. Hence a change in the absolute value of the cross-section 
can indeed be used to look at the sensitivity of the signal to \rpv\
couplings.   As a matter of fact such an  analysis~\cite{D0DY} of the 
differential distribution  in the dimuon mass of the dimuon pair 
cross-section  has been used by the D0 collaboration to
put limits on the Quark-Electron Compositeness scale, using the Tevatron data.
Recall that the analysis here does include the effect of the
higher order corrections on the SM contribution to the $\mu^+ \mu^-$  pairs.

Note that  $ \Delta \eta (= \eta_+ - \eta_-$), the difference 
of the lepton rapidities is directly related to the scattering 
angle in the subprocess center-of-mass frame.  Hence 
one expects to see the difference between an $s$--channel and a
$t$--channel process in this distribution.  This is borne out 
strikingly in the second of Fig.~\ref{fig:distribs}. The symmetry 
about $\Delta \eta = 0$ is 
dictated by the fact that the initial state is symmetric. And finally, 
one considers $\eta_{\rm pair}$ which is a measure of the Lorentz 
boost of the center-of-mass frame, or in other words, the mismatch 
of the quark and antiquark energies. 

To quantify our comparison of the differential distributions in the 
two cases (SM vs. \rpv), we devise a $\chi^2$ test. We divide the 
$M$--$\eta_{\rm pair}$--$\Delta \eta$ hypervolume into equal sized 
bins and compare the number of signal ($N_{n}^{\mathrm{SM} + \not{R}}$)
and background ($N_{n}^{\mathrm{SM}}$)   events in each bin 
for a given integrated luminosity. We then define a $\chi^2$
test of discrimination 
\be
   \chi^{2} = \sum_{n} \frac{(N_{n}^{{\mathrm{SM}}+ \not{R}} 
                               - N_{n}^{{\mathrm{SM}}}       )^{2}}
	{N_{n}^{\mathrm{SM}} + (\epsilon N_{n}^{\mathrm{SM}})^2 }
	\label{eq:chisq}
\ee
where $\epsilon$ is a measure of the systematic errors, arising
mainly from the uncertainties in luminosity measurement and
quark densities. 
To be specific, we use a uniform grid of ($20 \GeV, 0.15, 0.15$)
and choose two representative values for $\epsilon$, namely $5 \%$ 
and $15 \%$.

\begin{figure}[htb]
\begin{center}
\epsfxsize=8.5cm
\epsffile{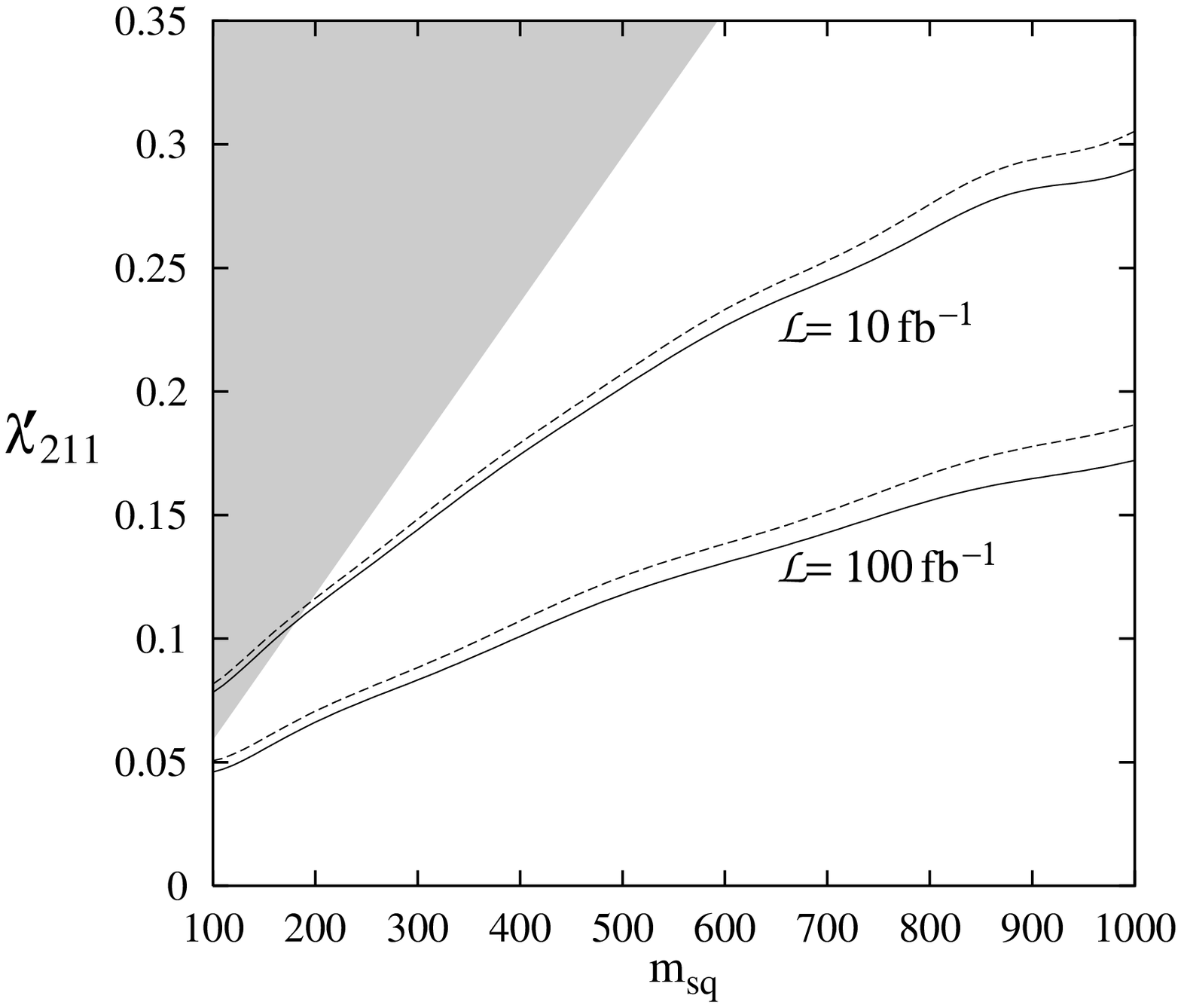}
\end{center}
\caption{\em The 95\% C.L.
	exclusion curves that may be obtained at the LHC for
	two different values of integrated luminosity. 
	In each case, the lower curve corresponds to a systematic 
	uncertainty (see eq.\protect\ref{eq:chisq}) $\epsilon= 0.05$ while 
	the upper corresponds to $\epsilon= 0.15$. The shaded region
        indicates the region excluded by low energy measurements.
	}
    \label{fig:excl}
\end{figure}

In Fig.~\ref{fig:excl}, we present the exclusion plots in the 
($m_{\rm sq}, \lambda'_{211}$) plane that can be achieved at the 
LHC for a given luminosity. The interpretation is simple. The area
above the respective curves can be ruled out at 95\% C.L. As expected, 
for $\epsilon = 0.15$ we do slightly worse than for the case of 
$\epsilon = 0.05$. And similarly, we do better with larger luminosity. 
What is most interesting, however, is that we do significantly better 
than the bounds obtained~\cite{BGH} from low-energy experiments, {\em viz.}
\[ \dis
	\lambda'_{211} < 0.059 \; (100 \GeV / m_{\tilde d_R}) \ .
\]
The LHC bound clearly does not scale as the mass of the squark. This is 
easily understood from a perusal of the first of Figs.~\ref{fig:distribs}. 
For smaller squark masses, the \rpv\ contribution is also peaked at 
smaller values of $M$. However, the SM contribution is also much bigger at 
such $M$. Consequently, the relative deviations are smaller and so are the 
contributions to the $\chi^2$. For larger squark masses, the \rpv\ 
contributions concentrate at larger $M$ and hence the relative deviations 
are larger too. This is also the reason why our analysis shows much more
sensitivity to the \rpv\ couplings than the  analysis of 
Ref.~\cite{Joanne_rizzo}.

It is clear that a similar analysis can also be performed to study the
sensitivity of LHC experiments to effects of Leptoquarks/compositeness
etc.  which would give rise to four-fermion currents with unusual
chiral structure.


\setcounter{figure}{0}
\setcounter{table}{0}
\setcounter{section}{0}
\setcounter{equation}{0}
\setcounter{footnote}{0}
\clearpage

\begin{center}
{\large \bf Neutrino masses, R-parity violating supersymmetry and 
collider signals} \\
\vspace*{3mm}
{\sc A.K. DATTA, B. MUKHOPADHYAYA, S. ROY and F. VISSANI}\\
\end{center}

\vspace*{3mm}

In a scenario with R-parity violating supersymmetry one can have masses
for the neutrinos. The seed of this lies in the superpotential,
where, upon admission of lepton number violation, one can write the
following terms over and above those of the minimal SUSY standard model (MSSM):

\begin{equation}
W_{\not R} = \lambda_{ijk} L_i L_j E_k^c +
\lambda_{ijk}' L_i Q_j D_k^c +
\lambda_{ijk}''U_i^c D_j^c D_k^c + \epsilon_i L_i H_2
\end{equation}

\noindent

If one considers the framework where R-parity is violated through a
term of the type $\epsilon_i L_iH_2$, then one neutrino acquires a
mass at the tree level through mixing with neutralinos. We have argued
that this may give rise to the mass splitting corresponding to the
$\nu_{\mu} {\rightarrow} \nu_{\tau}$ oscillation, which provides an
explanation of the atmospheric muon neutrino deficit recorded by the
superkamiokande (SK) experiment \cite{R1}.  The other two neutrinos
remain massless at the tree level.  However, the $\lambda$-and
$\lambda^{'}$-type terms can at the same time induce mass terms for
all the three generations at the one-loop level, which can be
responsible for a smaller mass splitting between the the first two
neutrino generation, thereby explaining the solar neutrino problem.

The relevant parameters in this scenario, in addition to those
contained in the MSSM, are the $\epsilon_i$'s mentioned above and
vacuum expectation values (vev) for the scalar neutrinos, which are
unavoidable consequences of the former. These can be lumped into one
`basis-independent' parameter, namely, the vev of the sneutrino
corresponding to the state which acquires the tree-level mass.  The SK
results restrict this vev to be of the order of 100 keV. It is also
possible to constrain the parameter space of the soft bilinear SUSY
breaking terms of this theory from large angle mixing and neutrino
mass-squared difference as demanded by the atmospheric $\nu_\mu$
data. In addition, non-trivial restrictions on the hierarchy of the
$\epsilon$-parameters of different flavors or between the
$\epsilon$'s and $\mu$ follow if we demand that flavor-changing
neutral currents be suppressed \cite{R2}. Moreover, if one wants to
solve the solar neutrino problem in this scenario, either through MSW
solution or vacuum oscillation, then the $\lambda$ and $\lambda^{'}$
couplings also get restricted to values $\le 10^{-5}$ for the MSW
solution, and to much smaller values for the `just-so' solution.

Apart from the neutrino-neutralino mixing mentioned above, such a
scenario induces mixing in the chargino-charged lepton sector as well.
These mixings result in some couplings like $\tilde \chi^0_1 \tau W$,
$\tilde \chi^0_1 \nu Z$ etc., which are characteristic of bilinear
R-parity violation.

If the lightest neutralino ($\tilde \chi^0_1$) is heavier than the $Z$, then
we have the following additional decay channels of $\tilde \chi^0_1$ \cite{R4}:
\begin{equation}
\tilde \chi^0_1 \rightarrow \nu_l Z, \ \ \ \  l=e, \mu, \tau  
\end{equation}
\begin{equation}
\tilde \chi^0_1 \rightarrow l W, \ \ \ \ l=e, \mu, \tau  
\end{equation}

This kind of decay will not be possible with only the trilinear
R-violating interactions. Also, if the $\lambda$ and $\lambda^{'}$
terms have to be responsible for loop-induced solutions to the solar
neutrino problem, then they are restricted to such values that
three-body decays of the lightest neutralino triggered by them are
dominated by the two-body decays as and when they are allowed. In such
a case, demanding maximal mixing between the second and third
generations (as required by the SK data), one would expect to get
equal numbers of muons and taus in decays in the channel $\tilde
\chi^0_1 \rightarrow l W$ \cite{R5}.  This can be a remarkable feature
in collider searches of the lightest neutralino. Another important
observation is that, thanks to the small R-violating couplings
necessitated by the small tree-level neutrino mass, the decay length
of the lightest neutralino can be as large as 1 - 10 mm or so. This
implies a decay gap that can be observed in collider experiments.

Among other signals that have been investigated \cite{R6}, the signal
in the form of \emph{equal} number of \emph{like sign} $\mu$ and
$\tau$ events accompanied by \emph{like sign on-shell} $W$'s may be
seen at the upgraded Tevatron, LHC or future generation $e^+e^-$
colliders when a pair of lightest neutralinos ($\widetilde{\chi}_1^0$)
is produced directly or via cascades. Each neutralino may decay
directly to $\mu$ or $\tau$ (with BR $\sim 40 \%$) \cite{R5} along
with \emph{on-shell} $W$'s. We emphasize that this is a consequence of
neutralino-neutrino mixing triggered exclusively by the bilinear
R-parity violating terms in the superpotential.  A very interesting
possibility, essentially due to the Majorana nature of neutralinos, is
that of having like-sign dilepton events accompanied by two W's of
identical charge \cite{R7}.  Of course, for such a signal to be
distinct, one requires the associated $W$'s to be \emph{on-shell}. In
other words, such a signal is typical of a situation where
${\chi^0_1}$ is heavier than the W.  Otherwise, 3-body decays of the
lightest neutralino take place, and these can have contributions from
trilinear R-violating couplings as well.

The interesting feature is that there is almost no Standard Model
background for such a signal. Taming the MSSM background, too, is
unlikely to pose any serious difficulty, so long as the W's can be
identified \cite {R7}. One may further exploit the large decay gap
($\sim$ few mm) \cite{R5} in the lightest neutralino decay for
strengthening the signal.

\setcounter{figure}{0}
\setcounter{table}{0}
\setcounter{section}{0}
\setcounter{equation}{0}
\setcounter{footnote}{0}
\clearpage

\begin{center}
{\large \bf Introduction to 
GMSB phenomenology at TeV colliders} \\

\vspace{0.3cm} 

{\sc  S.~AMBROSANIO\footnote{E-mail: {\tt ambros@mail.cern.ch}} } \\

\end{center}
\vspace*{3mm}
\begin{abstract}
\nn
A brief account is given of models with gauge--mediated
supersymmetry breaking, as a general introduction for the following
two contributions.
\end{abstract}

\section{Introduction to Gauge-Mediated SUSY Breaking}
\label{sec:intro}
\noindent 
Since no superpartners have been detected at collider experiments so far,
supersymmetry (SUSY) cannot be an exact symmetry of Nature.
The requirement of ``soft'' supersymmetry breaking~\cite{StevePrimer} 
alone is not sufficient to reduce the free parameters
to a number suitable for predictive phenomenological studies. 
Hence, motivated theoretical hypotheses on the nature of SUSY breaking 
and the mechanism through which it is transmitted to the visible sector 
of the theory [here assumed to be the one predicted by the minimal SUSY 
extension of the standard model (MSSM)] are highly desirable. 
If SUSY is broken at energies of the order of the Planck mass and the SUSY 
breaking sector communicates with the MSSM sector through gravitational 
interactions only, one falls in the supergravity-inspired (SUGRA) scheme.
The most recognized alternative to SUGRA is based instead on the hypothesis
that SUSY breaking occurs at relatively low energy scales and is mediated 
mainly by gauge interactions (GMSB)~\cite{oldGMSB,newGMSB,GR-GMSB}. 
A good theoretical reason to consider such a possibility is that it 
provides a natural, automatic suppression of the SUSY contributions 
to flavor-changing neutral currents and CP-violating processes. 
A pleasant consequence is that, at least in the simplest versions of GMSB, 
the MSSM spectrum and other observables depend on just a handful of 
parameters, typically

\be
M_{\rm mess}, \; N_{\rm mess}, \; \Lambda, \; \tgb, \; {\rm sign}(\mu),
\label{eq:pars}
\ee

\noindent
where $M_{\rm mess}$ is the overall messenger scale; $N_{\rm mess}$ is the 
so-called messenger index, parameterizing the structure of the messenger
sector; $\Lambda$ is the universal soft SUSY breaking scale felt by the
low-energy sector; $\tgb$ is the ratio of the vacuum expectation 
values of the two Higgs doublets; sign($\mu$) is the ambiguity
left for the SUSY higgsino mass after conditions for correct 
electroweak symmetry breaking (EWSB) are imposed (see e.g.  
Refs.~\cite{GMSBmodels1,GMSBmodels2,AKM-LEP2,AB-LC}).

The phenomenology of GMSB (and, more generally, of any theory with
low-energy SUSY breaking) is characterized by the presence of a very
light gravitino $\G$~\cite{Fayet},
\be 
m_{3/2} \equiv m_{\G} = \frac{F}{\sqrt{3}M'_P} \simeq 
\left(\frac{\sqrt{F}}{100 \; {\rm TeV}}\right)^2 2.37 \; {\rm eV},  
\label{eq:Gmass}
\ee

\noindent
where $\sqrt{F}$ is the fundamental scale of SUSY breaking, 100 TeV is a 
typical value for it, and $M'_P = 2.44 \times 10^{18}$ GeV is the reduced 
Planck mass.
Hence, the $\G$ is always the lightest SUSY particle (LSP) in these theories. 
If $R$-parity is assumed to be conserved, any produced MSSM particle will 
finally decay into the gravitino. Depending on $\sqrt{F}$, the interactions 
of the gravitino, although much weaker than gauge and Yukawa interactions, 
can still be strong enough to be of relevance for collider physics. 
As a result, in most cases the last step of any SUSY decay chain is 
the decay of the next-to-lightest SUSY particle (NLSP), which can  
occur outside or inside a typical detector or even close to the interaction 
point. The pattern of the resulting spectacular signatures is determined 
by the identity of the NLSP and its lifetime before decaying into the $\G$,

\be 
c \tau_{\rm NLSP} \simeq \frac{1}{100 {\cal B}}
\left(\frac{\sqrt{F}}{100 \; {\rm TeV}}\right)^4 
\left(\frac{m_{\rm NLSP}} {100 \; {\rm GeV}}\right)^{-5},
\label{eq:NLSPtau}
\ee 

\noindent
where ${\cal B}$ is a number of order unity depending on the nature
of the NLSP.

The identity of the NLSP [or, to be more precise, the identity of the 
sparticle(s) having a large branching ratio (BR) for decaying into the 
gravitino and the relevant SM partner] determines four main scenarios 
giving rise to qualitatively different phenomenology:

\begin{description} 

\item[Neutralino NLSP scenario:] Occurs whenever 
$m_{\NI} < (m_{\tauu} - m_{\tau})$. Here typically a decay of the $\NI$  
to $\G\gamma$ is the final step of decay chains following
any SUSY production process. As a consequence, the main inclusive signature 
at colliders is prompt or displaced photon pairs + X + missing energy. 
$\NI$ decays to $\G \Z$ and other minor channels may also be relevant 
at TeV colliders.    

\item[Stau NLSP scenario:] Defined by 
$m_{\tauu} < {\rm Min}[m_{\NI}, m_{\lR}] - m_{\tau}$, 
features $\tauu \to \G \tau$ decays, producing $\tau$ pairs or 
charged semi-stable $\tauu$ tracks or decay kinks + X + missing energy.
Here and in the following, $\ell$ stands for $e$ or $\mu$. 

\item[Slepton co-NLSP scenario:] When 
$m_{\lR} < {\rm Min}[m_{\NI}, m_{\tauu} + m_{\tau}]$, 
$\lR \to \G \ell$ decays are also open with large BR. In addition to the 
signatures of the stau NLSP scenario, one also gets $\ell^+\ell^-$
pairs or $\lR$ tracks or decay kinks. 

\item[Neutralino-stau co-NLSP scenario:] If 
$| m_{\tauu} - m_{\NI} | < m_{\tau}$ and $m_{\NI} < m_{\lR}$, 
both signatures of the neutralino NLSP and stau NLSP scenario are present
at the same time, since $\NI \leftrightarrow \tauu$ 2--body 
decays are not allowed by phase space. 

\end{description}

Note that in the GMSB parameter space the relation $m_{\lR} > m_{\tauu}$ 
always holds. 
Also, one should keep in mind that the classification above is only 
valid as an indicative scheme in the limit $m_e$, $m_\mu \to 0$, neglecting 
also those cases where a fine-tuned choice of $\sqrt{F}$ and the sparticle 
masses may give rise to competition between phase-space suppressed decay 
channels from one ordinary sparticle to another and sparticle decays to the 
gravitino~\cite{AKM2}. 

In this report, two important aspects of the GMSB phenomenology 
at TeV colliders will be treated:
\begin{description}
\item[(A)] The consequences of the GMSB hypothesis on the light Higgs spectrum
      using the most accurate tools available today for model generation and 
      $\mh$ calculation;
\item[(B)] Studies and possible measurements at the LHC with the ATLAS detector
      in the stau NLSP or slepton co-NLSP scenarios, with focus 
      on determining the fundamental SUSY breaking scale $\sqrt{F}$.
\end{description}

For this purpose, we generated about 30000 GMSB models under
well defined hypotheses, using the program {\tt SUSYFIRE}~\cite{SUSYFIRE}, 
as described in the following section. 

\section{GMSB Models}
\label{sec:models}

\noindent 
In the GMSB framework, the pattern of the MSSM spectrum is simple,
as all sparticle masses are generated in the same way and scale 
approximately with a single parameter $\Lambda$, which sets
the amount of soft SUSY breaking felt by the visible sector. 
As a consequence, scalar and gaugino masses are related to each
other at a high energy scale, which is not the case in other SUSY 
frameworks, e.g.\ SUGRA. Also, it is possible to impose other 
conditions at a lower scale to achieve EWSB and  
further reduce the dimension of the parameter space. 

To build our GMSB models, we adopt the usual phenomenological approach,
in particular following Ref.~\cite{AB-LC}, where problems relevant 
for GMSB physics at TeV colliders were also approached.
We do not specify the origin of the SUSY higgsino mass $\mu$, nor do we 
assume that the analogous soft SUSY breaking parameter $B\mu$ vanishes at the 
messenger scale. Instead, we impose correct EWSB to trade $\mu$
and $B\mu$ for $M_Z$ and $\tgb$, leaving the sign of $\mu$ undetermined. 
However, we are aware that to build a satisfactory GMSB model one should also 
solve the latter problem in a more fundamental way, perhaps by providing a 
dynamical mechanism to generate $\mu$ and $B\mu$, possibly with values
of the same order of magnitude. This might be accomplished radiatively 
through some new interactions. However, in this case the other
soft terms in the Higgs potential, namely $m^2_{H_{1,2}}$, will be also  
affected and this will in turn change the values of $|\mu|$ and $B\mu$ 
coming from EWSB conditions~\cite{GR-GMSB,GMSBmodels1,GMSBmodels2}. 
Within the study {\bf (A)}, we are currently considering some ``non-minimal'' 
possibilities for GMSB models that to some extent take this problem into 
account, and we are trying to assess the impact on the light Higgs mass. 
We do not treat this topic here, but refer to~\cite{AHW} for further details.

To determine the MSSM spectrum and low-energy parameters, we solve 
the renormalization group (RG) evolution with the following 
boundary conditions at the $M_{\rm mess}$ scale,

\bea
M_a & = & N_{\rm mess} \Lambda 
g\left(\frac{\Lambda}{M_{\rm mess}}\right) \frac{\alpha_a}{4\pi}, 
\; \; \; (a=1, 2, 3) \nonumber \\
\tilde{m}^2 & =  & 2 N_{\rm mess} \Lambda^2 
f\left(\frac{\Lambda}{M_{\rm mess}}\right) 
\sum_a \left(\frac{\alpha_a}{4\pi}\right)^2 C_a,
\label{eq:bound}
\eea 

\noindent
respectively for the gaugino and the scalar masses. 
In Eq.~(\ref{eq:bound}), $g$ and $f$ are the one-loop and two-loop functions 
whose exact expressions can be found e.g. in Ref.~\cite{AKM-LEP2}, 
and $C_a$ are the quadratic Casimir invariants for the scalar fields.
As usual, the scalar trilinear couplings $A_f$ are assumed to vanish
at the messenger scale, as suggested by the fact that they (and not
their squares) are generated via gauge interactions with the messenger 
fields at the two loop-level only. 

To single out the interesting region of the GMSB parameter space, we
proceed as follows.  Barring the case where a neutralino is the NLSP
and decays outside the detector (large $\sqrt{F}$), the GMSB
signatures are very spectacular and are generally free from SM
backgrounds. Keeping this in mind and being interested in GMSB
phenomenology at future TeV colliders, we consider only models where
the NLSP mass is larger than 100 GeV, assuming that searches at LEP
and the Tevatron, if unsuccessful, will in the end exclude a softer
spectrum in most cases.  We require that $M_{\rm mess} > 1.01
\Lambda$, to prevent an excess of fine-tuning of the messenger masses,
and that the mass of the lightest messenger scalar be at least 10
TeV. We also impose $M_{\rm mess} > M_{\rm GUT} \; {\rm
exp}(-125/N_{\rm mess})$, to ensure perturbativity of gauge
interactions up to the GUT scale. Further, we do not consider models
with $M_{\rm mess} \gtap 10^{5} \Lambda$. As a result of this and
other constraints, the messenger index $N_{\rm mess}$, which we assume
to be an integer independent of the gauge group, cannot be larger
than~8. To prevent the top Yukawa coupling from blowing up below the
GUT scale, we require $\tgb > 1.2$ (and in some cases $> 1.5$). This is
also motivated by the current bounds from SUSY Higgs searches at
LEP~II~\cite{mhiggsmSUGRA}.  Models with $\tgb \gtap 55$ (with a mild
dependence on $\Lambda$) are forbidden by the EWSB requirement and
typically fail to give $m_A^2 > 0$.

To calculate the NLSP lifetime relevant to our study {\bf (B)}, one needs 
to specify the value of the fundamental SUSY breaking scale $\sqrt{F}$ on a
model-by-model basis. Using perturbativity arguments, for each given 
set of GMSB parameters, it is possible to determine a lower bound
according to Ref.~\cite{AKM-LEP2}, 

\be 
\sqrt{F} \ge \sqrt{F_{\rm mess}} \equiv \sqrt{\Lambda M_{\rm mess}} > \Lambda.
\label{eq:sqrtFmin}
\ee
On the contrary, no solid arguments can be used to set an upper limit 
on $\sqrt{F}$ of relevance for collider physics, although some 
semi-qualitative cosmological arguments are sometimes evoked.

In order to generate our model samples using {\tt SUSYFIRE}, we used
logarithmic steps for $\Lambda$ (between about 45 TeV/$N_{\rm mess}$
and about 220 TeV/$\sqrt{N_{\rm mess}}$, which corresponds to
excluding models with sparticle masses above $\sim 4$ TeV), $M_{\rm
mess}/\Lambda$ (between 1.01 and $10^5$) and $\tgb$ (between 1.2 and
about 60), subject to the constraints described above.  {\tt SUSYFIRE}
starts from the values of SM particle masses and gauge couplings at
the weak scale and then evolves up to the messenger scale through
RGEs. At the messenger scale, it imposes the boundary conditions
(\ref{eq:bound}) for the soft sparticle masses and then evolves the
full set of RGEs back to the weak scale. The decoupling of each
sparticle at the proper threshold is taken into account.  Two-loop
RGEs are used for gauge couplings, third generation Yukawa couplings
and gaugino soft masses. The other RGEs are taken at the one-loop
level. At the scale $\sqrt{m_{\stopu}m_{\stopd}}$, EWSB conditions are
imposed by means of the one-loop effective potential approach,
including corrections from stops, sbottoms and staus.  The program
then evolves up again to $M_{\rm mess}$ and so on.  Three or four
iterations are usually enough to get a good approximation for the MSSM
spectrum.

\clearpage
\setcounter{section}{0}
\setcounter{figure}{0}
\setcounter{table}{0}
\setcounter{equation}{0}

\begin{center}

{\large \bf The light Higgs boson spectrum in GMSB models} \\

\vspace*{0.3cm} 
{\sc S.~AMBROSANIO, S.~HEINEMEYER and G.~WEIGLEIN} \\
\end{center}

\vspace*{0.3cm}

\begin{abstract}
\nn
This study deals with the characteristics of the light Higgs 
boson spectrum allowed by the (minimal and non-minimal) GMSB framework.
Today's most accurate GMSB model generation and two-loop 
Feynman-diagrammatic calculation of $\mh$ have been combined. 
The Higgs masses are shown in dependence of various model parameters
at the messenger and electroweak scales. In the minimal model, an upper
limit on $\mh$ of about 124 GeV is found for $\mt = 175$~GeV.

\end{abstract}

\section{Introduction}
\label{sec:A1}

\noindent
Within the MSSM, the masses of the $\cp$-even neutral Higgs bosons are 
calculable in terms of the other low-energy parameters. The mass of the
lightest Higgs boson, $\mh$, has been of particular interest, as
it is bounded to be smaller than the $\Z$~boson mass at the tree level. 
The \onel\ results~\cite{mhiggs1l,mhiggsf1l,mhiggsf1ldab,pierce} 
for $\mh$ have been supplemented in the
last years with the leading \twol\ corrections, performed in the
renormalization group (RG)
approach~\cite{mhiggsRG1,mhiggsRG2}, in the effective
potential approach~\cite{mhiggsEP} and most recently in
the Feynman-diagrammatic (FD)
approach~\cite{mhiggsletter,mhiggslong,mhiggslle}. 
The \twol\ corrections have turned out to be sizeable. They can
lower the \onel\ results by up to 20\%.
These calculations predict an upper bound on $\mh$ of about 
$\mh \le 130$ GeV for an unconstrained MSSM with $\mt = 175$ GeV 
and a common SUSY mass scale $\msusy \le 1$ TeV.

As discussed in the Introduction \cite{ambro1}, the GMSB scenario
provides a relatively simple set of constraints and thus constitutes a
very predictive and readily testable realization of the MSSM.  The
main goal of the present analysis is to study the spectrum of the
lightest neutral $\cp$-even Higgs boson, $\mh$, within the GMSB
framework.  Particular emphasis is given to the maximal value of $\mh$
achievable in GMSB after an exhaustive scanning of the parameter
space.  Our results are discussed in terms of the GMSB constraints on
the low-energy parameters and compared to the cases of a
SUGRA-inspired or an unconstrained MSSM.

\section{Calculation of $\mh$}
\label{sec:A2}

\noindent
We employ the currently most accurate calculation to evaluate $\mh$,
based on the FD approach as given in
Refs.~\cite{mhiggsletter,mhiggslong,mhiggslle}.  The most important
radiative corrections to $\mh$ arise from the top and scalar top
sector of the MSSM, with the input parameters $\mt$, the masses of the
scalar top quarks, $\mste$, $\mstz$, and the $\Stop$-mixing angle,
$\tst$. Here we adopt the conventions introduced in
Ref.~\cite{mhiggslong}.  The complete diagrammatic \onel\
result~\cite{mhiggsf1ldab} has been combined with the dominant \twol\
corrections of $\oaas$~\cite{mhiggsletter,mhiggslong} and with the
subdominant corrections of $\ogmzmts$~\cite{mhiggsRG1,mhiggsRG2}.
GMSB models are generated with the program {\tt SUSYFIRE}, according
to the discussion of \cite{ambro1}. For this study, we consider only
models with $\tgb > 1.5$~\cite{mhiggsmSUGRA_1} and $m_A > 80$
GeV~\cite{lepc}.  In addition, we always use $\mt = 175$ GeV. A change
of 1 GeV in $\mt$ translates roughly into a shift of 1 GeV (with the
same sign) in $\mh$ as well. Thus, changing $\mt$ affects our results
on $\mh$ in an easily predictable way.

The results of the $\mh$ calculation have been implemented in the 
{\tt Fortran} code \fh\ \cite{feynhiggs}. This program has been combined 
with {\tt SUSYFIRE}, which has been used to calculate the low 
energy parameters $\mste$, $\mstz$, $\tst$, $\mu$, $M_1$, $M_2$, $\mgl$, 
$\dots$ for each of the $\sim$30000 GMSB models generated. 
These have then been passed to \fh\ for the $\mh$ evaluation in a
coherent way. Indeed, we transfer the \msbar\ parameters in the 
{\tt SUSYFIRE} output to on-shell parameters before feeding
them into \fh. Compact expressions for the relevant transition 
formulas can be found in Refs.~\cite{m20,bse}.

Compared to an existing analysis in the GMSB framework~\cite{kaeding},
we use a more complete evaluation of $\mh$. This leads in particular
to smaller values of $\mh$ for a given set of input parameters in our 
analysis. Also, in Ref.~\cite{kaeding} although some GMSB scenarios with 
generalized messenger sectors were considered, the parameter space for the 
``minimal'' case with a unique, integer messenger index 
$N_{\rm mess} = N_1 = N_2 = N_3$ was not fully explored. 
Indeed, $\Lambda$ was in most cases limited to values smaller 
than 100 TeV and $M_{\rm mess}$ was fixed to $10^5$ TeV. Furthermore, 
partly as a consequence of the above assumptions, the authors did not 
consider models with $N_{\rm mess} > 4$, i.e. their requirements  
for perturbativity of the MSSM gauge couplings up to the GUT scale were 
stronger than ours. We will see in the following section that maximal 
$\mh$ values in our analysis are instead obtained for larger values of 
the messenger scale and the messenger index. 

\section{The Light Higgs Spectrum in GMSB}
\label{sec:A3}

\noindent 
In the following, we give some results in the form of scatter plots showing 
the pattern in GMSB for $\mh$, $m_A$ as well as other low-energy parameters
of relevance for the light Higgs spectrum.

\begin{figure}[h!]
\hspace{-0.5cm}
\begin{center}
\epsfxsize=0.49\textwidth 
\epsffile{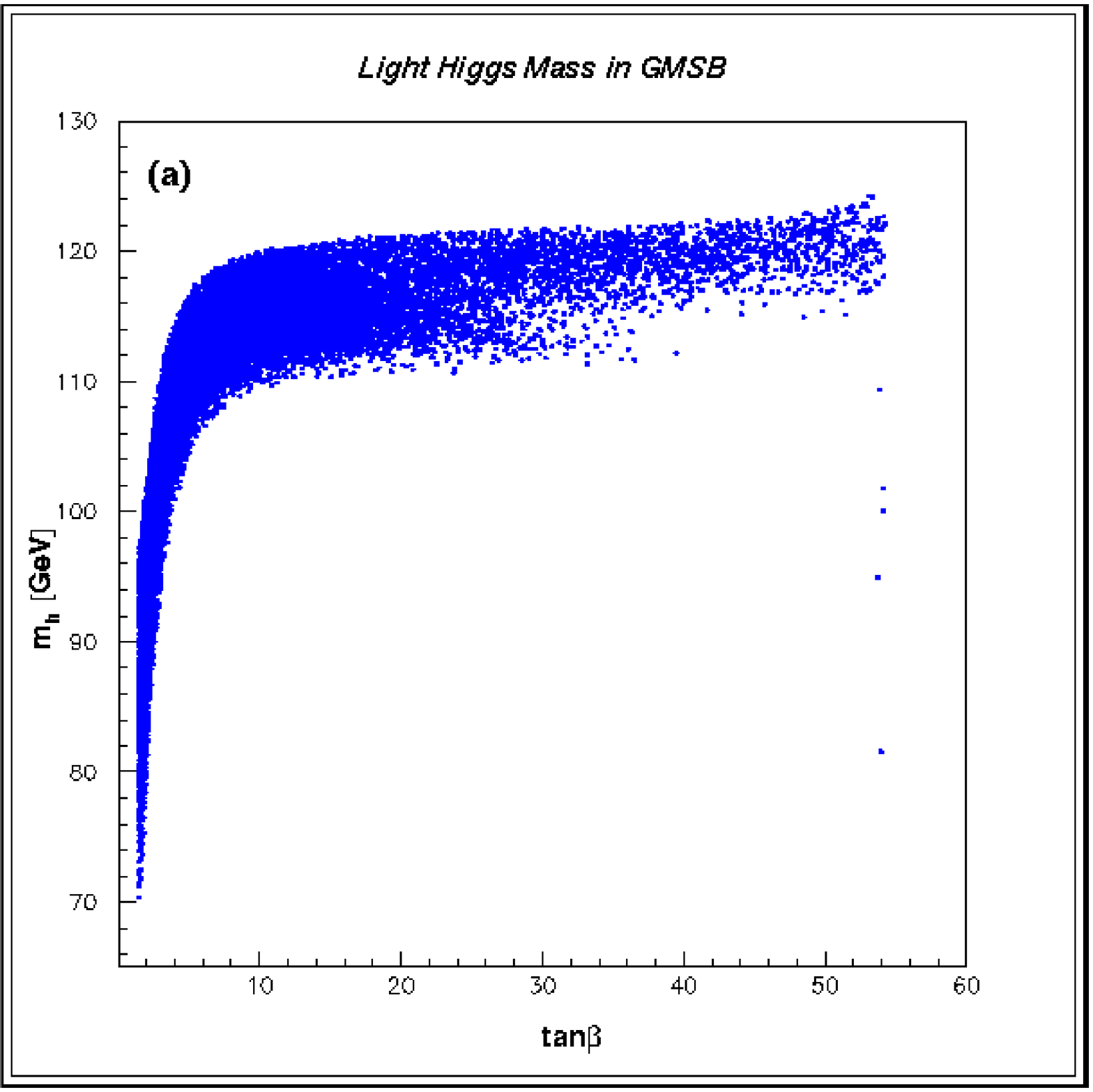}
\epsfxsize=0.49\textwidth 
\epsffile{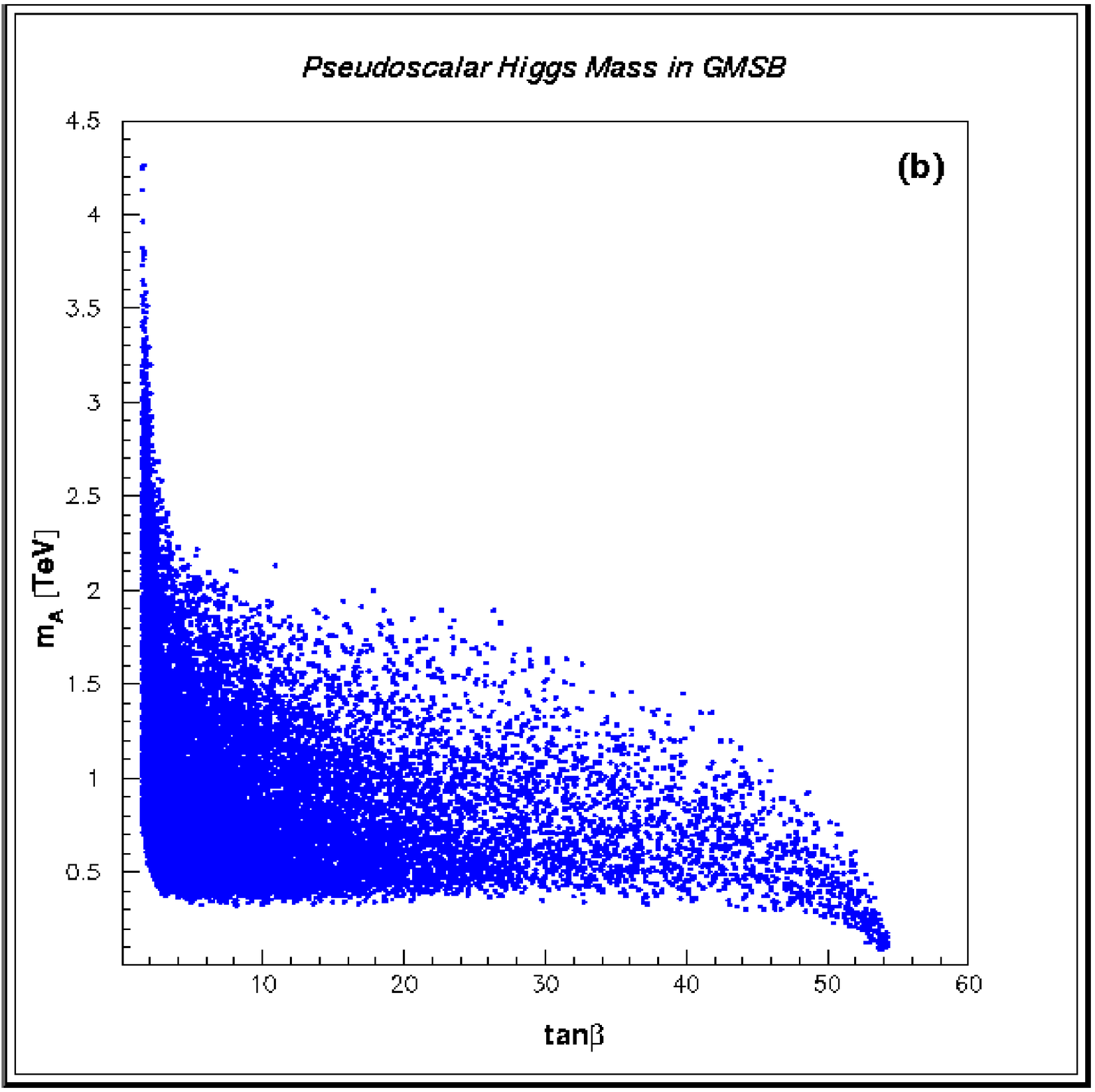}
\end{center}
\vspace*{-5mm}
\caption{\sl Scatter plots for the light scalar (a) and pseudoscalar (b)
Higgs masses as functions of $\tgb$. Only GMSB models with 
$\tgb > 1.5$, $m_A > 80$ GeV and $m_{\rm NLSP} > 100$ GeV are considered.
}
\label{fig:mh-mavstb}
\end{figure}

In Fig.~\ref{fig:mh-mavstb}(a), we show the dependence of $\mh$ on
$\tgb$, where only models with $\tgb > 1.5$, $m_A > 80$ GeV and $m_{\rm
NLSP} > 100$ GeV are considered, while $m_t$ is fixed to 175 GeV.  The
dependence is strong for small $\tgb \ltap 10$, while for larger $\tgb$
the increase of the lightest Higgs mass is rather mild.  The maximum
values for $\mh \simeq 124$ GeV are achieved for $\tgb > 50$.  It
should be noted that for very large $\tgb \gtap 52$, we also find a few
models with relatively small $\mh \ltap 100$ GeV. This is due to the
fact that in this case EWSB conditions tend to drive $m_A$ toward very
small values \cite{ambro1}. This is made visible by the scatter plot
in Fig.~\ref{fig:mh-mavstb}(b), where the pseudoscalar Higgs mass is
shown as a function of $\tgb$.  For such small values of $m_A$ and for
large $\tgb$, the relation $\mh \approx m_A$ holds. Thus small $\mh$
values are quite natural in this region of the parameter space. On the
other hand, one can see that extremely large values of $m_A \gtap 2$
TeV can only be obtained for small or moderate $\tgb \ltap 10$ GeV.  A
comparison between Fig~\ref{fig:mh-mavstb}(a) and (b) reveals that the
largest $\mh$ values $\gtap 123$ GeV correspond in GMSB to $m_A$
values in the 300--800 GeV range.  Indeed, it has been checked that
such large $\mh$ values are in general obtained in the FD calculation
for $300 \ltap m_A \ltap 1000$ GeV, see Ref.~\cite{mhiggslong}.

\begin{figure}[h!]
\begin{center}
\vspace*{-5mm}
\epsfxsize=0.55\textwidth 
\epsffile{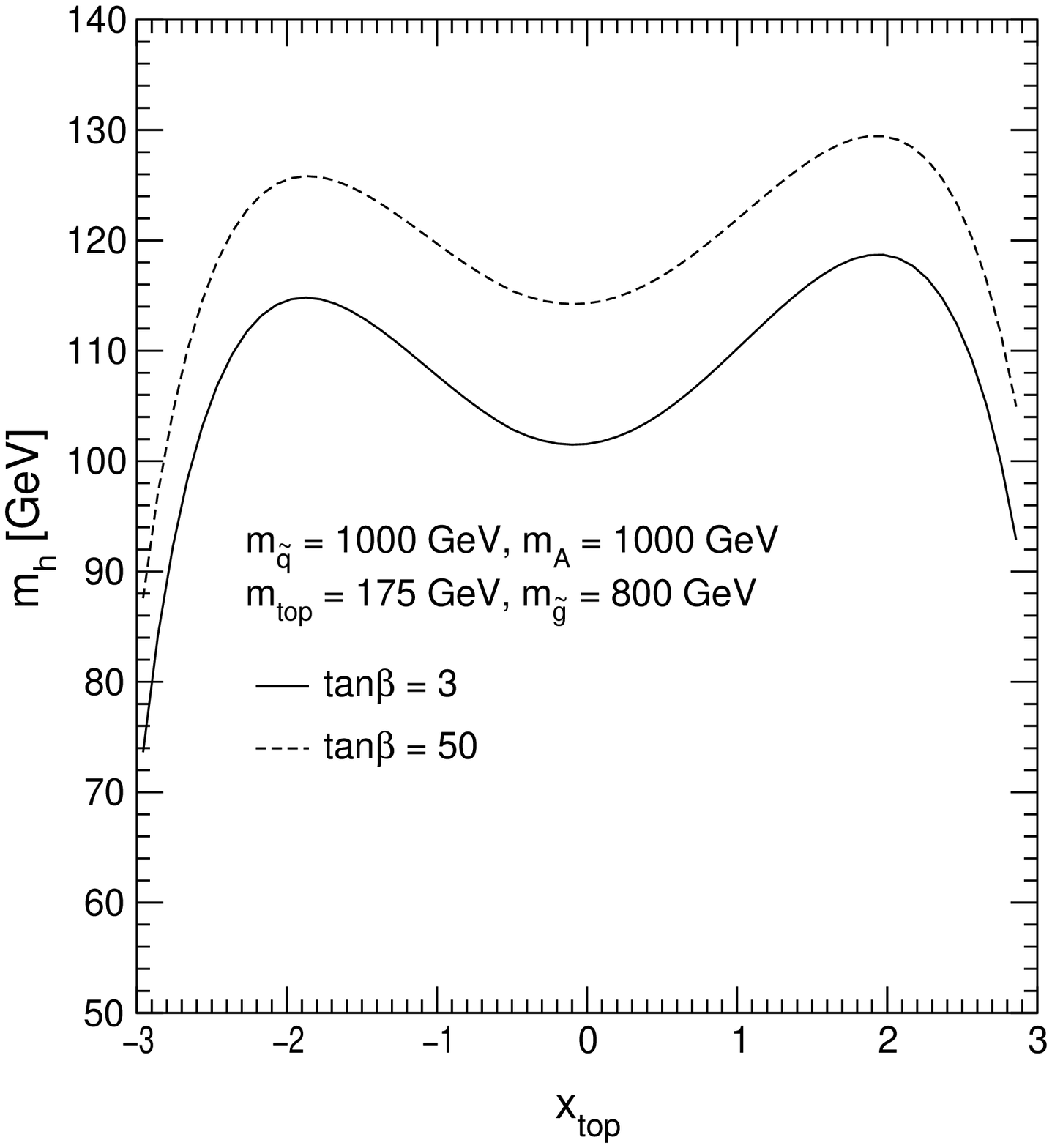}
\end{center}
\vspace*{-5mm}
\caption{\sl The light \cp-even Higgs boson mass is given as a function of 
$x_{\rm top}$ for $\tgb = 3, 50$, $m_A = 1000$ GeV, a common soft SUSY 
breaking scale for the squarks, $\msq = 1000$ GeV, and a gluino 
mass $\mgl = 800$ GeV.
}
\label{fig:mhvsxt}
\end{figure}

\begin{figure}[h!]
\begin{center}
\hspace{-0.5cm}
\epsfxsize=0.49\textwidth 
\epsffile{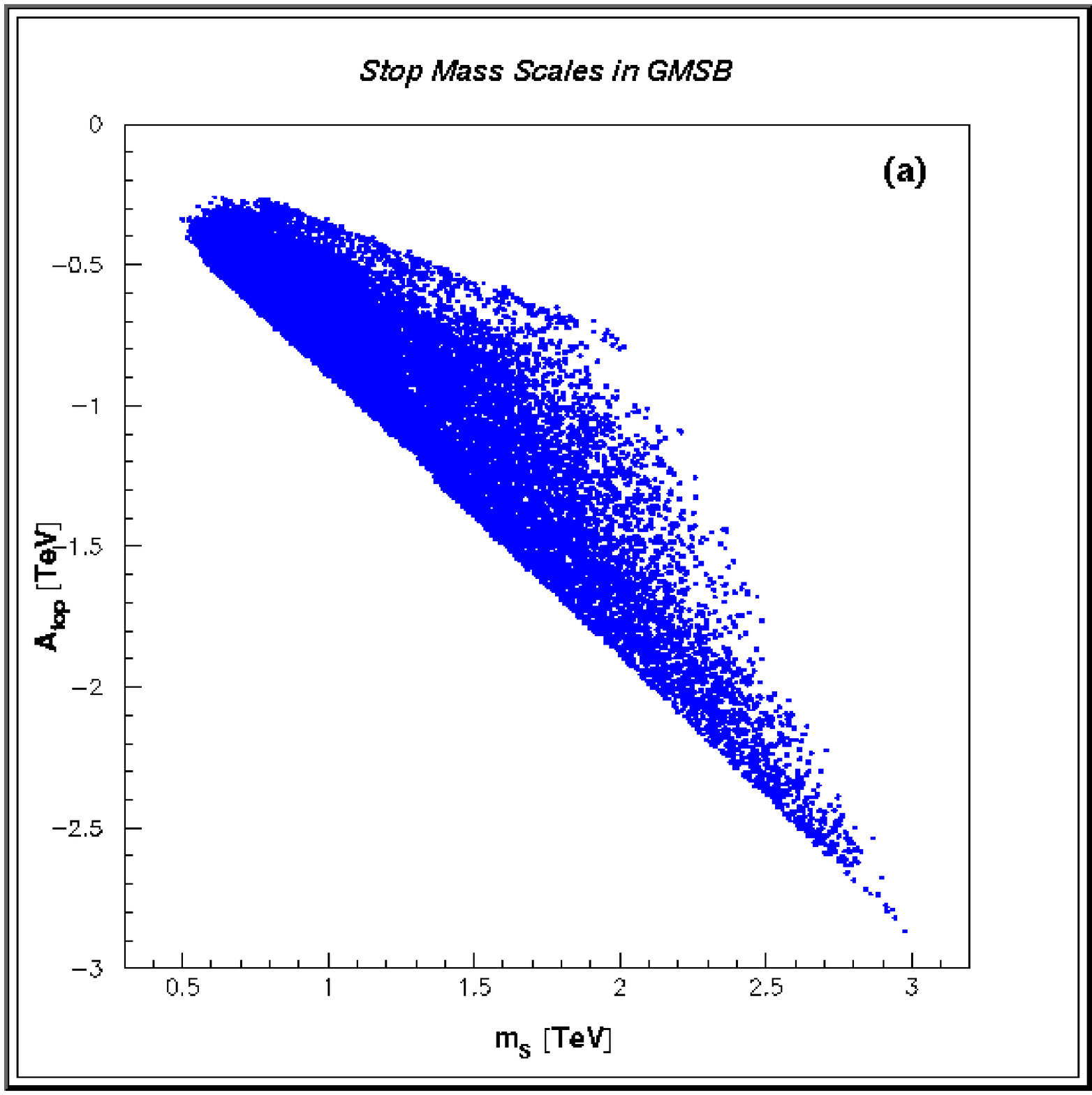}
\epsfxsize=0.49\textwidth 
\epsffile{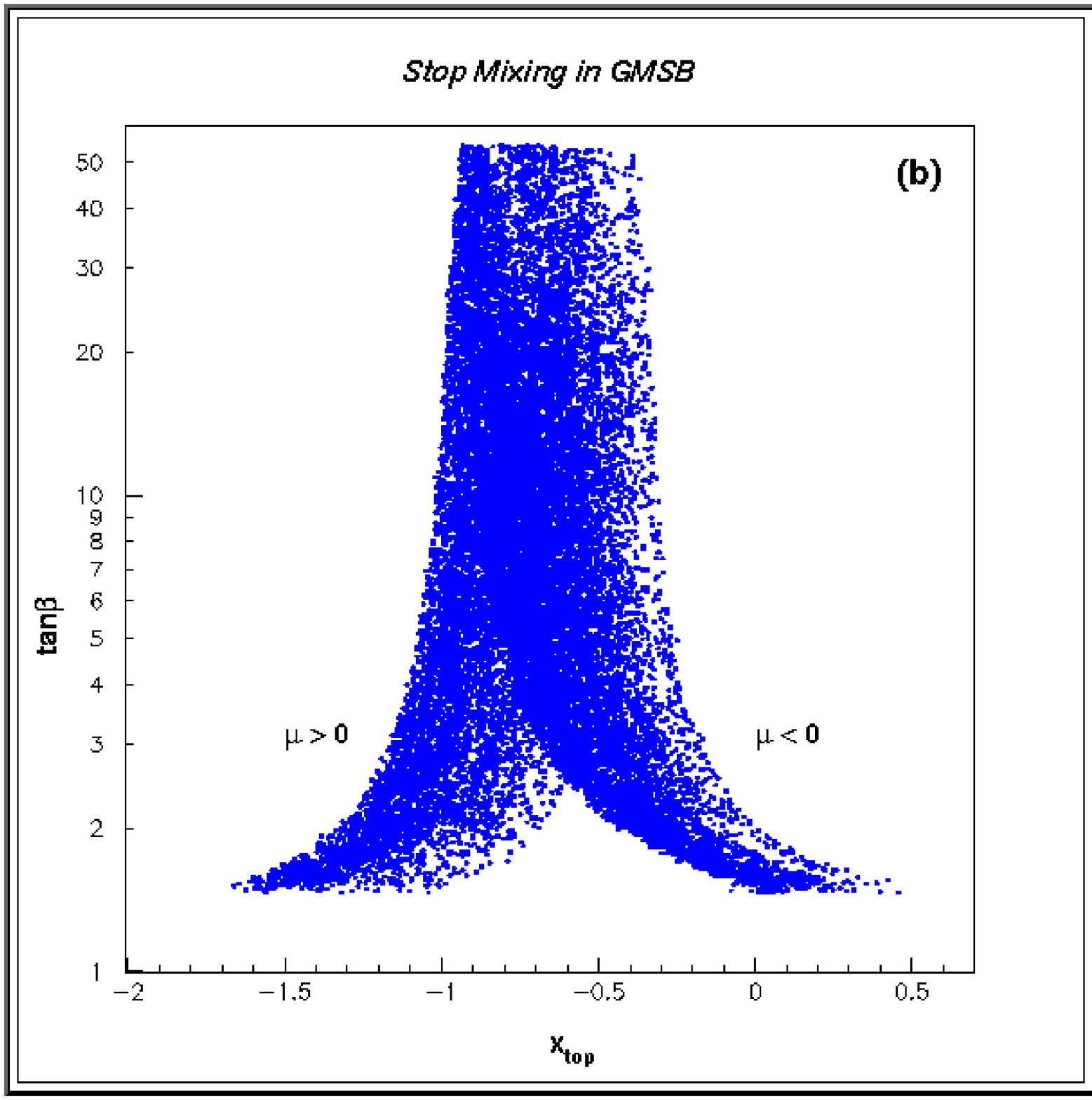}
\end{center}
\vspace*{-5mm}
\caption{\sl Scatter plots of $A_{\rm top}$ vs. $m_S$, the
mass scale appearing in the denominator of Eq.~(\ref{eq:stmix}) (a)
and $\tgb$ vs. $x_{\rm top}$ (b).
}
\label{fig:atvsmsxtvstb}
\end{figure}

In Fig.~\ref{fig:mhvsxt}, we show the dependence of the lightest Higgs 
boson mass on the stop mixing parameter $x_{\rm top}$ defined by 

\be 
x_{\rm top} \equiv \frac{A_{\rm top} - \mu/\tgb}{m_S}, \; \; \; \;
{\rm where} \; \; \; \; m_S = {\sqrt{(m_{\stopu}^2 + m_{\stopd}^2)/2}}.
\label{eq:stmix}
\ee

For equal soft SUSY breaking parameters in the stop sector with
the $D$-terms neglected, $x_{\rm top}$ corresponds to the ratio 
$X_t/M_S$ of the off-diagonal and diagonal entries in the stop mixing
matrix, see e.g. Ref.~\cite{bse}.

\begin{figure}[ht]
\begin{center}
\hspace{-0.5cm}
\epsfxsize=0.49\textwidth 
\epsffile{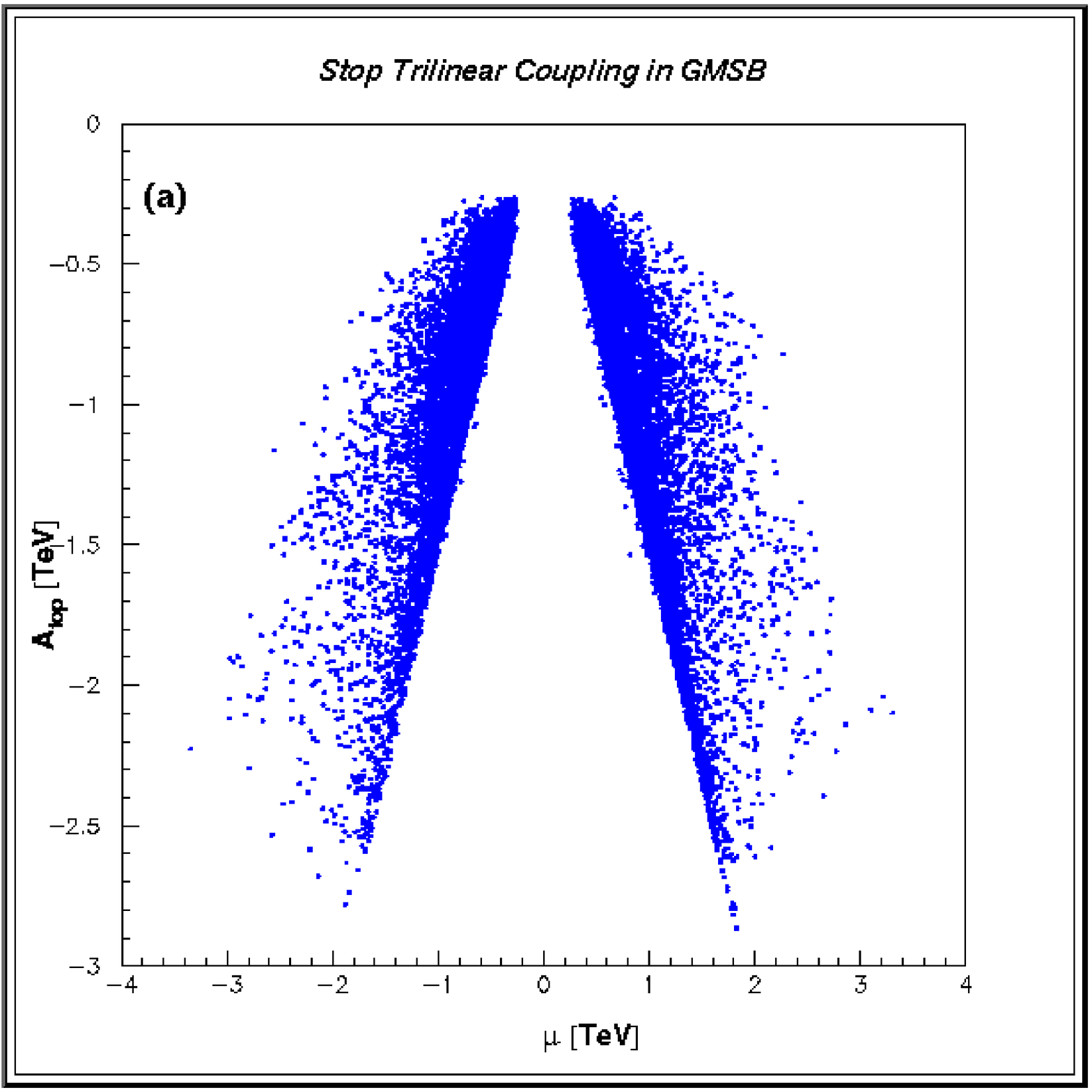}
\epsfxsize=0.49\textwidth 
\epsffile{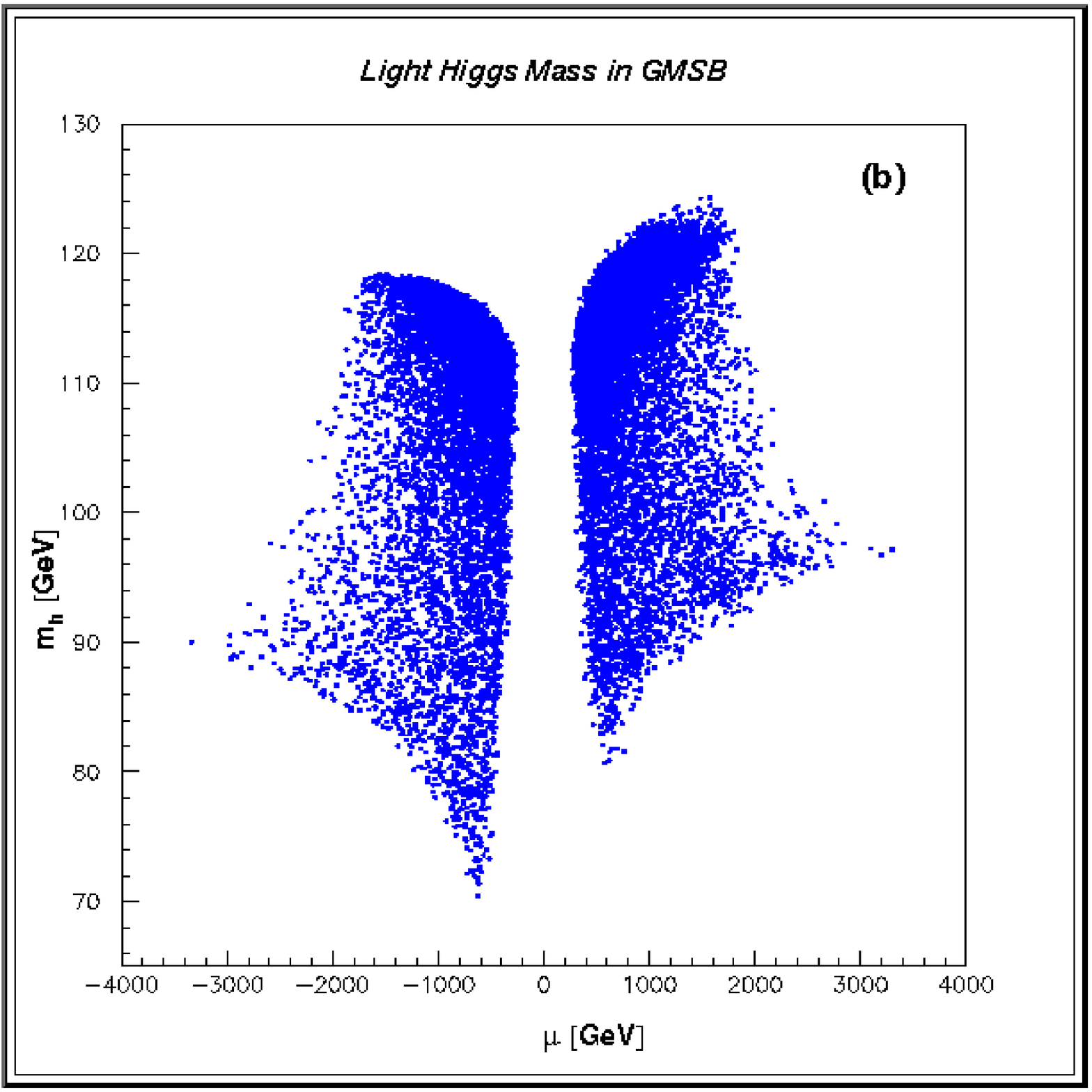}
\end{center}
\vspace*{-5mm}
\caption{\sl Scatter plots for $A_{\rm top}$ (a) and 
the light scalar Higgs mass (b) vs. the SUSY higgsino 
mass $\mu$ evaluated at the electroweak scale.
}
\label{fig:at-mhvsmu}
\end{figure}

Maximal $\mh$ values are obtained for $x_{\rm top} \approx \pm 2$, a
minimum is reached around $x_{\rm top} \approx 0$. Thus, for large
$\mh$ values a large numerator in Eq.~(\ref{eq:stmix}) is required.
From Fig.~\ref{fig:atvsmsxtvstb}(a), one can see that in GMSB only negative 
values of $A_{\rm top}$ are allowed at the electroweak scale, as a 
consequence of the fact that the trilinear couplings are negligible at 
the messenger scale. Due to the logarithmic dependence of $\mh$ on the stop 
masses, relatively large values of $|A_{\rm top}|$ are needed for large
$\mh$. In addition, large $\tgb$ is also required. 
From Fig.~\ref{fig:atvsmsxtvstb}(b) 
one can check that this leads to values of $x_{\rm top} \approx -0.95$, 
which can only be achieved for positive $\mu$. 
Fig.~\ref{fig:at-mhvsmu}(a) shows the dependence of $A_{\rm top}$
on $\mu$. Large values of $|A_{\rm top}|$ are only reached for large 
$|\mu|$ values. Therefore maximal $h$ masses are obtained for relatively 
large and positive $\mu$, as can be seen in 
Fig.~\ref{fig:at-mhvsmu}(b).\footnote{In general, for large values of 
$|\mu|$ and $\tgb$ the effects of the corrections from the $b$--$\tilde{b}$ 
sector can become important, leading to a decrease in $\mh$. For the 
GMSB models under consideration, however, this is not the case as a 
consequence of the relatively large $\tilde{b}$ masses.}

All these arguments about the combination of low energy parameters
needed for large $\mh$ in GMSB are summarized in Tab.~\ref{tab:one}.
where we report the 10 models in our sample that give rise to the
highest $\mh$ values. Together with $\mh$, Tab.~\ref{tab:one} shows
the corresponding input GMSB parameters (Eq.~(\ref{eq:pars}) in
\cite{ambro1}) as well as the values of the low energy parameters
mentioned above.

It is interesting to note that all the models shown in Tab.~1 feature
a large messenger index and values of the messenger scale not far from
the maximum we allowed while generating GMSB models. We could not
construct a single model with $\mh \gtap 122.5$ GeV having $N_{\rm
mess} < 6$ or $M_{\rm mess} < 10^5$ TeV, for $m_t = 175$ GeV. It is
hence worth mentioning here that our choice of imposing $M_{\rm
mess}/\Lambda < 10^5 \Rightarrow M_{\rm mess} \ltap 2 \times 10^{10}$
GeV does not correspond to any solid theoretical prejudice. On the
other hand it is true that $M_{\rm mess} \gtap 3 \times 10^8$ GeV
always corresponds to gravitino masses larger than $\sim 1$ keV, due
to Eqs.~(\ref{eq:Gmass}) and (\ref{eq:sqrtFmin}) in \cite{ambro1}.
The latter circumstance might be disfavored by cosmological
arguments~\cite{cosmo}. A curious consequence is that the GMSB models
with the highest $\mh$ belong always to the stau NLSP or slepton
co-NLSP scenarios.

Note also that restricting ourselves to GMSB models with 
$\Lambda < 100$ TeV, $M_{\rm mess} < 10^5$ TeV and $N_{\rm mess} \le 4$,
we find a maximal $\mh$ value of 122.2 GeV, for $\mt = 175$ GeV and 
$\tgb \sim 52$. This is to be compared with the one-loop result of 
Ref.~\cite{kaeding}, $\mh$(max) = 131.7, for $\tgb$ around 30 (the
assumed value of $\mt$ is not quoted).

Values for $\mh$ slightly larger than those we found here may also
arise from non-minimal contributions to the Higgs potential, in
connection with a dynamical generation of $\mu$ and $B\mu$
\cite{ambro1}. A treatment of this problem can be found in
Ref.~\cite{AHW_1}.

One should also keep in mind that our analysis still suffers from 
uncertainties due to unknown higher order corrections both in the RGEs
for GMSB model generation and in the evaluation of $\mh$ from low energy 
parameters. A rough estimate of these effects leads to shifts in $\mh$
not larger than 3 to 5 GeV.

\section{Conclusions}
\label{sec:A4}

\noindent
We conclude that in the minimal GMSB framework described above, values 
of $\mh \gtap 124.2$ GeV are not allowed for $\mt = 175$ GeV. This is 
almost 6 GeV smaller than the maximum value for $\mh$ one can achieve 
in the MSSM without any constraints or assumptions about the structure 
of the theory at high energy scales~\cite{mhiggslong,tbexcl,bench}.
On the other hand, the alternative mSUGRA framework allows values of 
$\mh$ that are $\sim 3$ GeV larger than in GMSB~\cite{mhiggsmSUGRA_1}. 
This makes the GMSB scenario slightly easier to explore via Higgs boson 
search. This result was expected in the light of the rather strong GMSB 
requirements, such as the presence of a unique soft SUSY breaking scale, 
the relative heaviness of the squarks and the gluino compared to 
non-strongly interacting sparticles, and the fact that the soft SUSY
breaking trilinear couplings $A_f$ get nonzero values at the electroweak 
scale only by RGE evolution.
Nevertheless, once the whole parameter space is explored, it is not true 
that mGMSB gives rise to $\mh$ values that are considerably smaller than in 
mSUGRA. Even smaller differences in the maximal $\mh$ might be present 
when considering non-minimal, complex messenger sectors~\cite{kaeding}
or additional contributions to the Higgs potential~\cite{GMSBmodels2_2,AHW_1}.
In any case, as for mSUGRA, current LEP~II or Tevatron data on Higgs boson 
searches are far from excluding mGMSB, and the upgraded Tevatron and 
the LHC will certainly be needed to deeply test any realistic SUSY model. 

\begin{table}[ht]
\begin{tabular}{|r||c|c|c|c|c|c|c|c|c|c|c|}  \hline
ID & $\mh$ & $N_{\rm mess}$ & $M_{\rm mess}$    & $\Lambda$ &
$\mu$ & $\tgb$ & $m_A$ & $m_{\stopu}$ 
& $m_{\stopd}$ & $A_{\rm top}$ & $x_{\rm top}$ \\
& GeV &        & $10^6$ TeV    & TeV  & GeV &             
& GeV & GeV  & GeV    & GeV    &  \\ \hline \hline
A1    & 124.2 & 7 & 1.00 & 72.7 & 1470 & 53.4 & 367 & 2320
      & 2510  & -2150 & -0.90 \\ \hline 

A2    & 124.2 & 8 & 4.48 & 66.7 & 1570 & 53.1 & 436 & 2400
      & 2600  & -2310 & -0.94 \\ \hline

A3    & 123.7 & 6 & 2.07 & 87.0 & 1580 & 52.8 & 485 & 2420
      & 2630  & -2240 & -0.90 \\ \hline 

A4    & 123.7 & 8 & 4.67 & 52.4 & 1270 & 52.9 & 373 & 1930 
      & 2100  & -1850 & -0.93 \\ \hline

A5    & 123.5 & 8 & 4.89 & 51.1 & 1250 & 52.7 & 388 & 1880
      & 2050  & -1810 & -0.93 \\ \hline

A6    & 123.5 & 6 & 2.54 & 67.1 & 1260 & 53.0 & 349 & 1910 
      & 2080  & -1760 & -0.89 \\ \hline

A7    & 123.4 & 8 & 4.62 & 61.6 & 1470 & 51.9 & 549 & 2230
      & 2430  & -2160 & -0.94 \\ \hline

A8    & 123.4 & 6 & 4.15 & 88.1 & 1630 & 51.9 & 609 & 2450  
      & 2670  & -2300 & -0.91 \\ \hline
 
A9    & 123.3 & 7 & 3.77 & 70.3 & 1490 & 51.8 & 567 & 2260
      & 2460  & -2150 & -0.92 \\ \hline

A10   & 123.3 & 8 & 3.74 & 72.1 & 1677 & 50.7 & 756 & 2580 
      & 2800  & -2500 & -0.94 \\ \hline
\end{tabular}
\caption{\sl The 10 GMSB models giving rise to the highest $\mh$ values
in our sample. For each model, together with the light Higgs mass, we show 
the values of the GMSB input parameters and other low energy parameters of
interest for calculating $\mh$.}
\label{tab:one}
\end{table}

\newpage

\setcounter{section}{0}
\setcounter{figure}{0}
\setcounter{table}{0}
\setcounter{equation}{0}

\begin{center}

{\large \bf Measuring the SUSY breaking scale at the LHC
in the slepton NLSP scenario of GMSB models} \\

\vspace*{0.3cm} 

{\sc S.~AMBROSANIO, B.~MELE, S.~PETRARCA, G.~POLESELLO and A.~RIMOLDI} \\

\end{center}

\vspace*{0.3cm}

\begin{abstract}

This study is focused on the measurement of the fundamental
SUSY breaking scale $\sqrt{F}$ at the LHC in the GMSB scenario where
a stau is the next-to-lightest SUSY particle (NLSP) and decays into a 
gravitino with $c\tau_{\rm NLSP}$ in the range 0.5~m to 1~km.
This implies the measurement of mass and lifetime of long lived sleptons.
The identification is performed by determining the time of flight in the 
ATLAS muon chambers. Accessible range and precision on $\sqrt{F}$ 
achievable with a counting method are assessed.

\end{abstract}

\section{Introduction}
\label{sec:B1}

\noindent
The fundamental scale of SUSY breaking $\sqrt{F}$ is perhaps the most
important quantity to determine from phenomenology in a SUSY theory.
In the mSUGRA framework, the gravitino mass sets the scale of the soft
SUSY breaking masses in the MSSM ($\sim 0.1-1$ TeV), so that
$\sqrt{F}$ is typically large $\sim 10^{10-11}$ GeV
(Eq.~(\ref{eq:Gmass}) in \cite{ambro2}). As a consequence, the
interactions of the $\G$ with the other MSSM particles $\sim F^{-1}$
are too weak for the gravitino to be of relevance in collider physics
and there is no direct way to access $\sqrt{F}$ experimentally. In
GMSB theories, the situation is completely different.  The soft SUSY
breaking scale of the MSSM and the sparticle masses are set by gauge
interactions between the messenger and low energy sectors to be $\sim
\alpha_{\rm SM}\Lambda$ (Eq.~(\ref{eq:bound}) in \cite{ambro2}), so
that typical $\Lambda$ values are $\sim 10-100$ TeV. On the other
hand, $\sqrt{F}$ is subject to the lower bound (\ref{eq:sqrtFmin}) in
\cite{ambro2} only, which tells us that values well below $10^{10}$
GeV and even as low as several tens of TeV are perfectly reasonable.
The $\G$ is in this case the LSP and its interactions are strong
enough to allow NLSP decays to the $\G$ inside a typical detector
size.  The latter circumstance gives us a chance for extracting
$\sqrt{F}$ experimentally through a measurement of the NLSP mass and
lifetime (Eq.~(\ref{eq:NLSPtau}) in \cite{ambro2}).

Furthermore, the possibility of determining $\sqrt{F}$ with good
precision opens a window on the physics of the SUSY breaking
(``secluded'') sector and the way this SUSY breaking is transmitted to
the messenger sector. Indeed, the characteristic scale of SUSY
breaking felt by the messengers (and hence the MSSM sector) given by
$\sqrt{F_{\rm mess}}$ in Eq.~(\ref{eq:sqrtFmin}) can also be
determined once the MSSM spectrum is known. By comparing the measured
values of $\sqrt{F}$ and $\sqrt{F_{\rm mess}}$ it might well be
possible to get information on the way the secluded and messenger
sectors communicate with each other. For instance, if it turns out that
$\sqrt{F_{\rm mess}} \ll \sqrt{F}$, then it is very likely that the
communication occurs radiatively and the ratio $\sqrt{F_{\rm mess}/F}$
is given by some loop factor. On the contrary, if the communication
occurs via a direct interaction, this ratio is just given by a
Yukawa-type coupling constant, with values $\ltap 1$, see
Refs.~\cite{GR-GMSB_2,AKM-LEP2_2}.

An experimental method to determine $\sqrt{F}$ at a TeV scale $\epem$
collider through the measurement of the NLSP mass and lifetime was
presented in Ref.~\cite{AB-LC_2}, in the neutralino NLSP scenario.
Here, we are concerned with the same problem, but at a hadron
collider, the LHC, and in the stau NLSP or slepton co-NLSP scenarios.
These scenarios provide a great opportunity at the LHC, since the
characteristic signatures with semi-stable charged tracks are
muon-like, but come from massive sleptons with $\beta$ significantly
smaller than 1. In particular, we perform our simulations in the ATLAS
muon detector, whose large size and excellent time resolution
\cite{TDR_am} allow a precision measurement of the slepton time of
flight from the production vertex out to the muon chambers and hence
of the slepton velocity.  Moreover, in the stau NLSP or slepton
co-NLSP scenarios, the knowledge of the NLSP mass and lifetime is
sufficient to determine $\sqrt{F}$, since the factor ${\cal B}$ in
Eq.(\ref{eq:NLSPtau}) of \cite{ambro2} is exactly equal to 1. This is
not the case in the neutralino NLSP scenario where ${\cal B}$ depends
at least on the neutralino physical composition and more information
and measurements are needed for extracting a precise value of
$\sqrt{F}$.

\section{Choice of the Sample Models and Event Simulation}
\label{sec:B2}

\noindent 
The two main parameters affecting the experimental measurement at the
LHC of the slepton NLSP properties are the slepton mass and momentum 
distribution. Indeed, at a hadron collider most of the NLSPs come
from squark and gluino production followed by cascade decays. 
Thus, the momentum distribution is in general a function of the whole 
MSSM spectrum. However, one can approximately assume that most of the 
information on the NLSP momentum distribution is provided by the squark 
mass scale $m_{\tilde q}$ only (in the stau NLSP scenario or slepton 
co-NLSP scenarios of GMSB, one generally finds 
$m_{\tilde{g}} \gtap m_{\tilde q}$). 
To perform detailed simulations, we select a representative set of 
GMSB models generated by {\tt SUSYFIRE}. We limit ourselves to models 
with $m_{\rm NLSP} > 100$ GeV, motivated by the discussion in 
\cite{ambro2}, and $m_{\tilde q} < 2$ TeV, in order to yield 
an adequate event statistics after a three-year low-luminosity run 
(corresponding to 30 fb$^{-1}$) at the LHC.
Within these ranges, we choose eight extreme points (four in the stau 
NLSP scenario and four in the slepton co-NLSP scenario) allowed by GMSB
in the ($m_{\rm NLSP}$, $m_{\tilde q}$) plane, in order to cover the 
various possibilities. 

\begin{table}[h!]
\begin{center}
\begin{tabular}{|r||r|r|r|r|c|} \hline
ID & $M_{\rm mess}$ (TeV) & $N_{\rm mess}$ 
   & $\Lambda$ (TeV)  & $\tgb$ & sign($\mu$)  \\ \hline\hline
B1 & 1.79$\times 10^4$ & 3 &  26.6  &  7.22  & --   \\ \hline
B2 & 5.28$\times 10^4$ & 3 &  26.0  &  2.28  & --   \\ \hline
B3 & 4.36$\times 10^2$ & 5 &  41.9  & 53.7~  & +    \\ \hline
B4 & 1.51$\times 10^2$ & 4 &  28.3  &  1.27  & --   \\ \hline
B5 & 3.88$\times 10^4$ & 6 &  58.6  & 41.9~  & +    \\ \hline
B6 & 2.31$\times 10^5$ & 3 &  65.2  &  1.83  & --   \\ \hline
B7 & 7.57$\times 10^5$ & 3 & 104~~  &  8.54  & --   \\ \hline
B8 & 4.79$\times 10^2$ & 5 &  71.9  &  3.27  & --   \\ \hline
\end{tabular}
\caption{\sl Input parameters of the sample GMSB models chosen for our
study.}
\label{tab:tbg}
\end{center}
\end{table}

In Tab.~\ref{tab:tbg}, we list the input GMSB parameters we used, while 
in Tab.~\ref{tab:tbg1} we report the corresponding values of the stau 
mass, the squark mass scale and the gluino mass. The ``NLSP'' column
indicates whether the model belongs to the stau NLSP or slepton co-NLSP
scenario. The last column gives the total cross section in pb for producing
any pairs of SUSY particles at the LHC. 

\begin{table}[h!]
\begin{center}
\begin{tabular}{|r||r|c|r|r|c|} \hline
ID &  $m_{\tilde\tau_1}$ (GeV) 
   & ``NLSP'' & $m_{\tilde q}$ (GeV) & $m_{\tilde g}$ (GeV) 
   & $\sigma$ (pb) \\ \hline \hline
B1 &  100.1 & $\tilde\tau$  &  577 &  631 & 42~~~~~ \\ \hline
B2 &  100.4 & $\tilde\ell$  &  563 &  617 & 50~~~~~ \\ \hline
B3 &  101.0 & $\tilde\tau$  & 1190 & 1480 & ~0.59~  \\ \hline
B4 &  103.4 & $\tilde\ell$  &  721 &  859 & 10~~~~~ \\ \hline
B5 &  251.2 & $\tilde\tau$  & 1910 & 2370 & ~0.023  \\ \hline
B6 &  245.3 & $\tilde\ell$  & 1290 & 1410 & ~0.36~  \\ \hline
B7 &  399.2 & $\tilde\tau$  & 2000 & 2170 & ~0.017  \\ \hline
B8 &  302.9 & $\tilde\ell$  & 1960 & 2430 & ~0.022  \\ \hline 
\end{tabular}
\caption{\sl Features of the sample GMSB model points studied.}
\label{tab:tbg1}
\end{center}
\end{table}

For each model, the events were generated with the {\tt ISAJET} 
Monte Carlo \cite{isajet_am} that incorporates the calculation of 
the SUSY mass spectrum and branching fraction 
using the GMSB parameters as input. We have checked that for 
the eight model points considered the sparticle masses 
calculated with {\tt ISAJET} are in good agreement with the output  
of {\tt SUSYFIRE}.

The generated events were then passed through {\tt ATLFAST} \cite{ATLFAST_am}, 
a fast particle-level simulation of the ATLAS detector. The {\tt ATLFAST}
package, however, was only used to evaluate the efficiency of the calorimetric
trigger that selects the GMSB events. The detailed response of the detector 
to the slepton NLSP has been parameterized for this work using the results  
of a full simulation study, as described in the next section.

\section{Slepton detection}
\label{sec:B3} 

The experimental signatures of heavy long-lived charged 
particles at a hadron collider have already been studied both 
in the framework of GMSB and in more general scenarios 
\cite{leandro,drtata, femoroi,marthom}.
The two main observables one can use to separate
these particles from muons are the high specific ionization
and the time of flight in the detector.

We concentrate here on the measurement of the time of flight, 
made possible by the timing precision ($\ltap 1$~ns) 
and the size of the ATLAS muon spectrometer.

It was demonstrated with a full simulation of the ATLAS muon
detector~\cite{gpar} that the $\beta$ of a particle can be measured
with a resolution that can be approximately parameterized as
$\sigma(\beta)/\beta^2=0.028$. The resolution on the transverse
momentum measurement for heavy particles is found to be comparable to
the one expected for muons. We have therefore simulated the detector
response to NLSP sleptons by smearing the slepton momentum and $\beta$
according to the parameterizations in Ref.~\cite{gpar}.

An important issue is the online selection of the SUSY events.
We have not made any attempt to evaluate whether the heavy sleptons 
can be selected using the muon trigger. For the event selection, we 
rely on the calorimetric $E_{T}^{\rm miss}$ trigger, 
consisting in the requirement of at least one hadronic jet with $p_T>50$~GeV,
and a transverse momentum imbalance calculated only from the energy 
deposition in the calorimeter larger than 50~GeV. We checked that 
this trigger has an efficiency in excess of 80\% for all the considered
models. 

A detailed discussion of the experimental assumptions underlying
the results presented here is given in Ref.~\cite{AMPPR}.

\section{Event Selection and Slepton Mass Measurement} 
\label{sec:B4} 

\noindent
In order to select a clean sample of sleptons, we apply the
following requirements: 

\begin{itemize}
\item
at least one hadronic jet with $P_T>50$~GeV and a calorimetric 
\mbox{$E_T^{\rm miss}>50$~GeV} (trigger requirement);
\item
at least one candidate slepton satisfying the following cuts: 
\begin{itemize}
\item
$|\eta|<$2.4 to ensure that the particle is in the
acceptance of the muon trigger chamber, and therefore 
both coordinates can be measured;
\item
$\beta_{\rm meas}<0.91$, where $\beta_{\rm meas}$ is the $\beta$ of the
particle measured with the time of flight in the precision chambers;
\item
The $P_T$ of the slepton candidate, after the energy loss in the
calorimeters has been taken into account, must be larger than
10~GeV, to ensure that the particle traverse all of the muon stations.
\end{itemize}
\end{itemize} 

If we consider an integrated luminosity of 30~fb$^{-1}$, 
a number of events ranging from a few hundred for the models with
2~TeV squark mass scale to a few hundred thousand for a 500~GeV
mass scale survive these cuts and can be used for measuring the 
NLSP properties.

From the measurements of the slepton momentum and of particle $\beta$,
the mass can be determined using the standard relation 
$m=p\frac{\sqrt{1-\beta^2}}{\beta}$. 
For each value of $\beta$ and momentum, the measurement error is
known and it is given by the parameterizations in Ref.~\cite{gpar}.
Therefore, the most straightforward way of measuring the mass 
is just to use the weighted average of all the masses 
calculated with the above formula. 

In order to perform this calculation, the particle momentum is needed,
which implies measuring the $\eta$ coordinate. In fact, with the precision 
chambers only one can only measure the momentum components  
transverse to the beam axis. 

The measurement of the second coordinate must be provided by the 
trigger chambers, for which only a limited time window around the beam 
crossing is read out, therefore restricting the $\beta$ range where 
this measurement is available. Hence, we have evaluated the achieved 
measurement precision for two different $\beta$ intervals: $0.6<\beta<0.91$ 
and $0.8<\beta<0.91$ for the eight sample points.  
We found a statistical error well below the 0.1\% level for those model
points having $m_{\tilde q} < 1300$ GeV. Even for the three models 
(B5, B7, B8) with lower statistics ($m_{\tilde q} \simeq 2$ TeV), the 
error stays below the 0.4\% level. 

Many more details, tables and figures about this part of our study 
can be found in Ref.~\cite{AMPPR}. 

\section{Slepton Lifetime Measurement}
\label{sec:B5}

\noindent 
The measurement of the NLSP lifetime at a high energy $e^+e^-$ collider
was studied in detail in Ref.~\cite{AB-LC_2} for the neutralino NLSP case.
Similar to that study, the measurement of the slepton NLSP lifetime we are 
interested in here can be performed by exploiting the fact that two NLSPs 
are produced in each event. 
One can therefore select $N_1$ events where a slepton is detected through 
the time-of-flight measurement described above, count the number of 
times $N_2$ when a second slepton is observed and use this information  
to measure the lifetime. 
Although in principle very simple, in practice this method requires 
an excellent control on all possible sources of inefficiency for 
detecting the second slepton.

\begin{figure}[h!]
\begin{center} 
\epsfxsize=0.7\textwidth
\epsffile{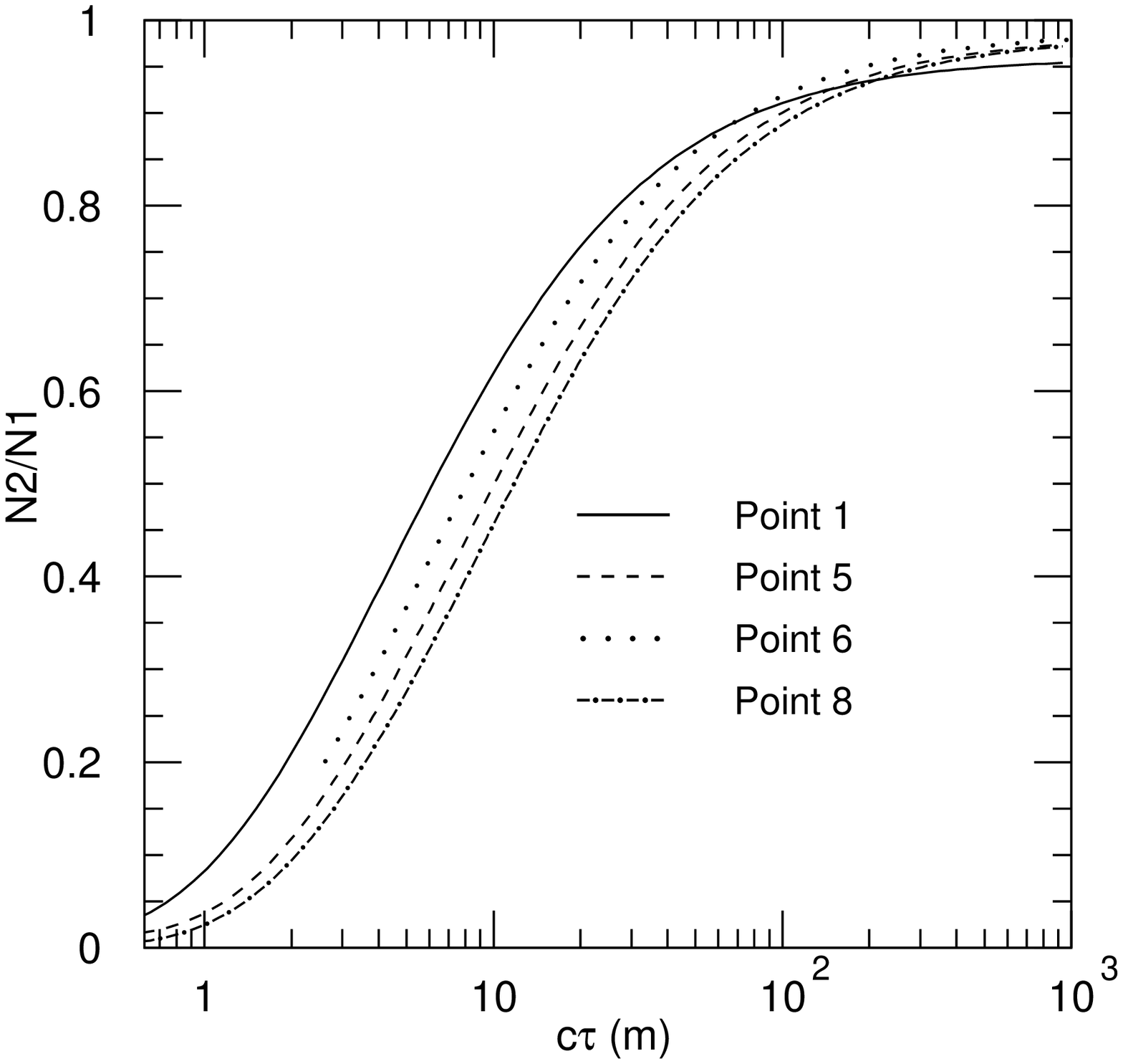}
\end{center}
\vspace*{-7mm}
\caption{\sl The ratio $R=N_2/N_1$ defined in the text as a function 
of the slepton lifetime $c\tau$. Only the curves corresponding to 
the model points B1, B5, B6, B8 are shown.} 
\label{fig:ratio}
\end{figure} 

We give here the basis of the method, without mentioning the experimental 
details. We provide an estimate of the achievable statistical error for 
the models considered and a parameterization of the effect on the lifetime
measurement of a generic systematic uncertainty on the slepton efficiency.
In case the sparticle spectrum and BRs can be measured from the SUSY events, 
as e.g. shown in Ref.~\cite{ihfp}, an accurate simulation of all the SUSY 
production processes can be performed, and the results from this section are 
representative of the measurement precision achievable in a real experiment.

Another method based on the same principles, but assuming minimal knowledge 
of the SUSY spectrum, is described in Ref.~\cite{AMPPR}, where 
a detailed estimate of the achievable systematic precision is given.

We define $N_1$ starting from the event sample 
defined by the cuts discussed in Sec.~\ref{sec:B4}, 
with the additional requirement that, 
for a given value of the slepton lifetime, 
at least one of the produced sleptons decays at 
at a distance from the interaction vertex $>10$~m,
and is therefore reconstructed in the muon system.
For the events thus selected, we define $N_2$ as the 
subsample where a second particle with a transverse momentum 
$>10$~GeV is identified in the muon system. The search for the
second particle should be as inclusive as possible, in order to minimize
the corrections to the ratio. In particular, the cut $\beta_{\rm meas}<0.91$ 
is not applied, but  particles with a mass measured from $\beta$ and momentum 
incompatible with the measured slepton mass are rejected. This leaves
a background of high momentum muons in the sample that can be statistically
subtracted using the momentum distribution of electrons. 
The ratio

\be
R=\frac{N_2}{N_1}
\label{eq:ratio}
\ee

\noindent
is a function of the slepton lifetime. Its dependence on the NLSP
lifetime $c\tau$ in meters in shown in Fig.~\ref{fig:ratio} for four
among our eight sample models. The curves for the model points not
shown are either very similar to one of the curves we show or are
mostly included between the external curves corresponding to points B1
and B8, thus providing no essential additional information.  Note that
the curve for model 6 starts from $c\tau = 2.5$~m and not from $c\tau
= 50$~cm, as for the other models. This is due to the large value of
$M_{\rm mess}$ (cfr. Tab.~\ref{tab:tbg}), determining a minimum NLSP
lifetime allowed by theory which is macroscopic in this case
(Eqs.~(\ref{eq:NLSPtau}) and (\ref{eq:sqrtFmin}) in \cite{ambro2}).

\begin{figure}[h!] 
\epsfxsize = 1.\textwidth 
\vspace*{-5mm}
\epsffile{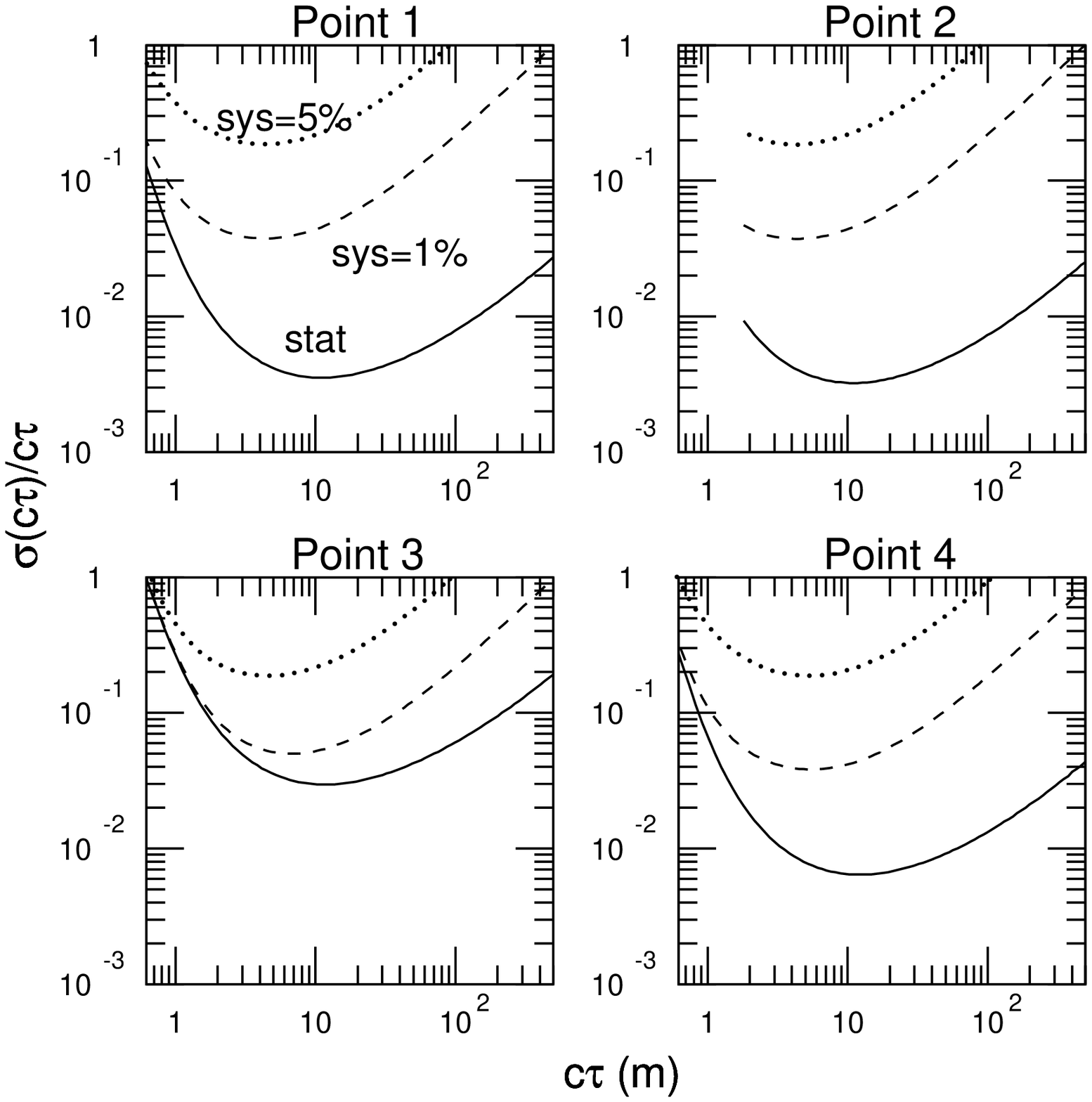}
\vspace*{-10mm}
\caption{\sl Fractional error on the measurement of the 
slepton lifetime $c\tau$, for model sample points B1 to B4.
We assume an integrated luminosity of 30~fb$^{-1}$. 
The curves are shown for three different assumptions on the 
fractional systematic error on the $R$ measurement: 
statistical error only (full line), 1\% systematic error (dashed line),
5\% systematic error (dotted line).
}
\label{fig:sigctau1}
\end{figure}

The probability for a particle of mass $m$, momentum $p$ and proper 
lifetime $\tau$ to travel a distance $L$ before decaying is
given by the expression

\be
P(L)=e^{-mL/pc\tau}.
\label{eq:prob} 
\ee

$N_2$ is therefore a function of the momentum distribution
of the slepton, which is determined by the details of the SUSY spectrum.
One therefore needs to be able to simulate the full SUSY cascade decays
in order to construct the $c\tau$--$R$ relationship. 

\begin{figure}[h!]
\epsfxsize = 1.0\textwidth
\vspace*{-5mm}
\epsffile{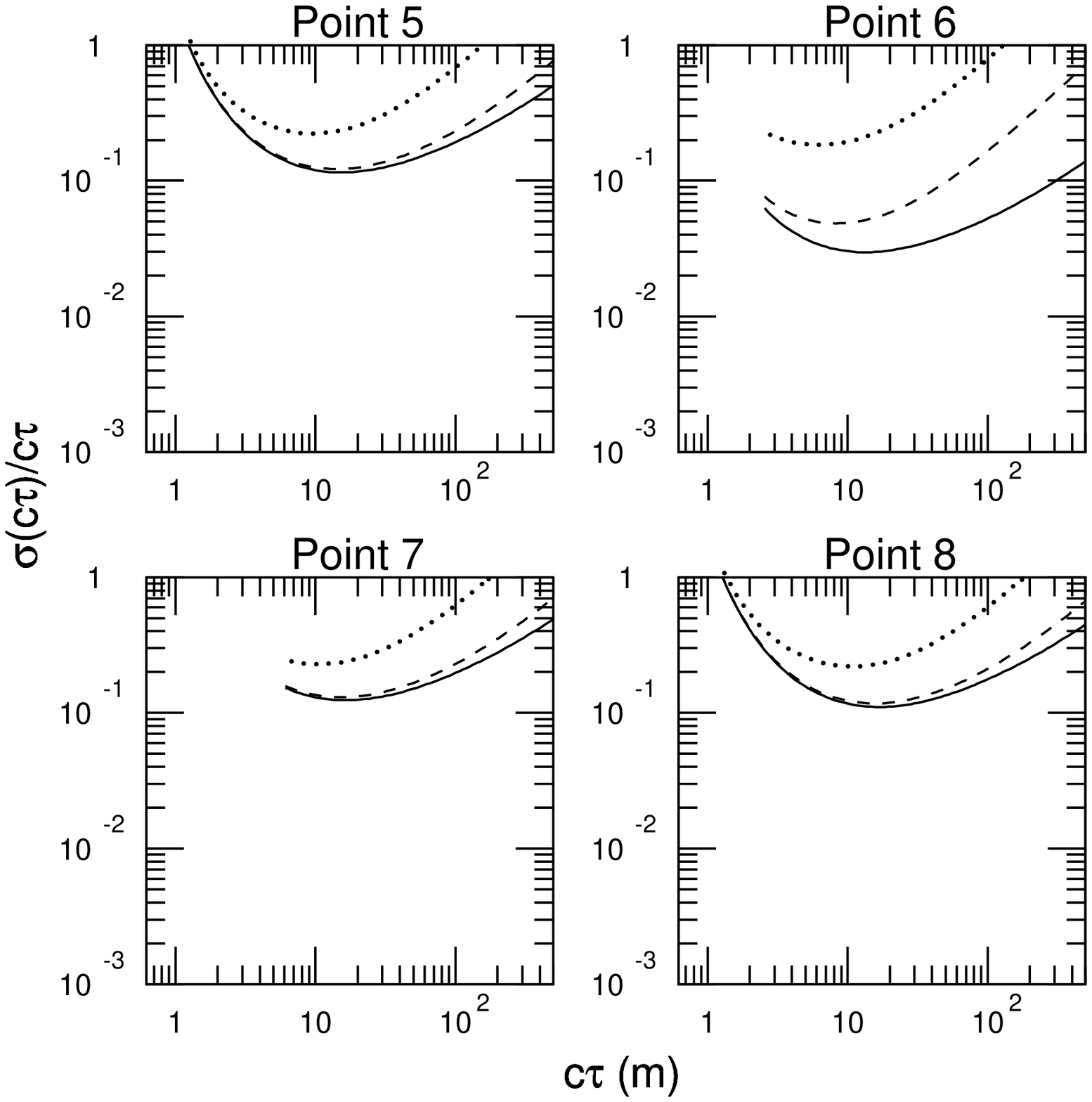}
\vspace*{-10mm}
\caption{\sl The same as in Fig.~\ref{fig:sigctau2}, but for 
the model sample points B5 to B8.
}
\label{fig:sigctau2}
\end{figure}

The statistical error on $R$ can be evaluated as

\be
\sigma(R)=\sqrt{\frac{R(1-R)}{N_1}}.
\label{eq:errat} 
\ee

Relevant for the precision with which the SUSY breaking scale can be 
measured is instead the error on the measured $c\tau$. This can be 
extracted from the curves shown in Fig.~\ref{fig:ratio} and can be 
evaluated as

\be
\sigma({\mathrm c}\tau)=\sigma(R)/
\left[\frac{\partial R(c\tau)}{\partial c\tau}\right].
\label{eq:erctau}
\ee

The measurement precision calculated according to this formula is shown 
in Figs.~\ref{fig:sigctau1} and \ref{fig:sigctau2} for the eight sample 
points, for an integrated luminosity of 30~fb$^{-1}$. 
The full line in the plots is the error on $c\tau$ considering the statistical
error on $R$ only. The available statistics is a function 
of the strongly interacting sparticles' mass scale. 
Even if a precise $R$--$c\tau$ relation can be built from the 
knowledge of the model details, there will be a systematic uncertainty 
in the evaluation of the losses in $N_2$, because of sleptons produced outside
the $\eta$ acceptance, or absorbed in the calorimeters, or escaping the 
calorimeter with a transverse momentum below the cuts. 
The full study of these uncertainties is in progress. 

\begin{figure}[h!]
\epsfxsize = 1.0\textwidth
\vspace*{-5mm}
\epsffile{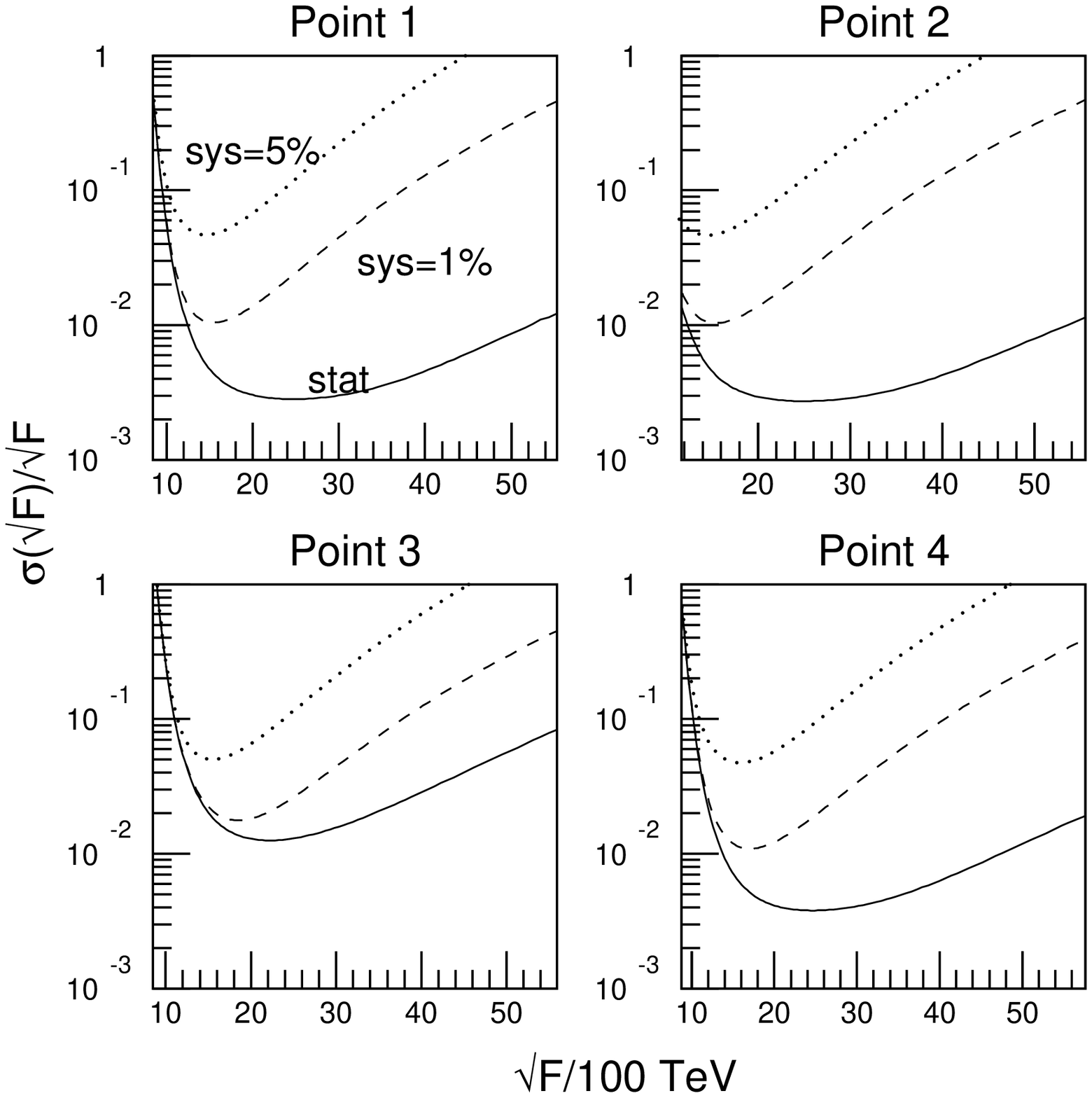}
\vspace*{-10mm}
\caption{\sl Fractional error on the measurement of the 
SUSY breaking scale $\sqrt{F}$ for model sample points B1 to B4.
We assume an integrated luminosity of 30~fb$^{-1}$. 
The curves are shown for the three different assumptions on the fractional 
systematic error used in Figs.~\ref{fig:sigctau1} and \ref{fig:sigctau2}.
}
\label{fig:sigsqrtf1}
\end{figure}

At this level, we just parameterize the systematic error as a term 
proportional to $R$, added in quadrature to the statistical error. 
We choose two values, $1\% R$ and $5\% R$, and propagate the error to 
the $c\tau$ measurement. 
The results are represented by the dashed and dotted lines in
Figs.~\ref{fig:sigctau1} and \ref{fig:sigctau2}. 

For the models with squark mass scales up to 1200~GeV,
assuming a 1\% systematic error on the measured ratio, 
a precision better than 10\% on the $c\tau$ measurement 
can be obtained for lifetimes between 0.5--1~m and 50--80~m. 
If the systematic uncertainty grows to 5\%, the 10\% precision
can only be achieved in the range 1--10~m. If the mass scale goes up to 2~TeV,
even considering a pure statistical error only, a 10\% precision
is not achievable. However a 20\% precision is possible  
over $c\tau$ ranges between 5 and 100~m, assuming a 1\% systematic error.

\begin{figure}[h!]
\epsfxsize = 1.0\textwidth
\vspace*{-5mm}
\epsffile{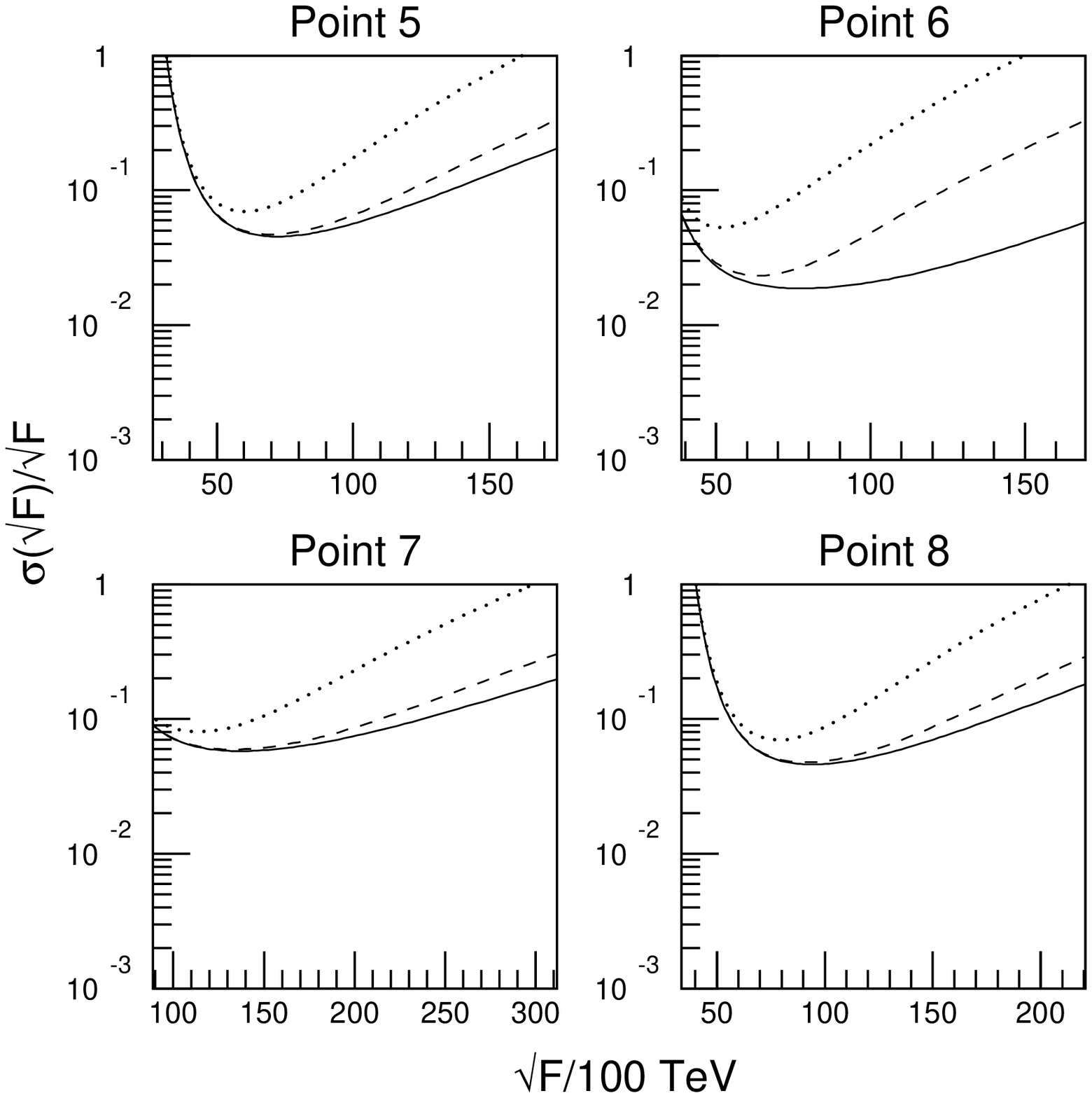}
\vspace*{-10mm}
\caption{\sl The same as in Fig.~\ref{fig:sigsqrtf1}, but for model
sample points B5 to B8.
}
\label{fig:sigsqrtf2}
\end{figure}

Note that the curves corresponding to the model points B2, B6 and B7 
do not start from $c\tau = 50$ cm, but from the theoretical lower 
limit on $c\tau$ of 1.8, 2.5 and 6.1 meters, respectively.

\section{Determining the SUSY Breaking Scale $\sqrt{F}$}
\label{sec:B6}

\noindent 
Using the measured values of $c\tau$ and the NLSP mass, the SUSY
breaking scale $\sqrt{F}$ can be calculated from Eq.(\ref{eq:NLSPtau})
in \cite{ambro2}, where ${\cal B} = 1$ for the case where the NLSP is
a slepton. From simple error propagation, the fractional uncertainty
on the $\sqrt{F}$ measurement can be obtained adding in quadrature one
fourth of the fractional error in $c\tau$ and five fourths of the
fractional error on the slepton mass.

In Figs.~\ref{fig:sigsqrtf1} and \ref{fig:sigsqrtf2}, we show the
fractional error on the $\sqrt{F}$ measurement as a function of
$\sqrt{F}$ for our three different assumptions on the $c\tau$ error.
The uncertainty is dominated by $c\tau$ for the higher part of the
$\sqrt{F}$ range and grows quickly when approaching the lower limit on
$\sqrt{F}$. This is because very few sleptons survive and the
statistical error on both $m_{\tilde \ell}$ and $c\tau$ gets very
large.  If we assume a 1\% systematic error on the ratio $R$ from
which $c\tau$ is measured (dashed lines in Figs.~\ref{fig:sigsqrtf1}
and \ref{fig:sigsqrtf2}), the error on $\sqrt{F}$ is better than 10\%
for $1000 \ltap \sqrt{F} \ltap 4000$~TeV for model points B1--B4 with
higher statistics. For points B5--B8, in general one can explore a
range of higher $\sqrt{F}$ values with a small relative error,
essentially due to the heaviness of the decaying NLSP in these models.
Note also that the theoretical lower limit (\ref{eq:sqrtFmin}) in
\cite{ambro2} on $\sqrt{F}$ is equal to about 1200, 1500, 3900, 8900
TeV respectively in model points B2, B5, B6, B7, while it stays well
below 1000 TeV for the other models.

\section{Conclusions}
\label{sec:B7}

\noindent 
We have discussed a simple method to measure at the LHC with the ATLAS 
detector the fundamental SUSY breaking scale $\sqrt{F}$ in the GMSB scenarios 
where a slepton is the NLSP and decays to the gravitino with a lifetime
in the range 0.5~m $\ltap c\tau_{\rm NLSP} \ltap 1$~km. This method requires
the measurement of the time of flight of long lived sleptons 
and is based on counting events with one or two identified NLSPs.  
It relies on the assumptions that a good knowledge of the MSSM sparticle 
spectrum and BRs can be extracted from the observation of the SUSY events 
and that the systematic error in evaluating the slepton losses can be kept
below the few percent level.
We performed detailed, particle level simulations for eight representative 
GMSB models, some of them being particularly hard due to low statistics. 
We found that a level of precision of a few 10's \% on the SUSY breaking 
scale measurement can be achieved in significant parts of the 
$1000 \ltap \sqrt{F} \ltap 30000$~TeV range, for all models considered. 
More details as well as a full study of the systematics associated with this 
procedure and another less ``model-dependent'' method to measure $\sqrt{F}$ 
is presented in detail in Ref.~\cite{AMPPR}.

\setcounter{figure}{0}
\setcounter{table}{0}
\setcounter{section}{0}
\setcounter{equation}{0}
\setcounter{footnote}{0}
\newpage

\def\wtil{\widetilde}

\def\ignore#1{}
\def\sgn{\mathop{\rm sgn}}
\def\pion{\pi}
\def\gev{\, {\rm GeV}}
\def\etmiss{\slashchar{E}_T}
\def\tq{{\tilde q}}
\def\tg{{\tilde g}}
\def\tchi{{\tilde\chi}}
\def\lsp{{\tilde\chi_1^0}}
\def\tell{{\tilde\ell}}
\def\ttop{{\tilde t}}
\def\tb{{\tilde b}}
\def\tG{{\tilde G}}
\def\Meff{M_{\rm eff}}
\def\mgrav{m_{3/2}}
\def\fbi{{\rm fb^{-1}}}
\def\cm{{\rm cm}}

\def\slashchar#1{\setbox0=\hbox{$#1$}           
   \dimen0=\wd0                                 
   \setbox1=\hbox{/} \dimen1=\wd1               
   \ifdim\dimen0>\dimen1                        
      \rlap{\hbox to \dimen0{\hfil/\hfil}}      
      #1                                        
   \else                                        
      \rlap{\hbox to \dimen1{\hfil$#1$\hfil}}   
      /                                         
   \fi}                                         %

\begin{center}

{\large\bf Anomaly mediated SUSY breaking at the LHC} \\
\vspace*{3mm}

{\sc F.E. PAIGE and J. WELLS} \\
\end{center}
\vspace*{3mm}

\begin{abstract}
Anomaly Mediated SUSY Breaking models are reviewed. Possible signatures
at the LHC for one case of the minimal realistic model are examined.

\end{abstract}
\section{Introduction}

The signatures for SUSY at the LHC depend very much on the SUSY masses,
which presumably result from spontaneous SUSY breaking. It is not
possible to break SUSY spontaneously using just the MSSM fields; instead
one must do so in a hidden sector and then communicate the breaking
through some interaction. In supergravity models, the communication is
through gravity. In gauge mediated models it is through gauge
interactions; the gravitino is then very light and can play an important
role. Simple examples of both have been discussed previously. A third
possibility is that the hidden sector does not have the right structure
to provide masses through either mechanism; then the leading
contributions come from a combination of gravity and anomalies. This is
known as Anomaly Mediated SUSY Breaking (AMSB), and it predicts a different
pattern of masses and signatures.

\section{Anomaly-Mediated Supersymmetry Breaking}

In the supersymmetric standard model there exist AMSB contributions
to the soft mass parameters that arise via the superconformal 
anomaly~\cite{Randall:1998uk, Giudice:1998xp}.
The effect
can be understood by recognizing several important features of
supersymmetric theories.  First, supersymmetry
breaking can be represented by a chiral superfield 
$\Phi=1+m_{3/2}\theta^2$ which also acts as a compensator
for super-Weyl transformations. Treating $\Phi$ as a spurion, one can
transform a theory into a super-conformally invariant theory.  
Even if a theory is superconformal at the outset (i.e., no
dimensionful couplings), the spurion $\Phi$ is employed since the
quantum field theory requires a regulator that implies scale 
dependence (Pauli-Villars
mass, renormalization scale in dimensional reduction, etc.).  
To preserve scale invariance the renormalization scale parameter $\mu$ in
a quantum theory then becomes $\mu/\sqrt{\Phi^\dagger\Phi}$.  It is
the dependence of the regulator on $\Phi$ that induces supersymmetry
breaking contributions to the scalars and gauginos.

The anomaly induced masses can be derived straightforwardly
for the scalar masses.  The K\" ahler kinetic terms depend on wave function
renormalization as in the following superfield operator,
\be
 \int d^2\theta d^2\bar \theta Z_Q
\left( \frac{\mu}{\sqrt{\Phi^\dagger\Phi}}
 \right) Q^\dagger Q.
\ee
Taylor expanding $Z$ around $\mu$ and projecting out the $FF^\dagger$ terms
yields a supersymmetry breaking mass for the scalar field $\tilde Q$:
\be
\label{squarkmass}
m_{\tilde Q}^2 = -\frac{1}{4}\frac{d^2\ln Z_Q}{d(\ln \mu )^2} m^2_{3/2}
  = -\frac{1}{4}\left(\frac{\partial\gamma_Q}{\partial g}\beta_g
   +\frac{\partial \gamma_Q}{\partial y}\beta_y\right)m_{3/2}^2.
\ee
Similar calculations can be done for the gauginos and the $A$ terms:
\beq
M_i & = & -\frac{g_i^2}{2}\frac{dg_i^{-2}}{d\ln \mu} m_{3/2} 
    =\frac{\beta_{g_i}}{g_i}m_{3/2}, \\
A_y & = & \frac{1}{2}\sum_a \frac{d\ln Z_{Q_a}}{d\ln\mu}m_{3/2}
   = -\frac{\beta_y}{y}m_{3/2}
\label{gauginomass}
\eeq
where the sum over $a$ includes all fields associated with the Yukawa
coupling $y$ in the superpotential.

There are several important characteristics 
of the AMSB spectrum to note. First,
the equations for the supersymmetry breaking contributions are
scale invariant.  That is, the value of the soft masses at any scale is
obtained by simply plugging in the gauge couplings and Yukawa couplings 
at that scale into the above formulas.  Second, the masses are related to the
gravitino mass by a one loop suppression.  In AMSB 
$M_i\sim m_{3/2}\alpha_i /4\pi$, whereas in SUGRA $M_i\sim m_{3/2}$.
While the AMSB contributions are always present in a theory independent of
how supersymmetry breaking is accomplished, they may be highly
suppressed compared to standard hidden sector models.  Therefore, for
AMSB to be the primary source of scalar masses, one needs to assume or
arrange that supersymmetry breaking is not directly communicated from a
hidden sector.  This can be accomplished, for example, by assuming 
supersymmetry breaking on a distant brane~\cite{Randall:1998uk}.
Finally, the squared masses
of the sleptons are 
negative (tachyonic)
because $\beta_g>0$ for $U(1)$ and $SU(2)$ gauge groups.  This
problem rules out the simplest AMSB model based solely on 
eqs.~\ref{squarkmass}-\ref{gauginomass}.

Given the tachyonic slepton problem, it might seem most rational to
view AMSB as a good idea that did not quite work out.  However, there
are many reasons to reflect more carefully on AMSB.  As already mentioned
above, AMSB contributions to scalar masses are always present if
supersymmetry is broken.  Soft masses in the MSSM come for free, whereas
in all other successful theories of supersymmetry breaking a communication
mechanism must be detailed.  In particle, hidden sector models require
singlets to give the gauginos an acceptable mass.  
In AMSB, singlets are not necessary.
Also, there may be small variations on the AMSB idea that can
produce a realistic spectrum and can have important 
phenomenological consequences.  This is our motivation for writing 
this note. 

\section{Two realistic minimal models of AMSB: mAMSB and DAMSB}

As we discussed in the introduction,
the pure AMSB model gives negative squared masses for the
sleptons, thus breaking electromagnetic gauge invariance, so some
additional contributions must be included. The simplest assumption
that solves this problem is to add
at the GUT scale
a single universal scalar
mass $m_0^2$ to all the sfermions' squared masses.
We will call this model mAMSB.
The description and many phenomenological implications
of this model are given in Refs.~\cite{GGW,Feng:1999hg}.
The parameters of the model after the usual
radiative electroweak symmetry breaking are then
$$
m_0,\quad \mgrav,\quad \tan\beta,\quad \sgn\mu=\pm .
$$
This model has been implemented in ISAJET~7.48~\cite{ISAJET_we}; a
pre-release version of ISAJET has been used to generate the events for
this analysis.

For this note the AMSB parameters were chosen to be
$$
m_0=200\,\GeV,\quad \mgrav=35\,\TeV,\quad \tan\beta=3,\quad \sgn\mu=+
$$
For this choice of parameters the slepton squared masses are positive at
the weak scale, but they are still negative at the GUT scale. This means
that charge and color might be broken (CCB) at high temperatures in the
early universe. However, at these high energies there are also large
finite temperature effects on the mass, which are positive (symmetry
restoration occurs at higher $T$). In fact, a large class of SUSY
models with CCB minima naturally fall into the correct SM minimum when
you carefully follow the evolution of the theory from high T to today.
If CCB minima are excluded at all scales, then the value of $m_0$ must
be substantially larger, so the sleptons must be quite heavy.

	The masses from ISAJET~7.48 for this point are listed in
Table~\ref{mtable}. The mass spectrum has some similarity to that for
SUGRA Point 5 studied previously~\cite{HPSSY,TDR_we}: the gluino and squark
masses are similar, and the decays $\tchi_2^0 \to \tell\ell$ and
$\tchi_2^0 \to \lsp h$ are allowed. Thus, many of the techniques
developed for Point 5 are applicable here. But there are also important
differences. In particular, the $\tchi_1^\pm$ is nearly degenerate with
the $\lsp$, not with the $\tchi_2^0$.  The mass splitting between
the lightest chargino and the lightest neutralino must be calculated
as the difference between the lightest eigenvalues
of the full one-loop neutralino and chargino mass matrices.
The mass splitting is always above $m_{\pion^\pm}$, thereby allowing
the two-body decays 
$\chi^\pm_1\to\chi^0_1+\pion^\pm$~\cite{GGW,Feng:1999fu}.  Decay lifetimes
of $\chi^\pm$ are always less than 10~cm over mAMSB parameter space, and
are often less than 1~cm.

Another unique feature of the spectrum is the near degeneracy of
the $\tilde\ell_L$ and $\tilde\ell_R$ sleptons.  The mass splitting 
is~\cite{GGW}
\beq
m^2_{\tell_L}-m^2_{\tell_R}\simeq 0.037 \left( -m^2_Z\cos 2\beta
  +M^2_2\ln \frac{m_{\tell_R}}{m_Z}\right).
\eeq
There is no symmetry requiring this degeneracy, but rather it
is an astonishing accident and prediction of the mAMSB model.

It is instructive to compare the masses
from ISAJET with those calculated in Ref.~\cite{GGW} to provide
weak-scale input to ISAJET. These masses are listed in the right hand
side of Table~\ref{mtable}. Since the agreement is clearly adequate for
the purposes of the present study, no attempt has been made to
understand or resolve the differences. It is clear, however, that if
SUSY is discovered at the LHC and if masses or combinations of masses
are measured with the expected precision, then more work is needed
to compare the LHC results with theoretical models in a sufficiently
reliable way.

\begin{table}[t]
\caption{Masses of the SUSY particles, in GeV, for the mAMSB model with
$m_0=200\,\GeV$, $m_{3/2}=35\,\TeV$, $\tan\beta=3$, and $\sgn\mu=+$ from
ISAJET (left side) and from Ref.~\cite{GGW} (right side) using the
ISAJET sign conventions. \label{mtable}}
\begin{center}
\begin{tabular}{cccc|cccc} \hline \hline
Sparticle  & mass \qquad &Sparticle & mass &
Sparticle  & mass \qquad &Sparticle & mass \\ \hline
$\wtil g$  		& 815 	& 			&	&
$\wtil g$  		& 852 	& 			&	\\
$\wtil \chi_1^\pm$	& 101 	& $\wtil \chi_2^\pm$	& 658	&
$\wtil \chi_1^\pm$	& 98 	& $\wtil \chi_2^\pm$	& 535	\\
$\wtil \chi_1^0$	& 101	& $\wtil \chi_2^0$	& 322	&
$\wtil \chi_1^0$	& 98	& $\wtil \chi_2^0$	& 316	\\
$\wtil \chi_3^0$	& 652	& $\wtil \chi_4^0$	& 657	&
$\wtil \chi_3^0$	& 529	& $\wtil \chi_4^0$	& 534	\\
$\wtil u_L$		& 754	& $\wtil u_R$		& 758	&
$\wtil u_L$		& 760	& $\wtil u_R$		& 814	\\
$\wtil d_L$		& 757	& $\wtil d_R$		& 763	&
$\wtil d_L$		& 764	& $\wtil d_R$		& 819	\\
$\wtil t_1$		& 516	& $\wtil t_2$		& 745	&
$\wtil t_1$		& 647	& $\wtil t_2$		& 778	\\
$\wtil b_1$		& 670	& $\wtil b_2$		& 763	&
$\wtil b_1$		& 740	& $\wtil b_2$		& 819	\\
$\wtil e_L$		& 155	& $\wtil e_R$		& 153	&
$\wtil e_L$		& 161	& $\wtil e_R$		& 159	\\
$\wtil \nu_e$		& 137 	& $\wtil \nu_\tau$	& 137	&
$\wtil \nu_e$		& 144 	& $\wtil \nu_\tau$	& 144	\\
$\wtil \tau_1$		& 140	& $\wtil \tau_2$	& 166	&
$\wtil \tau_1$		& 152	& $\wtil \tau_2$	& 167	\\
$h^0$			& 107	& $H^0$			& 699	&
$h^0$			& 98	& $H^0$			& 572	\\
$A^0$			& 697	& $H^\pm$		& 701	&
$A^0$			& 569	& $H^\pm$		& 575	\\
\hline \hline
\end{tabular}
\end{center}
\end{table}

Another variation on AMSB is deflected AMSB (DAMSB).  The idea is
based on Ref.~\cite{Pomarol:1999ie} who demonstrated that realistic
sparticle spectrums with non-tachyonic sleptons can be induced if
a light {\it modulus} field $X$ (SM singlet) is coupled to heavy, non-singlet 
vector-like messenger fields $\Psi_i$ and $\bar\Psi_i$:
$$
W_{\rm mess} = \lambda_\Psi X\Psi_i\bar\Psi_i.
$$
To ensure gauge coupling unification we identify $\Psi_i$ and $\bar\Psi_i$
as $5+\bar 5$ representations of $SU(5)$.
When the messengers are integrated out at some
scale $M_0$, the beta
functions do not match the AMSB masses, and the masses are 
deflected from the AMSB renormalization group trajectory.  The subsequent
evolution of the masses below $M_0$ induces positive mass squared for
the sleptons, and a reasonable spectrum can result.  Although there
may be additional significant parameters associated with the generation
of the $\mu$ and $B_\mu$ term in the model, we assume for this discussion
that they do not affect the spectra of the MSSM fields.  The values of
$\mu$ and $B_\mu$ are then obtained by requiring that the conditions
for EWSB work out properly.

The parameters of DAMSB are
$$
m_{3/2},\quad n,\quad M_0,\quad \tan\beta,\quad \sgn\mu=\pm
$$
where $n$ is the number of $5+\bar 5$ messenger multiplets, and $M_0$ is
the scale at which the messengers are integrated out.
Practically, the spectrum is obtained by imposing the boundary conditions
at $M_0$, and then using SUSY soft mass renormalization group equations to
evolve these masses down to the weak scale.  Expressions for
the boundary conditions
can be found in Refs.~\cite{Pomarol:1999ie,rsw}, and details on
how to generate the low-energy spectrum are given in~\cite{rsw}.
The resulting spectrum of superpartners
is substantially different from that of mAMSB.  The most characteristic
feature of the DAMSB spectrum is the near proximity of all superpartner
masses.  In Table~\ref{DAMSB} we show the
spectrum of a model with $n=5$, $M_0=10^{15}\gev$, and $\tan\beta =4$
as given in~\cite{rsw}.
The LSP is the lightest neutralino, which is a Higgsino.  (Actually,
the LSP is the fermionic component of the modulus $X$, but the decay
of $\chi_1^0$ to it is much greater than collider time scales.)
All the gauginos and squarks are between $300\gev$
and $500\gev$, while the sleptons and higgsinos are a bit
lighter ($\sim 150\gev$ to $\sim 250\gev$) 
in this case.  

\begin{table}[t]
\caption{Masses of the SUSY particles, in GeV, for the DAMSB model with
$n=5$, $M_0=10^{15}\, \GeV$, and $\tan\beta=4$ from Ref.~\cite{rsw}.
\label{DAMSB}}
\begin{center}
\begin{tabular}{cccc} \hline \hline
Sparticle  & mass \qquad &Sparticle & mass \\ \hline
$\wtil g$               & 500   &                       &       \\
$\wtil \chi_1^\pm$      & 145   & $\wtil \chi_2^\pm$    & 481   \\
{$\wtil \chi_1^0$}  & 136   & $\wtil \chi_2^0$      & 152   \\
$\wtil \chi_3^0$        & 462   & $\wtil \chi_4^0$      & 483   \\
$\wtil u_L$             & 432   & $\wtil u_R$           & 384   \\
$\wtil d_L$             & 439   & $\wtil d_R$           & 371   \\
$\wtil t_1$             & 306   & $\wtil t_2$           & 454   \\
$\wtil b_1$             & 371   & $\wtil b_2$           & 406   \\
$\wtil e_L$             & 257   & $\wtil e_R$           & 190   \\
$\wtil \nu_e$           & 246   & $\wtil \nu_\tau$      & 246   \\
$\wtil \tau_1$  & 190   & $\wtil \tau_2$        & 257   \\
$h^0$                   & 98    & $H^0$                 & 297   \\
$A^0$                   & 293   & $H^\pm$               & 303   \\
\hline \hline
\end{tabular}
\end{center}
\end{table}

In summary, we have outlined
two interesting directions to pursue in modifying
AMSB to make a realistic spectrum.  The first direction we call
mAMSB, and is constructed by adding a common scalar mass to the
sfermions at the GUT scale to solve the negative squared slepton mass
problem of pure AMSB.  The other direction that we outlined is deflected
anomaly mediation that is based on throwing the scalar masses off the
pure AMSB renormalization group trajectory by integrating out heavy
messenger states coupled to a modulus.  The spectra of the two approaches
are significantly different, and we should expect the LHC signatures
to be different as well.  In this note, we study the mAMSB carefully
in a few observables to demonstrate how it is distinctive from
other, standard approaches to supersymmetry breaking, such as mSUGRA
and GMSB.

\section{LHC studies of the example mAMSB model point}

\begin{figure}[t]
\begin{center}
\dofigs{3in}{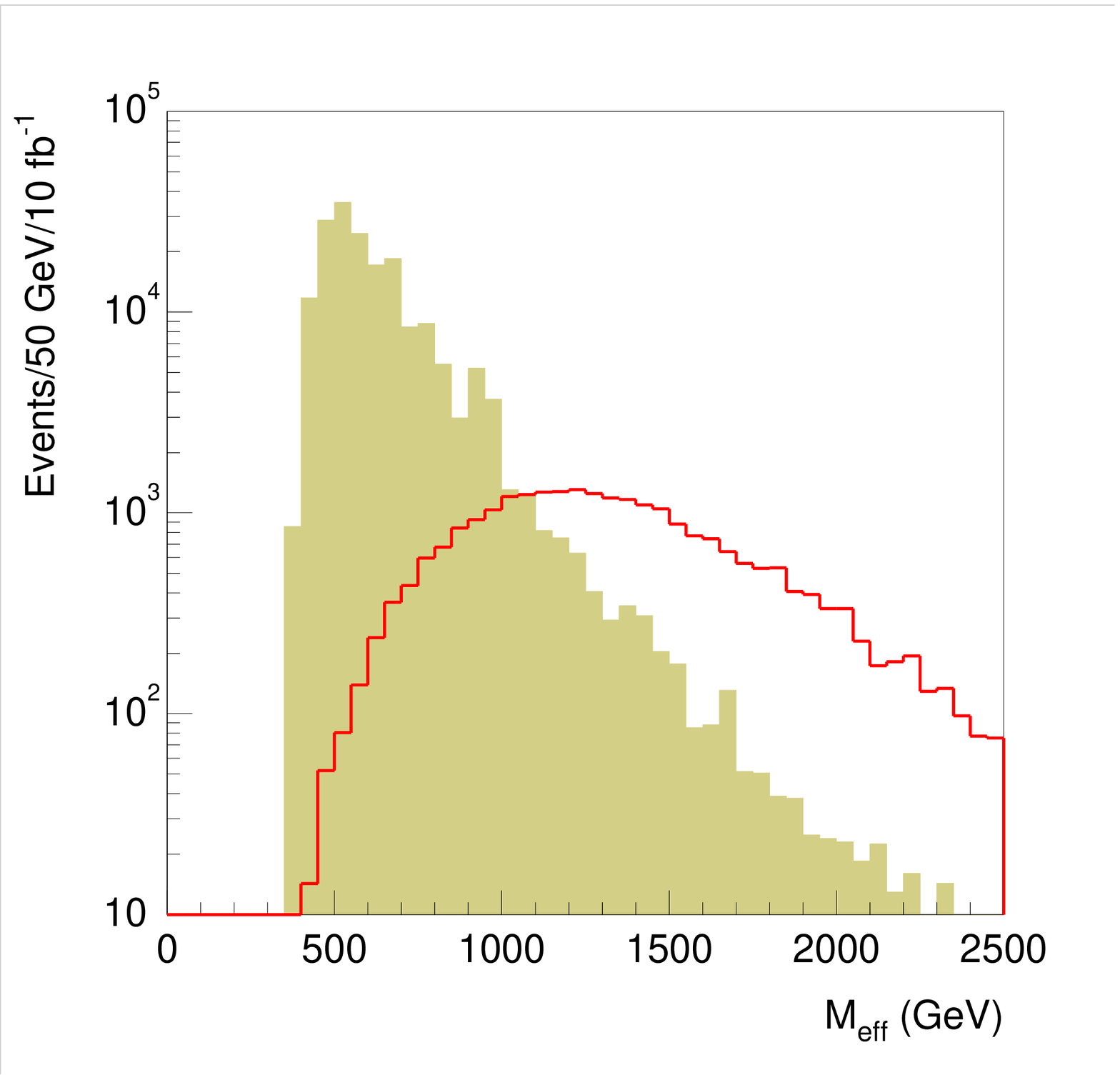}{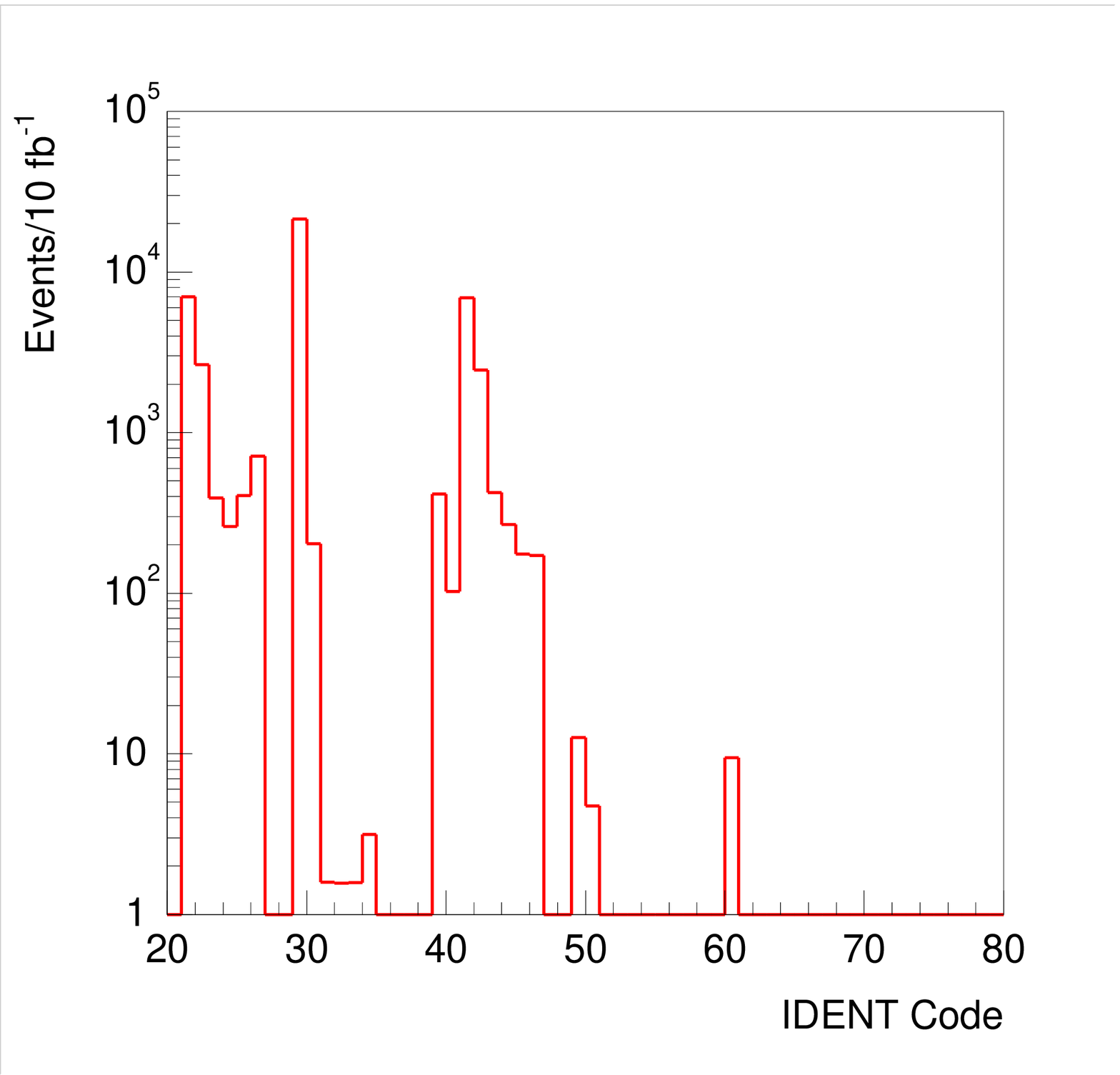}
\caption{Left: Effective mass distribution for signal (curve) and
Standard Model background (shaded). Right: ISAJET IDENT codes for all
produced particles contributing to $\Meff$ distribution after cuts. The
dominant contributions are $\tq_L$ (21--26), $\tg$ (29), and $\tq_R$
(41--46). \label{a1hmeff}}
\end{center}
\end{figure}

We now turn to a study of the example mAMSB spectra presented
in Table~1.
A sample of $10^5$ signal events was generated; since the total
signal cross section is $16\,{\rm nb}$, this corresponds to an
integrated LHC luminosity of $6\,\fbi$. All distributions shown in this
note are normalized to $10\,\fbi$, corresponding to one year at low
luminosity at the LHC. Events were selected by requiring
\begin{itemize}
\item At least four jets with $p_T>100,50,50,50\,\GeV$;
\item $\etmiss > \min(100\,\GeV,0.2\Meff)$;
\item Transverse sphericity $S_T>0.2$;
\item $\Meff > 600\,\GeV$;
\end{itemize}
where the ``effective mass'' $\Meff$ is given by the scalar sum of the
missing $E_T$ and the $p_T$'s of the four hardest jets,
$$
\Meff = \etmiss + p_{T,1} + p_{T,2} + p_{T,3} + p_{T,4} .
$$
Standard model backgrounds from gluon and light quark jets, $t\bar t$,
$W+{\rm jets}$, $Z + {\rm jets}$, and $WW$ have also been generated,
generally with much less equivalent luminosity. The $\Meff$
distributions for the signal and the sum of all backgrounds with all
except the last cut are shown in Figure~\ref{a1hmeff}. The ISAJET IDENT
codes for the SUSY events contributing to this plot are also shown. It
is clear from this plot that the Standard Model backgrounds are small
with these cuts, as would be expected from previous 
studies~\cite{HPSSY,TDR_we}.

\begin{figure}[t]
\begin{center}
\dofigs{3in}{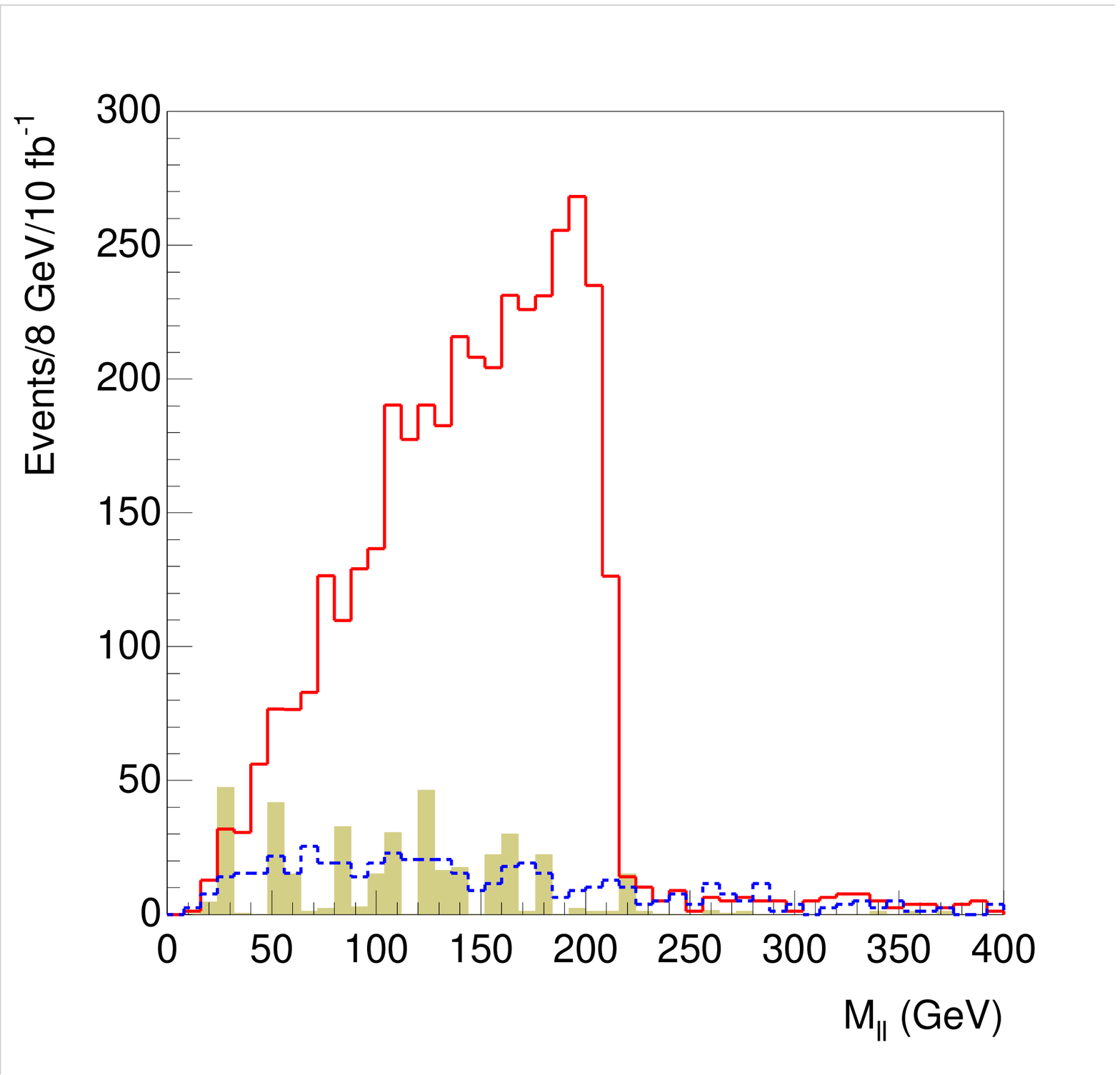}{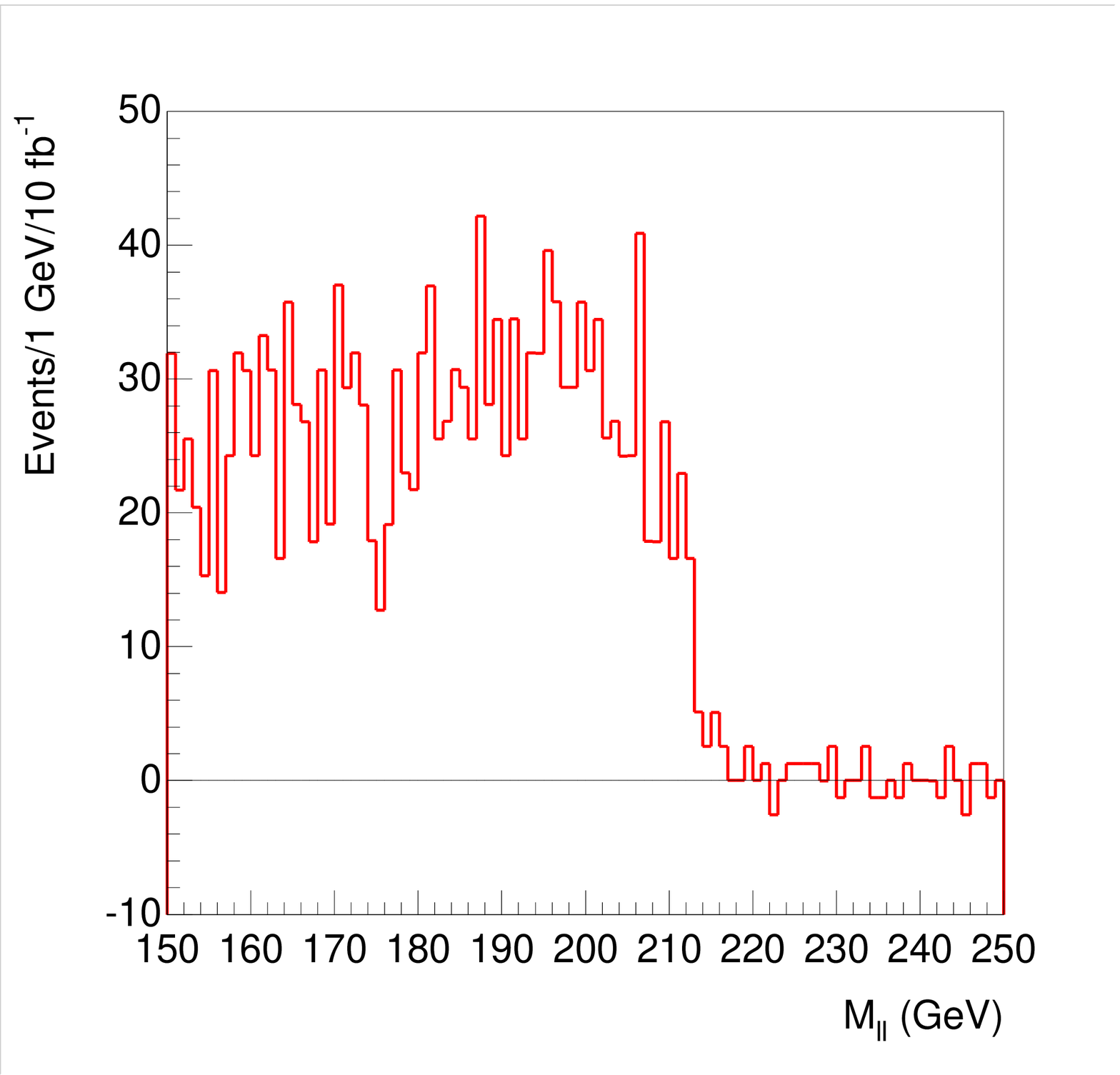}
\caption{Mass distribution for opposite sign dileptons. Left:
Distributions for same flavor signal (solid curve), opposite flavor
signal (dashed curve), and Standard Model same flavor background
(shaded). Right:  $e^+e^- + \mu^+\mu^- - e^\pm\mu^\mp$ distribution for
signal on a finer scale. \label{a1hmll}}
\end{center}
\end{figure}

	The mass distribution for $\ell^+\ell^-$ pairs with the same and
opposite flavor is shown in Figure~\ref{a1hmll}. The opposite-flavor
distribution is small, and there is a clear endpoint in the same-flavor
distribution at
$$
M_{\ell\ell}^{\rm max} = \sqrt{(M_{\tchi_2^0}^2 - M_{\tell}^2)
(M_{\tell}^2 - M_\lsp^2)\over M_{\tell}^2} = 213.6, 215.3\,\GeV
$$
corresponding to the endpoints for the decays $\tchi_2^0 \to
\tell_{L,R}^\pm \ell^\mp \to \lsp \ell^+\ell^-$. This is similar to what
is seen in SUGRA Point 5, but in that case only one slepton contributes.
It is clear from the $e^+e^- + \mu^+\mu^- - e^\pm\mu^\mp$ dilepton
distribution with finer bins shown in the same figure that the endpoints
for $\tell_R$ and $\tell_L$ cannot be resolved with the expected ATLAS
dilepton mass resolution. More work is needed to see if the presence of
two different endpoints could be inferred from the shape of the edge of
the dilepton distribution.

\begin{figure}[t]
\begin{center}
\dofigs{3in}{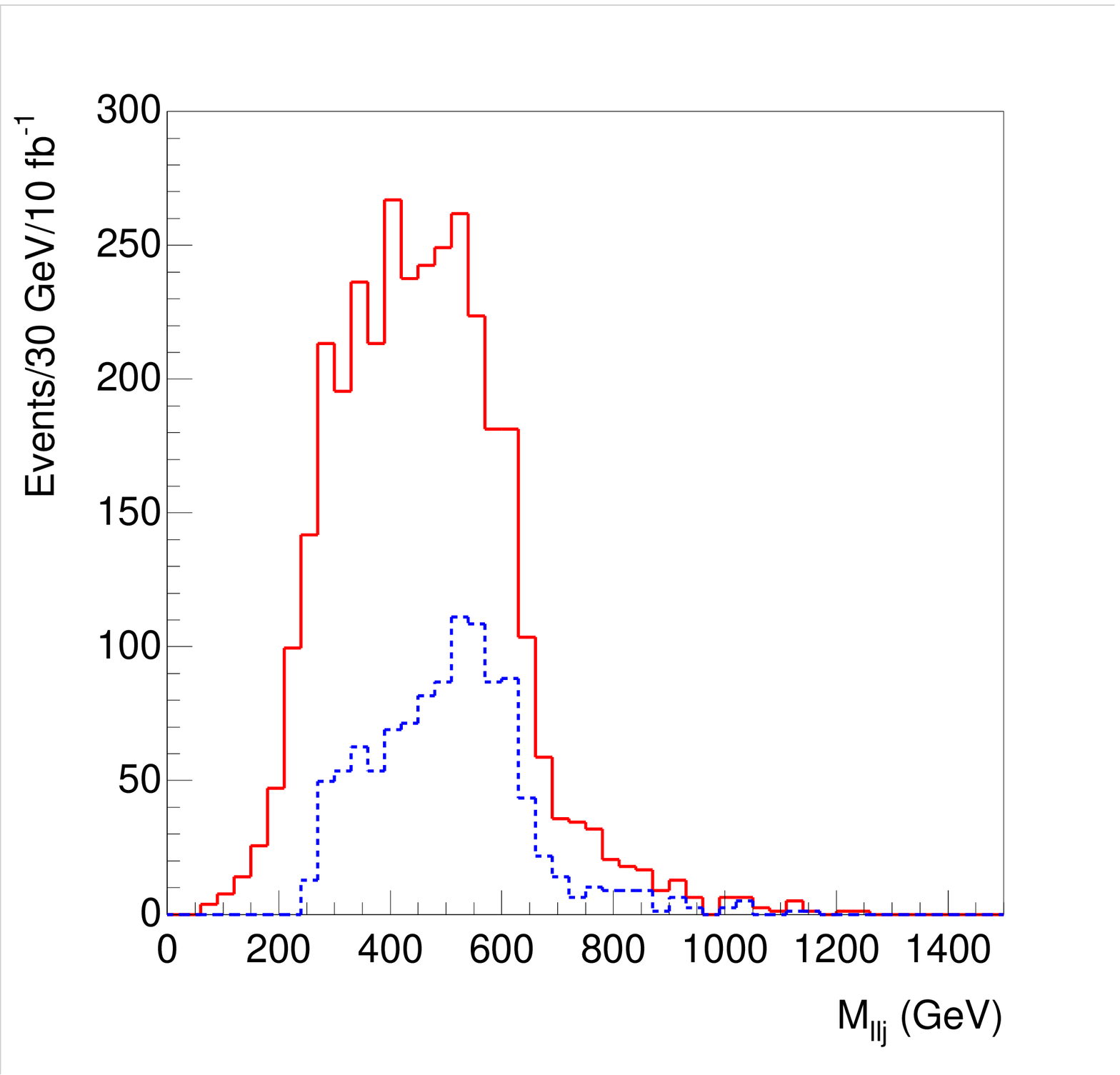}{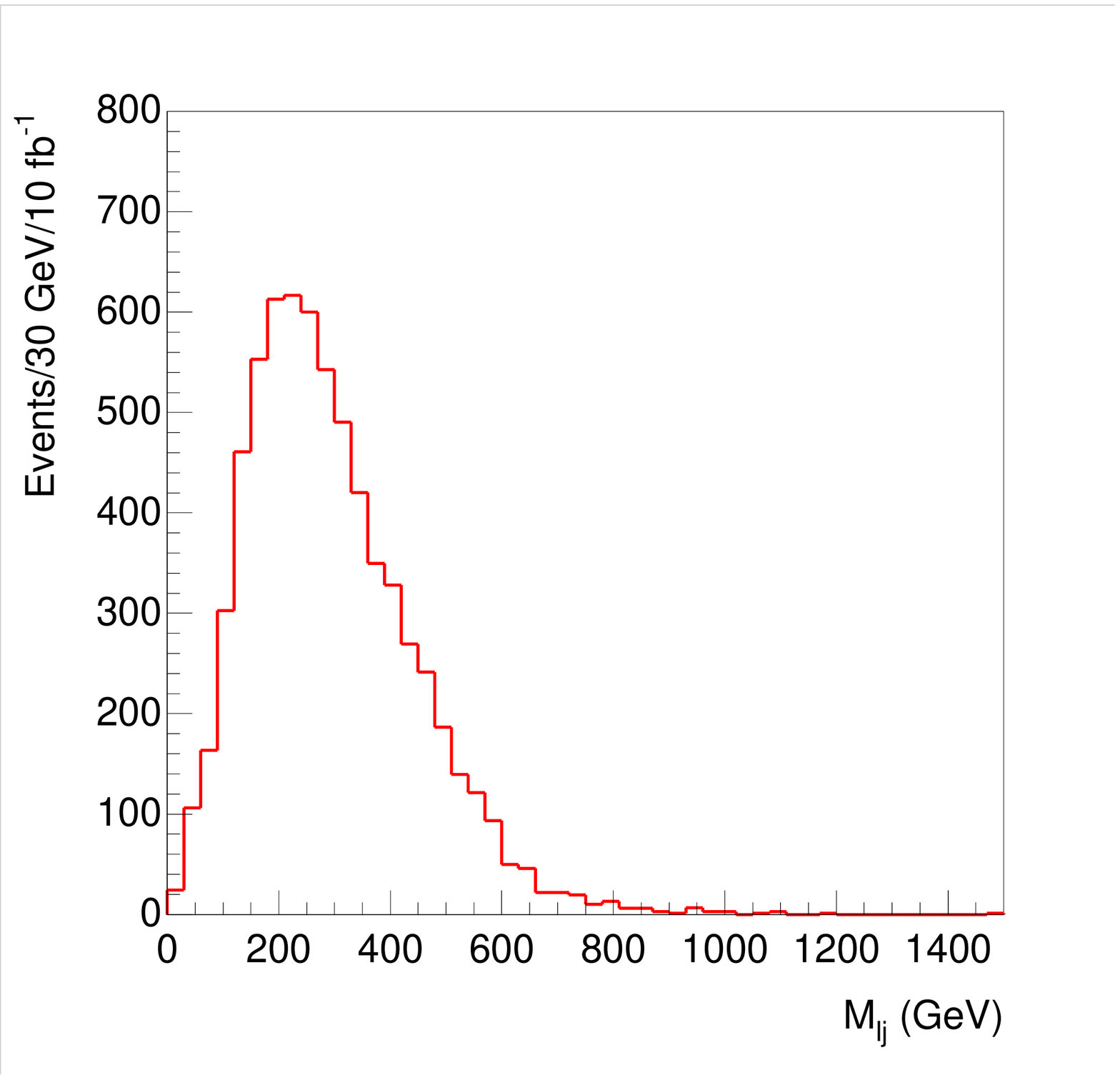}
\caption{Mass distribution for $e^+e^- + \mu^+\mu^- - e^\pm\mu^\mp$
events combined with one of the two hardest jets. Left: $\ell^+\ell^-j$
mass distribution (solid) and same with $M_{\ell\ell}>175\,\GeV$
(dashed).  Right: $\ell^\pm j$ mass distribution for the one of the two
hardest jets that gives the smaller $\ell\ell j$ mass.
\label{a1hmllj}}
\end{center}
\end{figure}

\begin{figure}[t]
\begin{center}
\dofigs{3in}{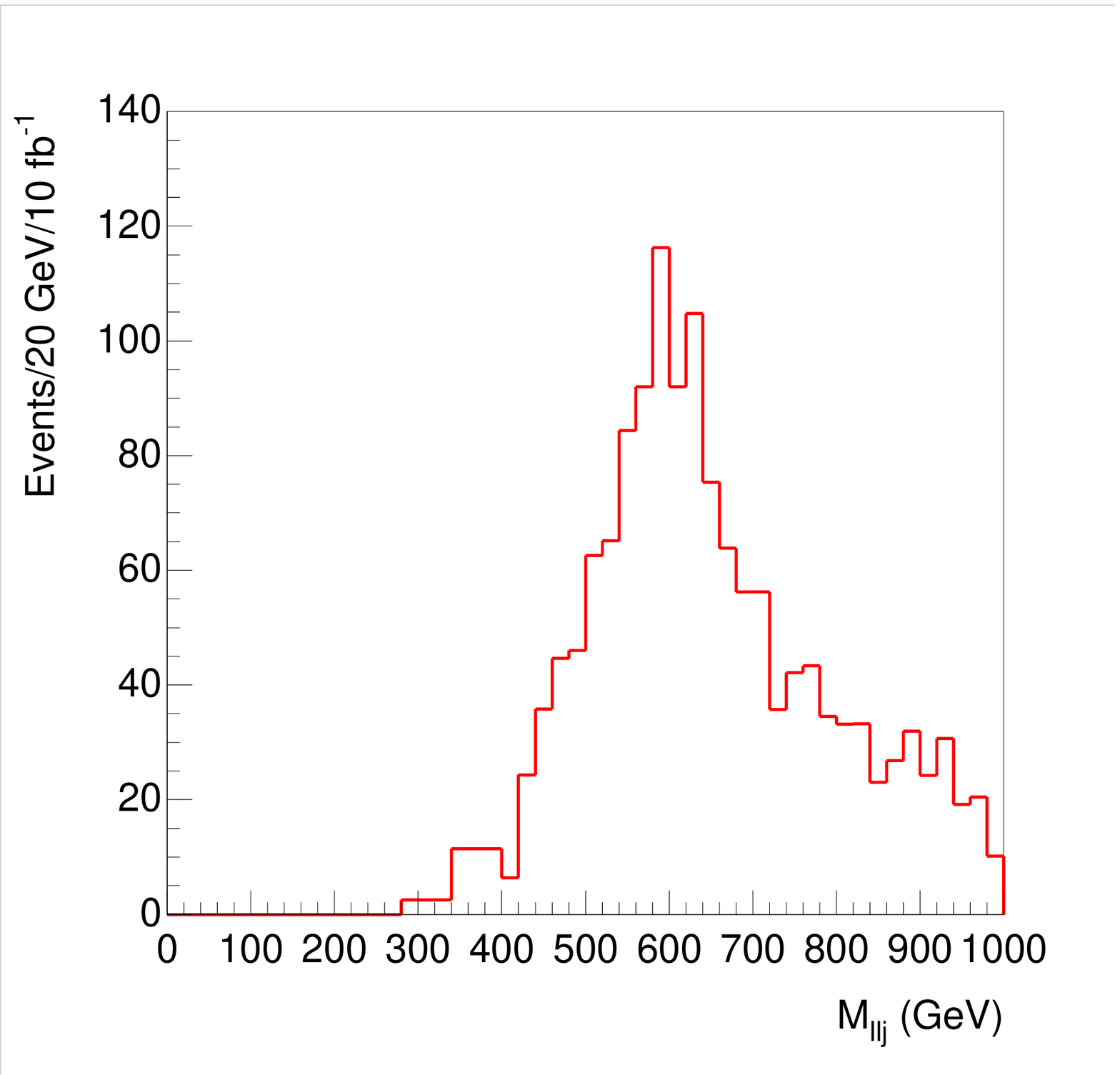}{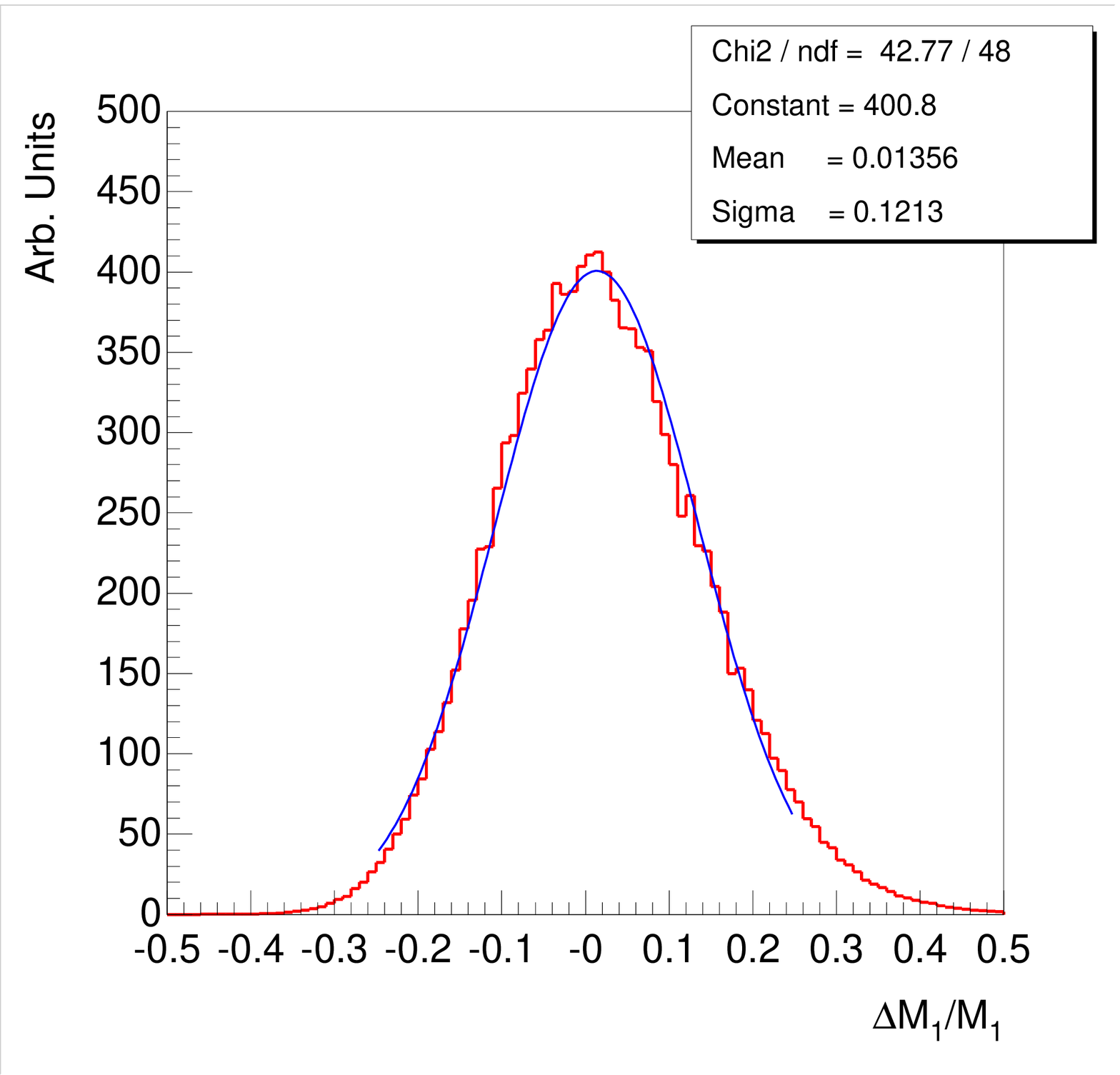}
\caption{Left: Lower edge from larger $\ell\ell j$ mass combining $e^+e^- +
\mu^+\mu^- - e^\pm\mu^\mp$ with one of the two hardest jets. Right:
Model-independent fit for $\lsp$ mass.\label{a1hmlljbk}}
\end{center}
\end{figure}

	Since the main source for $\tchi_2^0$ is $\tq_R \to \tchi_2^0 q$,
information on the squark masses can be obtained by combining the
leptons from $\tchi_2^0 \to \tell \ell$ decays with one of the two
hardest jets in the event, since the hardest jets are generally products
of the squark decays. Figure~\ref{a1hmllj} shows the distribution for
the smaller of the two $\ell^+\ell^- j$ masses formed with the two
leptons and each of the two hardest jets in the event. The dashed curve
in this figure shows the same distribution for $M_{\ell\ell}>175\,\GeV$,
for which the backgrounds are smaller. Both distributions should have
endpoints at the kinematic limit for $\tq_R \to \tchi_2^0 \to \tell \ell
\to \lsp \ell \ell$,
$$
\left[{(M^2_{\tq_R} - M^2_{\tchi_2^0})(M^2_{\tchi_2^0} -M^2_{\lsp}) \over
M^2_{\tchi_2^0}}\right]^{1/2} = 652.9\,\GeV .
$$
Figure~\ref{a1hmllj} also shows the $\ell^\pm j$ mass distribution
formed with each of the two leptons combined with the jet that gives the
smaller of the two $\ell\ell j$ masses. This should have a 3-body
endpoint at
$$
\left[{(M^2_{\tq_R} - M^2_{\tchi_2^0})(M^2_{\tchi_2^0} -M^2_{\tell}) \over
M^2_{\tchi_2^0}}\right]^{1/2} = 605.4\,\GeV .
$$
The branching ratio for $\tb_1 \to \tchi_2^0 b$ is very small, so the
same distributions with $b$-tagged jets contain only a handful of events
and cannot be used to determine the $\tb_1$ mass.

	The decay chain $\tq_R \to \tchi_2^0 q \to \tell_{L,R}^\pm
\ell^\mp q \to \lsp \ell^+\ell^- q$ also implies a lower limit on the
$\ell\ell q$ mass for a given limit on $z=\cos\theta^*$ or equivalently
on the $\ell\ell$ mass. For $z>0$ (or equivalently $M_{\ell\ell} >
M_{\ell\ell}^{\rm max}/\sqrt{2}$) this lower limit is
\begin{equation}
\begin{array}{lcl}
(M_{\ell\ell q}^{\rm min})^2 &=& \frac{1}{4 M_2^2 M_e^2} \times \\
&& \Biggl[-M_1^2 M_2^4 + 3 M_1^2 M_2^2 M_e^2 - M_2^4 M_e^2 - M_2^2 M_e^4
- M_1^2 M_2^2 M_q^2 - \\
&&\quad M_1^2 M_e^2 M_q^2 + 3 M_2^2 M_e^2 M_q^2 - M_e^4 M_q^2 +
(M_2^2-M_q^2)\times \\
&&\quad \sqrt{(M_1^4+M_e^4)(M_2^2 + M_e^2)^2 +
2 M_1^2 M_e^2 (M_2^4 - 6 M_2^2 M_e^2 + M_e^4)}\Biggr]\\
M_{\ell\ell q}^{\rm min}&=& 376.6\,\GeV\nonumber
\end{array}
\end{equation}
where $M_q$, $M_2$, $M_e$, and $M_1$ are the (average) squark,
$\tchi_2^0$, (average) slepton, and $\lsp$ masses. To determine this
lower edge, the larger of the two $\ell\ell j$ masses formed from two
opposite-sign leptons and one of the two hardest jets is plotted in
Figure~\ref{a1hmlljbk}. An endpoint at about the right value can clearly
be seen.

	The $\ell^+\ell^-$, $\ell^+\ell^-q$, $\ell^\pm q$, and lower
$\ell^+\ell^-q$ edges provide four constraints on the four masses
involved. Since the cross sections are similar to those for SUGRA
Point~5, we take the errors at high luminosity to be negligible on the
$\ell^+\ell^-$ edge, 1\% on the $\ell^+\ell^-q$ and $\ell^\pm q$ upper
edges, and 2\% on the $\ell^+\ell^-q$ lower edge. Random masses were
generated within $\pm50\%$ of their nominal values, and the $\chi^2$ for
the four measurements with these errors were used to determine the
probability for each set of masses. The resulting distribution for the
$\lsp$ mass, also shown in Figure~\ref{a1hmlljbk}, has a width of
$\pm11.7\%$, about the same as for Point~5; the errors for the other
masses are also comparable. Of course, the masses being measured in this
case are different: for example the squark mass is the average of the
$\tq_R$ rather than the $\tq_L$ masses.

\begin{figure}[t]
\begin{center}
\dofig{3in}{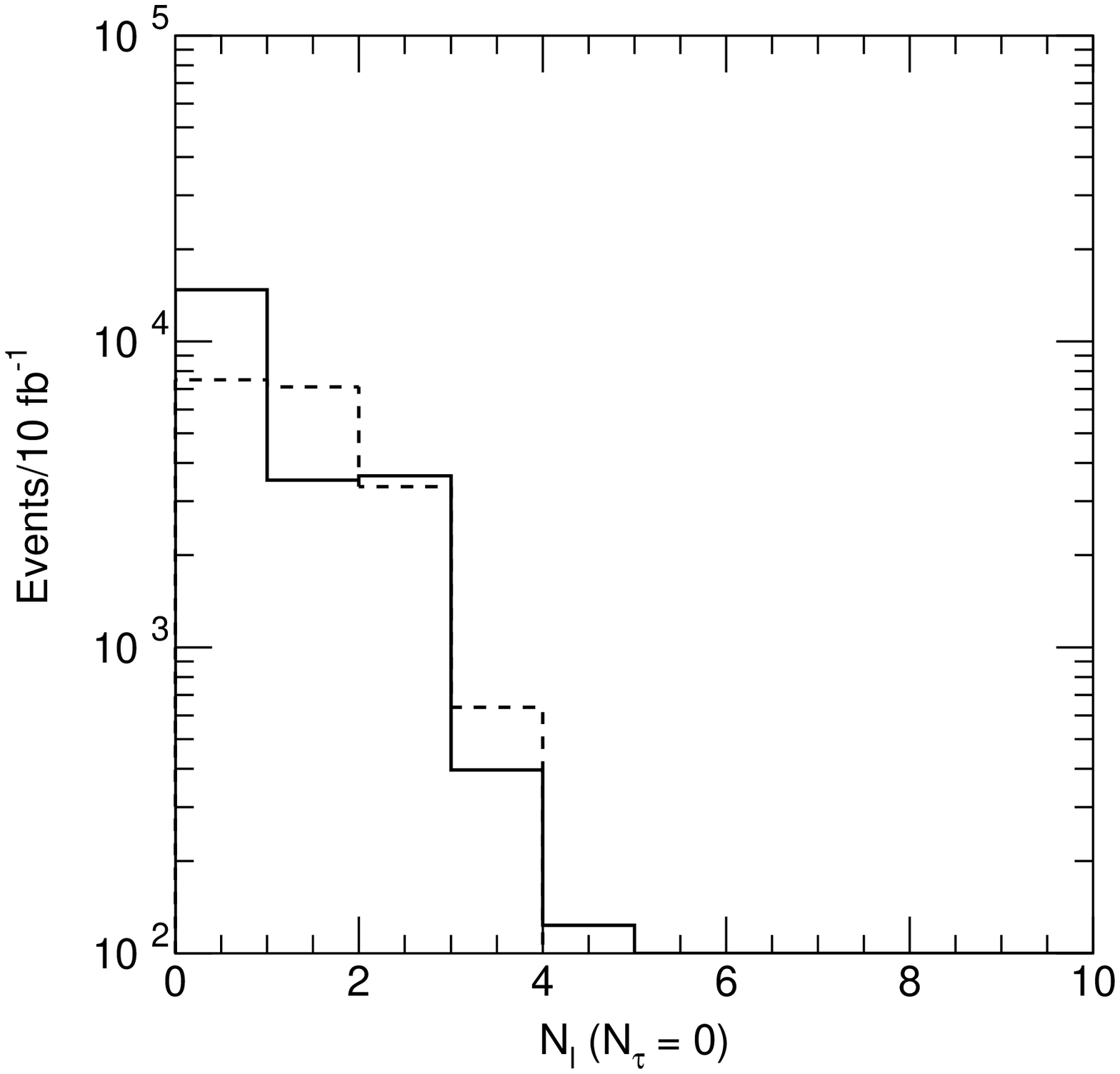}
\caption{Lepton multiplicity with a hadronic $\tau$ veto. Solid: AMSB
model. Dashed: Same but with $M_1 \leftrightarrow
M_2$. \label{w1hnlep}}
\end{center}
\end{figure}

	The leptons from $\tchi_1^\pm \to \lsp \ell^\pm \nu$ are very
soft. This implies that the rate for events with one or three leptons or
for two leptons with opposite flavor are all suppressed.
Figure~\ref{w1hnlep} shows as a solid histogram the multiplicity of
leptons with $p_T>10\,\GeV$ and $|\eta|<2.5$ for the AMSB signal with a
veto on hadronic $\tau$ decays. The same figure shows the distribution
for a model with the same weak-scale mass parameters except that the
gaugino masses $M_1$ and $M_2$ are interchanged. This model has a wino
$\tchi_1^\pm$ approximately degenerate with the $\tchi_2^0$ rather than
with the $\lsp$. Clearly the AMSB model has a much smaller rate for
single leptons and a somewhat smaller rate for three leptons; these
rates can be used to distinguish AMSB and SUGRA-like models.

\begin{figure}[t]
\begin{center}
\dofigs{3in}{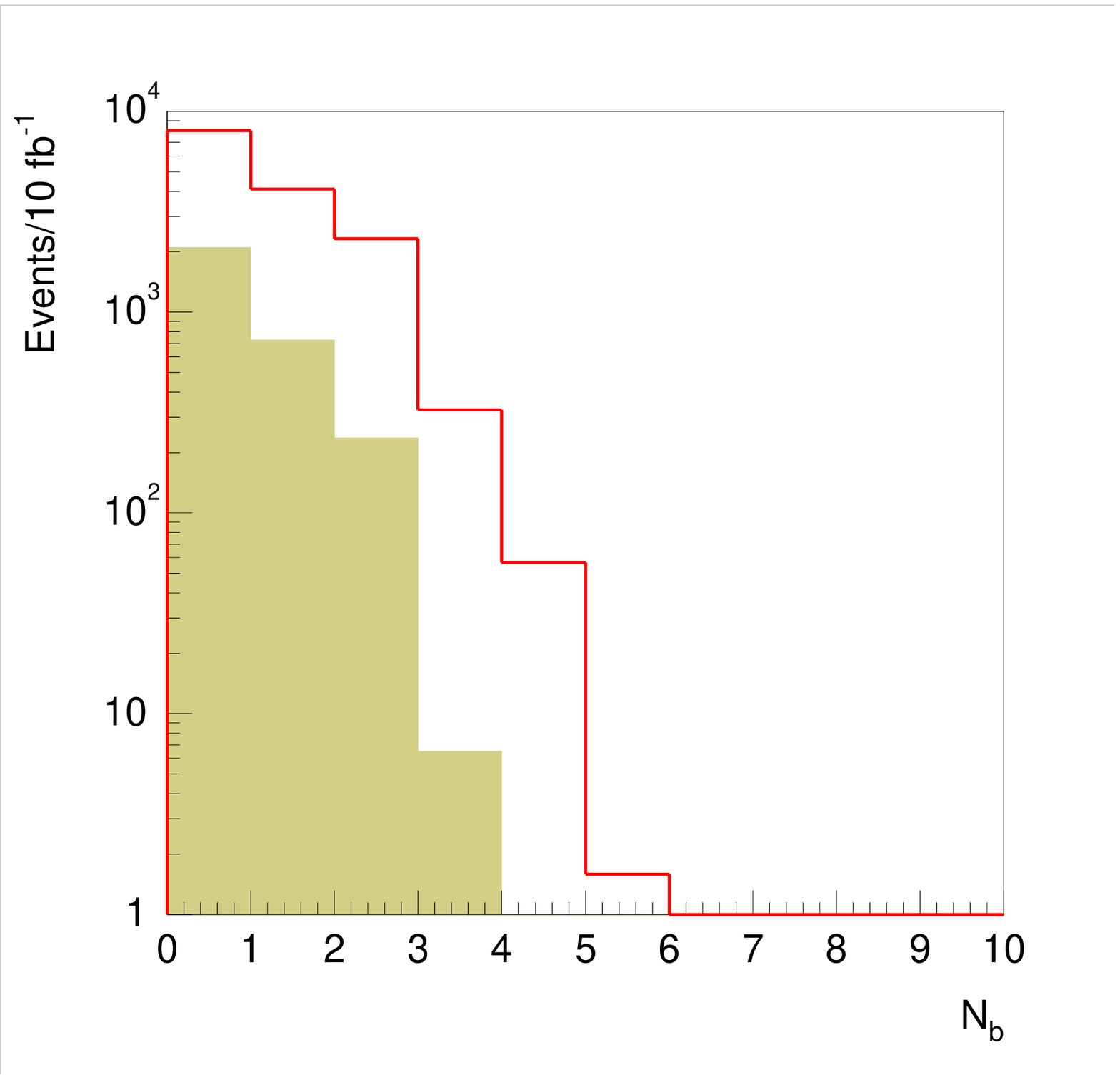}{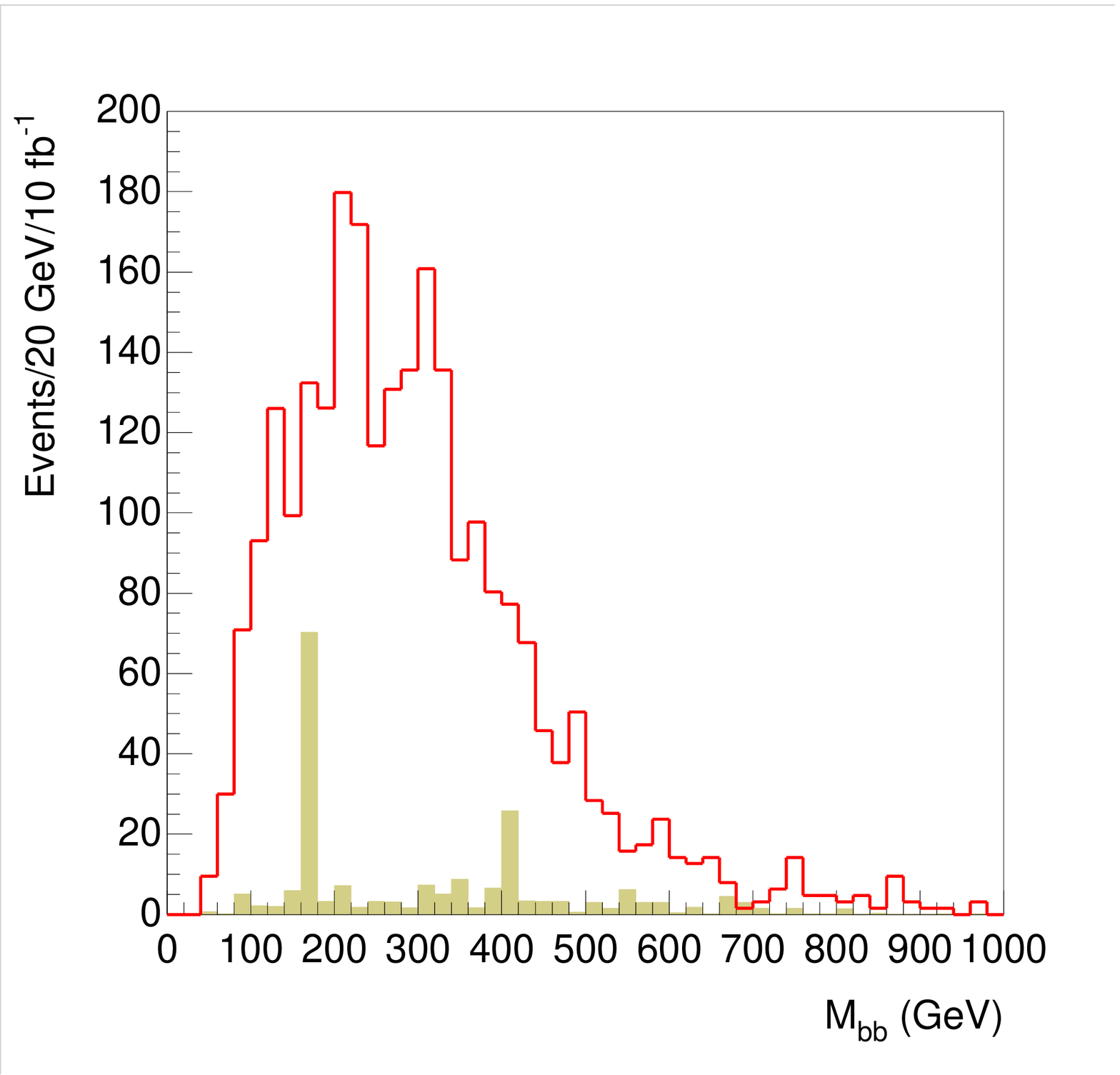}
\caption{Left: Multiplicity of $b$ jets for signal (curve) and Standard
Model background (shaded). Right: Smallest mass for pairs of $b$ jets
for signal (curve) and Standard Model background (shaded). Both plots
have $\Meff>1200\,\GeV$ in addition to the standard cuts and include a
$b$ tagging efficiency $\epsilon_b=60\%$. \label{a1hmbb}}
\end{center}
\end{figure}

	While the decay $\tchi_2^0 \to \lsp h$ is kinematically allowed,
the branching ratio is only about 0.03\%. Other sources of $h$ in SUSY
events are also quite small, so in contrast to SUGRA Point~5 there is no
strong $h \to b \bar b$ signal. However, there is a fairly large
branching ratio for $\tg \to \tb \bar b, \ttop \bar t$ with $\tb \to
\lsp b$, $\ttop \to \chi_1^+ b$, giving two hard $b$ jets and hence
structure in the $M_{bb}$ distribution. For this analysis $b$ jets were
tagged by assuming that any $B$ hadron with $p_{T,B}>10\,\GeV$ and
$|\eta_B|<2$ is tagged with an efficiency $\epsilon_B = 60\%$; the jet
with the smallest
$$
R = \sqrt{(\Delta\eta)^2 + (\Delta\phi)^2}
$$
was then taken to be $b$ jets. The two hardest jets generally come from
the squarks. To reconstruct $\tg \to \tb \bar b$ one of the two hardest
jets, tagged as a $b$, was combined with any remaining jet, also tagged
as a $b$. In addition to the standard multijet and $\etmiss$ cuts, a cut
$\Meff>1200\,\GeV$ was made to reduce the Standard Model background. The
resulting distributions for the $b$ jet multiplicity and for the
smallest $bb$ dijet mass are shown in Figure~\ref{a1hmbb}. The dijet
mass should have an endpoint at the kinematic limit for $\tg \to \tb_1
\bar b \to \lsp b \bar b$,
$$
M_{bb}^{\rm max} = \sqrt{(M_{\tg}^2 - M_{\tb}^2)
(M_{\tb}^2 - M_\lsp^2)\over M_{\tb}^2} = 418.7\,\GeV .
$$
While the figure is roughly consistent with this, the endpoint is not
very sharp; more work is needed to assign an error and to understand the
high mass tail. There should also be a $b\bar t$ endpoint resulting from
$\tg \to \ttop \bar t$, $\ttop \to \tchi_1^+ b$, with $M_{\tchi_1^+}
\approx M_\lsp$ and essentially invisible. Of course $m_t$ has to be
kept in the formula. This would be an apparent strong flavor violation
in gluino decays and so quite characteristic of these models.
Reconstructing the top is more complicated, so this has not yet been
studied.

\begin{figure}[t]
\begin{center}
\dofigs{3in}{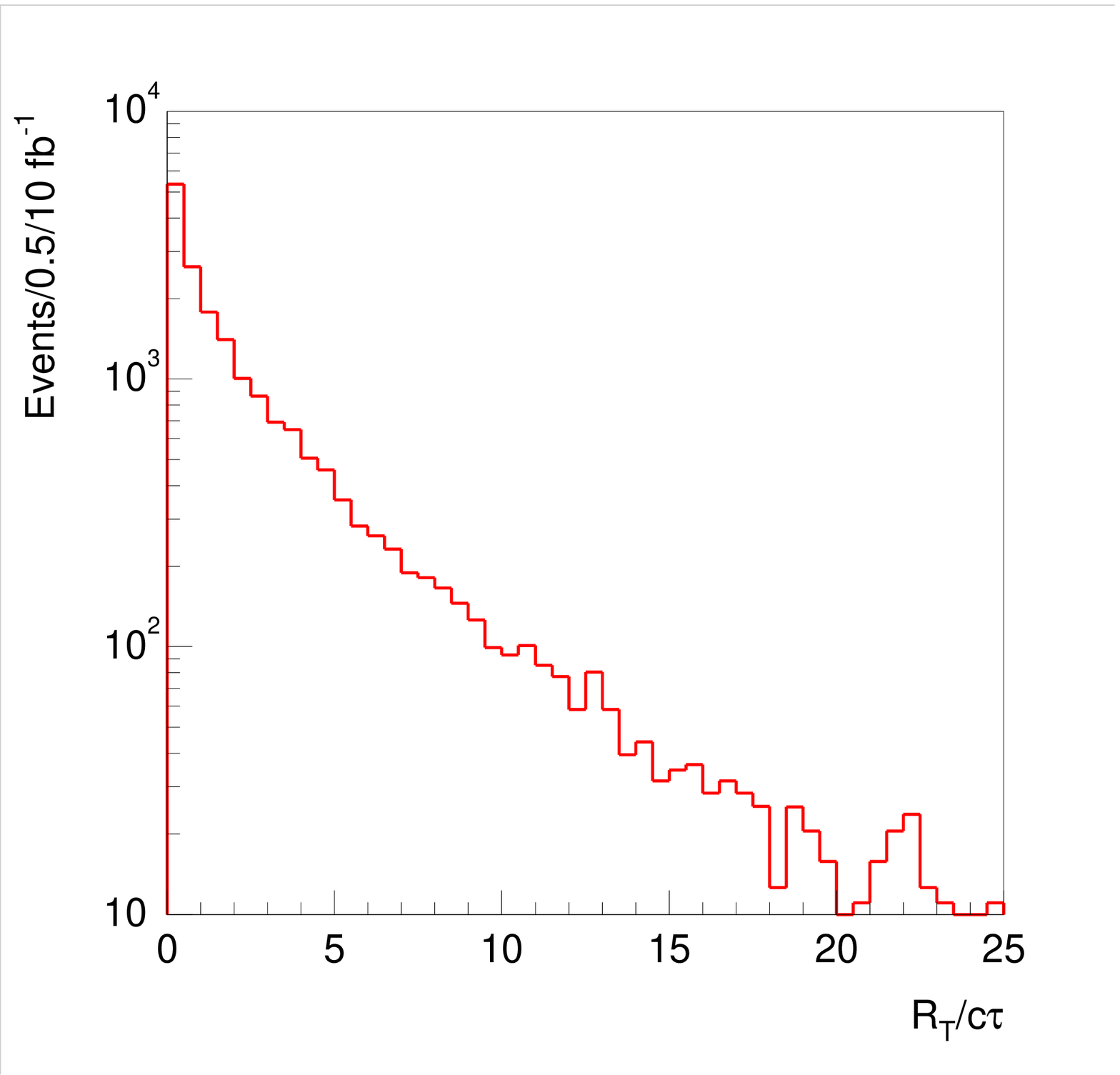}{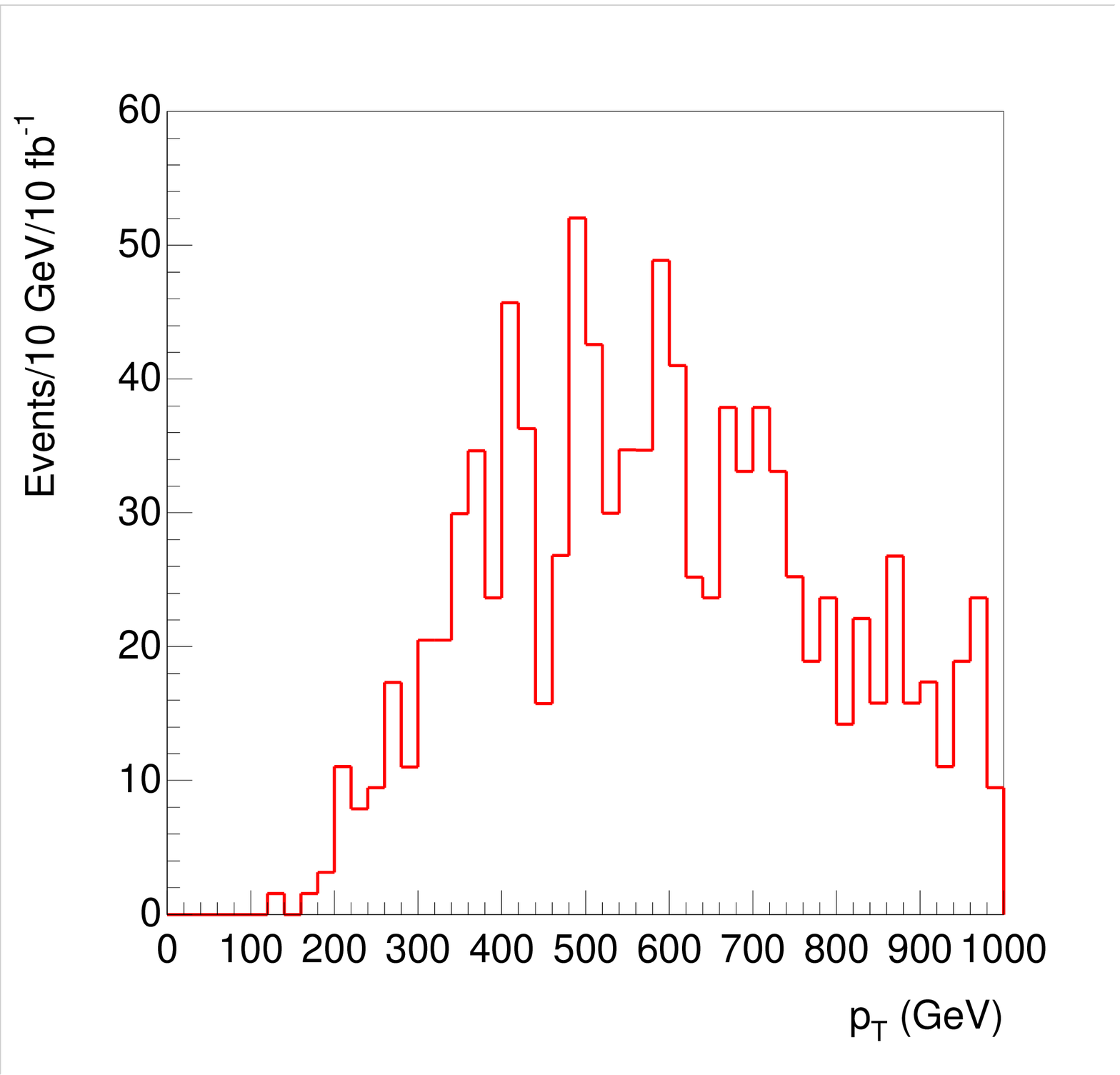}
\caption{Left: Radial track length distributions for $\tchi_1^\pm$ in
the barrel region, $|\eta|<1$. Right: $p$ distribution for $\tchi_1^\pm$
with $R_T > 10c\tau$. \label{a1hrwino}}
\end{center}
\end{figure}

	The splitting between the $\tchi_1^\pm$ and $\lsp$ is very small
in AMSB models. ISAJET gives a splitting of $0.189\,\GeV$ for this point
and $c\tau = 2.8\,\cm$, with the dominant decay being the two-body mode
$\tchi_1^\pm \to \lsp \pi$ via a virtual $W$. Ref.~\cite{GGW} gives a
somewhat smaller value of $\mu$ and so a smaller splitting. The lifetime
is of course quite sensitive to the exact splitting. Since the pion or
electron is soft and so difficult to reconstruct, it seems better to
look for the tail of long-lived winos. The signature is an isolated
stiff track in a fraction of the events that ends in the tracking volume
and produces no signal in the calorimeter or muon system.
Figure~\ref{a1hrwino} shows the radial track length $R_T$ distribution
in units of $c\tau$ for winos with $|\eta|<1$ and the (generated)
momentum distribution for those with $R_T>10c\tau$. Note that the ATLAS
detector has three layers of pixels with very low occupancy at radii of
4, 11, and 14~cm and four double layers of silicon strips between 30 and
50~cm. It seems likely that the background for tracks that end after the
pixel layers would be small. 

	It is instructive to compare this signature to that for GMSB
models with an NLSP slepton. Both models predict long-lived charged
particles with $\beta<1$. In the GMSB models, two NLSP sleptons occur in
every SUSY event, and they decay into a hard $e$'s, $\mu$'s, or $\tau$'s
plus nearly massless $\tG$'s. In the AMSB models, only a fraction of the
SUSY events contain long-lived charged tracks, and these decay into a
soft pion or electron plus an invisible particle. A detailed tracking
simulation should be done for both cases.

\bigskip
\noindent
{\bf Acknowledgements:}
This work was supported in part by the U.S. Department of Energy
under Contract DE-AC02-98CH10886.  We also acknowledge the support of the
Les Houches Physics Center, where part of this work was done.


\setcounter{figure}{0}
\setcounter{table}{0}
\setcounter{section}{0}
\setcounter{equation}{0}
\setcounter{footnote}{0}
\clearpage


\begin{thebibliography}{99}

\bibitem{bounds}
M.\ Carena, R.L.\ Culbertson, S.\ Eno, H.J.\ Frisch and S.\ Mrenna,
Rev.~Mod.~Phys.~{\bf 71} (1999) 937.

\bibitem{sqgl_sp} W.\ Beenakker, R.\ H\"opker, M.\ Spira and P.M.\ Zerwas,
\prl{74}{95}{2905}, \zp{C69}{95}{163}, and Nucl.\ Phys.\ {\bf B492}
(1995) 51.

\bibitem{stops} W.\ Beenakker, M.\ Kr\"amer, T.\ Plehn, M.\ Spira and P.M.\
Zerwas, Nucl.\ Phys.\ {\bf B515} (1998) 3.

\bibitem{gaunlo} T.\ Plehn, Ph.D.\ Thesis, University Hamburg 1998,
hep-ph/9803319; W.\ Beenakker, M.\ Klasen, M.\ Kr\"amer, T.\ Plehn, M.\
Spira and P.M.\ Zerwas, Phys.~Rev.~Lett.~{\bf 83} (1999) 3780.

\bibitem{lo} G.L.~Kane and J.P.~Leveill\'e, \pl{B112}{82}{227};
  P.R.~Harrison and C.H.~Llewellyn Smith, \np{B213}{83}{223} [Err.
  \np{B223}{83}{542}];
  E.~Reya and D.P.~Roy, \pr{D32}{85}{645};
  S.~Dawson, E.~Eichten and C.~Quigg, \pr{D31}{85}{1581};
  H.~Baer and X.~Tata, \pl{B160}{85}{159}.

\bibitem{decouple} J.\ Collins, F.\ Wilczek and A.\ Zee, \pr{D18}{78}{242};
W.J.\ Marciano, \pr{D29}{84}{580}; P.\ Nason, S.\ Dawson and R.K.\ Ellis,
\np{B303}{88}{607}.

\bibitem{count} S.P.~Martin and M.T.~Vaughn, \pl{B318}{93}{331};
I.\ Jack and D.R.T.\ Jones, preprint LTH--400, hep-ph/9707278,
in 'Perspectives in Supersymmetry', ed.\ G.\ Kane, Singapore 1997.

\bibitem{prospino} W.~Beenakker, R.~H\"opker and M.~Spira, hep-ph/9611232.

\end{thebibliography}

\begin{thebibliography}{99}

\bibitem{reach_papers}  H.~Baer, C.-H.~Chen, F.~Paige and X.~Tata,
  Phys. Rev. {\bf D52} (1995) 2746; Phys. Rev. {\bf D53} (1996) 6241;
  CMS collaboration (S. Abdullin {\it et al.}), CMS NOTE 1998/006,
  hep--ph/9806366.

\bibitem{sqgl_ab}  
S.~Abdullin and F.~Charles, Nucl. Phys. {\bf B547} (1999) 60.

\bibitem{nkrasnik} S.~Bituykov and N.~Krasnikov, hep--ph/9901398,
 hep-ph/9907257.

\bibitem{ISAJET_ab} 
 F.E. Paige et al., {\it ISAJET 7.40: a Monte Carlo event generator for $pp$,
 $\bar p p$, and $e^+e^-$ reactions}, BNL-HET-98-39, Oct 1998,
 hep--ph/9810440.

\bibitem{PYTHIA_ab} T.~Sj\"{o}strand, Comp. Phys. Commun. {\bf 82} (1994) 74.

\bibitem{CompHEP}
  E.E. Boos et al, hep-ph/9503280;
  A.E. Pukhov et al, {\it CompHEP user's manual, v.3.3},
    Preprint INP MSU 98-41/542, 1998, hep-ph/9908288;
  A. Semenov, Comp. Phys. Commun. {\bf 115} (1998) 124.

\bibitem{GRACE}
T. Tanaka, T. Kaneko and Y. Shimizu,  Comp. Phys. Commun. {\bf 64} (1991) 149;
T. Ishikawa et al, {\it GRACE manual}, KEK Report 92-19, 1993;
H. Tanaka et al, Nucl. Instrum. Meth. {\bf A389} (1997) 295.

\end{thebibliography}

\begin{thebibliography}{99}

\bibitem{atlas}
I. Hinchliffe, F.E. Paige, M.D. Shapiro, J. Soderqvist and W. Yao, 
Phys. Rev. {\bf D55}, 5520 (1997).

\bibitem{pdflib}
H. Plothow-Besch, Comput. Phys. Commun. {\bf 75}, 396 (1993).

\bibitem{herwig_ko}
G. Corcella, I.G. Knowles, G. Marchesini, S. Moretti,
K. Odagiri, P. Richardson, M.H. Seymour and B.R. Webber,
hep--ph/9912396. 

\bibitem{grace}
T.Tanaka, T.Kaneko and Y.Shimizu,  Comp. Phys. Commun. {\bf 64}, 149
(1991); 
T.Ishikawa et al, {\it GRACE manual}, KEK Report 92-19, 1993;
H.Tanaka et al, Nucl. Instrum. Meth. {\bf A389}, 295 (1997).
\end{thebibliography}

\begin{thebibliography}{99}

\bibitem{cdr} 'Conceptual Design of a 500 GeV $e^+e^-$ Linear Collider',
  eds. R.~Brinkmann et al.,
  DESY~1997-048 and ECFA~1997-182.

\bibitem{lcphysics}
  E.~Accomando et al., Phys. Rep. 299 (1998) 1.

\bibitem{ecfadesy} $2^{nd}$ ECFA/DESY Study on Physics and Detectors for a 
  Linear $e^+e^-$ Collider,\\
  http://www.desy.de/conferences/ecfa-desy-lc98.html

\bibitem{mb} H.-U.~Martyn and G.A.~Blair, hep-ph/99104161.

\bibitem{pythia_ma} T.~Sj\"ostrand, Comp. Phys. Comm. 82 (1994) 74.

\bibitem{circe} T.~Ohl, Comp. Phys. Comm. 101 (1996) 269.


\end{thebibliography}

\begin{thebibliography}{40}

\bibitem{degk1}

ECFA/DESY LC Physics Working Group (E. Accomando et al.). Phys. Rept. 
{\bf 299}, 1 (1998) [hep--ph/9705442]. A recent study in the framwork
of minimal supergravity is H.--U. Martyn and G.A. Blair, hep--ph/9910416.

\bibitem{degk2}
See H. Eberl, S. Kraml, W. Majerotto, A. Bartl and W. Porod,
hep--ph/9909378, and references therein.

\bibitem{degk4}
E.g. M. Berggren, R. Keranen, H. Nowak and A. Sopczak, hep--ph/9911345.

\bibitem{degk5}
J.L. Feng and D.E. Finnell, Phys. Rev. {\bf D49}, 2369 (1994)
[hep--ph/9310211].

\bibitem{degk6}
Particle Data Group, Eur. Phys. J. {\bf C3}, 1 (1998).

\bibitem{degk8}
E.A. Kuraev and V.S. Fadin, Sov. J. Nucl. Phys. {\bf 41}, 466 (1985).

\bibitem{degk9}
M. Drees and O.J.P. \'Eboli, Eur. Phys. J. {\bf C10}, 337 (1999)
[hep--ph/9902391].

\bibitem{degk3}
K.--I. Hikasa and Y. Nakamura, Z. Phys. {\bf C70}, 139 (1996), E: {\bf
C71}, 356 (1996).

\bibitem{degk10}
S. Catani, Yu.L. Dokshitzer, M. Olsson, G. Turnock, and B.R. Webber,
Phys. Lett. {\bf B269}, 432 (1991).

\bibitem{degk11}
R. Brinkmann, talk at LCWS99, Sitges, Spain, May 1999.

\end{thebibliography}

\begin{thebibliography}{99}

\bibitem{gudi:spincorr} 
G. Moortgat-Pick \textit{et al}, Phys. Rev. {\bf D59}, 015016 (1999), 
hep--ph/9708481; 
S.Y. Choi \textit{et al.}, Eur. Phys. J. {\bf C7}, 123 (1999), hep--ph/9806279;
V.~Lafage \textit{et al.}, hep-ph/9810504; 
see also the talk by D.~Perret-Gallix at the International Workshop on
Linear Colliders 99, Sitges.

\bibitem{nous:susygen3} 
S.~Katsanevas and P.~Morawitz, Comp. Phys. Commun. {\bf 112}, 227 (1998);
N. Ghodbane, S. Katsanevas, P. Morawitz and E. Perez, \texttt{SUSYGEN3}\\
\texttt{http://lyoinfo.in2p3.fr/susygen/susygen3.html}.

\bibitem{Nojiri:stau}
K. Fujii, M.M. Nojiri and T. Tsukamoto, hep--ph/9511338;
A.~Bartl \textit{et al}, Z. Phys. {\bf C73}, 469 (1997), hep--ph/9603410.

\bibitem{poko:phases}
T.~Ibrahim and P.~Nath, Phys. Rev. {\bf D58}, 111301 (1998), hep--ph/9807501;
T.~Falk and K.A.~Olive, Phys. Lett. {\bf B439}, 71 (1998), hep--ph/9806236;
M.~Brhlik, G.J.~Good and G.L.~Kane, Phys. Rev. {\bf D59}, 115004
(1999), hep--ph/9810457;
S. Pokorski, J.~Rosiek and C.A.~Savoy, hep-ph/9906206.

\bibitem{EDM_E_EXP} 
E. Commins {\em et al.}, Phys. Rev. {\bf A50} (1994) 2960; 
K. Abdullah {\em et al.}, Phys. Rev. Lett. {\bf 65} (1990) 234.

\end{thebibliography}

\begin{thebibliography}{100}
\bibitem{Drein} 
H. Dreiner, published in {\em Perspectives on Supersymmetry },
ed. by G.L. Kane, World Scientific (1998), hep-ph/9707435.

\bibitem{susygen}
SUSYGEN 3.0/06, N. Ghodbane, S. Katsanevas, P. Morawitz and E. Perez,
lyoinfo.in2p3.fr/susygen/susygen3.html; N.~Ghodbane, hep-ph/9909499.

\bibitem{ATLFAST}
E. Richter-Was, D. Froidevaux and L. Poggioli, 'ATLFAST 2.0: a fast
simulation package for ATLAS', ATLAS Internal Note ATL-PHYS-98-131 (1998).

\bibitem{PYTHIA} 
T.\,Sj\"ostrand, Comp. Phys. Comm. {\bfseries 82}, 74 (1994).

\bibitem{ONETOP}
D.O. Carlson, S. Mrenna and C.P. Yuan, private communication;
D.O. Carlson  and C.P. Yuan, Phys. Lett. {\bf B306}, 386 (1993).

\bibitem{herwig}
H. Dreiner, P. Richardson, and M.H. Seymour, OUTP-99-26P, RAL-TR-1999-080,
hep-ph/9912407;
G. Corcella et al., Cavendish HEP 99/17 (1999) hep-ph/9912396;
G. Marchesini et al., Computer Phys. Commun. {\bf 67}, 465 (1992).

\bibitem{TDR} 
The ATLAS Collaboration, 'ATLAS Detector and Physics Performance Technical
Design Report', ATLAS TDR 15, CERN/LHCC/99-15 (1999).

\bibitem{lmgp}
L. Megner and G. Polesello, these proceedings.

\bibitem{aleph2}
The ALEPH collaboration,  Internal Note ALEPH~99-078, CONF~99-050 (1999),
contributed to HEP-EPS 99.

\end{thebibliography}

\begin{thebibliography}{99}

\bibitem{Dreiner:1994ba}
H. Dreiner, M. Guchait and D.P. Roy, Phys. Rev. {\bf D49}, 3270 (1994);
M. Guchait and D.P. Roy, Phys. Rev. {\bf D52}, 133 (1995), hep--ph/9412329.

\bibitem{Baer:1996va}
H.A. Baer,C-H. Chen, F.E. Paige and X.R. Tata, Phys. Rev. {\bf D53}, 6241
(1996), hep--ph/9512383;
S. Abdullin and F. Charles, Nucl. Phys. {\bf B547}, 60 (1999),
hep--ph/9811402.

\bibitem{Matchev:1999nb}
K.T. Matchev and D.M. Pierce, Phys. Rev. {\bf D60}, 075004 (1999),
hep--ph/9904282;
K.T. Matchev and D.M. Pierce, hep--ph/9907505;
J. Nachtman, D. Saltzberg and M. Worcester, CDF collab.,
hep--ex/9902010.

\bibitem{Baer:1999bq}
H.A. Baer, M. Drees, F.E. Paige, P. Quintana and X.R. Tata, 
hep--ph/9906233,

\bibitem{Corcella:1999qn}
G. Corcella et al., ``HERWIG 6.1 release note'', hep--ph/9912396;
G. Marchesini et al., Comput. Phys. Commun. {\bf 67}, 465 (1992),

\bibitem{Sjostrand:1994yb}
T. Sj\"ostrand, Comput. Phys. Commun. {\bf 82}, 74 (1994).

\bibitem{Campbell:1999ah}
J.M. Campbell and R.K. Ellis, Phys. Rev. {\bf D60}, 113006 (1999),
hep--ph/9905386.

\bibitem{Bonciani:1998vc}
R. Bonciani, S. Catani, M.L. Mangano and P. Nason, Nucl. Phys. {\bf
B529}, 424 (1998), hep--ph/9801375.

\bibitem{Dreiner:1999qz}
H. Dreiner, P. Richardson and M.H. Seymour, hep--ph/9912407.

\bibitem{Allanach:1999ic}
B.C. Allanach, A. Dedes and H.K. Dreiner, hep--ph/9906209,

\bibitem{Hirsch:1995zi}
M. Hirsch, H.V. Klapdor-Kleingrothaus and S.G. Kovalenko, Phys. Rev. Lett.
{\bf 75}, 17 (1995); Phys. Rev. {\bf D53}, 1329 (1996),
hep--ph/9502385;
K.S. Babu and R.N. Mohapatra, Phys. Rev. Lett. {\bf 75}, 2276 (1995),
hep--ph/9506354.

\bibitem{Barger:1989rk}
V. Barger, G.F. Giudice and T. Han, Phys. Rev. {\bf D40}, 2987 (1989).

\bibitem{Dreiner:1998gz}
H. Dreiner, P. Richardson and M.H. Seymour, hep--ph/9903419;
B.C. Allanach et al., hep--ph/9906224.

\end{thebibliography}

\begin{thebibliography}{100}

\bibitem{Drein_me} 
H. Dreiner, published in {\em Perspectives on Supersymmetry },
ed. by G.L. Kane, World Scientific (1998), hep--ph/9707435.

\bibitem{TDR_me}
The ATLAS Collaboration, `ATLAS Detector and Physics Performance Technical
Design Report', ATLAS TDR 15, CERN/LHCC/99-15 (1999).

\bibitem{andy}
L. Drage and M.A. Parker, `Measurement of the LSP mass in supersymmetric 
models with $R-$parity violation', Atlas internal note ATL-PHYS-2000-007.

\bibitem{connors}
A. Connors, Ph.D. Thesis, University of Birmingham (UK) 1998; 
http://www.ep.bham.ac.ud/publications/thesis.

\bibitem{isajet_me}
H. Baer, F.E. Paige, S.D. Protopopescu, and X. Tata, hep--ph/9305342
(1993); hep-ph/9804321 (1998).

\bibitem{gplm}
L. Megner and G. Polesello, in preparation.

\bibitem{herwig_me}
G. Corcella et al., Cavendish HEP 99/17 (1999) hep-ph/9912396; 
G. Marchesini, B.R. Webber, G. Abbiendi, I.G. Knowles, M.H. Seymour
and L. Stanco, Computer Phys.  Commun. {\bf 67} (1992) 465.

\bibitem{drein_me}
H. Dreiner, P. Richardson, and M.H. Seymour, these proceedings.

\bibitem{ATLFAST_me}
E. Richter-Was, D. Froidevaux, L. Poggioli, `ATLFAST 2.0: a fast
simulation package for ATLAS', ATLAS Internal Note ATL-PHYS-98-131 (1998).

\end{thebibliography}

\begin{thebibliography}{99} 

\bibitem{lavori}
H. Dreiner and G.G. Ross, \np{B365}{91}{597};  
A. Datta, J.M. Yang, B.L. Young and X. Zhang, \pr{D56}{97}{3107};
R.J. Oakes, K. Whisnant, J.M. Yang, B.L. Young and X. Zhang,  
\pr{D57}{98}{534};
P. Chiappetta, A. Deandrea, E. Nagy, S. Negroni, G. Polesello and
J.M. Virey, hep-ph/9910483. 
 
\bibitem{sjostrand} 
T. Sj\"ostrand, Comput. Commun. 82 (1994) 74. 
 
\bibitem{atlfast_de} 
E. Richter-W\c as, D. Froidevaux, L. Poggioli, ATLAS Internal Note Phys-No-79, 
1996. 
 
\bibitem{gdr}  
G. Bhattacharyya, Nucl. Phys. Proc. Suppl. {\bf 52A} (1997) 83;
H. Dreiner, hep-ph/9707435, in Perspectives on Supersymmetry, ed. G. Kane,
World Scientific;
Report of the $R$-parity group of GDR SUSY, hep-ph/9810232, 
available at http://www.in2p3.fr/susy/ and references therein;
B.C. Allanach, A. Dedes and H.K. Dreiner, \pr{D60}{99}{07501};
B. Allanach et al., in Physics at Run II: Workshop on Supersymmetry/Higgs, 
Batavia, IL, 19-21 Nov, 1998, hep-ph/9906224.
 
\bibitem{TDR_de} The ATLAS Collaboration, Detector and Physics Performance 
Technical Design Report, CERN/LHCC/99-14, ATLAS TDR 14, 25 may 1999. 
 
\bibitem{onetop} Onetop, C.P. Yuan, D. Carlson, S. Mrenna, Barringer, B.  
Pineiro and R. Brock,\\
http://www.pa.msu.edu/$\tilde{}$ brock/atlas-1top/EW-top-programs.html 
 
\bibitem{stelzer} 
T. Stelzer, Z. Sullivan and S. Willenbrock, \pr{D58}{98}{094021}. 

\end{thebibliography}

\begin{thebibliography}{99}


\newcommand{\ib}[3]{{\em ibid.\/} {\bf #1,} #2 (19#3)}                 %
\newcommand{\npbps}[3]{Nucl. Phys. B (Proc. Suppl.) {\bf #1,} #2 (19#3)} %

\bibitem{ff} 
P.  Fayet, \pl{B69}{489}{77}; G.  Farrar and P.  Fayet, 
\pl{B76}{575}{78}.

\bibitem{HERA_expl} D.  Choudhury and S.  Raychaudhuri, 
\pl{B401}{54}{97}; G.  Altarelli, \etal, \np{B506}{3}{97}; H.  Dreiner 
and P.  Morawitz, \np{B503}{55}{97}; J.  Kalinowski, \etal, 
\zp{74}{595}{97}; T.  Kon and T.  Kobayashi, \pl{B409}{265}{97}; G. 
Altarelli, G.F. Giudice and M. L. Mangano, \np{B506}{29}{97}; J. 
Ellis, S. Lola and K. Sridhar, \pl{B408}{252}{97}; M.  
Carena, D.  Choudhury, S.  Raychaudhuri and C.E.M. Wagner, 
\pl{B414}{92}{97}.
        
\bibitem{HERA}
C. Adloff, \etal\ (H1 Collaboration), \zp{74}{191}{97}; 
J. Breitweg, \etal\ (ZEUS Collaboration), \zp{74}{207}{97}.

\bibitem{hallsuz} See, \eg, L. J. Hall and M. Suzuki, \np{B231}{419}{84}.  

\bibitem{leptopar} L. Iba\~{n}ez and G. G. Ross, \pl{B260}{291}{91}, 
\np{B368}{3}{92}; S. Lola and G.G. Ross, \pl{B314}{336}{93};
P. Bin\'etruy, E. Dudas, S. Lavignac and C.A. Savoy, \pl{B422}{171}{98};
J. Ellis, S. Lola and G.G. Ross, \np{B526}{115}{98}.

\bibitem{BGH} V. Barger, G.F. Giudice and T. Han, {\it Phys. Rev.}
 {\bf D40} (1989) 2987;   
      G. Bhattacharyya and D.~Choudhury,
                        Mod. Phys. Lett. {\bf A10}, 1699 (1995);
       G. Bhattacharyya, \npbps{52A}{38}{97};  
       H. Dreiner, hep-ph/9707435 v2;
       R. Barbieri et al, hep-ph/9810232.
\bibitem{collider_decays} V.~Barger, M.S.~Berger, P.~Ohmann and 
R.J.~Phillips, \pr{50}{4299}{94}; 
H.~Baer, C.~Kao and X.~Tata, \pr{51}{2180}{95}.

\bibitem{b_c_s} G.~Bhattacharyya, D.~Choudhury and K.~Sridhar, 
	\pl{B349}{118}{95}.
\bibitem{Joanne_rizzo} J. Hewett and T. Rizzo, \pr{56}{5709}{97}.
\bibitem{products} Stringent limits on products of \rpar--violating 
couplings are given by K. Agashe and  M. Graesser, \pr{54}{4445}{95}; 
D.  Choudhury and P.  Roy, \pl{B378}{153}{96}; 
F.  Vissani and A.  Yu.  Smirnov, \pl{B380}{317}{96};
R.M. Godbole, S. Pakvasa, S.D. Rindani and X. Tata, hep-ph/9912315.

\bibitem{QCD} T.~Matsuura, S.C.~van~der~Marck and W.L.~van~Neerven,
    {\it Phys. Lett.} {\bf B211} (1988) 171; {\it Nucl. Phys.} {\bf B319}
    (1989) 570; R. Hamberg, W.L. Van Neerven and T. Matsuura, {\it Nucl. Phys.}
{\bf B359} (1991) 343.
 
\bibitem{CTEQ4} H.L. Lai, \etal\ (CTEQ Collaboration), \pr{55}{1280}{97}.

\bibitem{dittmar} M. Dittmar, \pr{55}{161}{97}.

\bibitem{D0DY} B. Abbott \etal\ (D0 Collaboration), \prl{82}{4769}{99}.

\end{thebibliography}

\begin{thebibliography}{19}
\bibitem{R1} 
SuperKamiokande Collaboration, Y. Fukuda \emph{et al.},
Phys. Rev. Lett. {\bf 81}, 1562 (1998); Phys. Lett.
{\bf B433}, 9 (1998); Phys. Lett. {\bf B436}, 33 (1998).

\bibitem{R2} 
A.K. Datta, B. Mukhopadhyaya and S. Roy, hep-ph/9905549.

\bibitem{R4} 
S. Roy and B. Mukhopadhyaya, Phys. Rev. {\bf D55}, 7020 (1997).

\bibitem{R5} 
B. Mukhopadhyaya, S. Roy and F. Vissani, Phys. Lett. {\bf B443}, 191 (1998).

\bibitem{R6} 
S.Y. Choi, E.J. Chun, S.K. Kang and J.S. Lee, hep-ph/9903465.
  
\bibitem{R7} 
A.K. Datta, B. Mukhopadhyaya and F. Vissani, work in progress.

\end{thebibliography}

\begin{thebibliography}{99}

\bibitem{StevePrimer}
  For a recent pedagogical review of supersymmetry and supersymmetry 
  breaking, see S.P.~Martin, ``A Supersymmetry Primer'', in ``Perspectives
  on Supersymmetry'', G.L.~Kane ed., World Scientific 1998, hep--ph/9709356 
  and references therein. 

\bibitem{oldGMSB}
    M.~Dine, W.~Fischler and M.~Srednicki, \NPB{189}{1981}{575};
    S.~Dimopoulos and S.~Raby, \NPB{192}{1981}{353};
    M.~Dine and W.~Fischler, \PLBold{110}{1982}{227};
    M.~Dine and M.~Srednicki, \NPB{202}{1982}{238};
    M.~Dine and W.~Fischler, \NPB{204}{1982}{346};
    L.~Alvarez--Gaum\'e, M.~Claudson and M.B.~Wise, \NPB{207}{1982}{96};
    C.R.~Nappi and B.A.~Ovrut, \PLBold{113}{1982}{175};
    S.~Dimopoulos and S.~Raby, \NPB{219}{1983}{479}.

\bibitem{newGMSB}
    M.~Dine and A.E.~Nelson, \PRD{48}{1993}{1277};
    M.~Dine and A.E.~Nelson, Y.~Shirman, \PRD{51}{1995}{1362};
    M.~Dine and A.E.~Nelson, Y.~Nir, Y.~Shirman, \PRD{53}{1996}{2658}.

\bibitem{GR-GMSB}
  For a recent review, see G.F.~Giudice and R.~Rattazzi, hep--ph/9801271.

\bibitem{GMSBmodels1}
    S.~Dimopoulos, S.~Thomas and J.D.~Wells, \PRD{54}{1996}{3283};
    \NPB{488}{1997}{39}.

\bibitem{GMSBmodels2}
    J.A.~Bagger, K.~Matchev, D.M.~Pierce and R.~Zhang, 
    \PRD{55}{1997}{3188}. 

\bibitem{AKM-LEP2} 
    S.~Ambrosanio, G.D.~Kribs and S.~P.~Martin, \PRD{56}{1997}{1761}. 

\bibitem{AB-LC}
    S.~Ambrosanio and G.A.~Blair, \EPJC{12}{2000}{287}.

\bibitem{Fayet}
    P.~Fayet, \PLBold{70}{1977}{461}; \PLBold{86}{1979}{272};
    \PLB{175}{1986}{471} and in ``Unification of the fundamental 
    particle interactions", eds.~S.~Ferrara, J.~Ellis,   
    P.~van Nieuwenhuizen (Plenum, New York, 1980) p.~587.

\bibitem{AKM2} 
  A relevant example is discussed in:
  S.~Ambrosanio, G.D.~Kribs and S.P.~Martin, \NPB{516}{1998}{55}.

\bibitem{SUSYFIRE}
  An updated, generalized and {\tt Fortran}-linked version of the 
  program used in Ref.~\cite{AKM-LEP2}. It generates minimal and 
  non-minimal GMSB and SUGRA models.
  For inquiries about this software package, please send e-mail to  
 {\tt ambros@mail.cern.ch}. 

\bibitem{AHW}
  S.~Ambrosanio, S.~Heinemeyer and G.~Weiglein, in preparation.

\bibitem{mhiggsmSUGRA} 
  A.~Dedes, S.~Heinemeyer, P.~Teixeira-Dias and G.~Weiglein,
  hep--ph/9912249, to appear in Jour. Phys. G.
  
\end{thebibliography}

\bigskip

\begin{thebibliography}{99}

\bibitem{mhiggs1l} 
  H.~Haber and R.~Hempfling, \PRL{66}{1991}{1815};
  J.~Ellis, G.~Ridolfi and F.~Zwirner, \PLB{257}{1991}{83};
                                       \PLB{262}{1991}{477}.

\bibitem{mhiggsf1l} 
  P.~Chankowski, S.~Pokorski and J.~Rosiek, \NPB{423}{1994}{437}.

\bibitem{mhiggsf1ldab} 
  A.~Dabelstein, \NPB{456}{1995}{25}; \ZPC{67}{1995}{495}.

\bibitem{pierce} 
  J.~Bagger, K.~Matchev, D.~Pierce and R.~Zhang, \NPB{491}{1997}{3}.

\bibitem{mhiggsRG1}  
  J.~Casas, J.~Espinosa, M.~Quir\'os and A.~Riotto, \NPB{436}{1995}{3}, 
  E: ibid {\bf B439} (1995) 466;
  M.~Carena, J.~Espinosa, M.~Quir\'os and C.~Wagner, \PLB{355}{1995}{209};
  M.~Carena, M.~Quir\'os and C.~Wagner, \NPB{461}{1996}{407}.

\bibitem{mhiggsRG2} 
  H.~Haber, R.~Hempfling and A.~Hoang, \ZPC{75}{1997}{539}.

\bibitem{mhiggsEP} 
  R.~Hempfling and A.~Hoang, \PLB{331}{1994}{99};
  R.-J.~Zhang, \PLB{447}{1999}{89}.

\bibitem{mhiggsletter} 
  S.~Heinemeyer, W.~Hollik and G.~Weiglein, \PRD{58}{1998}{091701};
  \PLB{440}{1998}{296}.

\bibitem{mhiggslong} 
  S.~Heinemeyer, W.~Hollik and G.~Weiglein, \EPJC{9}{1999}{343}.

\bibitem{mhiggslle} 
  S.~Heinemeyer, W.~Hollik and G.~Weiglein, \PLB{455}{1999}{179}.

\bibitem{ambro1}
S. Ambrosanio, These Proceedings.

\bibitem{mhiggsmSUGRA_1} 
  A.~Dedes, S.~Heinemeyer, P.~Teixeira-Dias and G.~Weiglein,
  hep--ph/9912249, to appear in Jour. Phys. G.

\bibitem{lepc} 
  P.~McNamara, ``Combined LEP Higgs search results up to 
  $\sqrt{s} = 196$ GeV'', talk given at the LEPC meeting, September 7, 1999.

\bibitem{feynhiggs} 
  S.~Heinemeyer, W.~Hollik and G.~Weiglein, 
  Comp. Phys. Comm. {\bf 124} (2000) 76. 

\bibitem{m20} 
  S.~Heinemeyer, W.~Hollik and G.~Weiglein, hep--ph/9910283.

\bibitem{bse} 
  M.~Carena, H.~Haber, S.~Heinemeyer, W.~Hollik, C.~Wagner and 
  G.~Weiglein, DESY 197-99, hep--ph/0001002.

\bibitem{kaeding} 
  T.A.~Kaeding and S.~Nandi, hep--ph/9906432.

\bibitem{cosmo}
  See e.g.: H.~Pagels and J.R.~Primack, \PRL{48}{1982}{223}. 

\bibitem{AHW_1}
  S.~Ambrosanio, S.~Heinemeyer and G.~Weiglein, in preparation.

\bibitem{tbexcl} 
  S.~Heinemeyer, W.~Hollik and G.~Weiglein, hep--ph/9909540.

\bibitem{bench} 
  M.~Carena, S.~Heinemeyer, C.~Wagner and G.~Weiglein, hep--ph/9912223.

\bibitem{GMSBmodels2_2}
    S.~Dimopoulos, S.~Thomas and J.D.~Wells, \PRD{54}{1996}{3283};
    \NPB{488}{1997}{39}.

\end{thebibliography}

\bigskip

\begin{thebibliography}{10}

\bibitem{ambro2}
S. Ambrosanio, These Proceedings.

\bibitem{GR-GMSB_2}
  For a recent review, see G.F.~Giudice and R.~Rattazzi, hep--ph/9801271.

\bibitem{AKM-LEP2_2} 
  S.~Ambrosanio, G.D.~Kribs and S.~P.~Martin, \PRD{56}{1997}{1761}. 

\bibitem{AB-LC_2}
    S.~Ambrosanio and G.A.~Blair, \EPJC{12}{2000}{287}.

\bibitem{TDR_am}
  The ATLAS Collaboration, ``ATLAS Detector and Physics Performance Technical
  Design Report'', ATLAS TDR 15, CERN/LHCC/99-15 (1999).

\bibitem{isajet_am}
  H.~Baer, F.E.~Paige, S.D.~Protopopescu and X. Tata, hep--ph/9305342; 
  hep--ph/9804321.

\bibitem{ATLFAST_am}
  E.~Richter-Was, D.~Froidevaux and L.~Poggioli, ``{\tt ATLFAST 2.0}: A fast
  simulation package for ATLAS'', ATLAS Internal Note ATL-PHYS-98-131 (1998).

\bibitem{leandro}
  A.~Nisati, S.~Petrarca and G.~Salvini, Mod. Phys. Lett. 
 {\bf A12} (1997) 2213.

\bibitem{drtata}
  M.~Drees and X.~Tata, \PLB{252}{1990}{695}.

\bibitem{femoroi}
  J.~L.~Feng and T.~Moroi, \PRD{58}{1998}{035001}.

\bibitem{marthom}
  S.P.~Martin and J.D.~Wells, \PRD{59}{1999}{035008}.

\bibitem{gpar}
  G.~Polesello and A.~Rimoldi, ``Reconstruction of Quasi-stable Charged
  Sleptons in the ATLAS Muon Spectrometer'', 
  ATLAS Internal Note ATL-MUON-99-06.

\bibitem{AMPPR}
  S.~Ambrosanio, B.~Mele, S.~Petrarca, G.~Polesello and A.~Rimodi, 
  in preparation.

\bibitem{ihfp}
  I.~Hinchliffe and F.E.~Paige, \PRD{60}{1999}{095002}.

\end{thebibliography}

\begin{thebibliography}{99}

\bibitem{Randall:1998uk}
L.~Randall and R.~Sundrum, hep-th/9810155.

\bibitem{Giudice:1998xp}
G.F.~Giudice, M.A.~Luty, H.~Murayama and R.~Rattazzi, JHEP {\bf 12}, 
027 (1998), hep--ph/9810442.

\bibitem{GGW} 
T. Gherghetta, G.F. Giudice, and J.D. Wells, hep--ph/9904378.

\bibitem{Feng:1999hg}
J.~L.~Feng and T.~Moroi, hep--ph/9907319.

\bibitem{ISAJET_we} 
H. Baer, F.E. Paige, S.D. Protopescu and X. Tata, hep--ph/9810440.

\bibitem{HPSSY} 
I. Hinchliffe, F.E. Paige, M.D. Shapiro, J. Soderqvist,
and W. Yao, Phys.\ Rev.\ {\bf D55}, 5520 (1997).

\bibitem{TDR_we} 
ATLAS Collaboration, {\sl ATLAS Detector and Physics
Performance Technical Design Report}, CERN/LHCC/99-15.

\bibitem{Feng:1999fu}
J.L.~Feng, T.~Moroi, L.~Randall, M.~Strassler and S.~Su,
Phys. Rev. Lett. {\bf 83}, 1731 (1999), hep--ph/9904250.

\bibitem{Pomarol:1999ie}
A.~Pomarol and R.~Rattazzi, JHEP {\bf 05}, 013 (1999), hep--ph/9903448.

\bibitem{rsw}
R.~Rattazzi, A.~Strumia and J.D.~Wells, hep--ph/9912390.

\end{thebibliography}
\end{document}